\RequirePackage{lineno}
\documentclass[twocolumn,english,amsart,showpacs,preprintnumbers,amsmath,amssymb,floatfix,nofootinbib]{revtex4-1}

\usepackage{enumitem}
\usepackage[T1]{fontenc}
\usepackage{adjustbox}
\usepackage[latin9,utf8]{inputenc}
\usepackage{xcolor}
\usepackage{array}
\usepackage{amstext}
\usepackage{graphicx}
\usepackage{epstopdf}
\usepackage{esint}
\usepackage{float}
 \usepackage{tabularx}
\usepackage{dcolumn}
\usepackage{bm}
\usepackage{subfigure}
\usepackage{color}
\usepackage{float}
\usepackage{multirow}
\usepackage{xspace}
\usepackage{adjustbox}

\usepackage{booktabs}
\usepackage{amsmath}
\usepackage{hhline,multirow,tabularx}  
\DeclareGraphicsExtensions{.pdf}
\usepackage{latexsym,amssymb}
\usepackage[mathscr]{eucal}
\usepackage{rotating}
\usepackage{appendix}
\usepackage{ulem}
\usepackage{tikz,xcolor}
\usepackage[colorlinks = true,
linkcolor = blue,
urlcolor  = blue,
citecolor = blue,
anchorcolor = blue,
linktocpage]{hyperref}
\definecolor{lime}{HTML}{A6CE39}
\DeclareRobustCommand{\orcidicon}{%
	\begin{tikzpicture}
		\draw[lime, fill=lime] (0,0)
		circle [radius=0.16]
		node[white] {{\fontfamily{qag}\selectfont \tiny ID}};
		\draw[white, fill=white] (-0.0625,0.095)
		circle [radius=0.007];
	\end{tikzpicture}
	\hspace{-2mm}
}
\foreach \x in {A, ..., Z}{%
	\expandafter\xdef\csname orcid\x\endcsname{\noexpand\href{https://orcid.org/\csname orcidauthor\x\endcsname}{\noexpand\orcidicon}}
}

\newcommand{\dzero}{D\O\xspace}

\newcommand{\x}{\ensuremath{x}\xspace}

\makeatletter
\begin{document}

\title{Parton distribution functions and QCD coupling constant from LHC and non-LHC data}

\author{Majid Azizi$^{1}$\orcidA{}}
\email{m.azizi@semnan.ac.ir}

\author{Ali Khorramian$^{1}$\orcidB{}}
\email{khorramiana@semnan.ac.ir}

\author{Saeid Paktinat Mehdiabadi$^{2}$\orcidC{}} 
\email{spaktinat@yazd.ac.ir}

\affiliation{
	$^{(1)}$Faculty of Physics, Semnan University, 35131-19111, Semnan, Iran\\
	$^{(2)}$Department of Physics, Yazd University, P.O. Box 89195-741, Yazd, Iran}

\date{\today}
\begin{abstract}
We present a new QCD analysis of parton distribution functions (PDFs) at the next-to-leading order (NLO) and next-to-next-to-leading order (NNLO). 
In the present paper, the role of special types of experimental measurements for extracting PDFs is investigated and simultaneously the values of the strong coupling constant are determined. Essential elements of this QCD analysis are: the HERA combined data as a base, heavy quark cross section measurement  together with Tevatron data on jet production and  H1 and ZEUS jet cross sections as ``non-LHC'' data, and especially the ``LHC'' data set for top-quark and jet cross sections. These different data sets  have allowed detailed information at low $x$, worth information on the nucleon's flavor and have played an important role for some PDFs at large-$x$ and strong coupling constant. Since the large-$x$ gluon PDF benefits from an accurate determination of quark PDFs, we have enough motivation to focus on the  top-quark production and jet cross section measurements from LHC to find the impact of these data on the gluon PDF and strong coupling constant. These experimental data have an impact on the relative uncertainties of PDF. The gluon PDF at large-$x$ and the values of $\alpha_s(M_Z^2)$ are affected significantly. 
Although, the main motivation of this paper is to focus on gluon PDF and also its uncertainty at large-$x$, but notice to heavy PDFs behavior in this region is also very important. We study the intrinsic charm (IC) using BHPS model with together  our  extracted  extrinsic charm PDF  at large $x$ which can be very worth for future experiment at the LHC.

\end{abstract}

\maketitle
\tableofcontents{}

\section{Introduction} \label{Itro}
Great advances have been made  in our knowledge of the parton substructure of the hadrons, with the accessibility of new high-energy scattering experimental data from different worldwide accelerators. Quantum Chromodynamic (QCD) based on quark-parton model has provided a reliable computational framework to describe the hadron interactions. 

Basically, in standard factorization, Parton Distribution Functions (PDFs) are specified using a comparison of theoretical QCD predictions with hard scattering experimental measurements covering kinematic wide range in the Bjorken variable $x$ and the momentum scale $Q^2$.
However, these PDFs have generally been applied for the description of hadron structures. Within the context of the
QCD quark-parton model, unpolarized and polarized PDFs  have been considerably successful to describe a wide kind of different processes. Very recently, the results of polarized QCD analysis of PDFs is reported in Refs. \cite{Khorramian:2020gkr,Kumano:2021fem,Zhou:2022wzm,Guo:2021aik}. It should be noted that the modern PDFs are expected to deliver an accurate information of the partons not only in a wide kinematic region of $x$ and $Q^2$, but to provide also information of  the strong coupling constant $\alpha_s$ as a nonperturbative parameter. The value for this quantity is often correlated with the PDFs and therefore, have to be extracted simultaneously in  a QCD fit \cite{Abramowicz:2015mha}.

In recent years various PDF sets \cite{Hou:2019efy,H1:2015ubc,Bailey:2020ooq,Ball:2021leu,Butterworth:2015oua} have been developed to be used at the Large Hadron Collider (LHC). These accurate PDFs have been applied for new physics searches \cite{Han:2021kes,Bhattacharya:2021moj,Bandyopadhyay:2020otm,Bauer:2017isx,Gao:2017yyd,Mangano:2004wq,Lin:2017snn} and also for the evaluation of PDF uncertainties on QCD precision  observables.  The PDF extracted results  using different experimental measurements for both unpolarized and polarized QCD analysis are also reported in Refs.~\cite{Abdolmaleki:2019tbb,Monfared:2014nta,Arbabifar:2013tma,Azizi:2018iiq,Aleedaneshvar:2017bgs,Rostami:2015iva,Zhang:2021shm,Liu:2022fvl,Shen:2021eir,Helenius:2021tof,DelDebbio:2020rgv,ATLAS:2021qnl,Karpie:2021pap,Radyushkin:2021fel,Radyushkin:2022qvt,Salimi-Amiri:2018had}.  
Very recently the nuclear parton distributions (nPDFs) from a global QCD analysis at NLO is also presented in Ref.~\cite{AbdulKhalek:2022fyi}.

 By utilizing more powerful particle accelerators such as the LHC and Tevatron, we have access to the data that belong to various processes such as top quark and  jets production and at different center of mass energies. 
New experimental data provide an opportunity to additionally improve the structure of the hadrons which are described by PDFs.

Deep-inelastic scattering
(DIS) experimental data  from the electron(positron)-proton ($e^{\pm}p$) collider and in fixed-target experiments at HERA \cite{Abramowicz:2015mha} have allowed a precise  valuable
information of PDFs at small values of $x$. In spite, these data provide no useful constraint on the gluon PDF at large $x$ which is relevant for  phenomenological applications at hadron-hadron colliders such as the LHC and Tevatron. In the present analysis, DIS final inclusive HERA combine cross section data \cite{Abramowicz:2015mha} can be considered as a base and common data. 

In addition to this kind of data as a base data set, other supplementary processes are also employed for PDF extraction. The final heavy 
flavour and HERA jet cross section data \cite{Chekanov:2002be,Chekanov:2006xr,Abramowicz:2010cka,Aktas:2007aa,Aaron:2009vs,Aaron:2010ac,Andreev:2014wwa,H1:2018flt} provide
a detailed information on PDFs, especially for gluon distribution at low value of $x$ and help to constrain the strange PDF in the proton. 
 Although, the heavier targets data are available, but to avoid any influence
of nuclear correction uncertainties, these kind of data in QCD analysis are not used by some groups.

The precision measurements on jet production  from high-energy $p p$ and
$p\bar p$ scattering \cite{Aaltonen:2008eq,Abazov:2008ae,Khachatryan:2016mlc,Aad:2013lpa,Aad:2011fc} are also necessary for PDF QCD analysis. 
 These kind of data have provided a complementary information on the nucleons flavor structure. Also they are important in detailed PDF validations in order to reduce their uncertainties and the value of strong coupling constant determination. 

\begin{figure*}[htb]
	\begin{center}
	    \includegraphics[scale = 0.8]{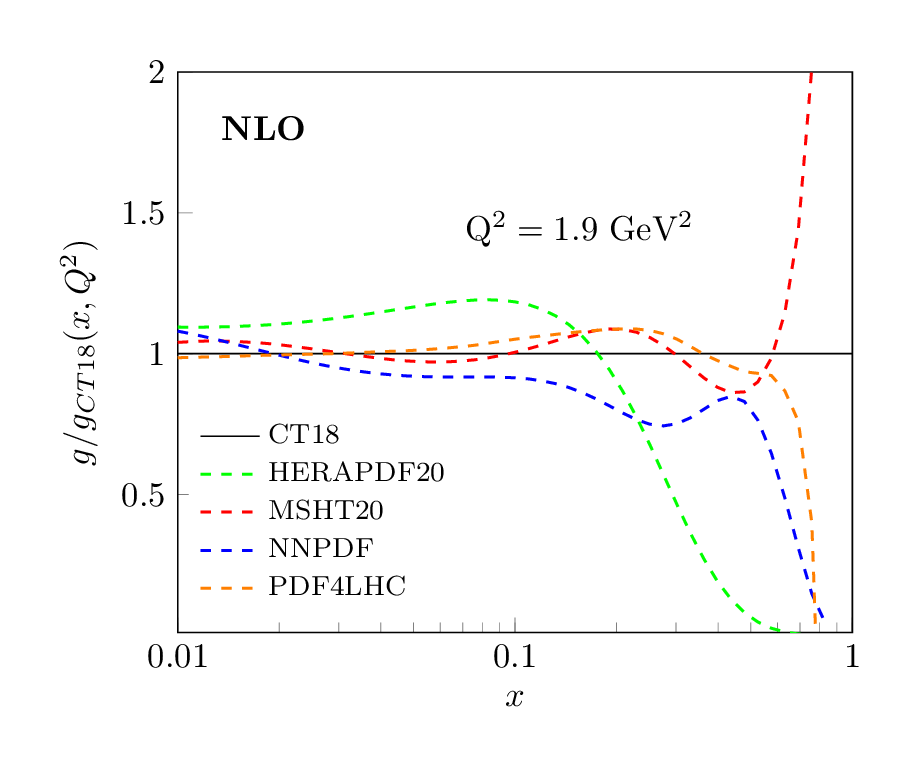}
	    \includegraphics[scale = 0.8]{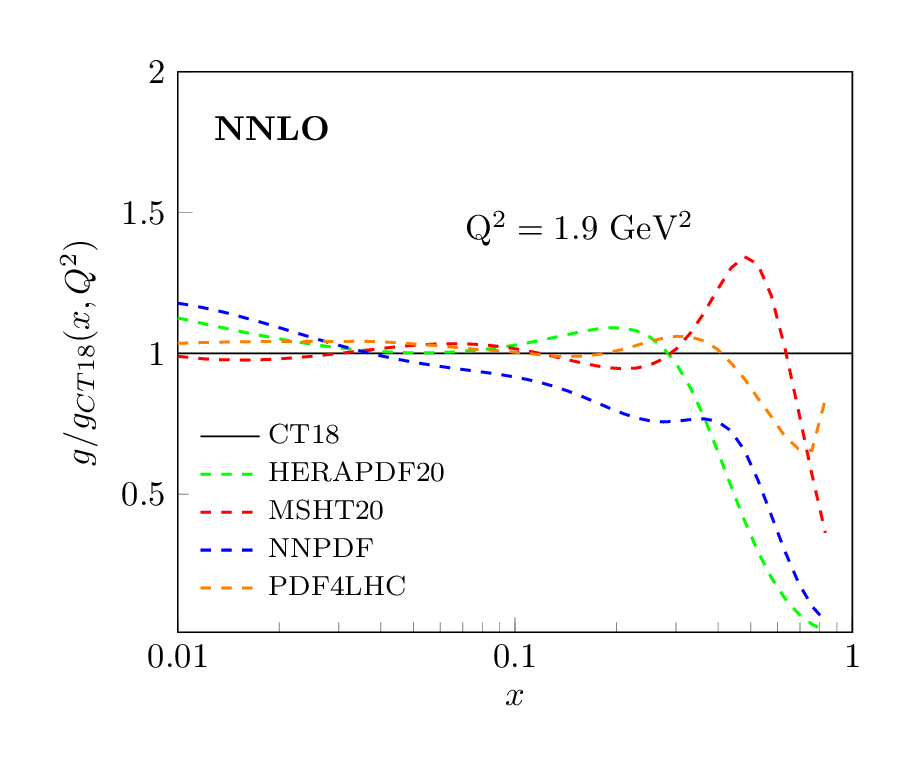}	    	        
                
	    		\caption{Gluon PDF ratios of the NLO (left) and NNLO (right) HERAPDF20 \cite{H1:2015ubc}, MSHT20 \cite{Bailey:2020ooq}, NNPDF  \cite{Ball:2021leu} and PDF4LHC \cite{Butterworth:2015oua} predictions to the CT18 \cite{Hou:2019efy} ones, $g/g_{CT18}(x,Q^2)$  as a function of $x$ at $Q^2$=1.9  GeV$^2$. }
		\label{fig:PDF-v-ratio}
	\end{center}
\end{figure*}

 To improve the gluon PDF situation with large $x$, in addition to above mentioned data sets commonly used to constrain PDFs, certain complementary processes such as top quark cross section data \cite{CMS:2012hkm, Khachatryan:2016yzq, Villalba:2021lkd, Khachatryan:2016mqs, Sirunyan:2018goh,Aaltonen:2013wca,D0:1998huz,ATLAS:2021xhc,Aaboud:2016zpd,ATLAS:2017fyu,ATLAS:2014ipf} may also be considered.
This kind of data have a good potential to constrain some PDFs at  large values of $x$ \cite{Czakon:2016olj}. Our motivation to include this kind of data in our analysis is that the gluon PDF is one of poorly known PDFs in the proton, particularly for $x>0.3$. In Fig. \ref{fig:PDF-v-ratio} the NLO and NNLO  gluon PDF ratios of the HERA \cite{H1:2015ubc}, MSHT20 \cite{Bailey:2020ooq}, NNPDF \cite{Ball:2021leu} and PDF4LHC \cite{Butterworth:2015oua} predictions to the CT18 \cite{Hou:2019efy} are compared. In this figure one can see the sizable difference between these extracted results for large $x$ values. 
So in the present study, we have enough motivation to provide a PDF set in presence of  top quark pair production and jet measurements extracted from hadron-hadron collisions to find the impact of these kind of data on central value and relative uncertainties of PDFs, especially the gluon PDF at large value of $x$ and the values of the strong coupling constant. In particular, these kind of data allow to reduce the uncertainties in the gluon PDF at a GeV scale especially for large $x$ values, as well as in strong coupling constant.

Although, the main motivation of our present study is to focus on gluon PDF and also its uncertainty at large $x$, but notice to heavy PDFs behavior in this region are also very important. As we will see, the accurate charm PDF and its uncertainty benefit from an accurate gluon determination and also the decrease of its uncertainty at large $x$. On the theoretical side, this may be very worth when one need to add the charm PDF (or extrinsic charm) to intrinsic charm (IC) which is dominant at large $x$.  We refer the reader to
Refs. \cite{BHPS1,BHPS2}, and references therein for a detailed
discussion of the  Brodsky, Hoyer, Peterson, and Sakai (BHPS) model. 
An important review of the intrinsic heavy PDF content of the proton
has been published in Ref.~\cite{Brodsky:2015fna}. Several phenomenological and theoretical studies have been reported using IC component in the proton
\cite{Brodsky:2020hgs,Lykasov:2020dri,Brodsky:2018zdh,Brodsky:2016tew,Dulat:2013hea,Pumplin:2005yf,Blumlein:2015qcn,Brodsky:2016fyh,Koshkarev:2016rci,Anjos:2001jr,Montano:1996nj,Bednyakov:2013zta}.  In Refs.~\cite{Chang:2014lea,An:2017flb}, the intrinsic light-quark sea in the proton using BHPS model are also studied.
A method to generate the intrinsic charm and bottom content of the proton is presented in Ref.~\cite{Lyonnet:2015dca}. Also a study of heavy flavor and Higgs production from intrinsic charm and bottom quarks is presented in Ref.~\cite{Brodsky:2007yz}. Recently, a new constraints on IC contribution using BHPS approach with the determination of  $\alpha_s(M_Z^2)$ and  $P_{c{\bar c}/p}$ IC probability values is reported in  Ref.~\cite{Abdolmaleki:2019tbb}.
Note that, the $c(x) $ and $\bar c(x)$ from intrinsic charm are very different.  This has very strong consequences for experiments at large $x$. This is marked contrast to the DGLAP and perturbative  $g \to  c + \bar c$ symmetric distributions. In Ref.~\cite{Sufian:2020coz}, the first lattice QCD calculation of the charm quark contribution to
study the constraints on the charm asymmetry  $c(x) \ne \bar c(x)$ in the nucleon from lattice gauge theory is presented.

In order to perform an accurate theoretical description, both 
the PDF evolution and the cross sections of hard scattering are necessary. For PDF evolution, open source codes such 
as QCDNUM \cite{Botje:2010ay} and QCD Pegasus \cite{Vogt:2004ns} are available in $x$ and Mellin $N$-space, respectively. To use the hard scattering cross sections of the different processes, different methods such as APPLGrid \cite{Carli:2010rw}, FastNLO \cite{Kluge:2006xs} and HATHOR \cite{Aliev:2010zk}  have been developed. Some groups have also reported 
their open source code for the theory predictions of the cross sections which are used in their analyses.
The \textsc{xFitter} \cite{Alekhin:2014irh} computational framework, previously known as 
HERAfitter \cite{HERAfitter}, as an open source package, provides a framework for PDF determination. In particular, \textsc{xFitter}  enables the
choice of theoretical options to  extract the PDF dependent 
cross section predictions and allows for a choice of different schemes for heavy PDFs. 
In our QCD analysis, we used  \textsc{xFitter} computational framework.
In Refs. \cite{Abdolmaleki:2019tbb,Tooran:2019cfz,Vafaee:2017nze}, \textsc{xFitter} is used to extract the PDFs and value of strong coupling constant.
Note that determination of strong coupling constant at NLO and NNLO concurrently with the parton distribution function  is one of the main purposes of this paper. 

The outline of this paper is as follows: In Sec.~II, we briefly discuss the experimental measurements used to constrain PDFs 
and stress the need to include only the necessary data sets in our analysis. In this section also all data sets, such as base,
non-LHC, and LHC data, which are included in the fit procedures are explained. Theoretical frameworks, computational setting and PDF parametrization utilized in the present analysis are described in Sec. III. In Sec. IV, we present and interpret the results of the fits and the impact of the different measurements on PDFs using different fits. The PDFs extracted from different fits are correlated with the strong coupling constant, therefore it is an important parameter to be determined  with the PDFs, simultaneously.  In addition, the reduction of gluon and charm PDF uncertainties at large $x$, as the most prominent case, determination of total charm PDFs considering BHPS model and Higgs boson cross section prediction to address PDF uncertainties are illustrated in Sec. IV. Finally, our conclusion and discussion is given in Sec. V.

\section{Description of experimental measurements for PDF analysis} \label{EXP}
Basically, a broad set of hard scattering cross sections from DIS, $pp$  and $p{\bar p}$ collisions, which provide the PDF information over a wide range of $x$ and $Q^2$ and for different flavor combinations, are used in this PDF QCD analysis. While the common PDF fits were based mostly on DIS process, in recent years other measurements and processes such as heavy flavor, inclusive jet, and also the top quark cross sections have proved a very important impact for constraining the PDFs \cite{Hou:2019efy,H1:2015ubc,Bailey:2020ooq,Ball:2021leu}.

	To have a precise constraint for all the PDFs in a wide region of $x$ and $Q^2$ (and not in especial region), we need to perform a PDF analysis using a completely global fit, including as much data as possible.

	One of the data sets which may include in a PDF  analysis is the fixed-target DIS data collected by some experiments (such as SLAC, BCDMS,  NMC, etc.) with proton and deuteron targets. These kinds of data are sensitive to some PDFs in specific regions. To add the fixed-target DIS data in a completely global fit, we have to add some additional parameters in some parton distribution due to the kinematic region of these data (low values of $Q^2$)  to have a better agreement in QCD fits.
	Another point is that the deuterium DIS data necessitate considering the differences between PDFs in the deuteron and those in the free proton and neutron. It is obvious that including fixed-target DIS data in the presence of all other data sets, which will be introduced in this section, impacts on some PDFs, especially valence PDFs, and consequently may impact on the gluon PDF at large $x$ via DGLAP evolution. We found that the gluon PDF will be somewhat different at low $x$ by including and excluding the fixed-target DIS data in the presence of all other data sets in the present analysis. 
	
	As we mentioned before,  to focus on data sets with
	impact on the high $x$ gluon, although the heavier targets data are available but to avoid any influence
	of nuclear correction uncertainties and adding some additional parameters, these kinds of data are not used in the present analysis.

	Although the impact of particular datasets on a particular PDF, such as the gluon PDF in a wide region of $x$ and $Q^2$, can only truly be assessed in a completely global fit, a study such as this analysis can give us some worth idea about the impact of particular data sets on some PDFs, especially the gluon PDF at large $x$.

In this analysis, we determine PDFs by including a selection of data sets (1891 data points) from the combined HERA I+II deep-inelastic scattering data along with other the  non-LHC and LHC  different data. The various PDF fits are performed to three different data sets: (i) inclusive HERA I+II cross section data as a base input data, (ii) non-LHC, (iii) LHC data set.
In this section an overview of the currently available experimental measurements reported by various collaborations that are included in our fit procedures, are presented.  At the following, the type of reaction, ranges of kinematic variables for all data sets also role of these experimental measurements in the extraction of parton distribution functions and strong coupling constant are mentioned. The details of utilized experimental data  are explained in Tables    \ref{AllData}, \ref{tab:II} and  \ref{tab:XsectionTable-ATLASand CMS}.

\subsection{ Inclusive HERA I+II cross sections}
In this subsection we discuss the significant role of final HERA I+II data, since this data have improved constraints on small $x$ sea quarks and gluon PDF. 

We use neutral and charged current combined final HERA run I+II data \cite{Abramowicz:2015mha} with the variety of beam energies. The H1 and ZEUS collaborations at HERA reported a combination of inclusive deep inelastic $e^{\pm}p$ scattering cross sections  for neutral and charged current interactions. The energies of proton beams  are 460, 575, 820 and 920 GeV and the energies of both $e^{\pm}$ beams are 27.5 GeV. The correspond center of mass energies related to each proton beam  are 225, 251, 300 and 320 GeV approximately and the value of integrated luminosity is about 1 fb$^{-1}$. These experimental results are available  as a function of $Q^2$ and $x$.  The ranges of these quantities  are $0.045 \leq Q^2 \leq 50000$ and $ 6 \times 10^{-7} \leq x \leq 0.65$ for neutral current (NC) interactions, and $200 \leq Q^2 \leq 50000$ and $ 1.3 \times 10^{-2} \leq x \leq 0.40$ for charged current (CC).
The inclusive HERA I+II cross section data is included as a base data set in our analysis to determine PDFs and strong coupling constant. For this kind of data set, we use the kinematic cut $Q^2 >$ 6.5 GeV$^2$ which is necessary to remove the higher twist terms effectively.

\subsection{ Non-LHC experimental data} 
In this part we introduce the non-LHC data set which we use in the present
analysis. In QCD fits to inclusive HERA I+II data only, the  shape of the gluon PDF  depends on the value of strong coupling $\alpha_s(M^2_Z)$. 
In fact, the uncertainty on the gluon PDF is increased for fits with free  $\alpha_s(M^2_Z)$ compared to fits with fixed $\alpha_s(M^2_Z)$ \cite{Abramowicz:2015mha}. In this regard, in addition of HERA I+II inclusive data, we include the DIS heavy-quark production measurements by H1 and ZEUS collaborations at HERA \cite{H1:2018flt}.
These kind of data were proven to be consistent with the inclusive data and also provide a constraint on sea and gluon PDF 
at low $x$ at NLO and NNLO \cite{Abramowicz:2015mha}.

On the other hand, attention to include the experimental data  on jet production cross section  would be very worth in PDF fits. Because, this kind of data provides an independent measurement of the gluon PDF. These data are also sensitive to $\alpha_s(M^2_Z)$, and gluon PDF at low $Q^2$ and to the valence PDF at high $Q^2$.
So, the jet data not only reduces the uncertainty on high $x$ gluon PDF with fixed  $\alpha_s(M^2_Z)$ in the fits, but allows us to determine gluon PDF and accurate $\alpha_s(M^2_Z)$ simultaneously. We choose H1 and ZEUS jet data \cite{Chekanov:2002be,Chekanov:2006xr,Abramowicz:2010cka,Aktas:2007aa,Aaron:2009vs,Aaron:2010ac,Andreev:2014wwa} in the non-LHC data set.

Finally,  in addition to mentioned data in above, other measurements which are also sensitive to PDF determination and strong coupling constant values, such as  \dzero and CDF data that contains jet experimental measurements \cite{Aaltonen:2008eq,Abazov:2008ae} and  top quark cross sections \cite{Aaltonen:2013wca,D0:1998huz} are included.

 At the following, we present the details of non-LHC data set: heavy quark production cross sections \cite{H1:2018flt},
the H1 and ZEUS jet data \cite{Chekanov:2002be,Chekanov:2006xr,Abramowicz:2010cka,Aktas:2007aa,Aaron:2009vs,Aaron:2010ac,Andreev:2014wwa},  \dzero and CDF data that contains top quark cross sections \cite{Aaltonen:2013wca,D0:1998huz} and jet experimental measurements \cite{Aaltonen:2008eq,Abazov:2008ae}, which are added to  the DIS HERA data \cite{Abramowicz:2015mha} as a base set.

\begin{enumerate}[label=(\roman*)]

\item{\bf Heavy flavor HERA cross sections:} 
In addition to the HERA I+II inclusive combination which we mentioned in previous subsection, the combined measurements of charm and beauty quark cross sections reported by H1 and ZEUS collaborations \cite{H1:2018flt} are also available in the kinematic ranges of $2.5 < Q^2 < 2000$ GeV$^2$ and $3\times10^{-5} < x < 5\times10^{-2}$. Taking into account the heavy quark cross section data in a PDF fit will cause restrict on the gluon distribution \cite{H1:2018flt} and also are valuable to obtain the  strong coupling constant value \cite{Abramowicz:2015mha}.

\item{\bf{{H1 jet cross sections:}}} Four different jet cross section experimental measurements \cite{Aktas:2007aa,Aaron:2009vs,Aaron:2010ac,Andreev:2014wwa} reported by H1 collaboration are included in our analysis for extracting PDFs.
These jet cross section measurements play a significant role in processes for determining the strong coupling constant.

The first inclusive jet cross section \cite{Aktas:2007aa} is neutral current deep-inelastic $e^+p$ scattering with the values of center of mass energy and integrated luminosity 319 GeV and 65.4 pb$^{-1}$, respectively. The jet cross sections are measured as a function of $Q^2$ and transverse energy ($E_T$) with the ranges of $150 \leq Q^2 \leq 15000$ GeV$^{2}$ and $7 \leq E_T \leq 50$ GeV. 

The second H1 inclusive jet cross section experimental data \cite{Aaron:2009vs} is from electron-proton DIS neutral current process, reported at  $\sqrt{s}=319$ GeV and the related integrated luminosity of experiment is equal to 395 pb$^{-1}$.  The inclusive jet cross sections are function of $Q^2$ and transverse momentum that the ranges covered by them, are $150 \leq Q^2 \leq 15000$ GeV$^{2}$ and 7$\leq p_T\leq$50. 

The third H1 inclusive jet cross section data \cite{Aaron:2010ac} extracted from positron-proton deep inelastic scattering corresponds to the center of mass energy and integrated luminosity values of 319 GeV and 43.5 pb$^{-1}$, respectively.  The jet experimental data is reported as a function of $Q^2$ and jet transverse momentum that their ranges are $5 \leq Q^2 \leq 100$ GeV$^{2}$ and $5 \leq p_T \leq 80$ GeV.

The fourth and last H1 experimental data \cite{Andreev:2014wwa} included is the inclusive jet, dijet and trijet double differential cross sections. The mentioned cross sections are measured in neutral current deep inelastic electron-proton scattering. The center of mass energy is equal to 319 GeV and the value of integrated luminosity is 351 pb$^{-1}$. The inclusive jet cross section data is reported as a function of $150 \leq Q^2 \leq 15000$ and  $7 \leq p_T \leq 50$. The dijet and trijet double differential cross sections are reported as a function of $Q^2$ and transverse momentum that the range of this variable for both dijet and trijet data is $5 \leq p_T \leq 50$ GeV. 

According to fit results in the literatures,  all inclusive jet cross sections  are important in particular for determining the value of strong coupling constant \cite{Aktas:2007aa, Aaron:2009vs, Aaron:2010ac, Andreev:2014wwa}.

\item{\bf{{ZEUS jet cross sections:}}} Three different jet experimental measurements reported by ZEUS collaboration \cite{Chekanov:2002be,Chekanov:2006xr,Abramowicz:2010cka} are used in our PDF fit. At the following the properties of these measurements are explained. 

The first one is inclusive jet differential cross section \cite{Chekanov:2002be} related to neutral current deep inelastic positron-proton scattering with center of mass energy of 300 GeV, that the $Q^2$ of virtual bosons are larger than 125 GeV$^2$ and the value of integrated luminosity related to this experiment is 38.6 pb$^{-1}$. This experimental data is presented as functions of jet transverse energy, jet pseudorapidity and $Q^2$ that the range of jet transverse energy is $8 \leq E_T \leq 100$  GeV and the range of jet pseudorapidity variable is from -2 to 1.8. Described jet cross section data is important in processes for determining the strong coupling constant \cite{Chekanov:2002be}. 

The second inclusive jet cross section measurement  presented by ZEUS collaboration \cite{Chekanov:2006xr} is reported at the center of mass energy of 318 GeV and with the integrated luminosity of 82 pb$^{-1}$. The inclusive jet cross sections measured in NC deep inelastic electron-proton scattering. These measurements are presented as functions of transverse energy and $Q^2$ that their ranges are $8 \leq E_T \leq 100$  GeV and $1.25\times10^{2}\leq Q^2 \leq10^{4}$ GeV$^2$, respectively, \cite{Chekanov:2006xr}. In particular, in a global fit, these data are able to constrain the gluon density \cite{Chekanov:2006xr}.

The third and last ZEUS experimental data \cite{Abramowicz:2010cka} that is included in fits of present QCD analysis is from inclusive dijet cross section measurement. The values of center of mass energy and integrated luminosity of the experiment are 319 GeV and 374 pb$^{-1}$, respectively. The dijet cross sections are reported as a function of $Q^2$ and the jet transverse energy, with the ranges of $125 \leq Q^2 \leq 20000$ GeV$^{2}$ and $8 \leq E_T \leq 60$ GeV, respectively. Also, invariant mass of the mentioned dijet system is larger than 20 GeV. In a fit procedure these dijet data have potential to improve the gluon uncertainty  at  high-$x$ region particularly and in addition are important for determining the value of strong coupling constant \cite{Abramowicz:2010cka}.

\item{\bf{{CDF jet cross sections:}}} The measurement of inclusive jet cross section is reported by CDF \cite{Aaltonen:2008eq} collaboration, using $p\bar p$ collisions at the center of mass energy of 1.96 TeV with the integrated luminosity  of  1.3 fb$^{-1}$. The measurement is presented as a function of jet rapidity and jet transverse momentum, with the covering ranges of $ \mid y \mid \leq  2.1$ and $62 \leq p_T \leq 700$ GeV. This experimental data in a global fit causes constraints at high-$x$ region on PDFs and particularly on gluon distribution and also can reduce the gluon uncertainty at high-$x$. It occurs when the measured jet cross sections have tendency to be lower than the central NLO predictions of perturbative QCD \cite{Aaltonen:2008eq}.

\item{\bf{{\dzero jet cross sections:}}} The inclusive jet cross section experimental data is reported by \dzero collaboration \cite{Abazov:2008ae} from $p\bar p$ collisions, as a function of jet rapidity and  jet transverse momentum
with the ranges of $\mid y \mid \leq  2.4$ and $50 \leq p_T \leq 600$ GeV, respectively. The values of center of mass energy and integrated luminosity related to this experiment are 1.96 TeV and  0.7 fb$^{-1}$, respectively. In researches for extracting PDFs  by considering these experimental data constraints on gluon PDF is highly expected \cite{Abazov:2008ae,Martin:2009iq}.

\item{\bf{{Total  $t \bar t$ Tevatron cross section:}}} The top quark total cross section measurements reported by \dzero and CDF  are also included in the fits.  The \dzero and CDF total top quark cross sections \cite{Aaltonen:2013wca,D0:1998huz} measured at both 1.8 and 1.96 TeV are  included in the fits to extract the PDFs and strong coupling constant. The details and values of these measurements are summarized in Tables \ref{tab:Data}.

\begingroup
\squeezetable

\begin{table*}[htp]
	
	\caption{\label{AllData} The list of all data sets: DIS HERA I+II, non-LHC and LHC data used in the present analysis. For each data set, we 
	indicate process, measurement, reference and the ranges of their kinematic cuts such as $x$, $y$, $Q^2$ [GeV$^2$], $p_T$ [GeV], $E_T$ [GeV], $\sqrt{s}$ [TeV] and $\mathcal{L}$ [fb$^{-1}$]. }

	\begin{center}
		\begin{ruledtabular}
		\begin{tabular}{l|c|l|c|lr}
			~Data set  & Process & ~Experiment   & Ref.  & ~Kinematic ranges and details\\
			\hline 
			
			\multicolumn{1}{l}{\bf{HERA I+II}}\\\hline	
			\hline 			
			& $e^{\pm}p \rightarrow \overset{(-)}{\nu} X$
			&  ~HERA I+II CC $e^+p$ &\cite{Abramowicz:2015mha}&~$3\times 10^{2}\leq Q^2 \leq3\times 10^{4}$, &$8.0\times 10^{-3}\leq x \leq$0.4  \\ 
			& & ~HERA I+II CC $e^-p$ &\cite{Abramowicz:2015mha}&~$3\times 10^{2}\leq Q^2 \leq3\times 10^{4}$, &$8.0\times 10^{-3}\leq x \leq$0.65\\ 
			&$e^{\pm}p \rightarrow e^{\pm}X   $ & ~HERA I+II NC $e^-p$ &\cite{Abramowicz:2015mha}&~$60\leq Q^2 \leq5\times 10^{4}$, &$8.0\times 10^{-4}\leq x \leq$0.65\\ 
			DIS $\sigma$	& & ~HERA I+II NC $e^-p$ 460~ &\cite{Abramowicz:2015mha} &~$6.5\leq Q^2 \leq8\times 10^{2}$, &$3.48\times 10^{-5}\leq x \leq$0.65\\ 
			& & ~HERA I+II NC $e^-p$ 575~ &\cite{Abramowicz:2015mha}&~$6.5\leq Q^2 \leq8\times 10^{2}$, &$3.48\times 10^{-5}\leq x \leq$0.65\\ 
			& & ~HERA I+II NC $e^+p$ 820~ &\cite{Abramowicz:2015mha}&~$6.5\leq Q^2 \leq3\times 10^{4}$, &$6.21\times 10^{-7}\leq x \leq$0.4\\ 
			& & ~HERA I+II NC $e^+p$ 920~ &\cite{Abramowicz:2015mha}&~$6.5\leq Q^2 \leq3\times 10^{4}$, &$5.02\times 10^{-6}\leq x \leq$0.65 \\
			\hline
			\multicolumn{1}{l}{\bf{Non-LHC}}\\\hline	
			\hline 				
			
			DIS heavy-quarks	&$e^{\pm}p\rightarrow e^{\pm} c  X  $ & ~H1-ZEUS Charm & \cite{H1:2018flt}&~2.5$\leq Q^2 \leq2\times10^{3}$, &   1$\times10^{-3}\leq x \leq$ 5$\times10^{-2}$ \\
			&$e^{\pm}p\rightarrow e^{\pm} b  X  $ & ~H1-ZEUS Beauty & \cite{H1:2018flt}&~2.5$\leq Q^2 \leq2\times10^{3}$, &   3$\times10^{-5}\leq x \leq$ 5$\times10^{-2}$ \\
			\hline
			
			&$e^{\pm}p \rightarrow j X \,$ &~H1 65.4 pb$^{-1}$ & \cite{Aktas:2007aa}&~$1.5\times10^{2}\leq Q^2 \leq1.5\times10^{4}$, &7$\leq E_T\leq$50\\
			& & ~H1 395 pb$^{-1}$ & \cite{Aaron:2009vs}& ~$1.5\times10^{2}\leq Q^2 \leq1.5\times10^{4}$, & 7$\leq p_T\leq$50\\ 
			& & ~H1 43.5 pb$^{-1}$ & \cite{Aaron:2010ac}&~$5\leq Q^2 \leq10^2$, & 5$\leq p_T\leq$80\\   
			& & ~H1 351 pb$^{-1}$ &  \cite{Andreev:2014wwa}& ~$1.5\times10^{2}\leq Q^2 \leq1.5\times10^{4}$, & 7$\leq p_T\leq$50\\

			Lepton-Hadron Jet	&$e^{\pm}p \rightarrow 2$-$j X \,$ & ~H1 dijets &  \cite{Andreev:2014wwa}&~$1.5\times10^{2}\leq Q^2 \leq1.5\times10^{4}$, & 5$\leq p_T\leq$50\\
			&$e^{\pm}p \rightarrow 3$-$j X \,$ & ~H1 trijets & \cite{Andreev:2014wwa}&~$1.5\times10^{2}\leq Q^2 \leq1.5\times10^{4}$, &  5$\leq p_T\leq$50\\
			& & ~ZEUS 300 GeV &  \cite{Chekanov:2002be}&~$1.25\times10^{2}\leq Q^2 \leq10^{4}$, &8$\leq E_T\leq$100\\  
			&$e^{\pm}p \rightarrow j X \,$ & ~ZEUS 318 GeV &\cite{Chekanov:2006xr}& ~$1.25\times10^{2}\leq Q^2 \leq10^{4}$, &8$\leq E_T\leq$100\\  
			&$e^{\pm}p \rightarrow 2$-$j X \,$ & ~ZEUS  dijet & \cite{Abramowicz:2010cka}&~$1.25\times10^{2}\leq Q^2 \leq2\times10^{4}$, & 8$\leq E_T\leq$60\\	
			\hline

			Hadron-Hadron Jet	&$h h \rightarrow j X \,$ & ~CDF & \cite{Aaltonen:2008eq}&~~$6.2\times10\leq p_T\leq7\times10^{2}$,&$ \mid y \mid \leq  2.1$\\ 
			& & ~\dzero &  \cite{Abazov:2008ae}&~$5\times10\leq p_T\leq6\times10^{2}$,&$ \mid y \mid \leq  2.4$\\	
			\hline 
			Hadron-Hadron Top	&$h h \rightarrow t\bar{t} \,$ & ~\dzero-CDF (total) &\cite{Aaltonen:2013wca} & ~$\sqrt{s} = 1.96$,&$\mathcal{L}=8.8$
			\\
			& & ~\dzero (total) &\cite{D0:1998huz} & ~$\sqrt{s} = 1.8$,&$\mathcal{L}=1.1\times10^{5}$
			\\
			\hline 
			\multicolumn{1}{l}{\bf{LHC}}\\
			\hline
			\hline 
		
			Hadron-Hadron Jet	&$h h \rightarrow j X \,$ & ~CMS &\cite{Khachatryan:2016mlc}& ~$7.4\times10\leq p_T\leq2.5\times10^{3}$,&$\mid y \mid \leq  3.0$
			\\
			&  & ~ATLAS &\cite{Aad:2013lpa}& ~$2\times10\leq p_T\leq4.3\times10^{2}$,&$\mid y \mid \leq  0.4$
			\\
			&  & ~ATLAS &\cite{Aad:2011fc}& ~$2\times10\leq p_T\leq1.5\times10^{3}$,&$\mid y \mid \leq  4.4$
			\\
			\hline 
			Hadron-Hadron Top	&$h h \rightarrow t\bar{t} \,$ &  ~CMS (total) &\cite{Khachatryan:2016yzq,Khachatryan:2016mqs,Sirunyan:2018goh,Villalba:2021lkd} & ~$5.02\leq\sqrt{s} \leq 13$,&$5\leq\mathcal{L}\leq3.04 \times 10^5$
			\\ 
			& & ~CMS (differential) &\cite{CMS:2012hkm} & ~$\sqrt{s} = 7$,&$0\leq p_T\leq400$\\
			& & ~ATLAS (total) &\cite{ATLAS:2021xhc,Aaboud:2016zpd,ATLAS:2017fyu} & ~$5.02\leq\sqrt{s} \leq 13$,&$3.2\leq\mathcal{L}\leq2.57 \times 10^5$\\
			& & ~ATLAS (differential) &\cite{ATLAS:2014ipf} & ~$\sqrt{s} = 7$,&$0\leq p_T\leq350$
			\\

		\end{tabular}
		\end{ruledtabular}
		
\end{center}

\end{table*}
\endgroup

\subsection{ LHC data}
We now present the inclusion of LHC data in the our analysis. This includes a variety of data on  jet  and top quark cross section. We will present more details of the fit quality and the PDFs extraction in the next sections, but first we present each of the different types of LHC data included in the present fit.

\item{\bf{{CMS and ATLAS jet cross sections:}}} The experimental measurements of double-differential jet cross sections ($pp \rightarrow jets + X$) are reported by CMS collaboration \cite{Khachatryan:2016mlc}, as a function of absolute jet rapidity ($\mid y \mid \leq  3.0$) and jet transverse momentum ($74 \leq p_T \leq 2500$ GeV) in the center of mass energy of 8 TeV and integrated luminosity of 19.7 fb$^{-1}$.  
These experimental high-$p_T$ double-differential jet cross section measurements can reduce significantly the uncertainties of gluon PDFs  and also are important for determining the value of strong coupling constant \cite{Khachatryan:2016mlc}.

Also, the inclusive jet cross section data are reported by ATLAS collaboration \cite{Aad:2013lpa}, with the center of mass energy of 2.76 TeV and integrated luminosity of 0.2 pb$^{-1}$. This experimental data presented as a function of jet rapidity  $\mid y \mid \leq  0.4$ and jet transverse momentum $20 \leq p_T \leq 430$ GeV. On the other hand, the ATLAS \cite{Aad:2011fc} jet cross section measurement at  $\sqrt{s} = 7$ TeV and  the integrated luminosity of 37 pb$^{-1}$ is included in the fit.
 This jet cross section data is a function of jet rapidity ($\mid y \mid \leq  4.4$) and jet transverse momentum ($20 \leq p_T \leq 1500$ GeV). The above mentioned ATLAS jet cross section data are able to provide important impact on gluon and sea quark distributions in the high $x$ region \cite{Aad:2013lpa}.

\begin{table*}
\caption{\label{tab:Data}The recent measurements of top quark pair production total cross section in different center of mass energies ($\sqrt{s}$) and integrated luminosities ($\mathcal{L}$) with corresponding uncertainties, reported by \dzero and CDF collaborations at Fermilab, and the ATLAS and CMS collaborations at LHC.}
   \begin{center}
   	\begin{ruledtabular}
        \begin{tabular}{ccccc}
            $\sqrt{s}$ [TeV] & ~~~ $\mathcal{L}$ [fb$^{-1}$] ~~~ &   Ref.    & $\sigma_{Exp.}^{\rm tot}(t\bar{t})$ [pb]   \\
            \hline 

                                     \multicolumn{4}{l}{{\bf  \dzero and CDF Experiments at Fermilab}}\\
            \hline                                       
             1.96   & 8.8   & \cite{Aaltonen:2013wca}  & 7.60 $\pm$ 0.41\\
             1.8   & $1.1 \times 10^5$   & \cite{D0:1998huz}  & 5.9 $\pm$ 1.2(stat.) $\pm$ 1.1(syst.)\\
            \hline             
            \multicolumn{4}{l}{{\bf  CMS Experiments at LHC}}\\
            \hline 
            
             5.02  & $3.04 \times 10^5$ &  \cite{Villalba:2021lkd} & 60.3 $\pm$ 5.0(stat.) $\pm$ 2.8(syst.) $\pm$ 0.9(lumi.)  \\
             7   & 5  &  \cite{Khachatryan:2016mqs} & 173.6 $\pm$ 2.1(stat.) $^{+4.5}_{-4}$(syst.) $\pm$ 3.8(lumi.)  \\
             7  & 5 &  \cite{Khachatryan:2016yzq} & 161.7 $\pm$ 6(stat.) $\pm$ 12(syst.) $\pm$ 3.6(lumi.)  \\
             8  & 19.6 &  \cite{Khachatryan:2016yzq} & 227.4 $\pm$ 3.8(stat.) $\pm$ 13.7(syst.) $\pm$ 6(lumi.) \\
             8   & 19.7 &  \cite{Khachatryan:2016mqs} & 244.9 $\pm$ 1.4(stat.) $^{+6.3}_{-5.5}$(syst.) $\pm$ 6.4(lumi.)  \\
             13  & 35.9  &  \cite{Sirunyan:2018goh} & 803 $\pm$ 2(stat.) $\pm$ 25(syst.) $\pm$ 20(lumi.)  \\             

            \hline
                         \multicolumn{4}{l}{{\bf  ATLAS Experiments at LHC}}\\
                         \hline

             5.02   & $2.57 \times 10^5$   & \cite{ATLAS:2021xhc}  & 66.0 $\pm$ 4.5(stat.) $\pm$ 1.6(syst.) $\pm$ 1.2(lumi.)$\pm$ 0.2(beam.)\\
             7   & 3.2   & \cite{Aaboud:2016zpd}  & 183 $\pm$ 3(stat.) $\pm$ 4(syst.) $\pm$ 4(lumi.)$\pm$ 3(beam.)\\
             8   & 3.2    & \cite{Aaboud:2016zpd}  & 243 $\pm$ 2(stat.) $\pm$ 5(syst.) $\pm$ 5(lumi.)$\pm$ 4(beam.)\\             
             8   & 20.2    & \cite{ATLAS:2017fyu}  & 248.3 $\pm$ 0.7(stat.) $\pm$ 13.4(syst.) $\pm$ 4.7(lumi.)\\
             13   & 3.2    & \cite{Aaboud:2016zpd}  & 818 $\pm$ 8(stat.) $\pm$ 27(syst.) $\pm$ 19(lumi.)$\pm$ 12(beam.)\\

        \end{tabular}
    \end{ruledtabular}
        		\label{tab:II}
    \end{center}
\end{table*}
\begingroup
\squeezetable
\begin{table}[h]
	\begin{ruledtabular}
	\begin{center}	
				\caption{\label{tab:Data1}Normalized differential cross-sections for $t \bar t$ production as a function of the $p_T$, reported by the CMS and ATLAS collaboration \cite{ATLAS:2014ipf,CMS:2012hkm}. The statistical and systematic uncertainties are also shown.\\}
						
		\begin{tabular}{c|c|c|c}
 			$p_T$ [GeV] & $ 1/\!\sigma\, d \sigma\!/\!d p_T$[GeV$^{-1}$] & Stat. [\%] & Sys. [\%] \\
            \hline        
            \multicolumn{4}{l}{{\bf CMS Experiment}}\\
            \hline			

			$   0$ to $ 60$ & $4.54 \cdot 10^{-3}$ &  $\pm$  2.5 & $\pm$   3.6 \\
			$  60$ to $100$ & $6.66 \cdot 10^{-3}$ & $\pm$   2.4 & $\pm$   4.9 \\
			$ 100$ to $150$ & $4.74 \cdot 10^{-3}$ &  $\pm$  2.4 &  $\pm$  3.2 \\
			$ 150$ to $200$ & $2.50 \cdot 10^{-3}$ &  $\pm$   2.6 &  $\pm$   5.1 \\
			$ 200$ to $260$ & $1.04 \cdot 10^{-3}$ &  $\pm$   2.9 &  $\pm$   5.5 \\
			$ 260$ to $320$ & $0.38 \cdot 10^{-3}$ & $\pm$   3.7 &  $\pm$   8.2 \\
			$ 320$ to $400$ & $0.12 \cdot 10^{-3}$ & $\pm$   5.8 &  $\pm$   9.5\\

                        \hline
			            \multicolumn{4}{l}{{\bf ATLAS Experiment}}\\
		            	\hline
			0 to   50 & $3.4 \cdot 10^{-3}$ & $\pm$     2.4 & $\pm$     5.1 \\ 
			50 to  100 &$ 6.7 \cdot 10^{-3}$ & $\pm$     1.2 & $\pm$     1.9 \\ 
			100 to  150 &$ 5.3 \cdot 10^{-3}$ & $\pm$     2.5 & $\pm$     2.6 \\ 
			150 to  200 &$ 2.6 \cdot 10^{-3}$ & $\pm$     2.0 & $\pm$     4.8 \\ 
			200 to  250 &$ 1.12 \cdot 10^{-3}$ & $\pm$     2.4 & $\pm$     4.8 \\ 
			250 to  350 &$ 0.32  \cdot 10^{-3}$ & $\pm$     3.5 & $\pm$     5.5 \\ 			 
		\end{tabular}	
		\label{tab:XsectionTable-ATLASand CMS}
	\end{center}
\end{ruledtabular}
\end{table}
\endgroup

\item{\bf{{Top quark cross sections:}}} The top quark cross section measurements reported by ATLAS and CMS are also included in the present analysis. The CMS top quark  data containing the differential cross sections are reported at the center of mass energy of 7 TeV  as a function of transverse momentum ($p_T$) \cite{CMS:2012hkm}. The CMS top quark total cross sections  are measured at various center of energies such as 5.02, 7, 8 and 13 TeV \cite{Khachatryan:2016yzq, Villalba:2021lkd, Khachatryan:2016mqs, Sirunyan:2018goh}. The ATLAS 
differential cross section data \cite{ATLAS:2014ipf} at the center of mass energies  7 TeV 
and the
top quark total cross section measurements \cite{ATLAS:2021xhc,Aaboud:2016zpd,ATLAS:2017fyu} at the center of mass energy of 5.02, 7, 8 and 13 TeV 
are also included in the final PDF fit. The details  of these measurements are summarized in Tables \ref{tab:Data} and \ref{tab:XsectionTable-ATLASand CMS}. According to the reported results in Ref.~\cite{Czakon:2013tha} the top quark cross section experimental data are able to create significant constraints on the gluon PDF.

We will see, choosing the complete data set as we shown in Table \ref{AllData}, with including LHC data along with the combined HERA and 
 non-LHC data set in the current analysis, the  LHC data will impact on gluon and charm PDF uncertainties at high-$x$ significantly, in respect to the fit of HERA data only and HERA with non-LHC data sets. 
\end{enumerate}

\section{Theoretical Framework} \label{Comput}

As previously mentioned, various types of experimental data  such as HEAR I+II, non-LHC and LHC data are included in our fit procedures. 
At the following section, most important computational aspects such as PDF parametrization and the minimization method which are considered for extracting PDF sets are explained.

The present QCD analysis for PDFs determination is performed in the \textsc{xFitter} framework.  The \textsc{xFitter} package \cite{Alekhin:2014irh}, updated version of HERAfitter \cite{HERAfitter},  is used to extract PDFs and the strong coupling constant with the main advantage of utilizing several computational programs simultaneously.

To perform a global fit, Dokshitser-Gribov-Lipatov-Altarelli-Parisi (DGLAP) \cite{Dokshitzer:1977sg,Gribov:1972ri,Altarelli:1977zs}  evolution equations 
derive PDFs at any scale relevant to comparisons with the experimental data from the parametrization of PDFs at an input scale of $Q_0^2$. For preparing the NLO and NNLO numerical solutions of DGLAP evolution equations of PDFs at higher scales $Q^2$>$Q_0^2$, the QCDNUM \cite{Botje:2010ay} package is utilized. In this regard, the parametric forms for the PDF, $ xf(x,Q_0^2)$ at the lower scale of QCD evolution are needed.   

To include the heavy quark contribution to any process, it is necessary to apply the variable flavor number scheme (VFNS), so that the different numbers of active
flavours ($N_f$) are adopted at different $Q^2$ scales. In this regard, the Thorne-Roberts variable-flavor number scheme \cite{Thorne:1997ga} is used to calculate the contributions of the heavy quarks.

The jet cross section data reported by ATLAS and CMS collaborations are available to include in \textsc{xFitter} framework. To include the jet experimental cross section data, APPLGrid \cite{Carli:2010rw} and FastNLO \cite{Kluge:2006xs} computational programs may be used in the fit procedures.  The grid file of ATLAS jet cross sections is obtained by APPLGrid and the grid file of CMS jet cross section is available by FastNLO  computational package.  Since the ATLAS jet cross section grid files are not available at NNLO, so using $k$-factor technique  is necessary to include these experimental data at NNLO. The HATHOR \cite{Aliev:2010zk} and FastNLO computational packages are also used to include the total and differential top quark cross sections.  Finally, the MINUIT \cite{James:1975dr} program is used for the minimization. 

In this analysis, the minimum cut value which implied on HERA I+II data is $Q_{min}^2=6.5$ GeV$^2$ and the invariant mass squared of 
$W^2$>15 GeV$^2$. We didn't apply any other cuts in this analysis. The value of initial QCD scale assumed to be $Q_0^2=1.9$  GeV$^2$ which is below $c$-quark mass threshold.  The masses of heavy quarks \cite{H1:2015ubc} considered to be $m_c = 1.43$ GeV, $m_b = 4.5$ GeV and $m_t = 173.5$ GeV.

The PDFs at the initial scale are represented by the following general form 
\begin{equation}
 xf_i(x, Q_0^2) = A_i x^{B_i} (1-x)^{C_i} (1 + D_i x + E_i x^2)\;,
\label{eqn:pdf}
\end{equation}
where $i$ indicates the flavor of the parton distribution and for each PDF flavor combination the parameters of PDF shape are different. 
The $x^{B_i}$ term affects  the low $x$ PDF behavior and the $(1-x)^{C_i}$ term, ensures that the parton distribution disappear in $x\to1$ elastic limit.
Also the last term in Eq.~\eqref{eqn:pdf} controls the behavior of the PDFs away from the extrapolation regions of $x\to 0$ and $x\to 1$.
The normalization parameters $A_i$ can be determined via QCD sum rules.

The parameterized parton distributions, $xf_i$,  are chosen to be the valence, light sea, and the gluon PDFs. The detailed parameterizations for each PDF at the initial scale are as followings \cite{Aad:2013lpa}
\begin{eqnarray}
xu_v(x, Q_0^2) &=  & A_{u_v} x^{B_{u_v}}  (1-x)^{C_{u_v}}\left(1+E_{u_v}x^2 \right), \nonumber\\
xd_v(x, Q_0^2) &=  & A_{d_v} x^{B_{d_v}}  (1-x)^{C_{d_v}},\nonumber\\
x\bar{U}(x, Q_0^2) &=  & A_{\bar{U}} x^{B_{\bar{U}}} (1-x)^{C_{\bar{U}}}, \nonumber\\
x\bar{D}(x, Q_0^2)&=&A_{\bar{D}} x^{B_{\bar{D}}} (1-x)^{C_{\bar{D}}}, \nonumber\\
xg(x, Q_0^2) &=   & A_g x^{B_g} (1-x)^{C_g} - A_g' x^{B_g'} (1-x)^{C_g'}\;,~~
\label{eq:xpar-general}
\end{eqnarray}
here, $xu_v$ and $xd_v$ are the valance quark distribution, $x\bar{U}$ and $x\bar{D}$ indicate the $u$-type and $d$-type anti-quark distribution respectively, where $x\bar{U} = x\bar{u}$ and $\bar{D} = x\bar{d} + x\bar{s}$, and the gluon distribution is represented by $xg$.

The QCD sum rules determine the normalization parameters. The parameters $A_{u_v}$, $A_{d_v}$ are fixed with considering the rule of quark counting and $A_{g}$ parameter is fixed using the momentum sum rule. The relation between $A_{\bar{U}}$ and $A_{\bar{D}}$ is $A_{\bar{U}}=A_{\bar{D}} (1-f_s)$ and the $B_{\bar{U}}$ and $B_{\bar{D}}$ parameters are considered to be equal, $B_{\bar{U}}=B_{\bar{D}}$.  The last two relations are set to guarantee that $x\bar{u} \to x\bar{d}$ as $x \to 0$. The distribution of strange-quark is assumed as an $x$-independent fraction ($f_s$) of the down-type sea, $x\bar{s} = f_s x\bar{D}$, at $Q^2_0$ and the value of $f_s$ chosen to be equal to 0.31 \cite{Aad:2013lpa}.

The $xg$ gluon PDF in  Eq.~\ref{eq:xpar-general}, is an exception from Eq.~\ref{eqn:pdf}, with subtracting an additional term of the
form $A_g' x^{B_g'} (1-x)^{C_g'}$, which it makes more flexibility of parametrization at low $x$ \cite{Abramowicz:2015mha} than a single power and positive values of $xg$  at large $x$ as suggested in Ref. \cite{Martin:2009iq}. Note that as the first term of gluon PDF can not be controlled by the single power $B_g$ as $x$ approaches zero, we need to consider the second term of the gluon PDF parametrization. It needs that the powers ${B_g}$ and ${B_g'}$ be different. This additional term is considered at NLO and NNLO.The  $C'_g$ parameter is fixed to 25 for all fit procedures of this QCD analysis.

	The $\chi^2$-function described here is a criterion to find out how a QCD model is compatible with the experimental measurements. In this analysis, we extract the best values of 14 independent free parameters by minimizing the $\chi^2$-function. In cases where all of the correlated uncertainties related to experimental measurements are available, the $\chi^2$-function, as implemented in the \textsc{xFitter}  framework, is given by \cite{Aaron:2009aa}: 
	
	\begin{equation}
		\chi^2 = \sum_i \frac{ [d_i -  t_i(1-\sum_j \beta_{j}^{i} s_j )]^2 }{\delta^2_{i,unc} t^2_i +\delta^2_{i,stat} d_i t_i } + \sum_j s^2_j\;,
		\label{eq:chi2}
	\end{equation}
	here $t_i$ and $d_i$ are theoretical predictions and experimental measurements of $i$-th data point, respectively.
	Also, $\beta_{j}^{i}$ are corresponding systematic uncertainties, and $s_j$ are the nuisance parameters associated with the correlated systematic uncertainties, and $j$ indicate sources of correlated systematic uncertainties. In above, $\delta^2_{i,stat}$ and $\delta^2_{i,unc}$ indicate relative statistical and uncorrelated systematic uncertainties.
	The MINUIT \cite{James:1975dr} package is used for minimizing the $\chi^2$-function in the \textsc{xFitter}  framework. 
	
	Determination of the uncertainty of free parameters is also very worthwhile in addition to extracting the central values of these parameters. The uncertainties of parton distributions functions estimated in Refs.~\cite{Hirai:2007cx,Pumplin:2001ct,Pumplin:2002vw,Martin:2002aw,Blumlein:2002qeu,Hirai:2006sr,Leader:2005ci,deFlorian:2005mw,Hirai:2004wq}. One of the most important approaches is the Hessian method. Although the details of this method 
	are presented in these references, due to utilizing it in the present QCD analysis, a review of the Hessian method is described.

	According to the Hessian approach, by considering $\zeta_i$ ($i$=1, 2, $\cdot \cdot \cdot$, $N$) as free parameters, where  $N$ is  the number of the parameters which are extracted from the fit procedure, the $\chi^2$-function can expand around the minimum $\chi^2$ point  $\hat{\zeta}$:

	\begin{equation}
		\Delta\chi^2 (\zeta) = \chi^2(\hat{\zeta}+\delta \zeta)-\chi^2(\hat{\zeta})
		=\sum_{i,j} H_{ij}\delta \zeta_i \delta \zeta_j \ ,
		\label{eq:chi2expand}
	\end{equation}
	here just leading quadratic term is considered, and $H_{ij}$ denote the  second Hessian derivative matrix. The confidence region is expressed by assigning a value to  $\Delta \chi^2$ in the parameter space. 
	
	If the number of parameters is one, $i.e.,$ $N=1$ the confidence level is $68\%$ for $\Delta \chi^2 = 1$.	
	Generally, in the case of $N \ne 1$,  the value of $\Delta \chi^2$ should be changed. It is known that the confidence level $P$ is related to the number of  parameters and $\Delta \chi^2$ as: 
	
	\begin{equation}
		P=\int_0^{\Delta \chi^2} \frac{1}{2\Gamma(N/2)} 
		\left(\frac{\omega}{2}\right)^{(\frac{N}{2})-1} 
		\exp\left(-\frac{\omega}{2} \right) d\omega \ ,
		\label{eq:dchi2}
	\end{equation}
	here $\Gamma(N/2)$ is the well-known Gamma function. The standard confidence level $P$=0.6826 is correspondent  to one $\sigma$ error range \cite{Leader:2005ci,deFlorian:2005mw,Hirai:2004wq}. The $\Delta \chi^2$  value is numerically calculated by choosing a confidence level value by using Eq.~\ref{eq:dchi2}. 
	In the present analysis, we have $N$=14, and by supposing the confidence level $P$=0.6826, consequently the value of $\Delta \chi^2$ will be 15.94. By running the  MINUIT subroutine \cite{James:1975dr}, the Hessian matrix is available.

	The uncertainty of a certain observable  (${\cal O}$), which is introduced at the input scale $Q_0^2$ can be reached by Hessian method. By having the $\Delta \chi^2$ value and derivatives of the observables with respect of the free parameters, the uncertainties for a given observable  (${\cal O}$)  are calculable using:
	%
	\begin{equation}
		[\delta {\cal O}_{i}]^{2}=\Delta \chi^{2} \sum_{j,k}
		\left(\frac{\partial {\cal O}_{i} (\zeta)}
		{\partial \zeta_j} \right)_{\hat\zeta}
		H_{jk}^{-1}	\left(\frac{\partial {\cal O}_{i}
			(\zeta)}{\partial \zeta_{k}} \right)_{\hat\zeta} \,.
	\end{equation}
	
	Therefore, by having $\Delta \chi^{2}$, and derivatives of ${\cal O}_{i}$ with respect to extracted parameters ($\zeta_i$), the above equation return the uncertainty ($\delta {\cal O}_{i}$) of desired observable. At any other scale of $Q^2$, the uncertainties can be calculated by evolving obtained gradient terms with DGLAP evolution kernel \cite{Dokshitzer:1977sg,Gribov:1972ri,Altarelli:1977zs}.

\section{Fit Results} \label{Res}
            
To investigate the impact of top quark and jet cross sections measurements at the LHC on PDFs, in the presence of HERA I+II combine data as a base set, and also non-LHC data sets, we divide our QCD fits into three different fits in the present analysis. The detailed information for each data set are summarized  in Table~\ref{AllData}. For this purpose, three different fits with the name of  ``Fit A'', ``Fit B'', and ``Fit C''  are introduced in below at both NLO and NNLO: 

\begin{itemize}
	
	\item \textbf{Fit~A}: In the first step, we include the DIS HERA I+II combine data set of Table~\ref{AllData}, as a base fit, without any additional data. In this fit we have 1092 data points. Therefore Fit A provides a proper base for illustrating the impact of additional data on various PDFs and  the strong coupling constant.
		
	\item
	\textbf{Fit~B}: In the second step, we added the non-LHC experimental data from Table~\ref{AllData} to Fit A. This is all same as Fit A, except now  all non-LHC data is added to DIS HERA I+II combine data. In this fit we have 1555 data points. 	
	\item
	\textbf{Fit~C}: As a final fit procedure, we include the LHC data from Table~\ref{AllData} to Fit B. In this fit we have 1891 data points and so contains the complete analysis.
	\end{itemize}
In all fits, there are 14 unknown parameters for PDFs and $\alpha_s(M^2_Z)$. 

\begin{table*}
	\caption{\label{tab:Chi2}The numerical results for the correlated $\chi^2$, log penalty $\chi^2$, total $\chi^2$ and the total $\chi^2$/ degree of freedom (dof)  of each data sets for different  Fits A, B and C at NLO and NNLO.}

	\begin{ruledtabular}
		\begin{tabular}{llcccccccc}

			Observable & Experiment & Ref. &   &&{$\chi^2$ / n-points}   \\ 
			\cline{4-9} 
			\multicolumn{9}{r}{{\bf \hspace{0. cm} Fit A \hspace{2.3 cm} Fit B \hspace{2.4 cm} Fit C~~~~~~~~~~}}\\			      
			\cline{4-9}
			& & & NLO &NNLO&  NLO & NNLO & NLO& NNLO     \\

			\hline
			
			& HERA I+II CC $e ^+ p$ &\cite{Abramowicz:2015mha}& 48 / 39  & 49 / 39 & 66 / 39 & 64 / 39& 63 / 39  & 62 / 39   \\ 
			& HERA I+II CC $e ^- p$  &\cite{Abramowicz:2015mha}& 56 / 42 &55 / 42& 53 / 42 & 53 / 42& 53 / 42   & 54 / 42 \\ 
			& HERA I+II NC $e ^- p$ &\cite{Abramowicz:2015mha}& 222 / 159 &223 / 159& 235 / 159& 238 / 159 & 237 / 159  & 239 / 159  \\ 
			{\bf HERA I+II}& HERA I+II NC $e ^+ p$ 820 GeV &\cite{Abramowicz:2015mha}& 66 / 66 &59 / 66 & 64 / 66& 60 / 66 & 64 / 66  &59 / 66  \\ 
			& HERA I+II NC $e ^+ p$ 920 GeV &\cite{Abramowicz:2015mha}& 403 / 348 &390 / 348& 411 / 348 & 388 / 348& 413 / 348  & 389 / 348 \\ 
			& HERA I+II NC $e ^- p$ 460 GeV &\cite{Abramowicz:2015mha}& 206 / 195 &209 / 195& 206 / 195 & 209 / 195& 206 / 195  &209 / 195 \\ 
			& HERA I+II NC $e ^- p$ 575 GeV &\cite{Abramowicz:2015mha}& 210 / 243 &210 / 243& 212 / 243 & 211 / 243& 212 / 243  & 210 / 243  \\
			\hline
			&   Charm H1-ZEUS& \cite{H1:2018flt}  & -& -& 20 / 30 & 25 / 30 & 20 / 30   & 25 / 30  \\
			&   Beauty H1-ZEUS& \cite{H1:2018flt}  & -& -& 14 / 27 & 23 / 27 & 14 / 27   & 23 / 27  \\				    
			&   H1 65.4 pb$^{-1}$& \cite{Aktas:2007aa} & -& -& 11 / 24 & 12 / 24 & 11 / 24   &11 / 24  \\ 
			&   H1 395 pb$^{-1}$& \cite{Aaron:2009vs}  & -& -& 14 / 24 & 19 / 24 & 14 / 24   & 18 / 24 \\  
			{\bf Non-LHC} &   H1 43.5 pb$^{-1}$& \cite{Aaron:2010ac}  & -& -& 18 / 28 & 19 / 28 & 18 / 28  & 20 / 28   \\  
			&   H1 351 pb$^{-1}$& \cite{Andreev:2014wwa}   & -& -& 25 / 24 & 28 / 24 & 25 / 24   & 28 / 24 \\ 
			&   H1   dijets & \cite{Andreev:2014wwa} & -& -& 42 / 24 & 39 / 24 & 43  / 24  & 38 / 24 \\ 
			&   H1  trijets& \cite{Andreev:2014wwa}  & -& -& 9 / 16 & 9 / 16 & 9 / 16   & 10 / 16  \\ 
			&   ZEUS 300 GeV& \cite{Chekanov:2002be} & -& -& 28 / 30 & 30 / 30 & 28 / 30   &29 / 30  \\
			&   ZEUS 318 GeV  & \cite{Chekanov:2006xr}& -& -& 23 / 30 & 23 / 30 & 23 / 30   & 23 / 30   \\
			&   ZEUS  dijet & \cite{Abramowicz:2010cka} & -& -& 16 / 22 & 17 / 22 & 16 / 22   & 16 / 22  \\ 				 
			&   CDF-Jet& \cite{Aaltonen:2008eq}  &  - & -& 121 / 72 & 119 / 72 & 128 / 72   & 130 / 72 \\ 
			&   \dzero-Jet & \cite{Abazov:2008ae}  & -& -& 75 / 110 & 82 / 110 & 76 / 110    & 85 / 110\\ 
			
			&     CDF-\dzero Combined-Top quark &\cite{Aaltonen:2013wca} &  -& -& 6 / 1 & 2 / 1 & 6 / 1   & 2 / 1   \\
			&  \dzero-Top quark  &\cite{D0:1998huz}   & -& -& 0.33 / 1 & 0.1 / 1 & 0.25 / 1  &0.1 / 1 \\  
			\hline  
			&   CMS-Jet &\cite{Khachatryan:2016mlc}  &  -& -& - & -  & 141 / 168   & 151 / 168  \\
			&   ATLAS-Jet 2.76 TeV &\cite{Aad:2013lpa}  &  -& -& - & -  & 40 / 54   & 41 / 54  \\
			&   ATLAS-Jet 7 TeV &\cite{Aad:2011fc}  &  -& -& - & -  & 42 / 90   & 44 / 90  \\
			&  CMS 7 TeV ($e ~ \mu ~ j$)  &\cite{Khachatryan:2016yzq}  & -& -& - & - & 1 / 1  &0.03 / 1 \\ 
			&     CMS 8 TeV ($e ~ \mu ~ j$) &\cite{Khachatryan:2016yzq} & -& -& - & - & 0.63 / 1   & 0.03 / 1 \\
			&     CMS 7 TeV ($b ~ j$) &\cite{Khachatryan:2016mqs} & -& -& - & - & 0.15 / 1    &0.55 / 1 \\
			&     CMS 8 TeV ($b ~ j$) &\cite{Khachatryan:2016mqs}  & -& -& - & - & 0.44 / 1   & 1 / 1 \\
			&     CMS 13 TeV &\cite{Sirunyan:2018goh} & -& -& - & - & 0.48 / 1   & 0.07 / 1  \\
			{\bf LHC} &     CMS 5 TeV &\cite{Villalba:2021lkd}  & -& -& - & - & 0.31 / 1   & 1 / 1 \\
			&     CMS 7 TeV differential &\cite{CMS:2012hkm}  & -& -& - & - & 23 / 7   & 12 / 7\\
			&     ATLAS 5 TeV &\cite{ATLAS:2021xhc}& -& -& - & - & 0.1 / 1   & 0.03 / 1  \\     
			&     ATLAS 7 TeV &\cite{Aaboud:2016zpd}& -& -& - & - & 8 / 1   & 3 / 1  \\ 
			&     ATLAS 8 TeV ($Z \rightarrow l^+ l^- $) &\cite{Aaboud:2016zpd} & -& - & -& - & 0.59 / 1   &0.2 / 1  \\ 
			&     ATLAS 8 TeV ($l ~ j $) &\cite{ATLAS:2017fyu}  &  -& -& - & - & 0.34 / 1   &0.22 / 1 \\
			&     ATLAS 13 TeV &\cite{Aaboud:2016zpd} & -& -& - & - & 1 / 1   & 0.06 / 1  \\
			&    ATLAS 7 TeV differential &\cite{ATLAS:2014ipf}  & -& -& - & - & 12 / 6   & 5 / 6 \\
			
			\hline
			\hline 
			Correlated $\chi^2$ & & & 66& 69& 155& 170& 225& 237  \\
			Log penalty $\chi^2$ &  & & -9.10& -9.29& -0.02& -8.13& +66& +37.00  \\
			Total $\chi^2$ & &  & 1268  &  1255 & 1824& 1832 & 2219   &2213 \\    
			Total $\chi^2$ / dof & &  & 1268/1078&  1255/1078& 1824/1541& 1832/1541 & 2219/1877   & 2213/1877 \\  
			& &  & =1.176&  =1.164& =1.184& =1.189 & =1.182   & =1.179 \\   
			
		\end{tabular}
	\end{ruledtabular}

\end{table*}

According to Table~\ref{tab:Chi2} the extracted value of $\chi^2$/dof  for Fit A is 1.176 and 1.164 at the NLO and NNLO, respectively. These values for Fit B, that the other additional non-LHC are included, are 1.184 and 1.189 at the NLO and NNLO, respectively. Such as this slight growth at NNLO in respect to NLO for Fit B is due to including heavy quarks (charm and beauty) data. Same as this result has been reported  in Ref.~\cite{H1:2018flt}.   In Fit C, which we added both  LHC and non-LHC data to HERA I+II combined data, the extracted NLO and NNLO values of  $\chi^2$/dof are 1.182 and 1.179, respectively. Obviously, in comparison of NLO and NNLO results for Fit C as a complete fit, there
are almost 0.25\% improvements for $\chi^2$ values and the fit quality. Although in the current analysis, Fit C contains the complete analysis, the comparison of this fit with other fit results would be
interested. According to Table~\ref{tab:Chi2} the total $\chi^2$/dof values at NLO are 1.175, 1.184  and 1.182 for Fit A, B, and C respectively, considering DIS HERA I+II combined, non-LHC and LHC data sets. These values are   1.164, 1.189  and 1.179 at NNLO.
According to Table~\ref{tab:Chi2}, the NNLO improvement of ~1\% of $\chi^2$/dof  in our base fit, $i.e.,$ Fit A, with respect to NLO and 
NNLO improvement of ~0.25 \% in Fit C,  with respect to NLO is obtained. Also the increase of ~0.42 \% of
$\chi^2$/dof  at NNLO with respect to our NLO fit in Fit B is obtained.

\begin{figure*}[htb]
	\begin{center}	
			    
		\includegraphics[scale = 0.4]{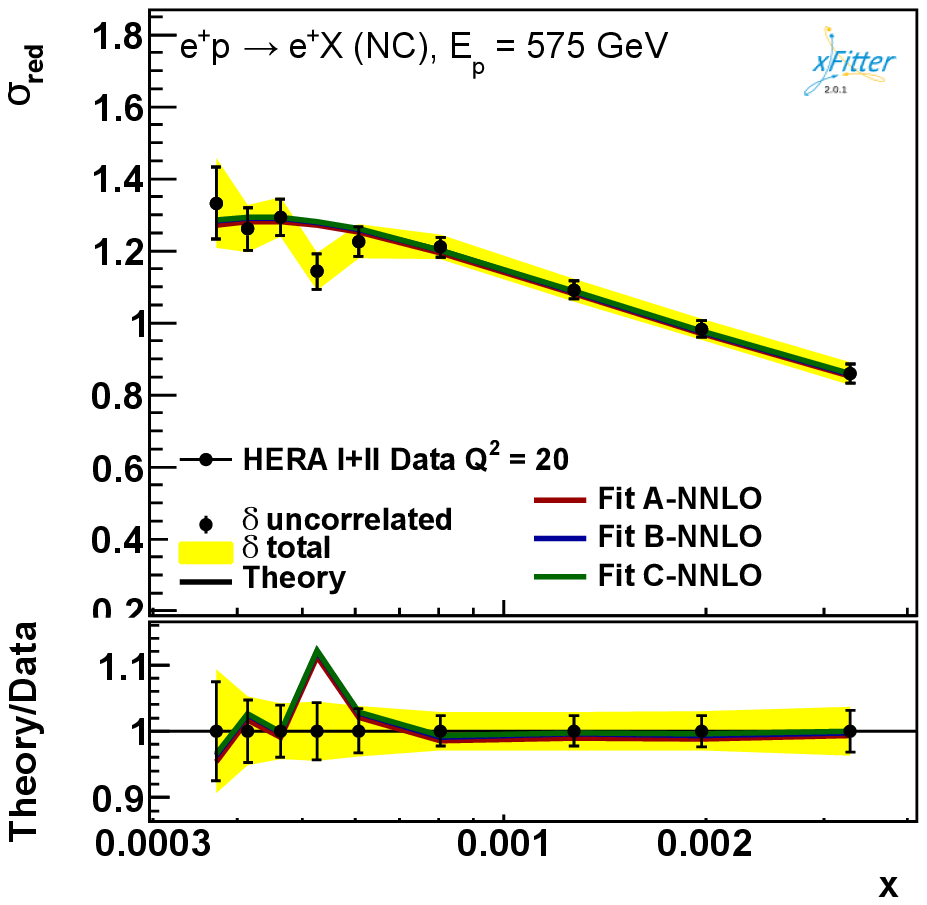}
		\includegraphics[scale = 0.4]{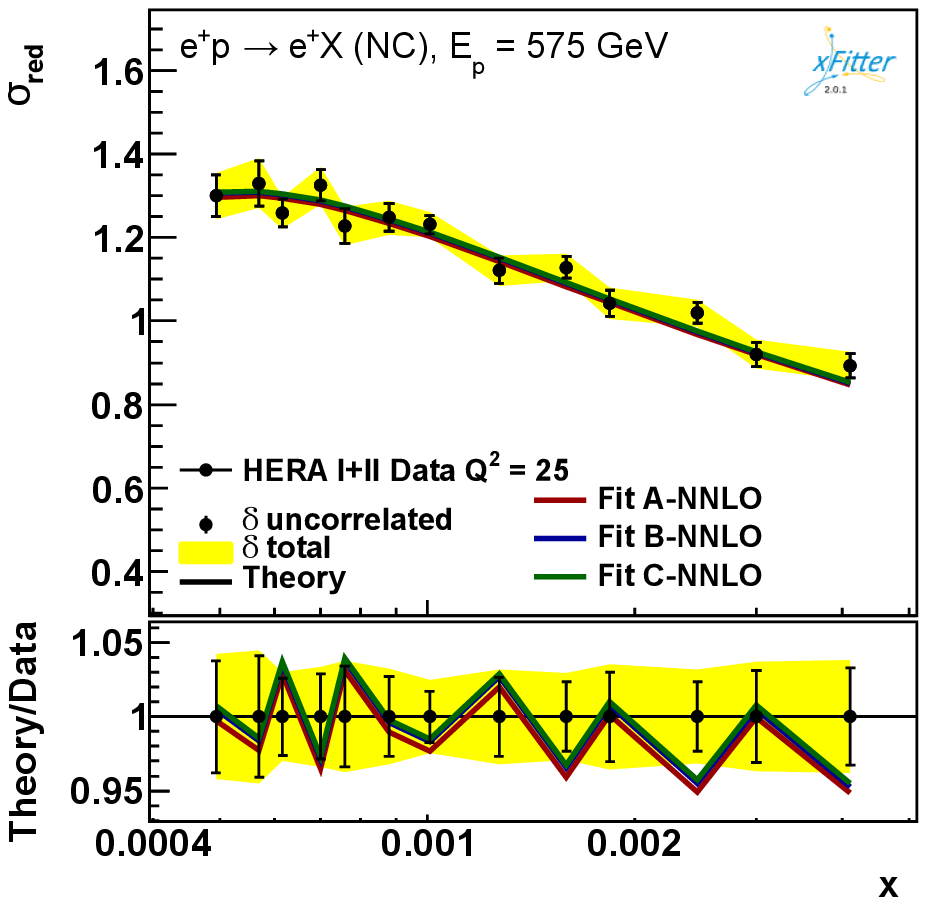}
		\includegraphics[scale = 0.4]{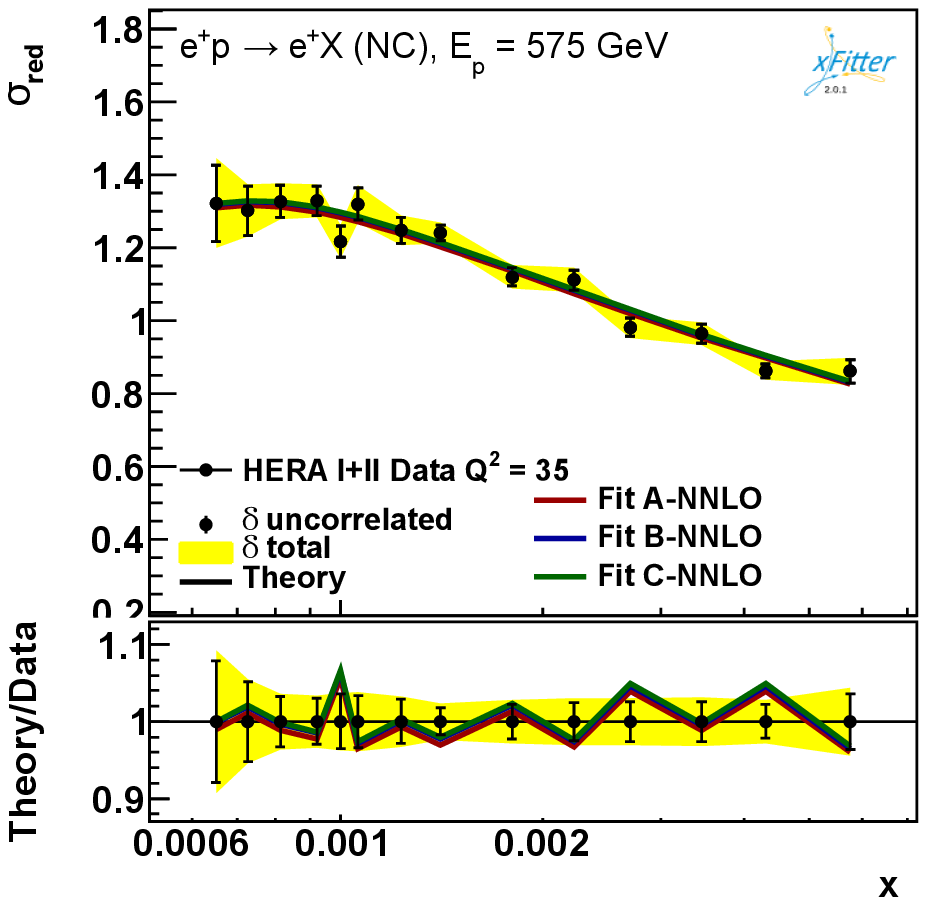}
		\includegraphics[scale = 0.4]{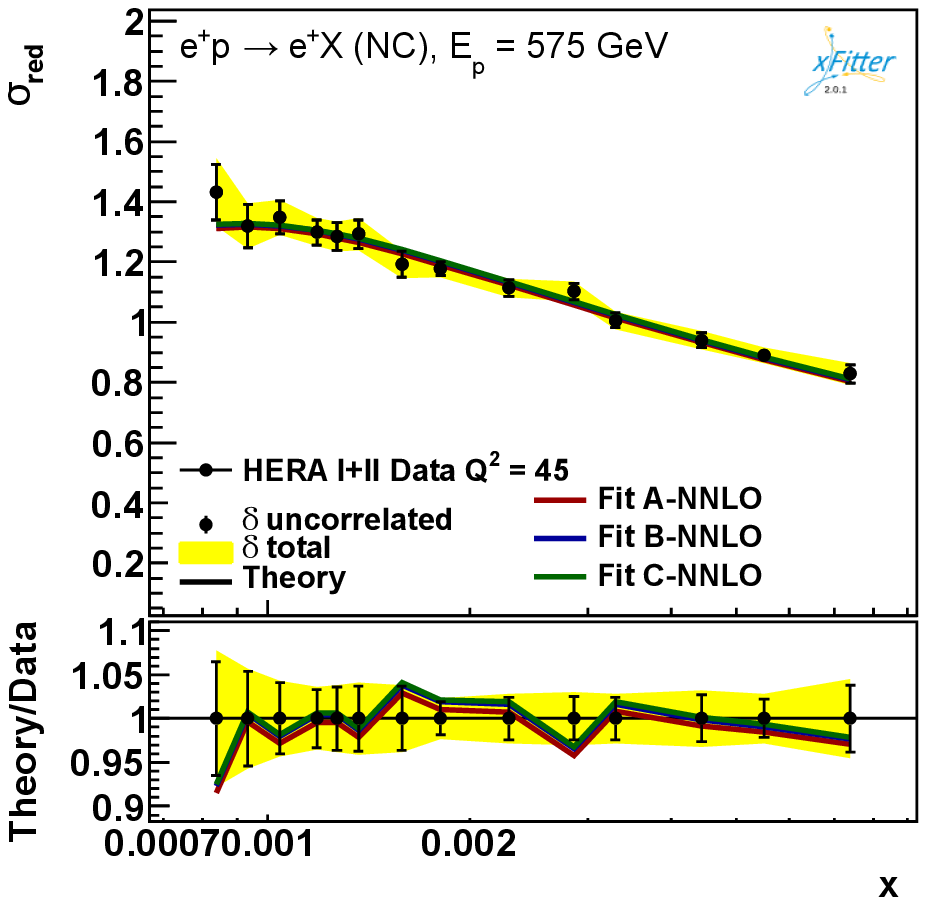}\\
		\includegraphics[scale = 0.4]{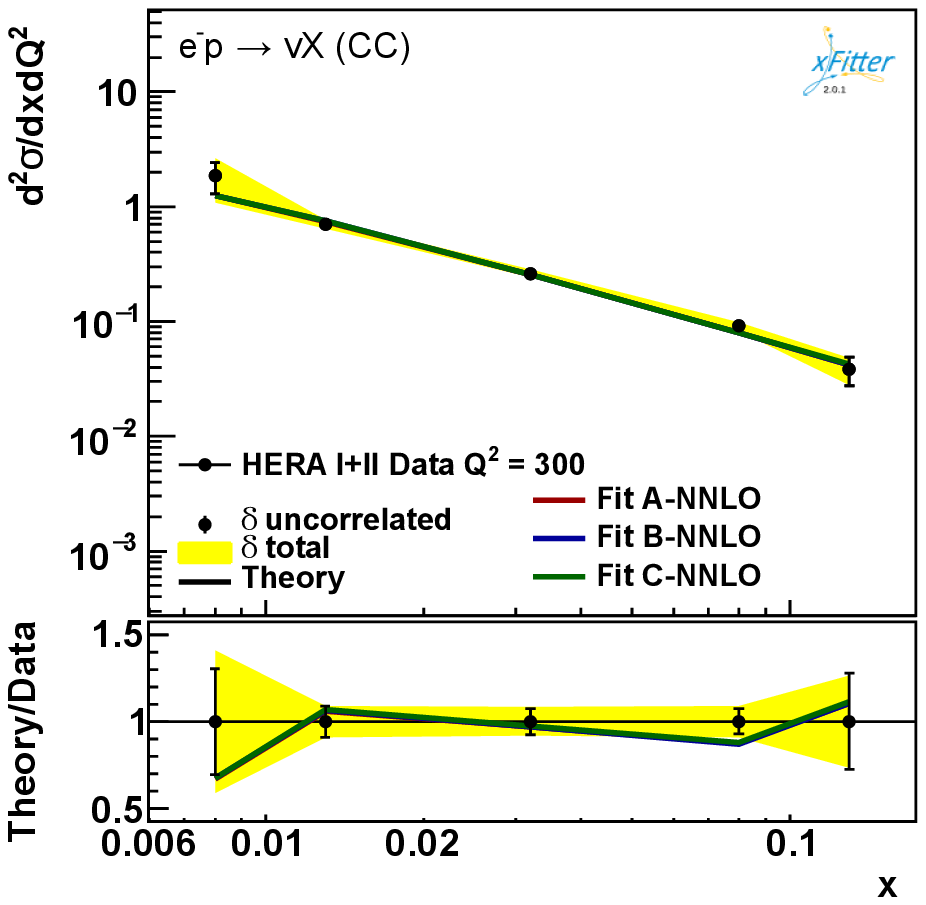}
        \includegraphics[scale = 0.4]{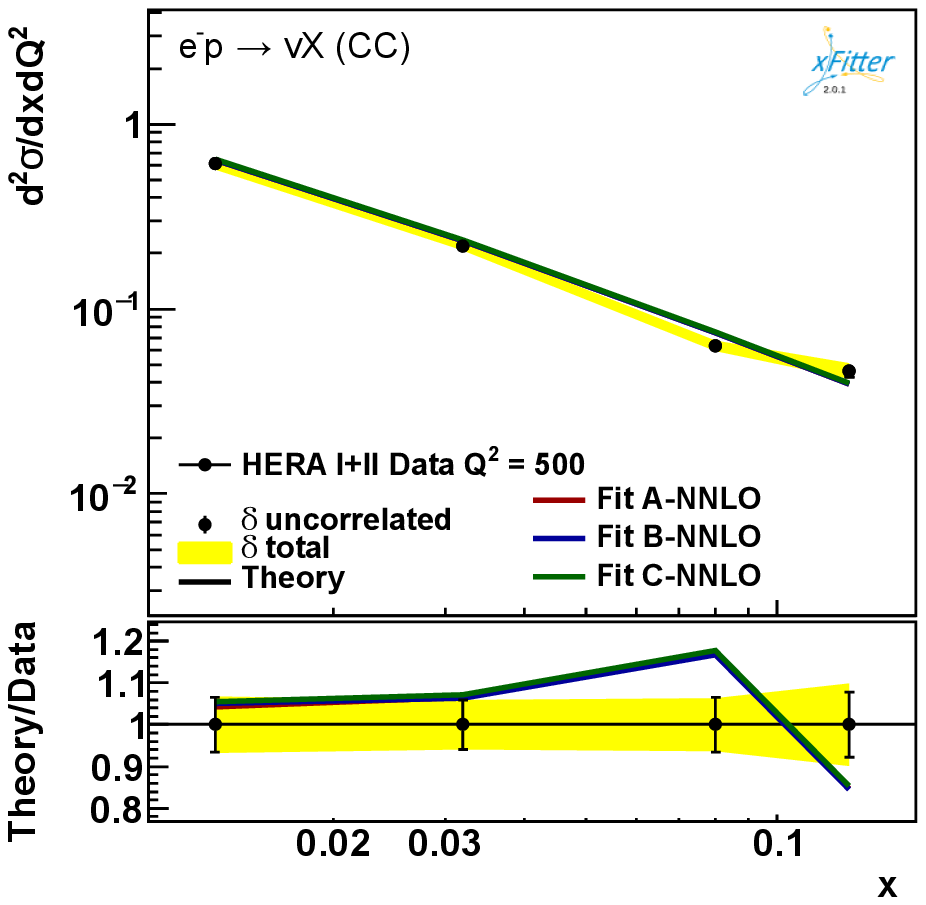}
        \includegraphics[scale = 0.4]{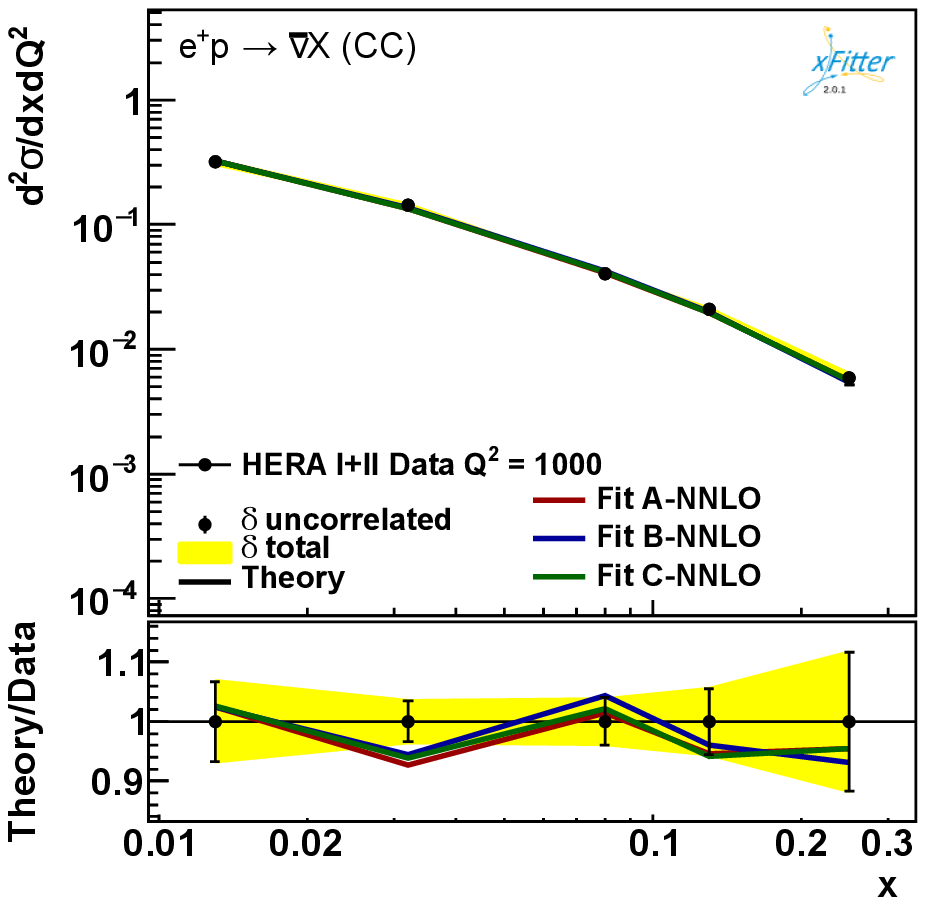}
		\includegraphics[scale = 0.4]{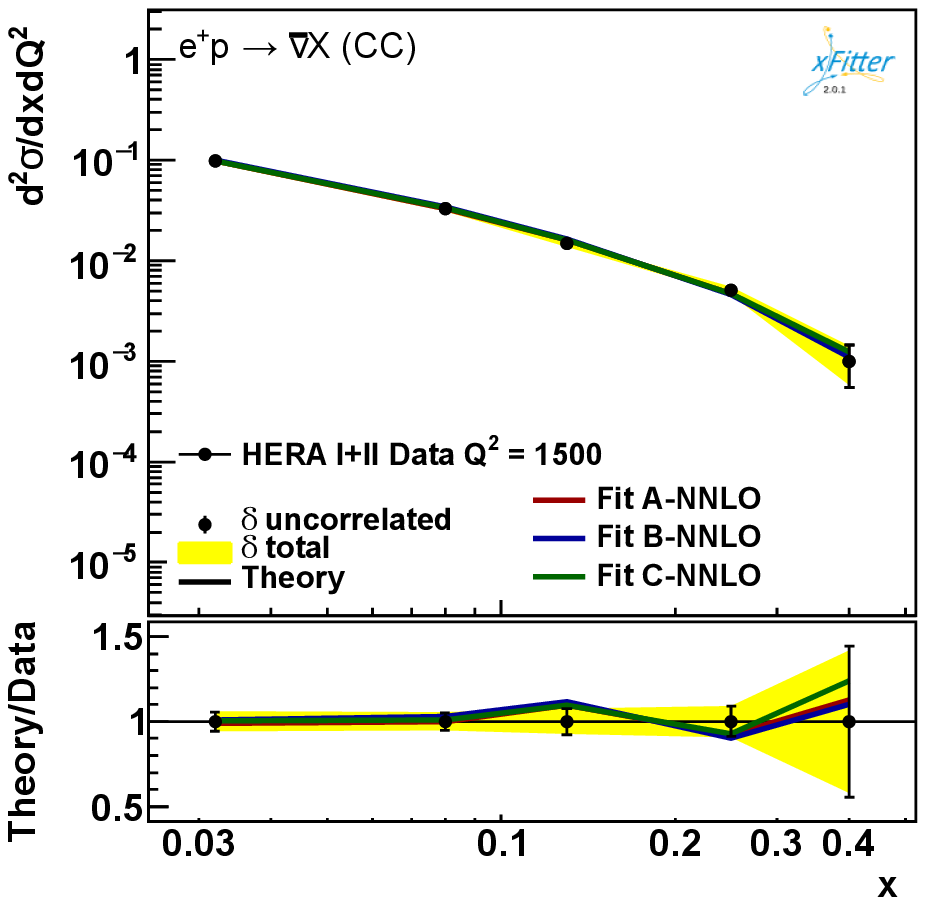}\\
		\includegraphics[scale = 0.4]{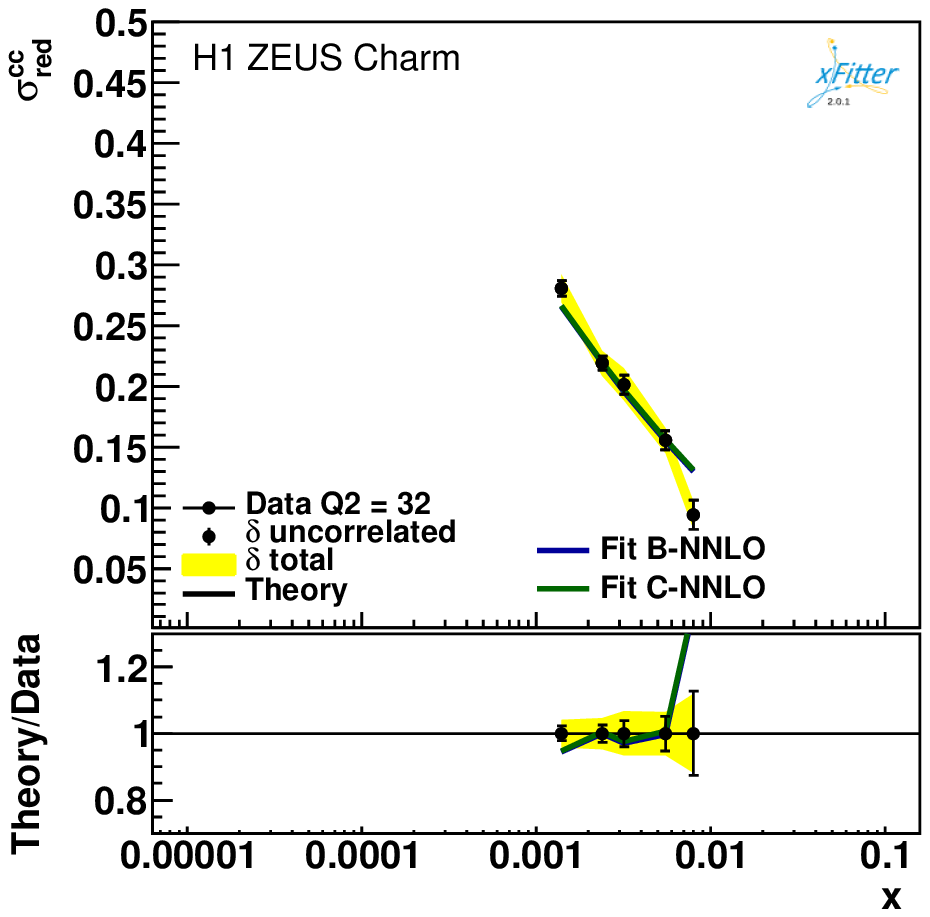}	
		\includegraphics[scale = 0.4]{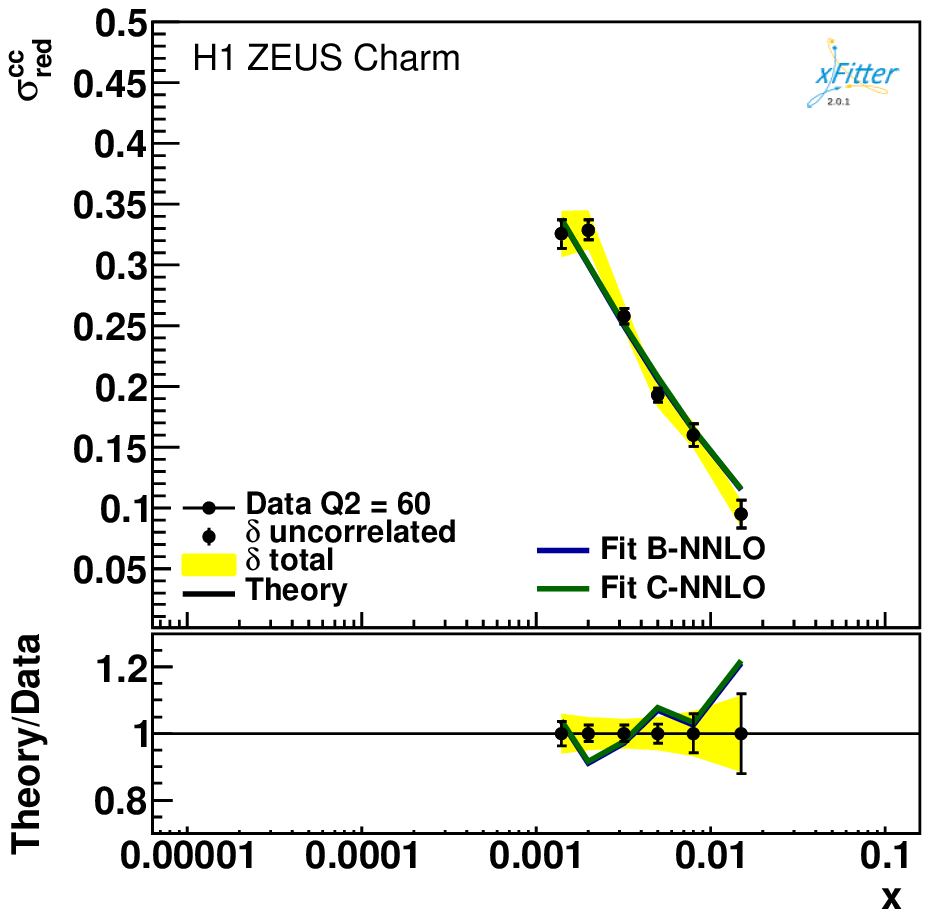}
		\includegraphics[scale = 0.4]{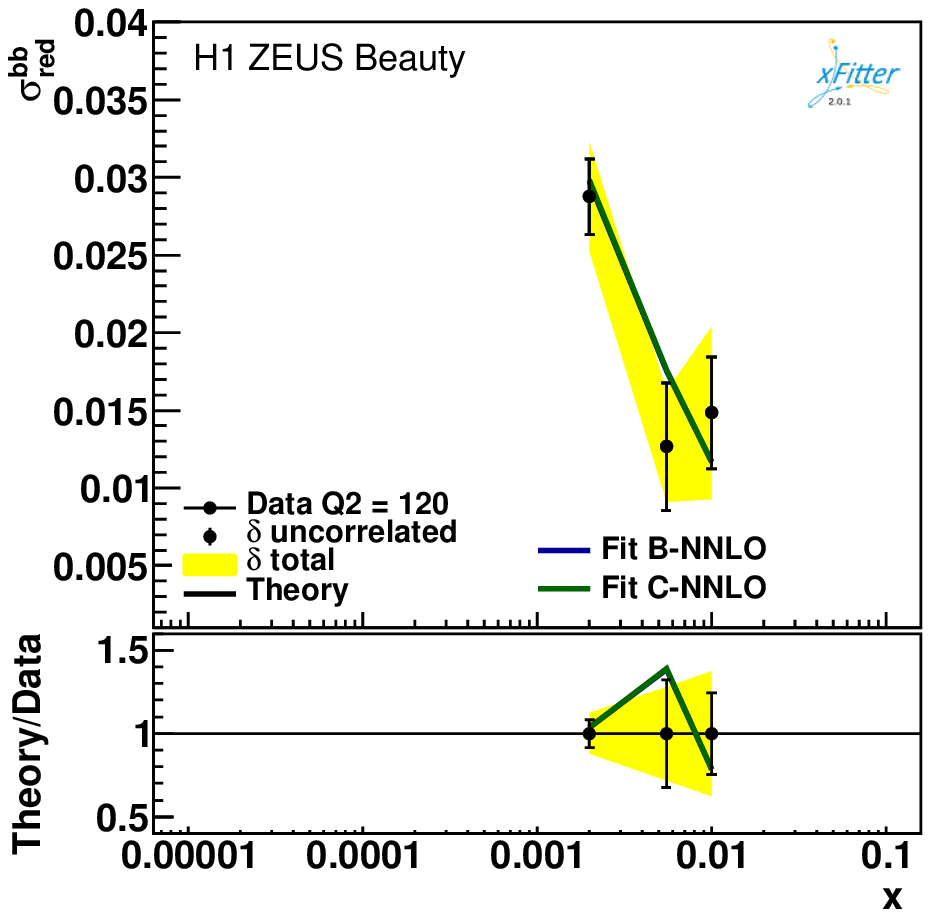}
		\includegraphics[scale = 0.4]{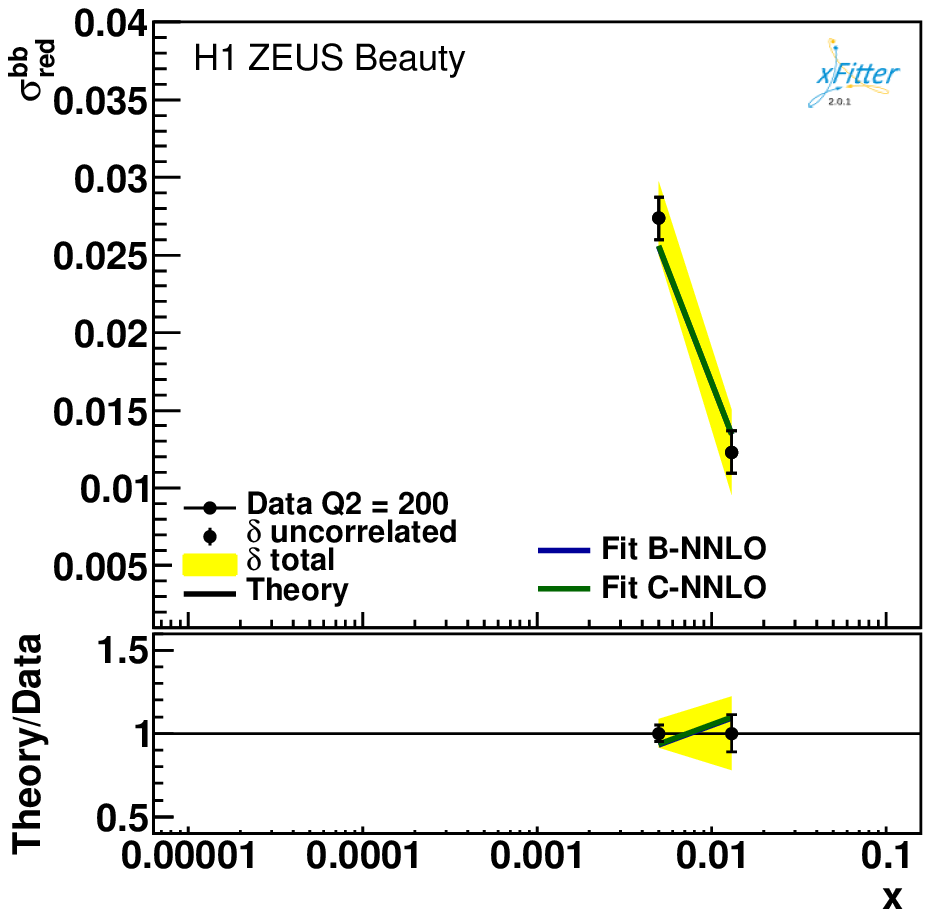} \\
		\includegraphics[scale = 0.4]{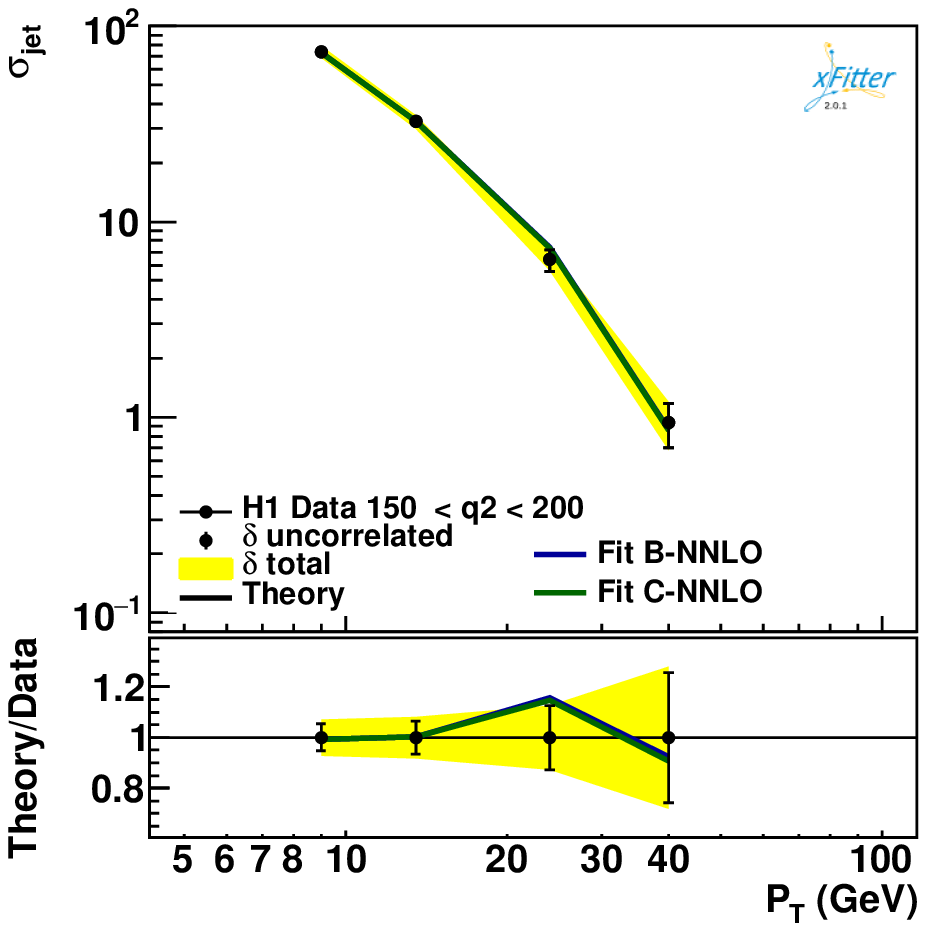}
		\includegraphics[scale = 0.4]{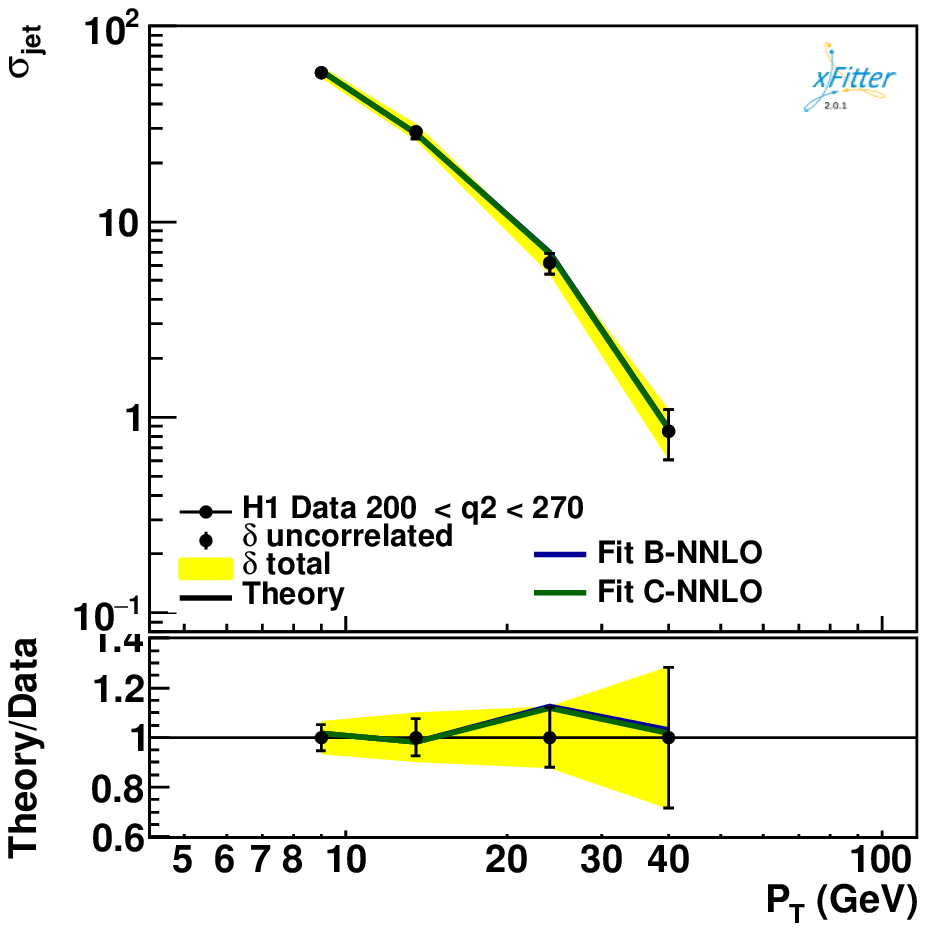}
		\includegraphics[scale = 0.4]{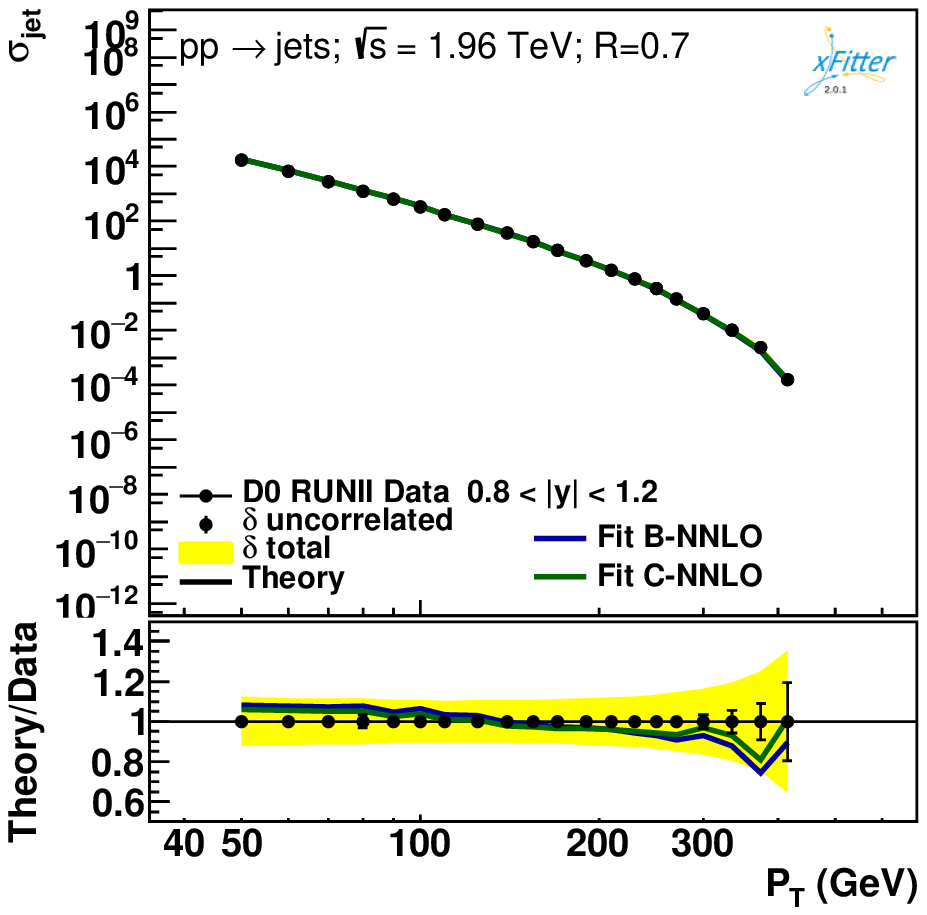}
		\includegraphics[scale = 0.4]{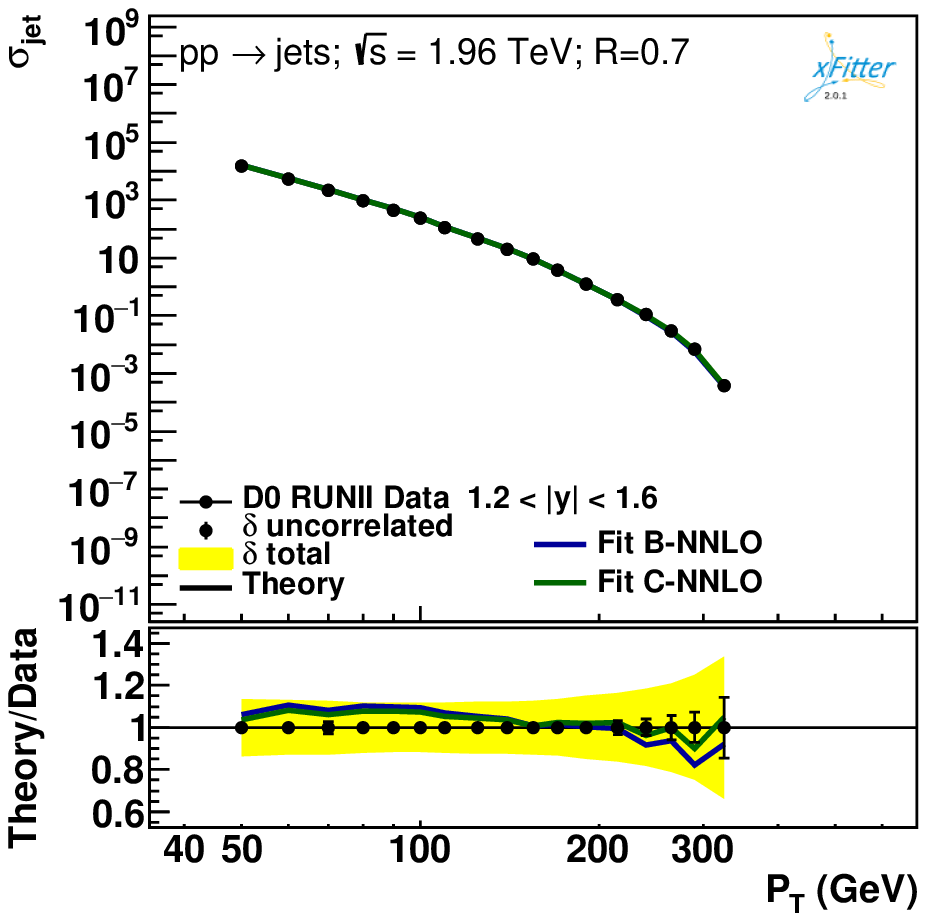}\\
		\includegraphics[scale = 0.4]{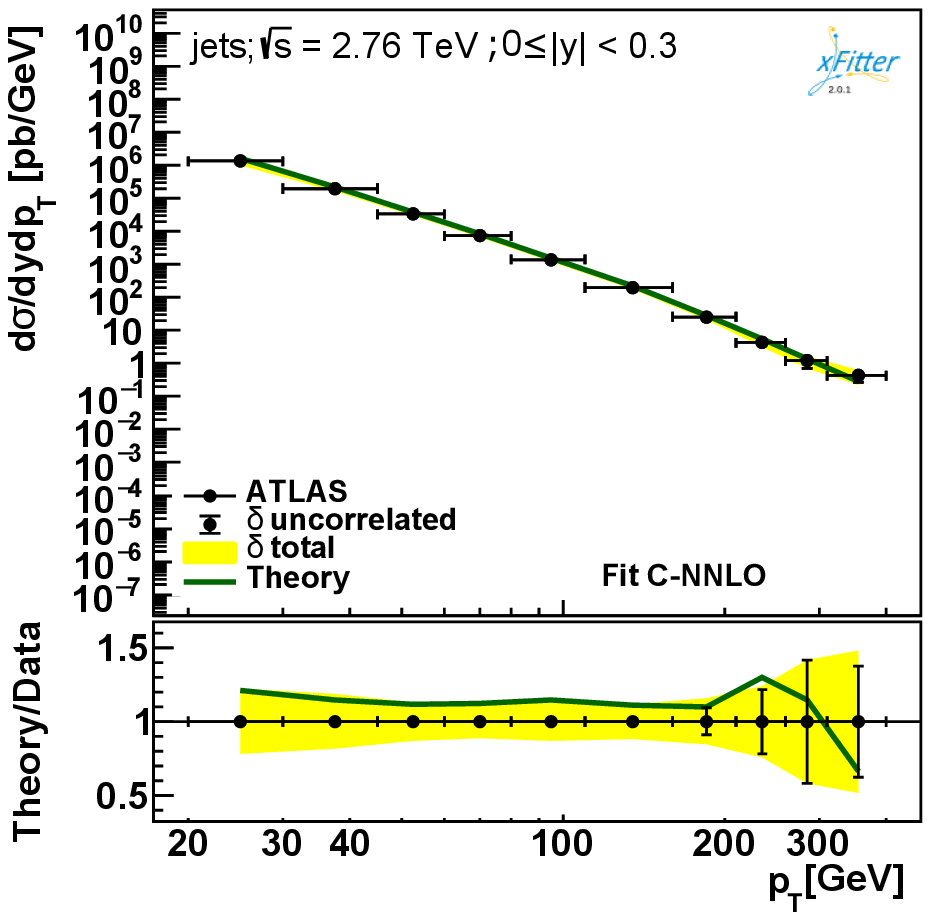}
		\includegraphics[scale = 0.4]{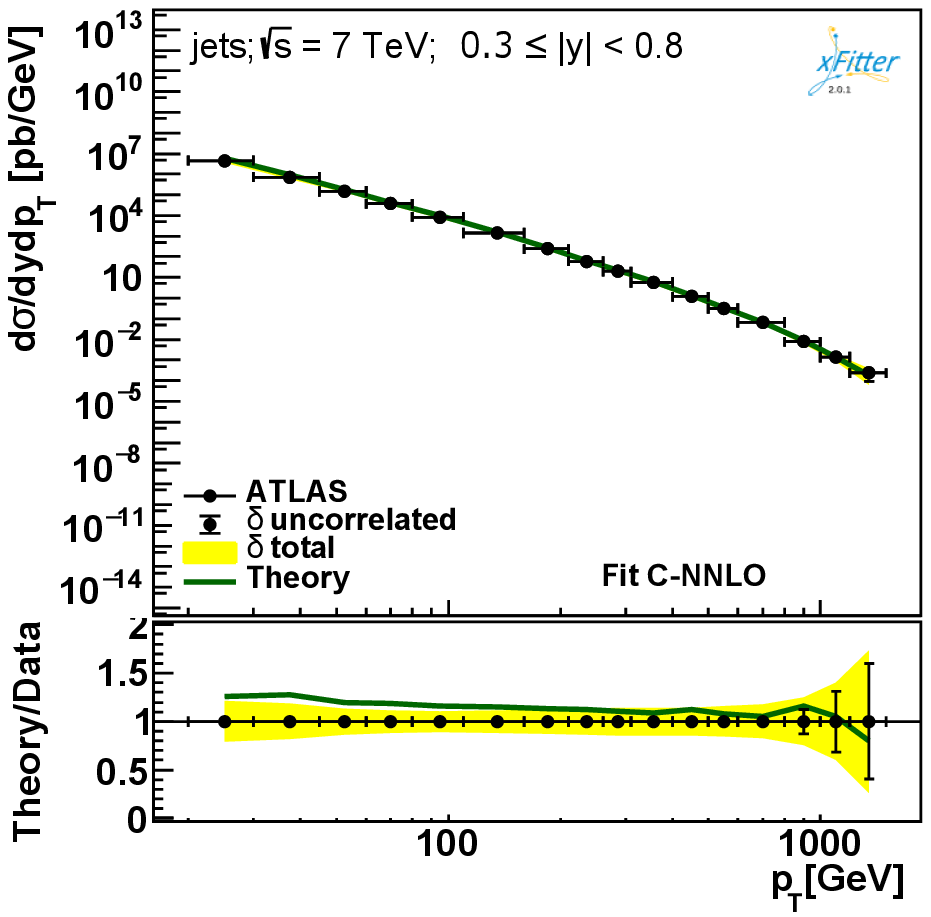}
		\includegraphics[scale = 0.4]{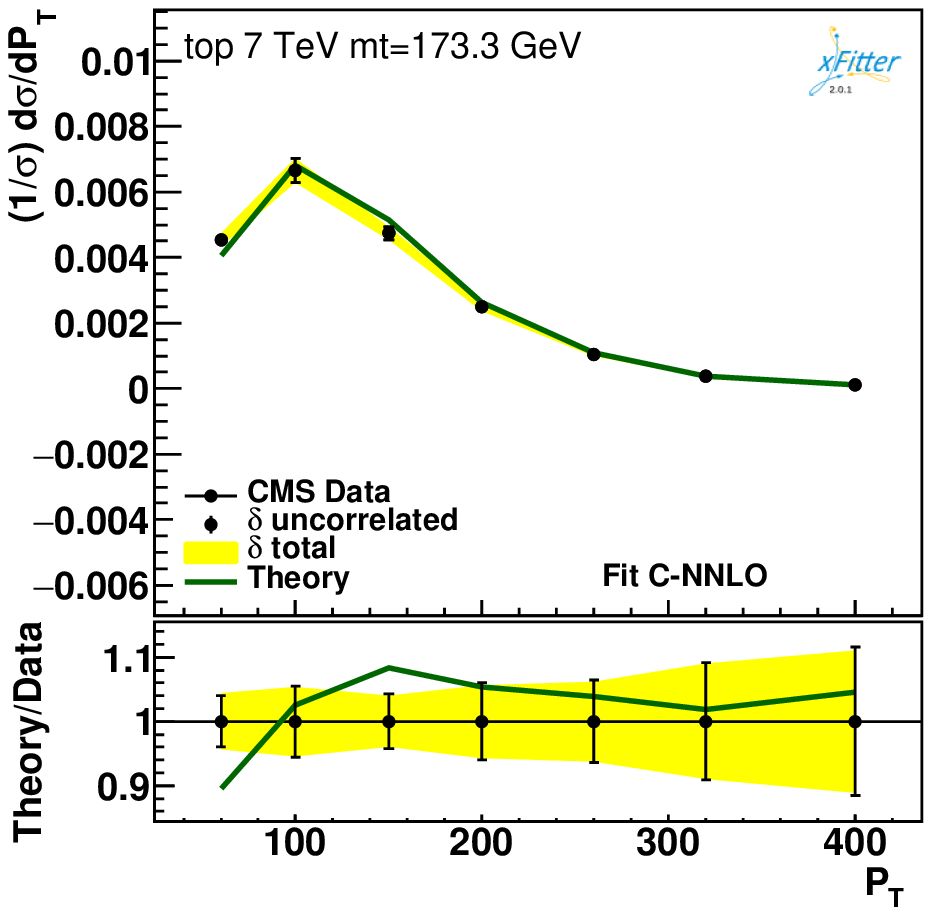}
		\includegraphics[scale = 0.4]{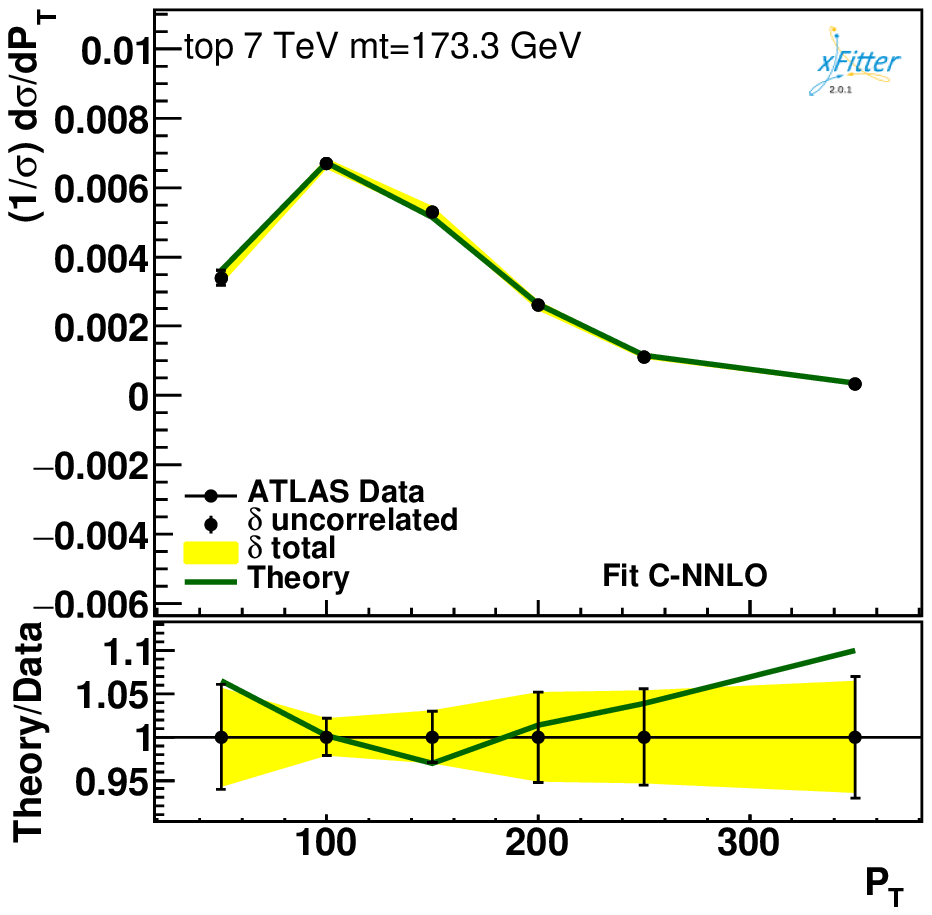}\\

		\caption{The results of pQCD  calculations for the cross sections of DIS processes, jet production and the differential cross section of top quark pair production, and their comparison with experimental measurements.}
		\label{fig:QCDfit1}
	\end{center}
\end{figure*}

Our main goal for considering three different fit procedures is investigating the specific impact of various types of cross sections experimental data, such as HERA I+II combined, non LHC and LHC data on PDFs and $\alpha_s(M^2_Z)$ simultaneously. In this regards, we need to
compare our results for our individual fits: Fits A, B, and C. In Fig.~\ref{fig:QCDfit1} some samples of our theoretical predictions  for various type of experimental data and their uncertainties at NNLO as a function of $x$ and $p_T$  are displayed.

In Table~\ref{tab:par}, the results  of numerical values for the PDF parameters which are described in section \ref{Comput} and also $\alpha_s(M^2_Z)$ extracted values  are summarized  for each fit procedures and for NLO and NNLO as well. 

\begin{table*}
	\caption{\label{tab:par} The numerical values and their uncertainties at the initial scale $Q_0^2=1.9$ GeV$^2$ extracted for parameters at NLO and NNLO related to Fits A, B and C.}
	\begin{center}
			\begin{tabular}{lcccccc}
				\hline
				\hline 
				\multicolumn{6}{r}{{\bf \hspace{. cm} Fit A \hspace{3.8 cm} Fit B \hspace{3.7 cm}  Fit C }}\\ 
								\cline{2-7}    
				Parameter~~~~~~~   & NLO & NNLO & NLO & NNLO & NLO & NNLO  \\     			
				\hline       
				\hline
				$B_{uv}$ & $0.733 \pm 0.015$& $0.731 \pm 0.034$& $0.625 \pm 0.016$& $0.610 \pm 0.016$&  $0.644 \pm 0.015$& $0.631 \pm 0.014$  \\ 
				$C_{uv}$ & $4.747 \pm 0.047$& $4.898 \pm 0.077$& $4.887 \pm 0.079$& $4.974 \pm 0.072$& $4.849 \pm 0.073$& $4.982 \pm 0.077$  \\   
				$E_{uv}$ & $10.10 \pm 0.14$& $11.3 \pm 1.5$& $15.1 \pm 1.3$& $17.9 \pm 1.4$& $13.3 \pm 1.2$& $15.9 \pm 1.1$  \\ 
				$B_{dv}$ & $0.838 \pm 0.040$& $0.869 \pm 0.085$& $0.765 \pm 0.061$& $0.704 \pm 0.066$&  $0.796 \pm 0.062$& $0.663 \pm 0.060$  \\ 
				$C_{dv}$ & $4.50 \pm 0.14$& $4.60 \pm 0.41$& $5.08 \pm 0.37$& $4.59 \pm 0.38$& $5.06 \pm 0.36$& $4.13 \pm 0.35$  \\ 
				$C_{\bar{U}}$ & $4.14 \pm 0.13$& $5.01 \pm 0.69$& $2.04 \pm 0.11$& $2.065 \pm 0.078$& $1.954 \pm 0.061$ & $1.972 \pm 0.047$  \\ 
				$A_{\bar{D}}$ & $0.2045 \pm 0.0098$& $0.231 \pm 0.014$& $0.1661 \pm 0.0079$& $0.1784 \pm 0.0069$& $0.1636 \pm 0.0070$& $0.1758 \pm 0.0065$  \\
				$B_{\bar{D}}$ & $-0.1293 \pm 0.0079$&  $-0.1271 \pm 0.0091$& $-0.1582 \pm 0.0077$& $-0.1597 \pm 0.0061$& $-0.1606 \pm 0.0070$& $-0.1624 \pm 0.0058$  \\
				$C_{\bar{D}}$  & $6.13 \pm 0.14$& $7.7 \pm 1.6$& $4.96 \pm 0.75$& $4.77 \pm 0.83$&$7.4 \pm 1.2$& $5.02 \pm 0.80$ \\   
				$f_s$ & $ 0.31 $~(Fixed)& $0.31 $~(Fixed)& $0.31 $~(Fixed)& $ 0.31 $~(Fixed)& $0.31 $~(Fixed)& $0.31 $~(Fixed) \\ 
				$B_g$ & $-0.345 \pm 0.060$& $-0.38 \pm 0.13$& $-0.696 \pm 0.055$& $-0.239 \pm 0.033$& $-0.680 \pm 0.046$& $-0.249 \pm 0.035$  \\   
				$C_g$ & $5.94 \pm 0.16$& $4.9 \pm 1.4$& $2.05 \pm 0.30$& $3.57 \pm 0.26$& $2.27 \pm 0.21$& $4.06 \pm 0.20$  \\
				$A'_g$ & $0.790 \pm 0.066$& $0.34 \pm 0.20$& $0.345 \pm 0.036$& $0.096 \pm 0.021$& $0.379 \pm 0.030$& $0.142 \pm 0.028$  \\ 
				$B'_g$ & $-0.419 \pm 0.048$& $-0.526 \pm 0.048$& $-0.709 \pm 0.047$& $-0.582 \pm 0.019$& $-0.696 \pm 0.039$& $-0.546 \pm 0.017$  \\
				$C'_g$ & $25.00$~(Fixed)& $25.00$~(Fixed)& $25.00$~(Fixed)& $25.00$~(Fixed)& $25.00$~(Fixed)& $25.00$~(Fixed)  \\ 

				$\alpha_s(M^2_Z)$  & $0.1167 \pm 0.0014$& $0.1134 \pm 0.0024$& $0.1198 \pm 0.0008$& $0.1187 \pm 0.0007$& $0.1188 \pm 0.0007$& $0.1179 \pm 0.0006$  \\
				\hline
				\hline   
			\end{tabular}
	\end{center}
	
\end{table*}

The extracted value of $\alpha_s(M^2_Z)$  from Fit A and Fit B at the NNLO are $0.1134\pm0.0024$ and $0.1187\pm0.0007$, respectively. These numerical values illustrate how non-LHC experimental data  impact on $\alpha_s(M^2_Z)$ when we compare it to  the world average value of
$\alpha_{s}(M_{Z}^{2})=0.1179\pm0.0085$~\cite{ParticleDataGroup:2020ssz}. This is a very reasonable result based on the fact that jet cross section data impact on $\alpha_{s}(M_{Z}^{2})$ significantly. This finding was already reported
in some previous QCD analysis. For more details, we refer the reader to the Ref. \cite{Abramowicz:2015mha}. According to Table~\ref{tab:par}  and for Fit C, by including LHC data we obtained $\alpha_s(M^2_Z)=0.1179\pm0.0006$  at the NNLO. It seems that LHC data in Table~\ref{tab:Chi2}  changes not only the central value of $\alpha_s(M^2_Z)$ but also its uncertainties in comparison to Fit A which we used the  HERA I+II combined data only.

\begin{figure*}[!htb]
	\begin{center}
		\includegraphics[scale = 0.4]{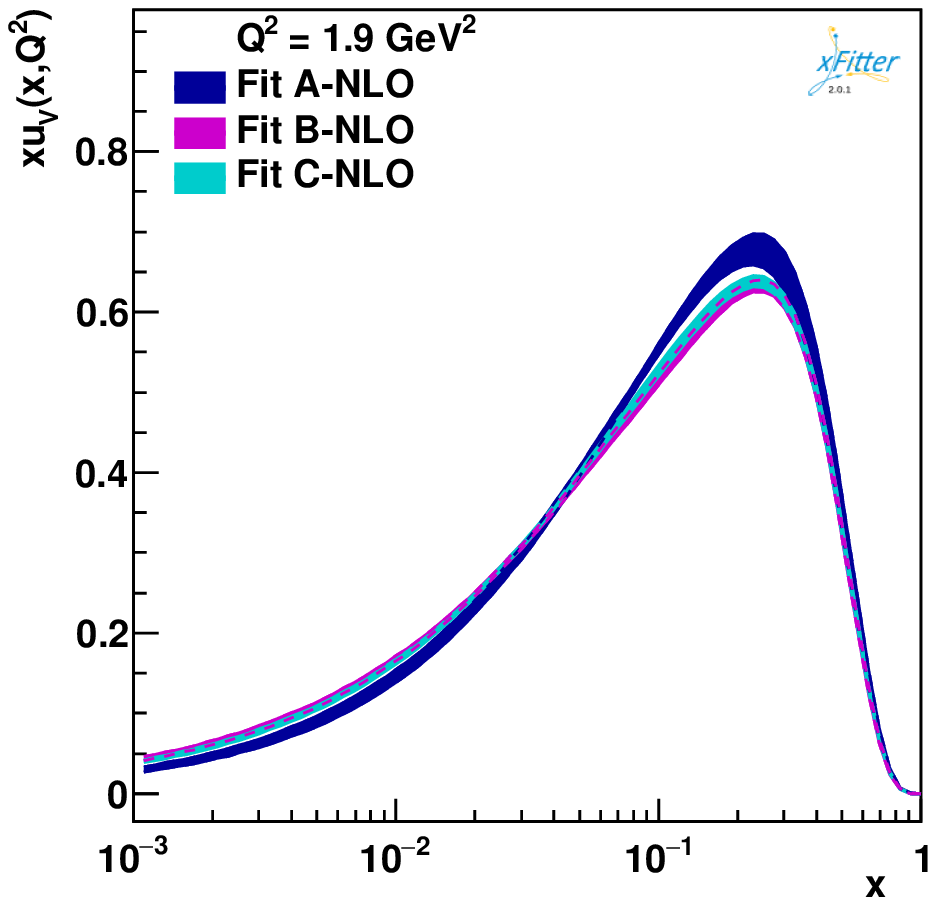}
		\includegraphics[scale = 0.4]{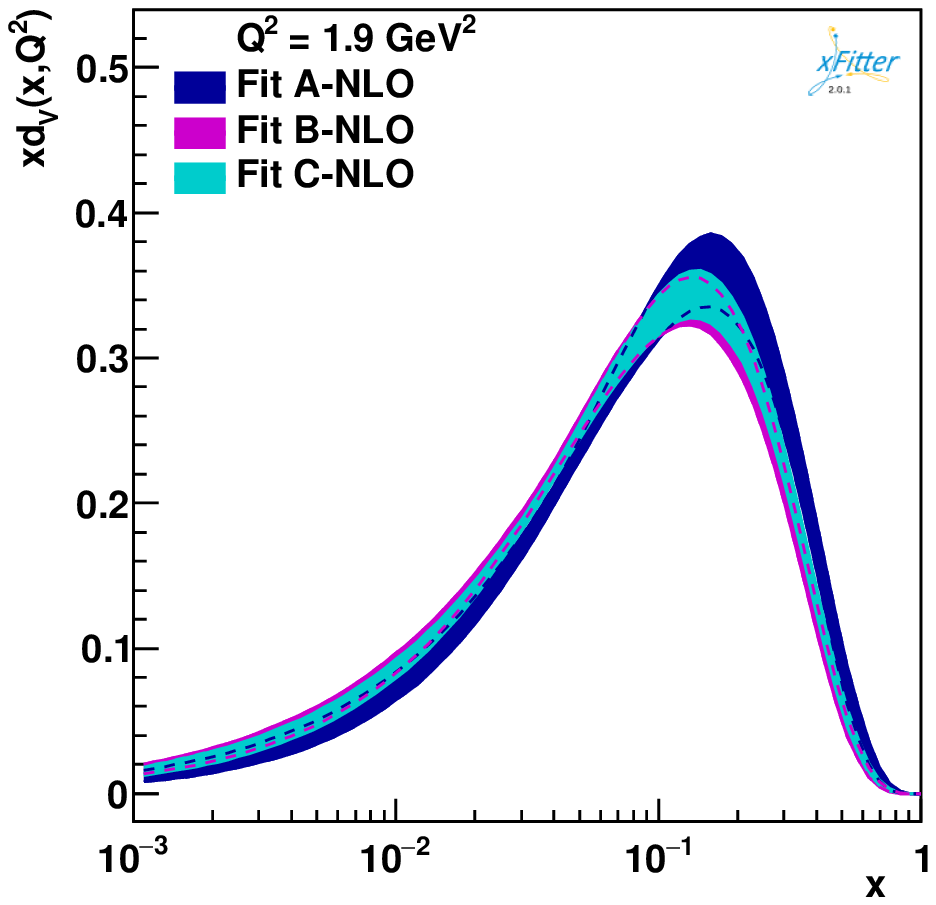}
		\includegraphics[scale = 0.4]{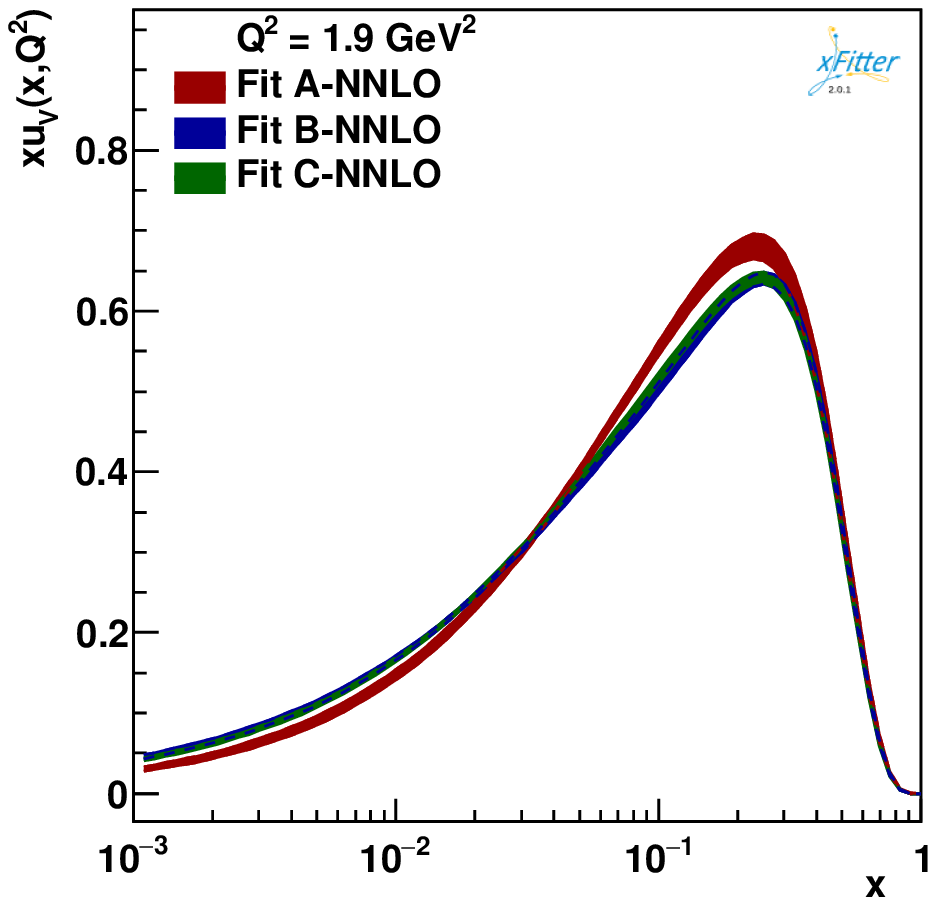}					    
		\includegraphics[scale = 0.4]{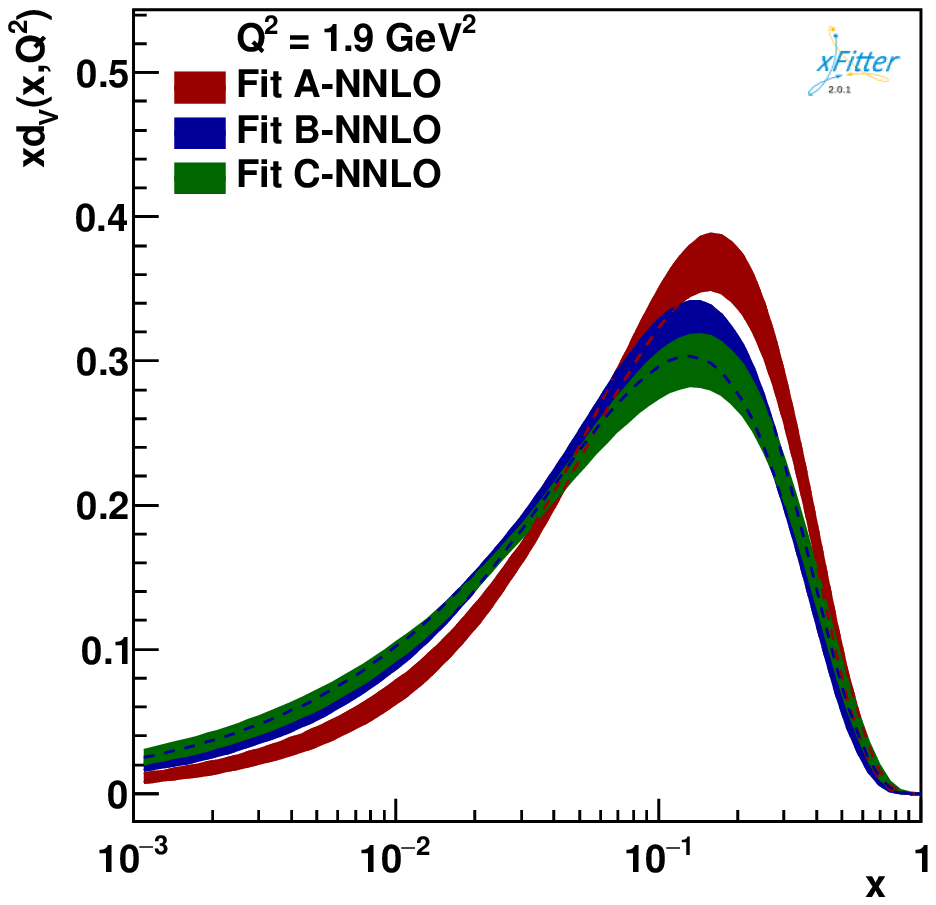}
			    		
		\includegraphics[scale = 0.4]{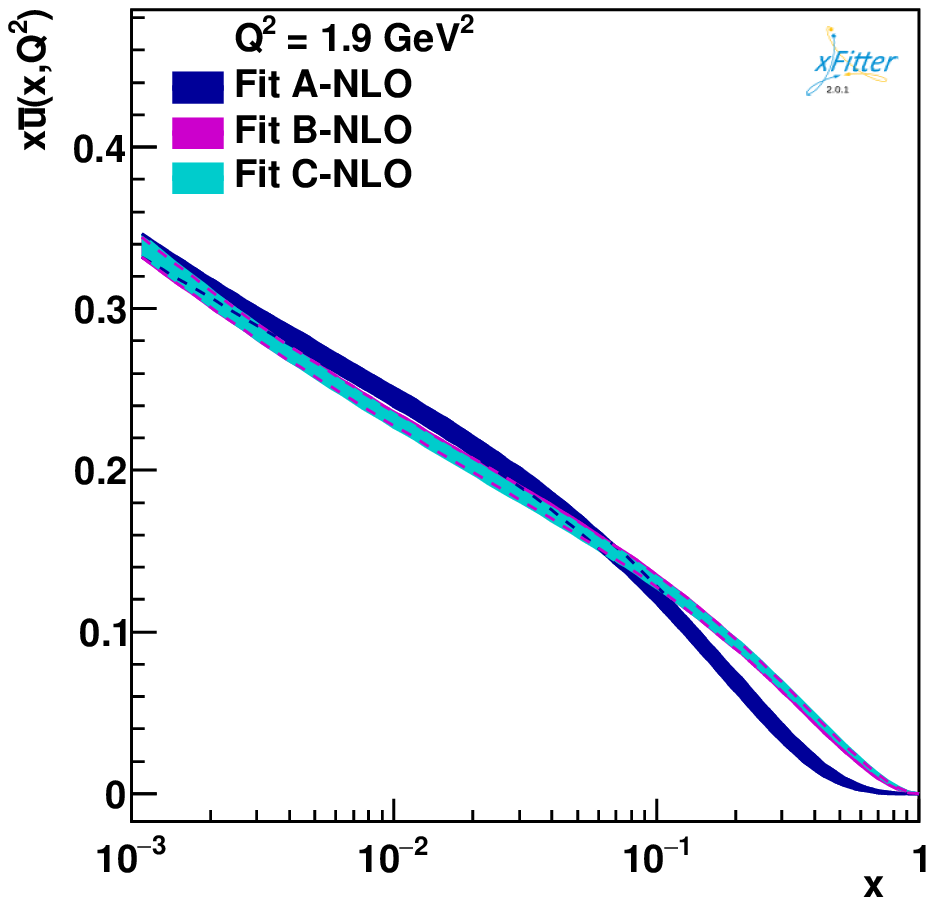}
		\includegraphics[scale = 0.4]{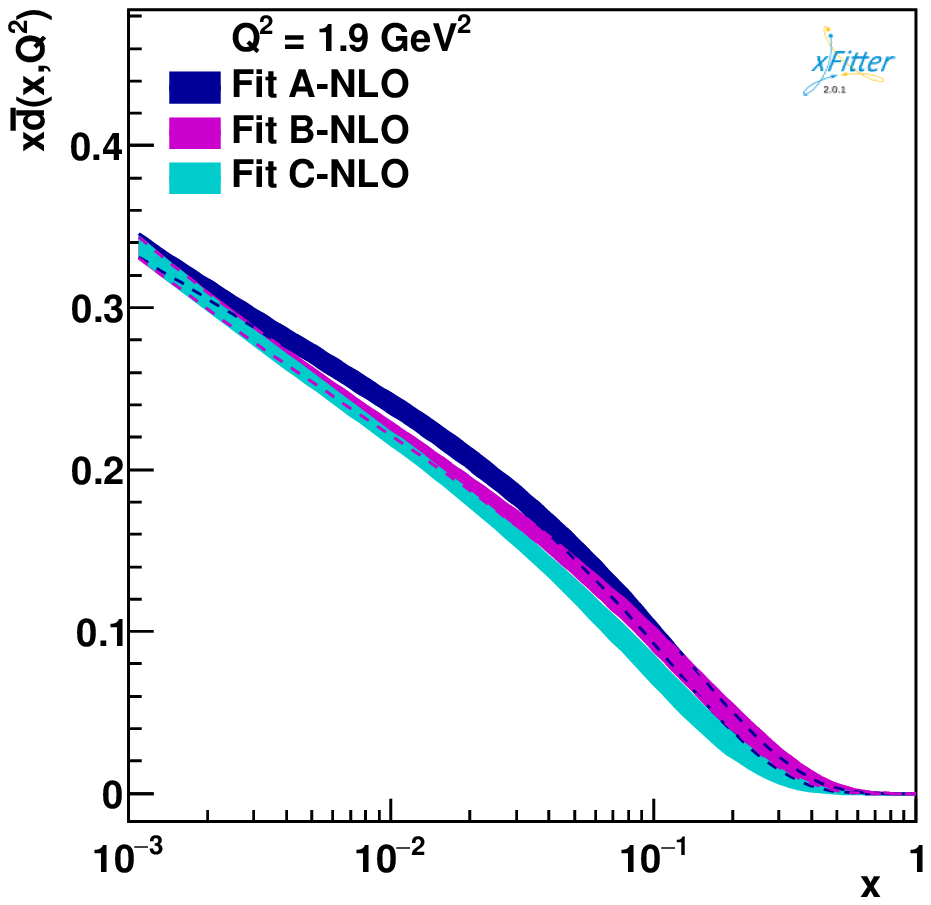}
		\includegraphics[scale = 0.4]{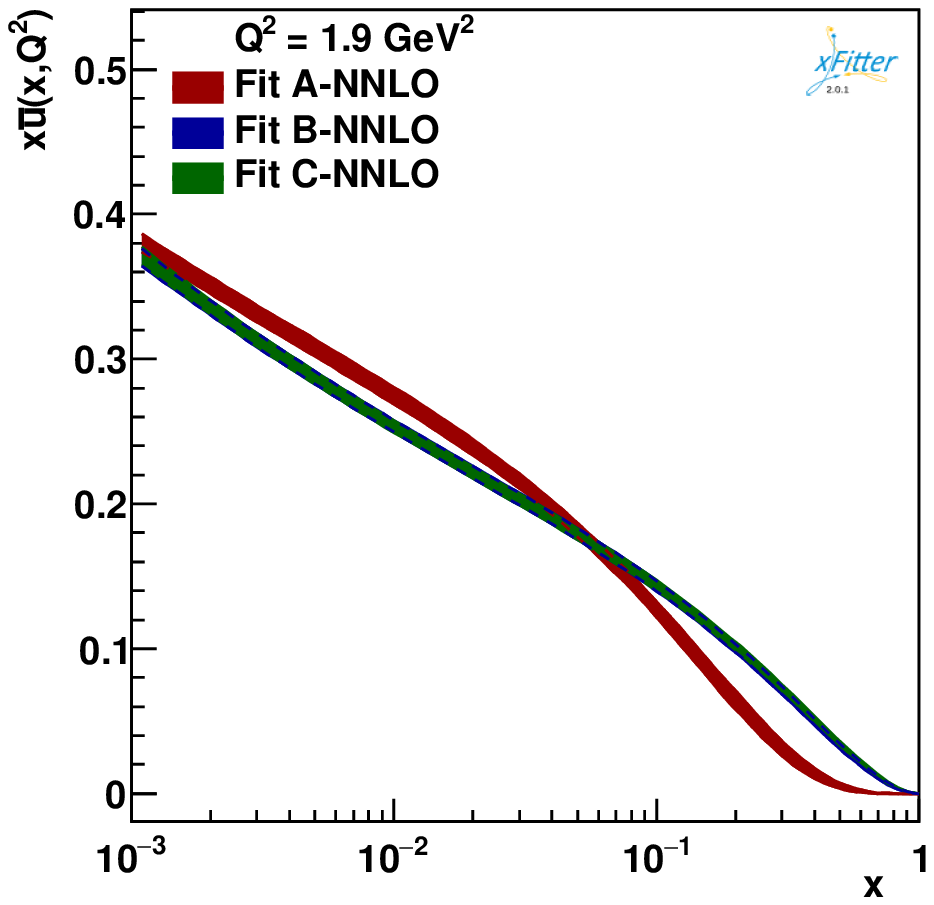}		    	
		\includegraphics[scale = 0.4]{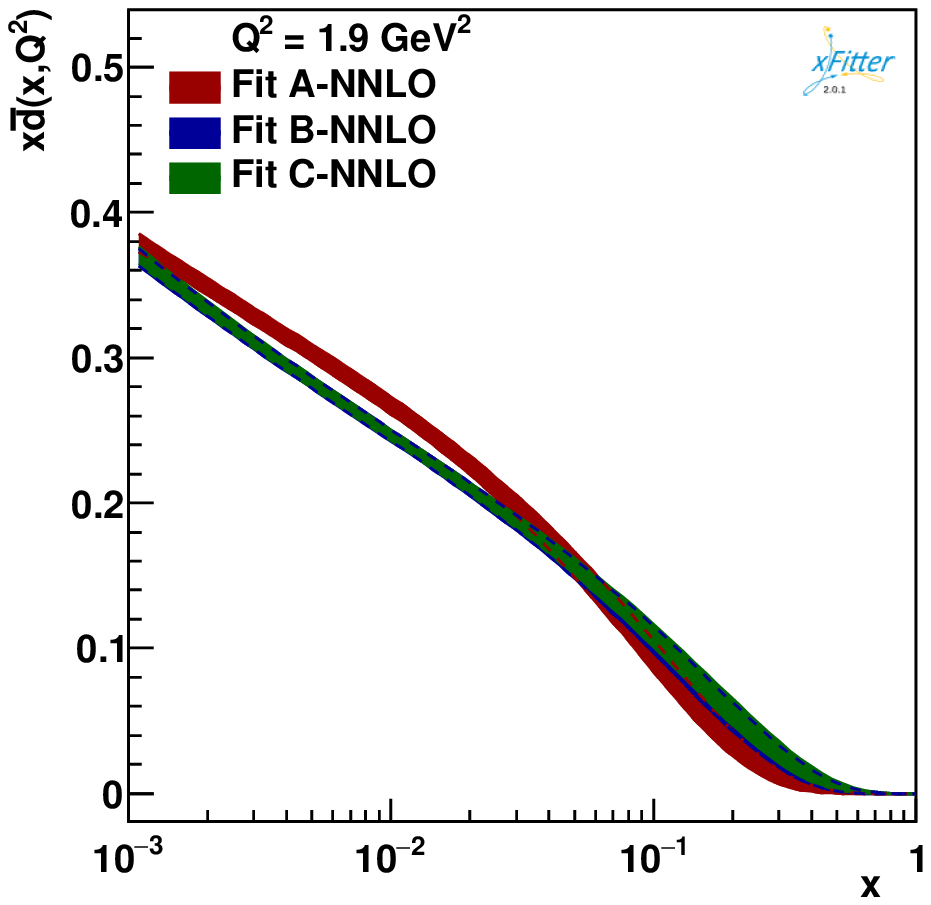}
			        
		\includegraphics[scale = 0.4]{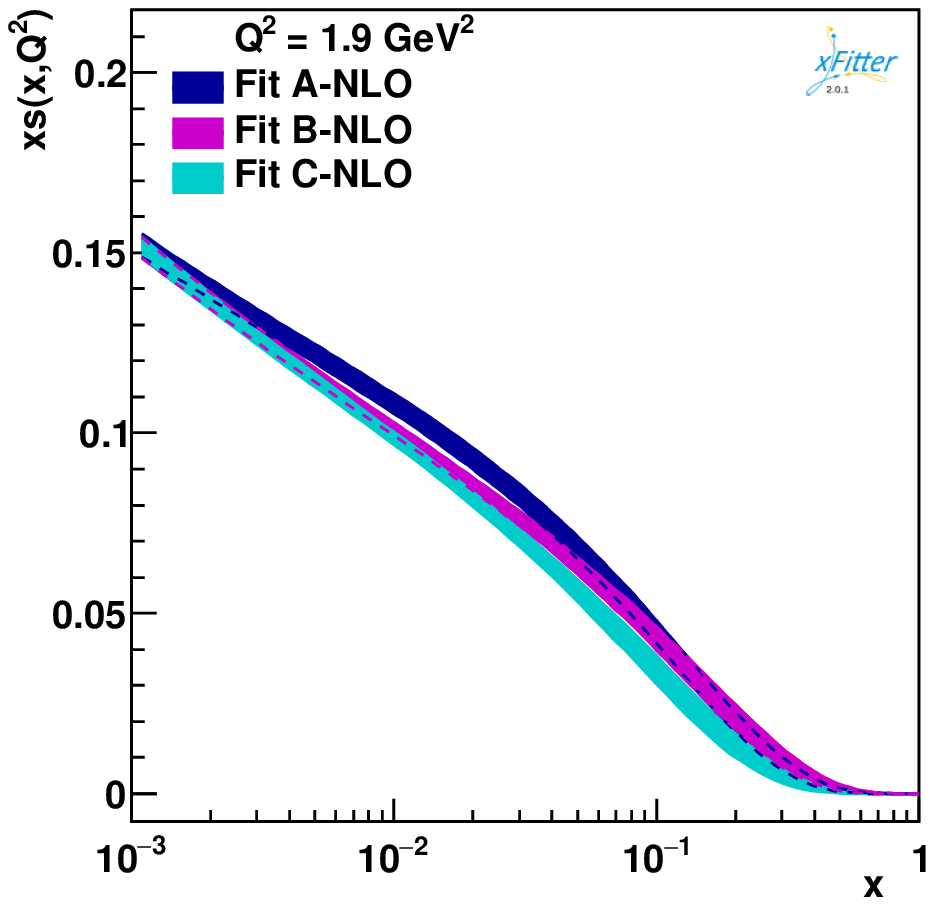}  
		\includegraphics[scale = 0.4]{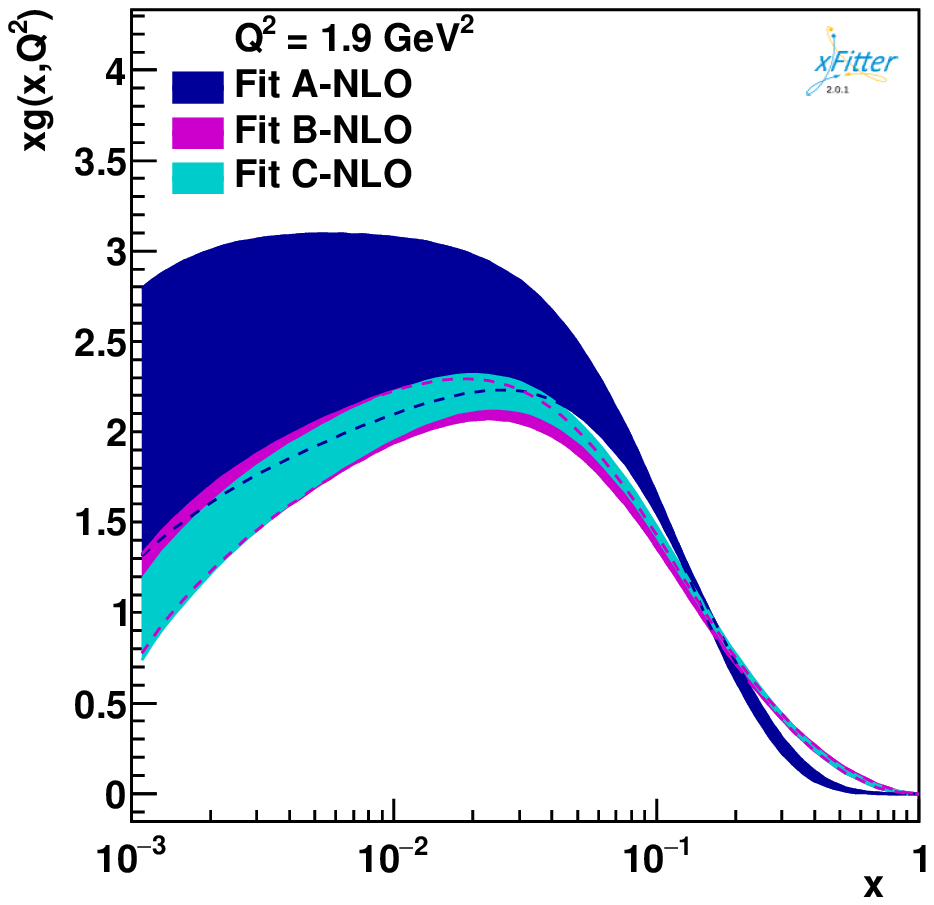} 
		\includegraphics[scale = 0.4]{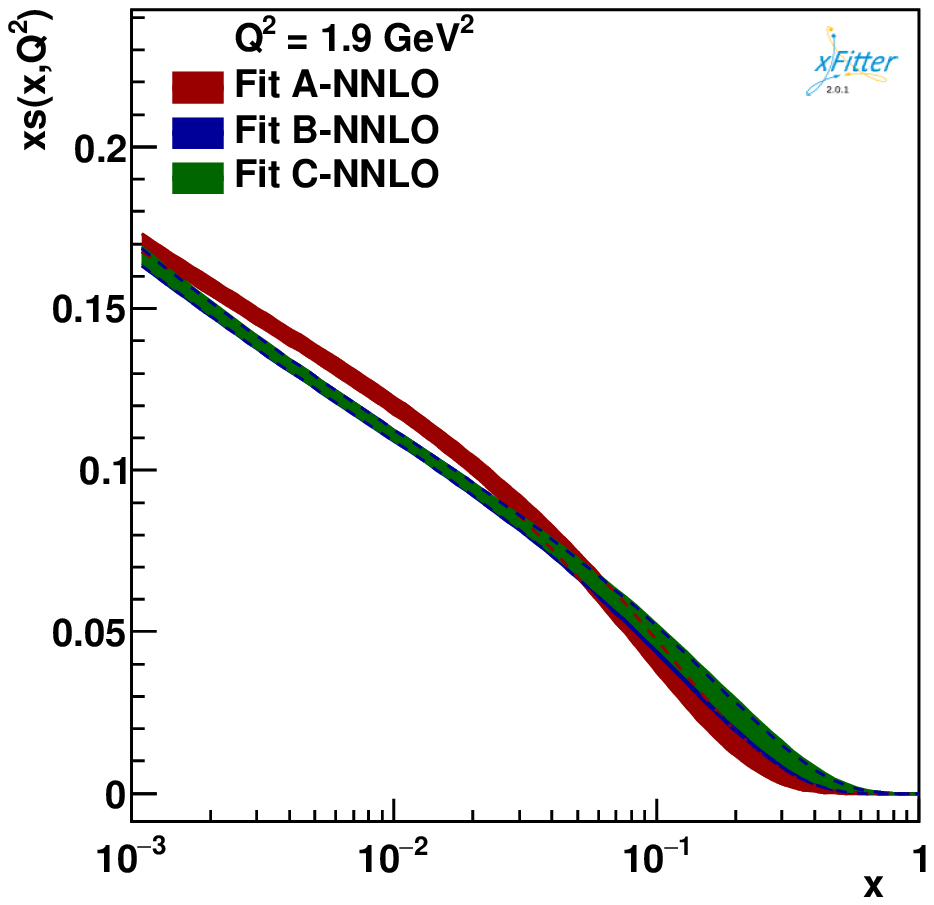} 			          
	          \includegraphics[scale = 0.4]{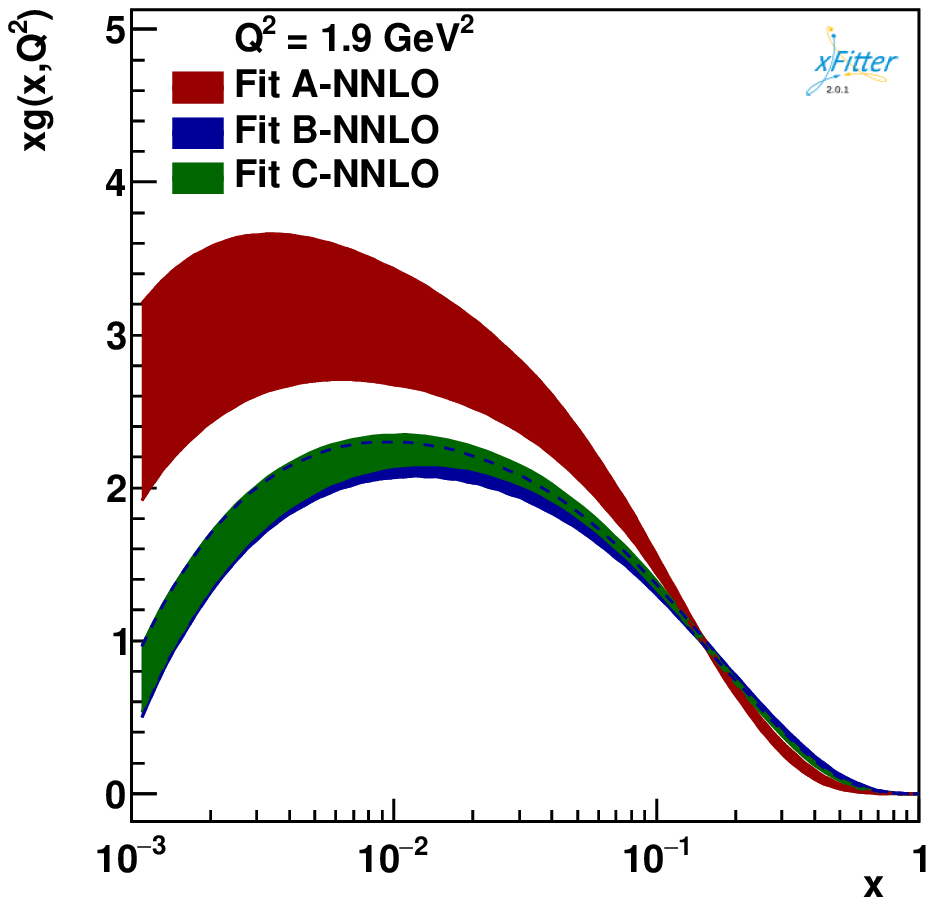}     
		
		\caption{The NLO and NNLO parton distribution of $xu_v$, $xd_v$, $x\bar{u}$, $x\bar{d}$, $xs$, 
		and $xg$ for Fit~A, Fit~B, and Fit~C extracted as a function of $x$ at $Q^2$=1.9 GeV$^2$. In the left panels, we present our results for NLO, whereas the right panels are for NNLO.}
		
		\label{fig:PDF-1.9-NLO+NNLO}
	\end{center}
\end{figure*}

 Figure. \ref{fig:PDF-1.9-NLO+NNLO} illustrates the NLO and NNLO QCD fit results for valence, sea and gluon PDFs as a function of $x$ at $Q^2$=1.9 GeV$^2$
for Fit~A, Fit~B, and Fit~C. In the left panels, we present our results for NLO, whereas the right panels are for NNLO. Although there are no significant changes in some PDFs in all fits at NLO (left panels), significant changes
in the central value of gluon PDF, their uncertainties, or both are observed from Fit~A to Fit~C. Such behaviors are also observed in NNLO from Fit~A to Fit~C (right panels).

\begin{figure*}[!htb]
	\begin{center}
		\includegraphics[scale = 0.4]{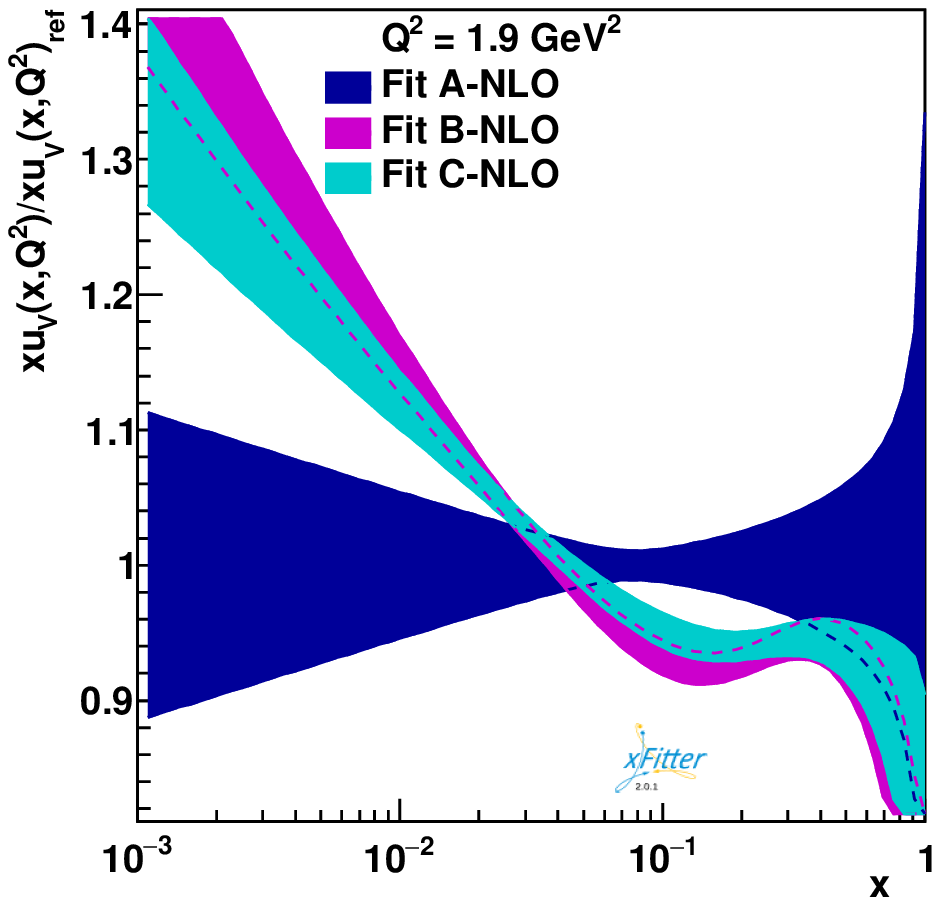}
		\includegraphics[scale = 0.4]{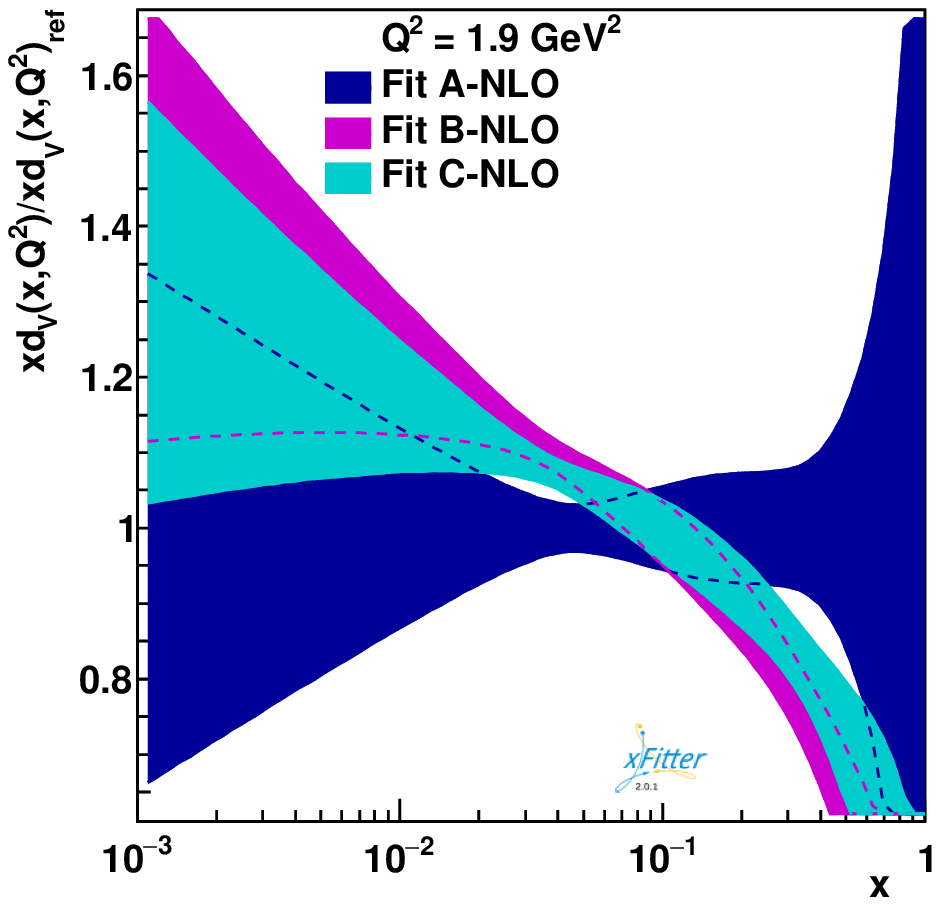}			
		\includegraphics[scale = 0.4]{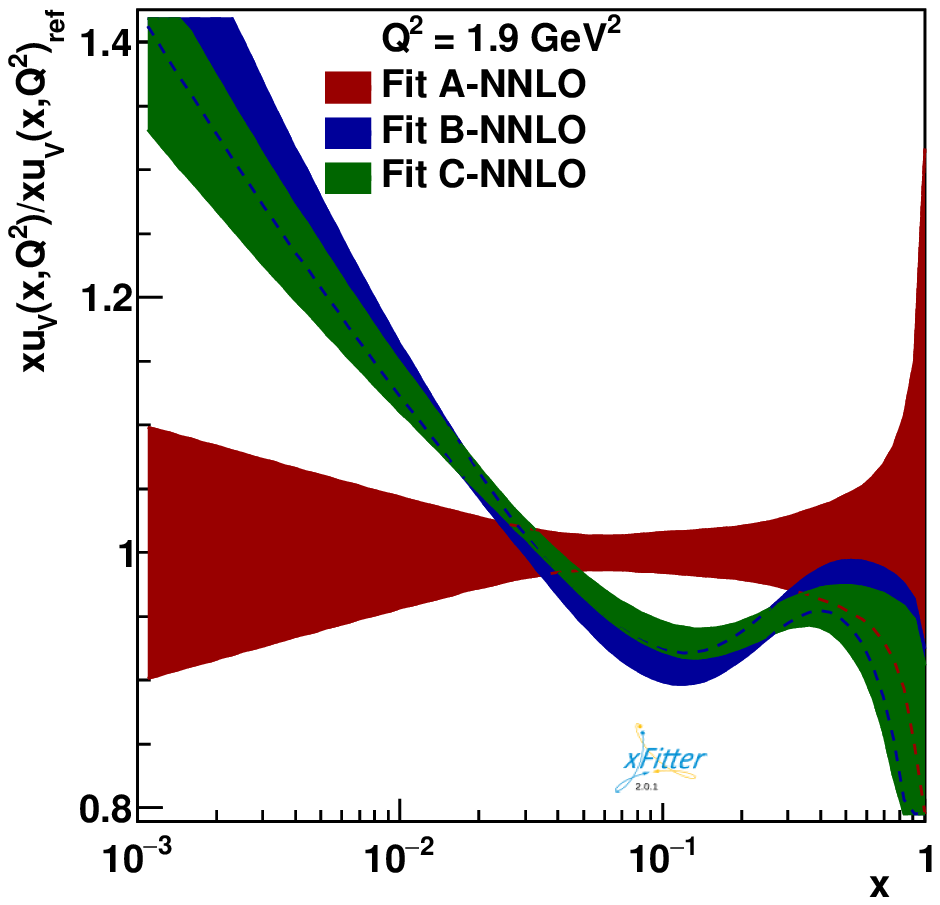}
		\includegraphics[scale = 0.4]{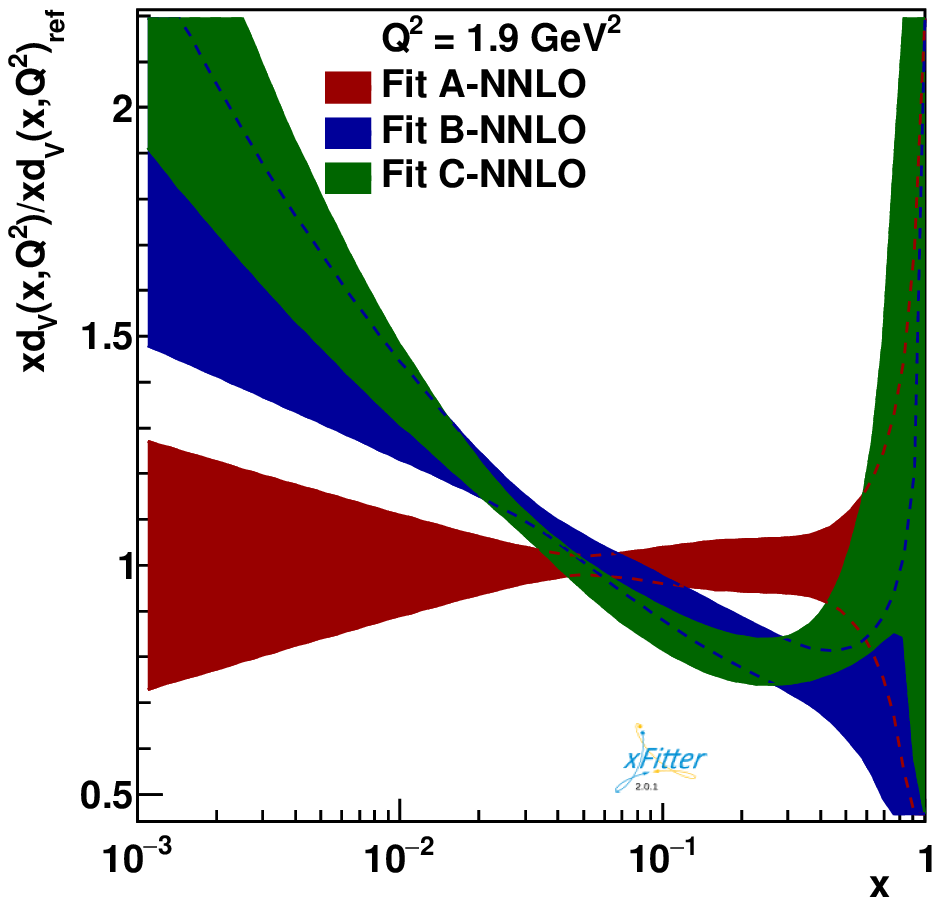}	
			    		
		\includegraphics[scale = 0.4]{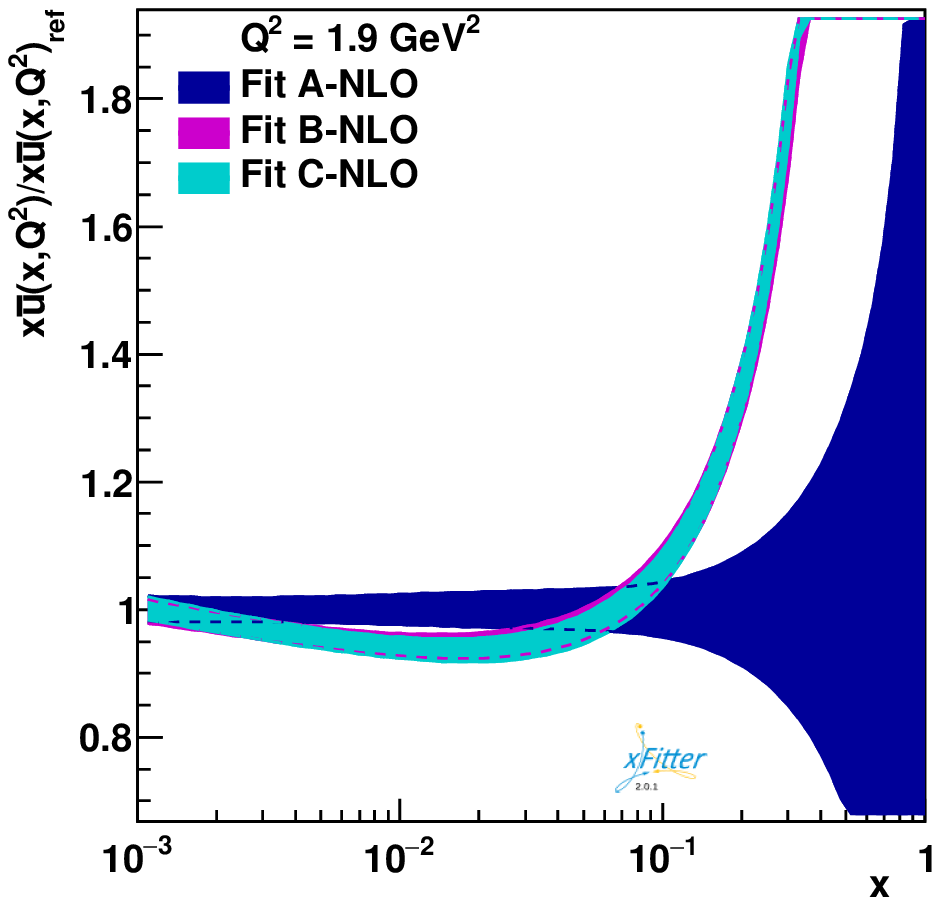}
		\includegraphics[scale = 0.4]{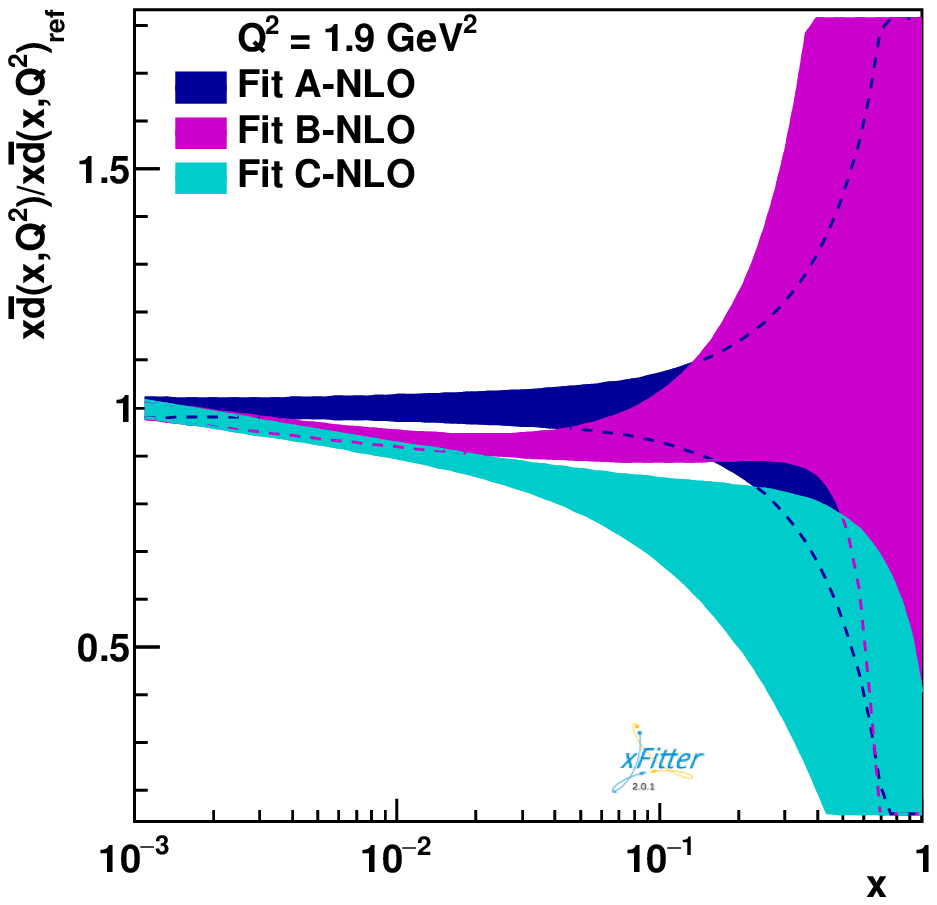}
		\includegraphics[scale = 0.4]{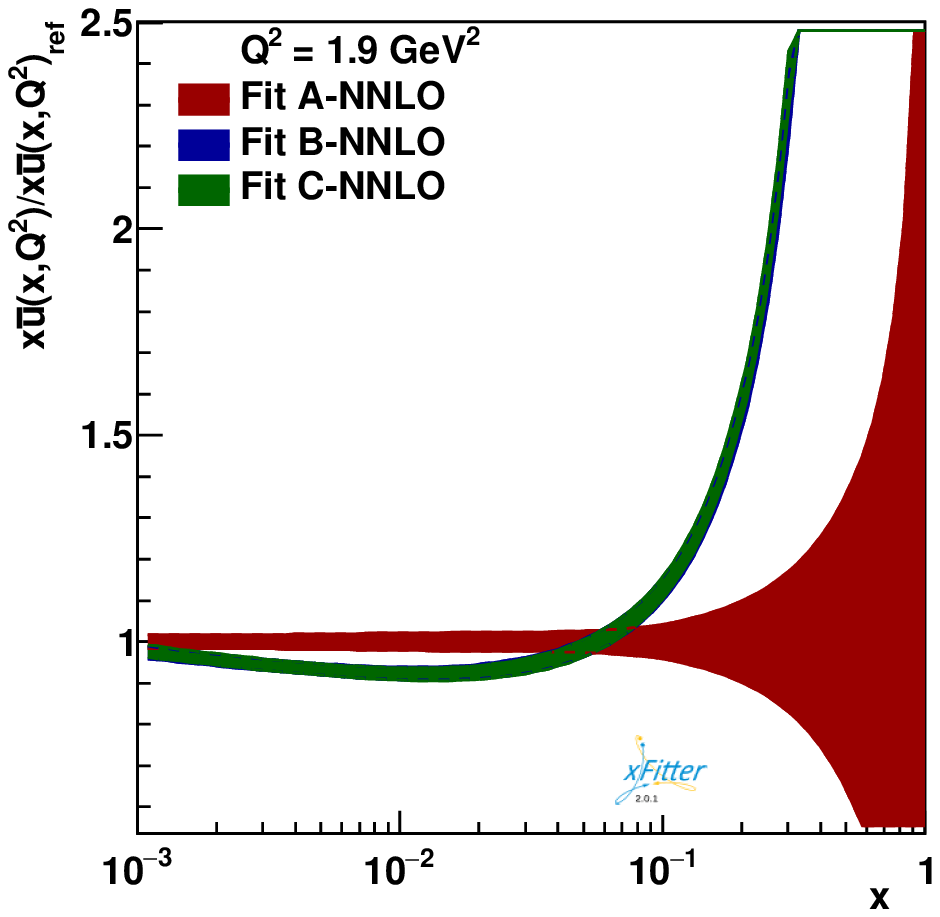}
		\includegraphics[scale = 0.4]{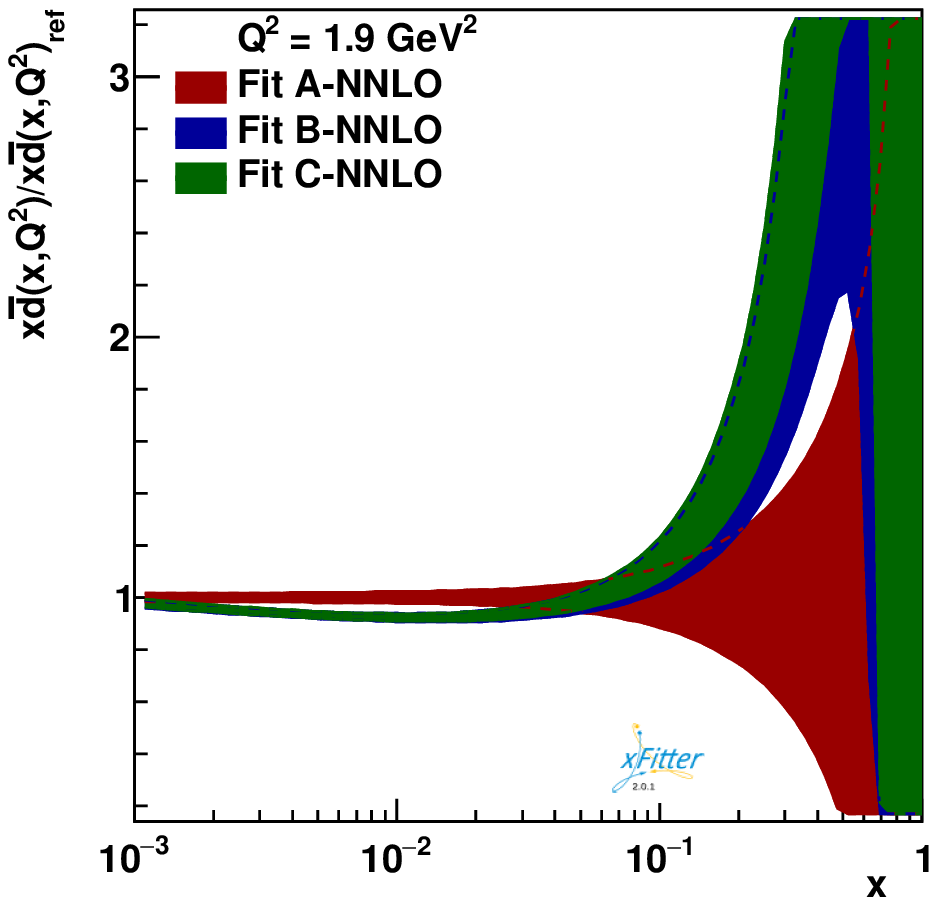}

		\includegraphics[scale = 0.4]{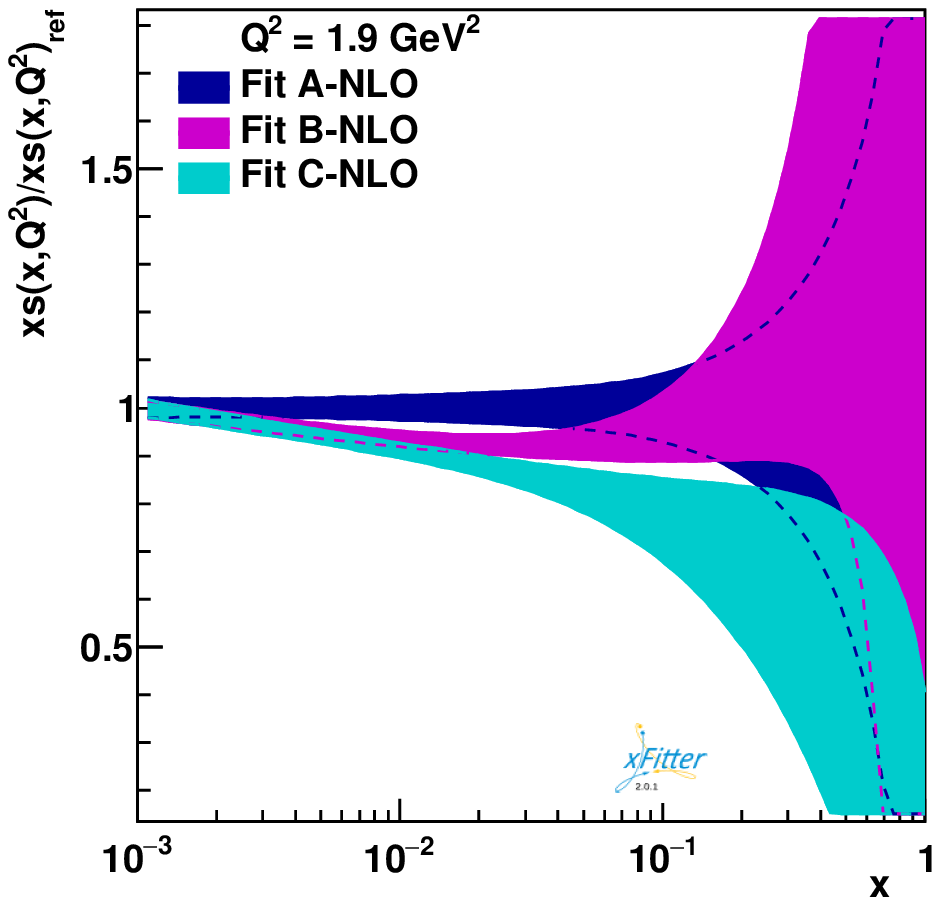} 
		\includegraphics[scale = 0.4]{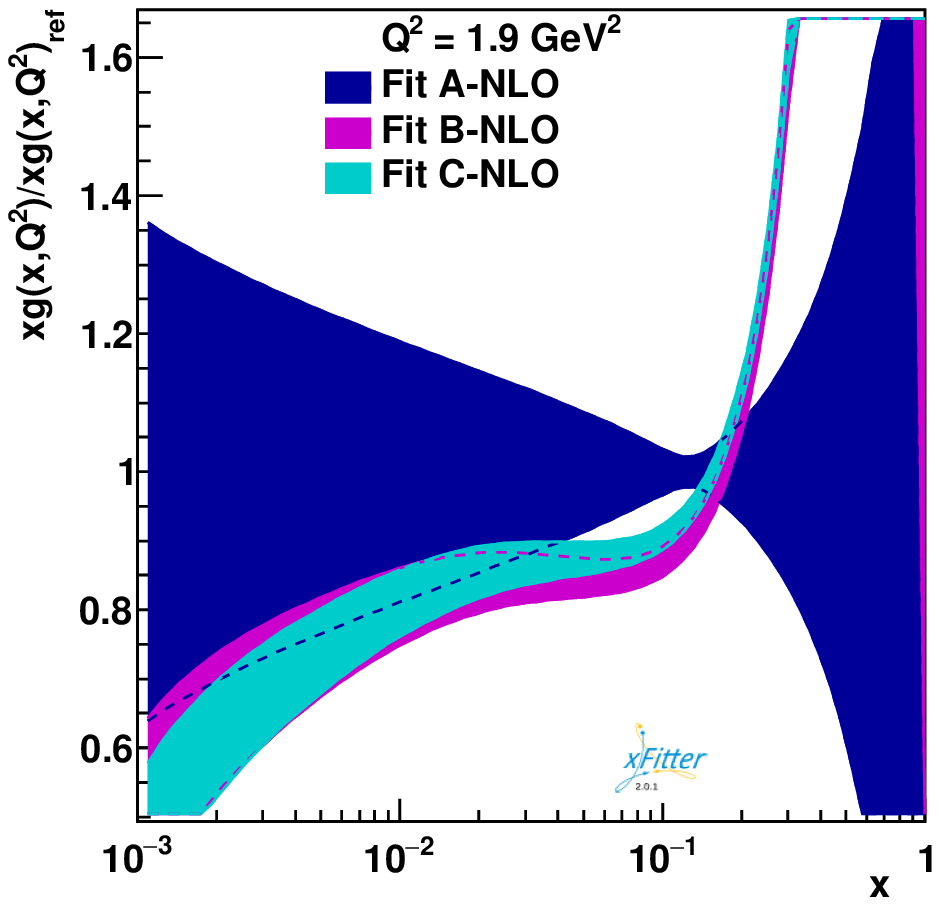}    
	    \includegraphics[scale = 0.4]{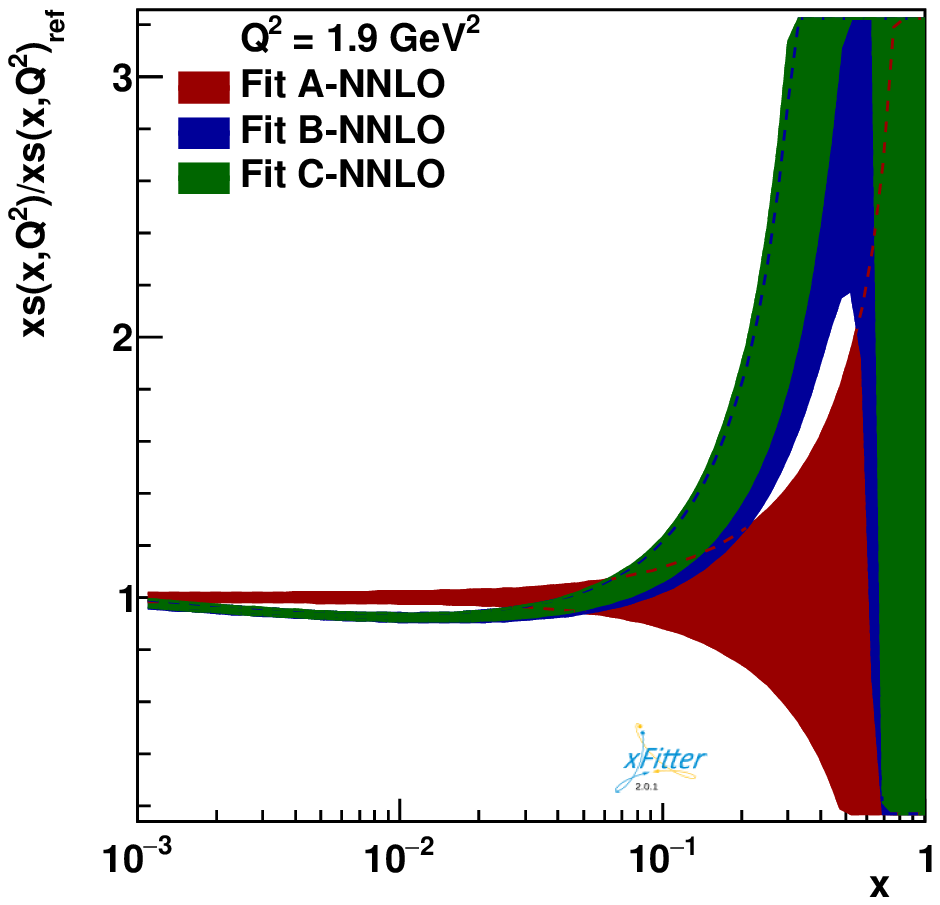}               
	    \includegraphics[scale = 0.4]{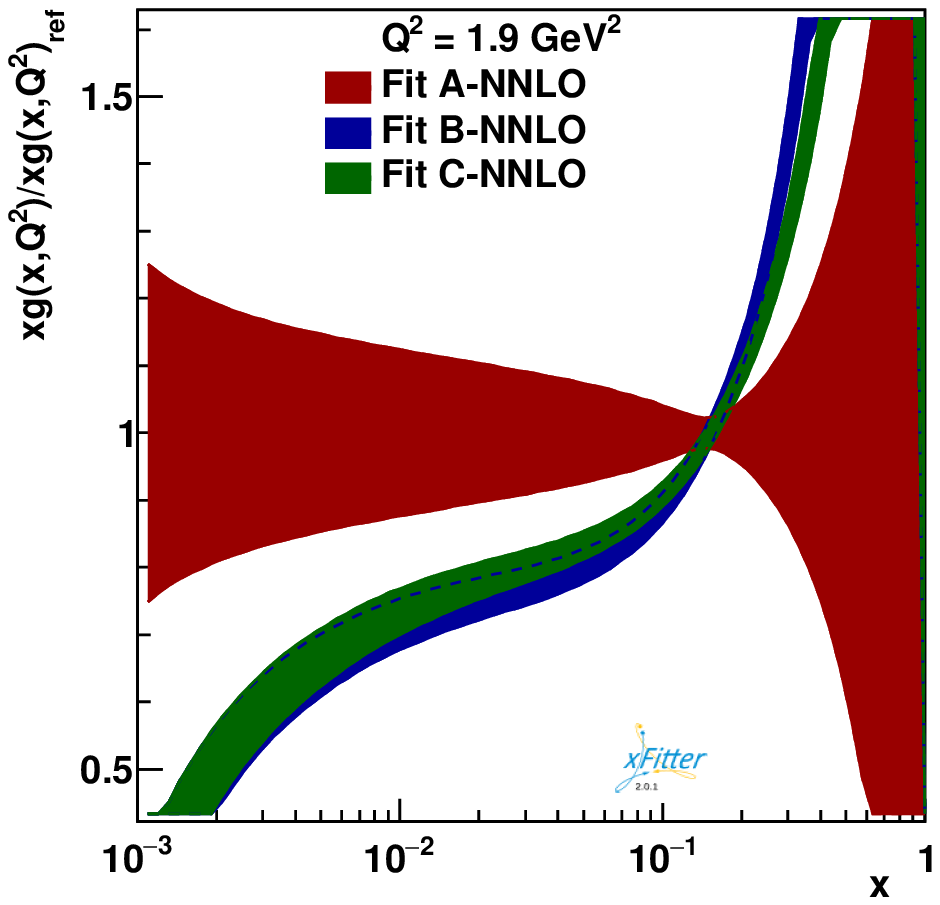}   
		
		\caption{The NLO and NNLO ratios of $xq(x,Q^2)/xq(x,Q^2)_{ref}$ at $Q^2$=1.9 GeV$^2$ for $q=u_v, d_v, \bar{u},\bar{d}, s$ and $g$ for Fit~A, Fit~B, and Fit~C, with respect to Fit~A. In the left panels, we present our results for NLO, whereas the right panels are for NNLO.}
		
		\label{fig:RURPDF-1.9-NLO+NNLO}
	\end{center}
\end{figure*}

To clarify the differences between our QCD fits from Fit~A to Fit~C, we present the ratios $xq(x,Q^2)/xq(x,Q^2)_{ref}$ with respect to the Fit~A at NLO (left panels) and also NNLO (right panels) in Fig. \ref{fig:RURPDF-1.9-NLO+NNLO}. According to this figure and for both NLO and NNLO analysis,  the changes of the central values of all PDFs, their uncertainties, or both are observed at $Q^2$=1.9 GeV$^2$.

\begin{figure*}[!htb]
	\begin{center}
		\includegraphics[scale = 0.4]{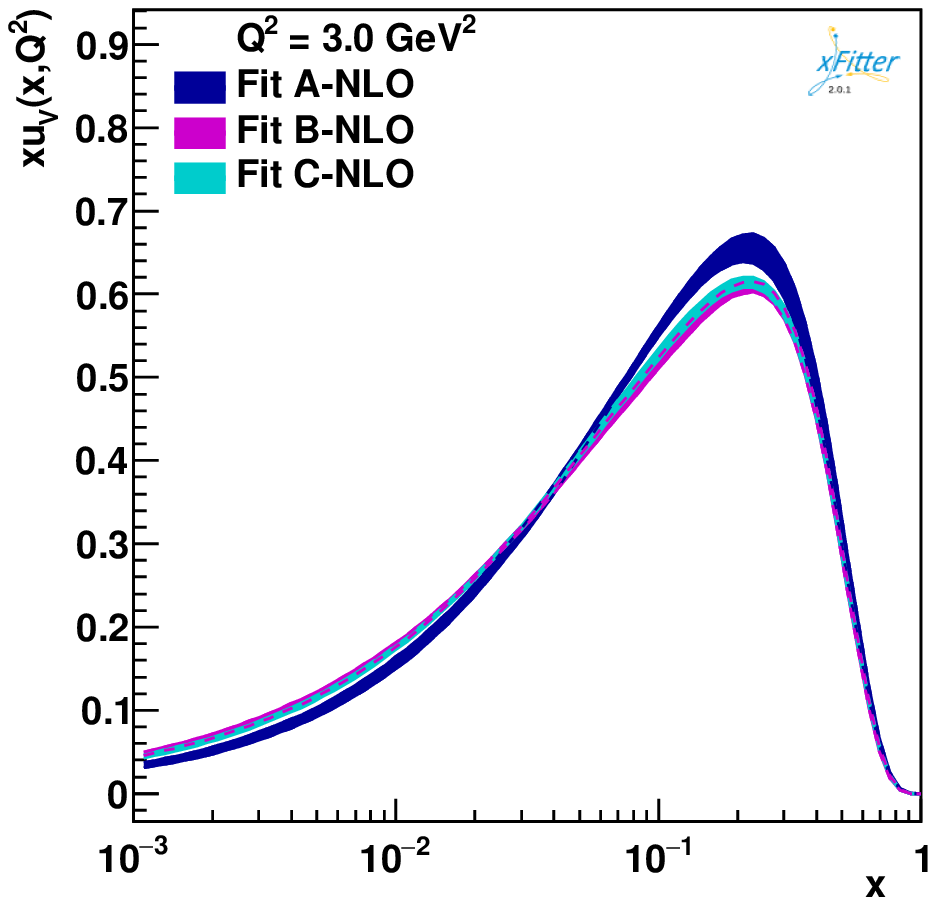}
		\includegraphics[scale = 0.4]{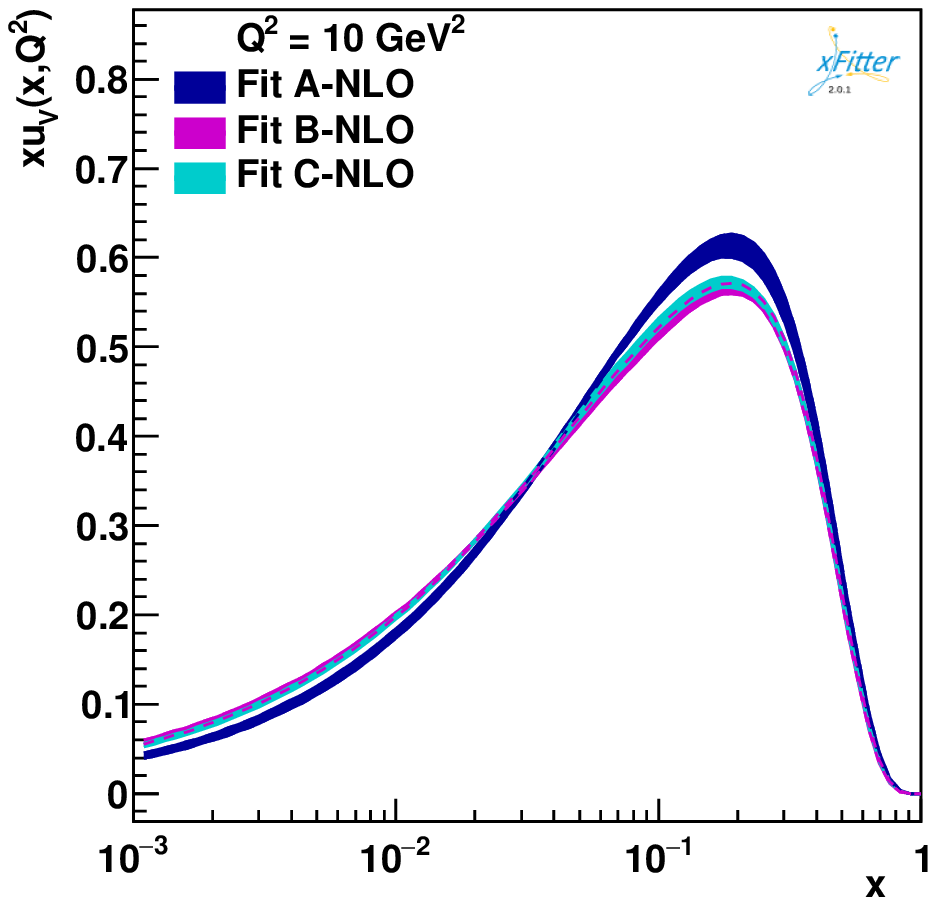}
		\includegraphics[scale = 0.4]{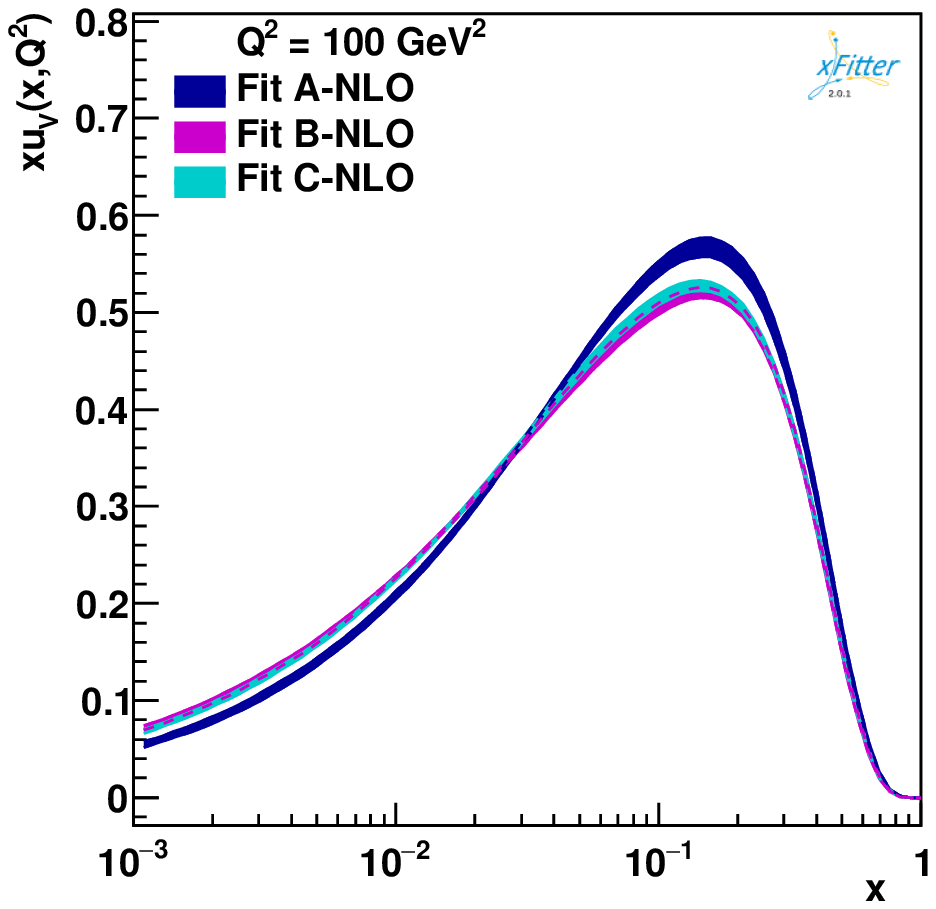}
		
		\includegraphics[scale = 0.4]{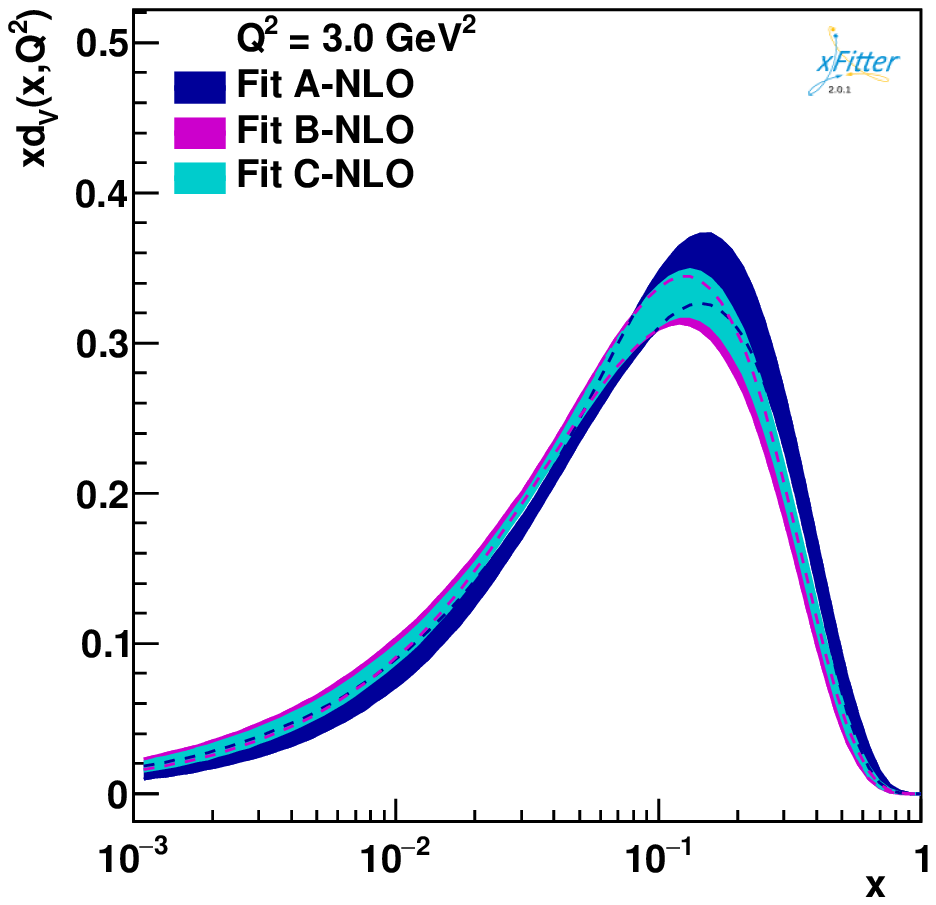}		
		\includegraphics[scale = 0.4]{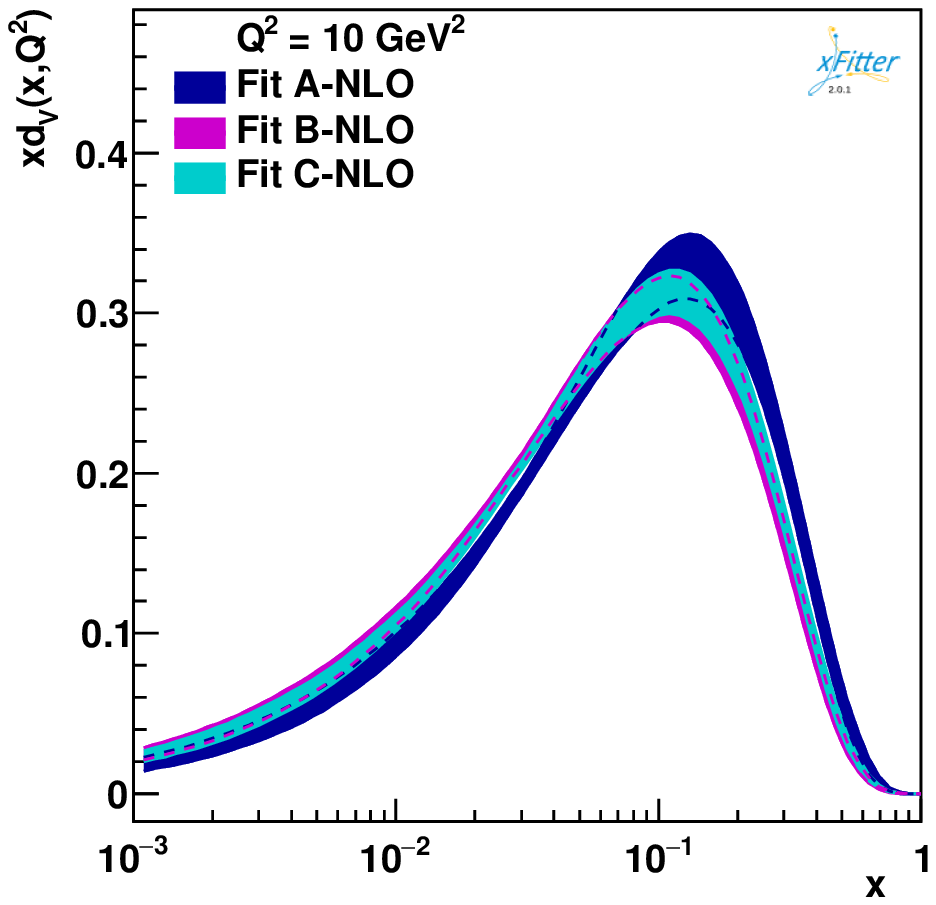}
		\includegraphics[scale = 0.4]{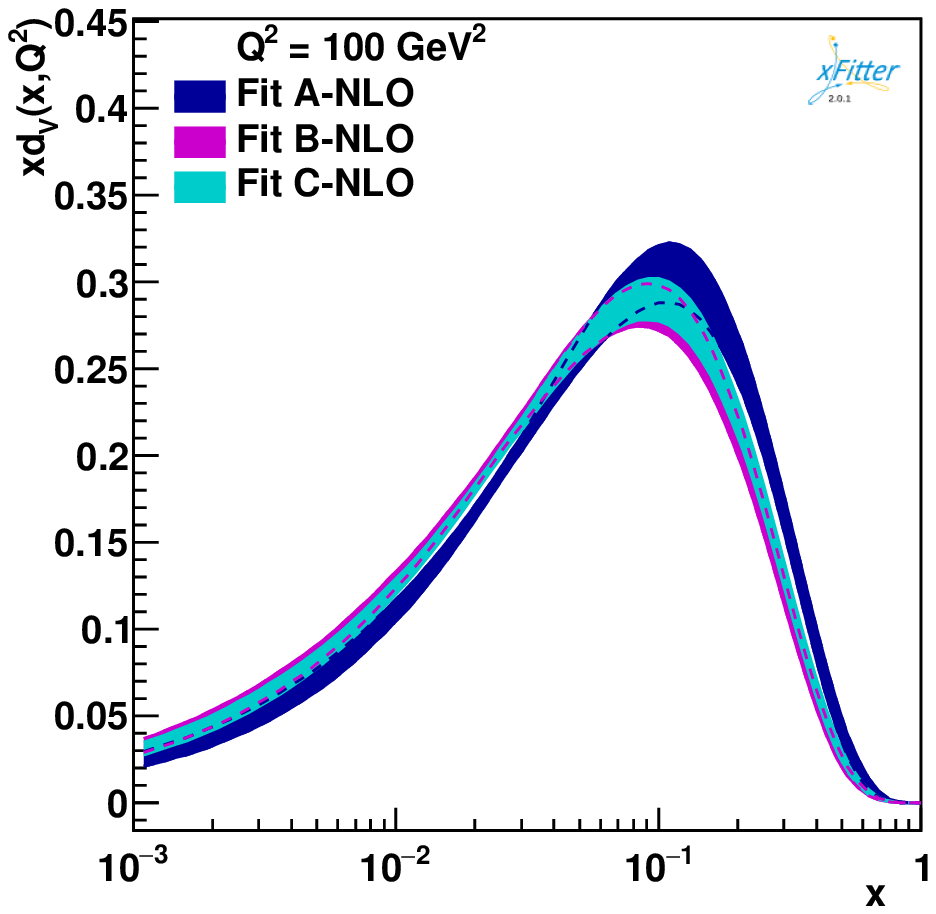}
		
		\includegraphics[scale = 0.4]{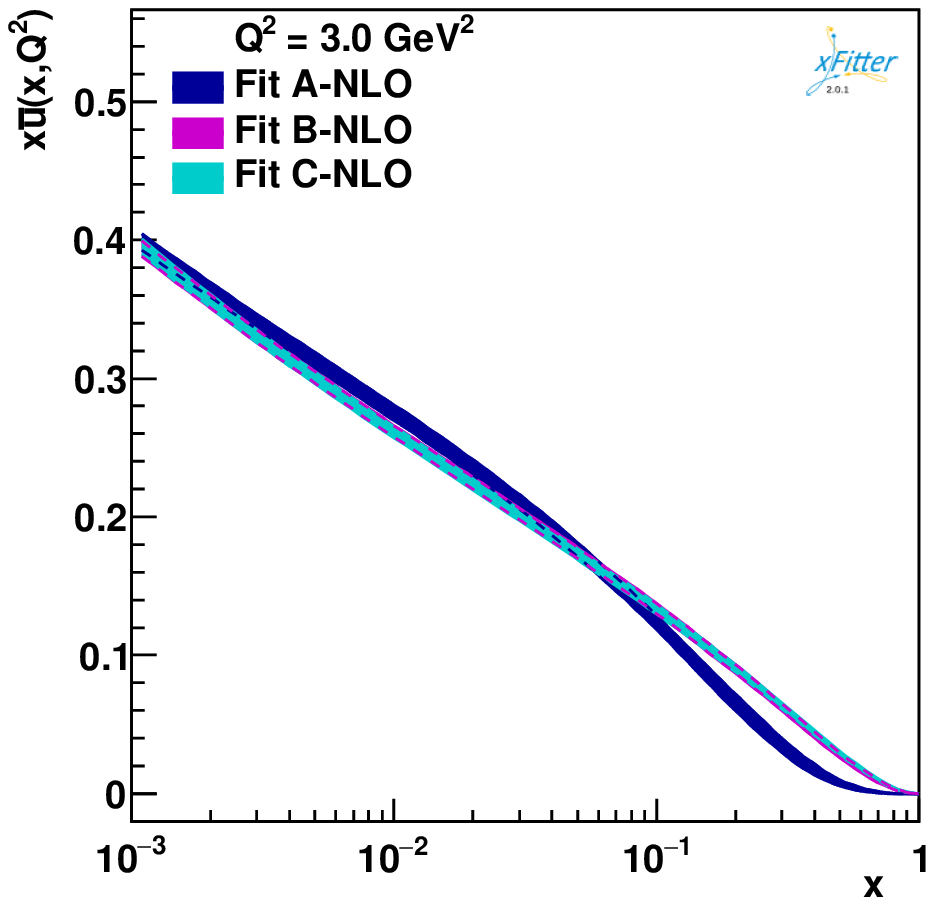}		
		\includegraphics[scale = 0.4]{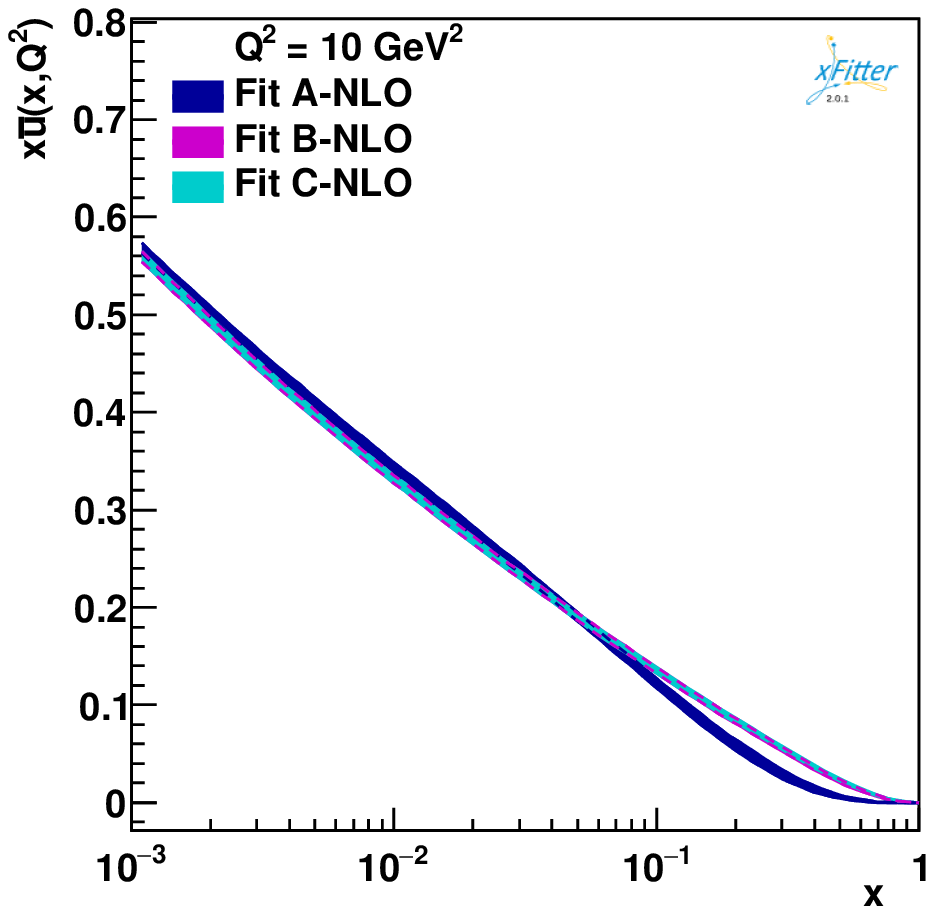}
		\includegraphics[scale = 0.4]{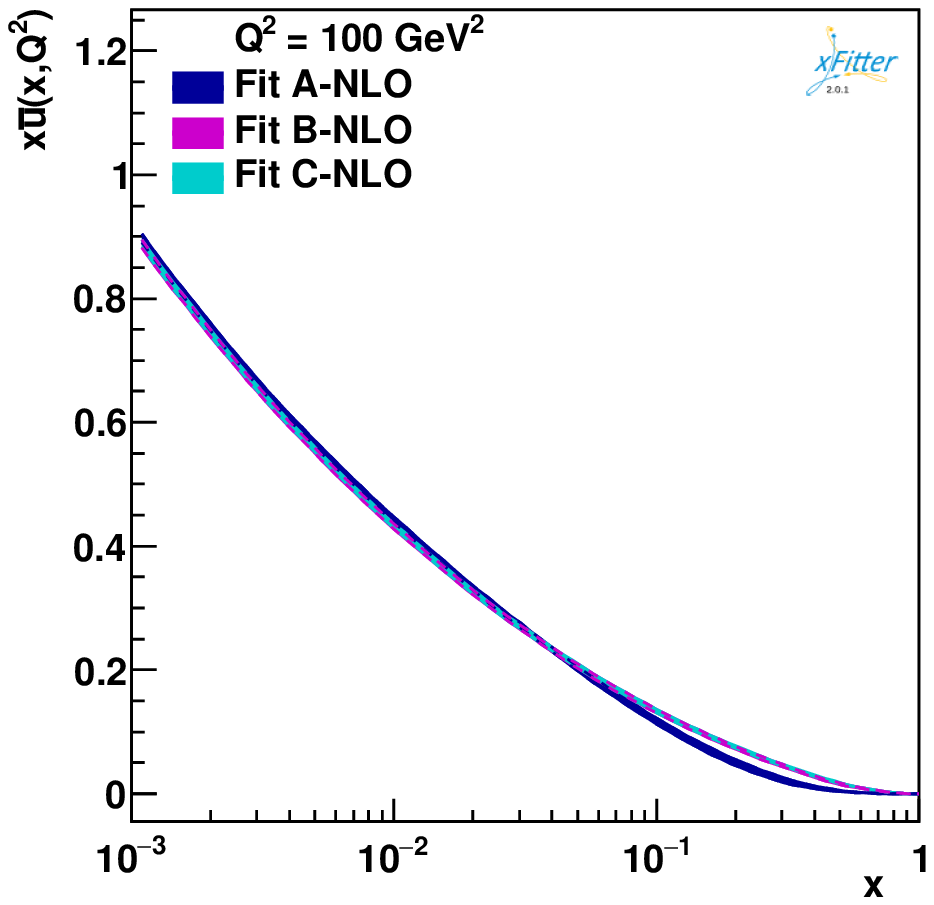}	 
		
		\includegraphics[scale = 0.4]{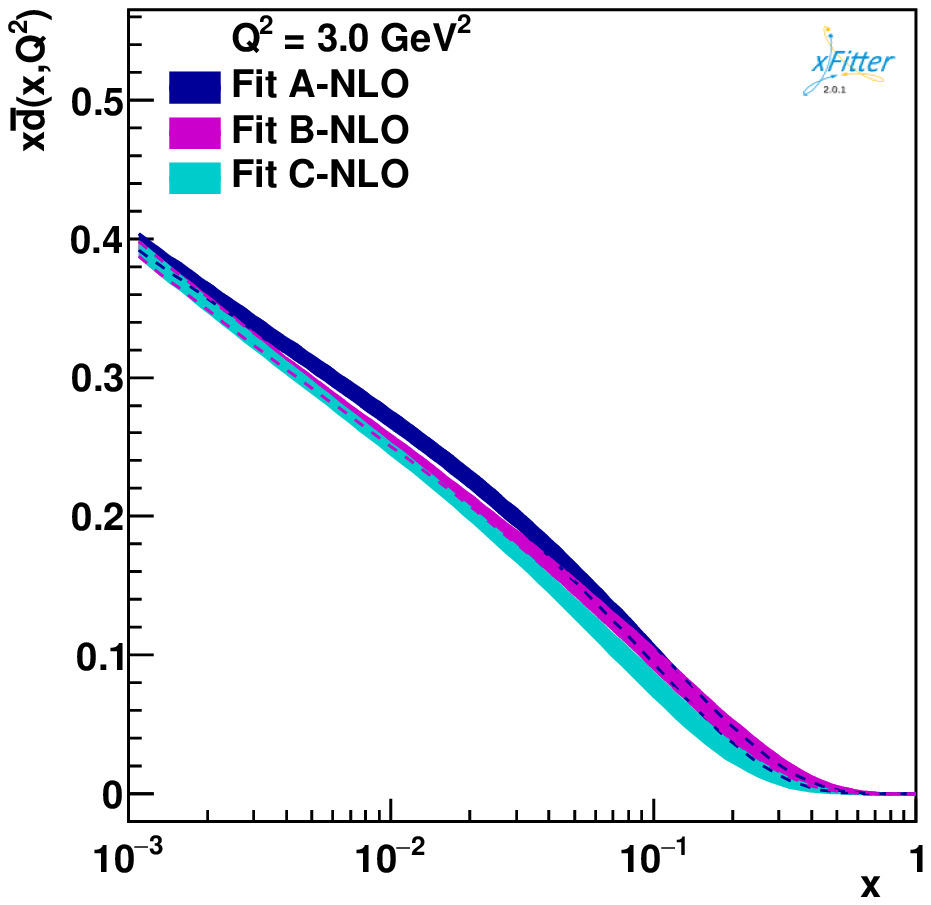}		
		\includegraphics[scale = 0.4]{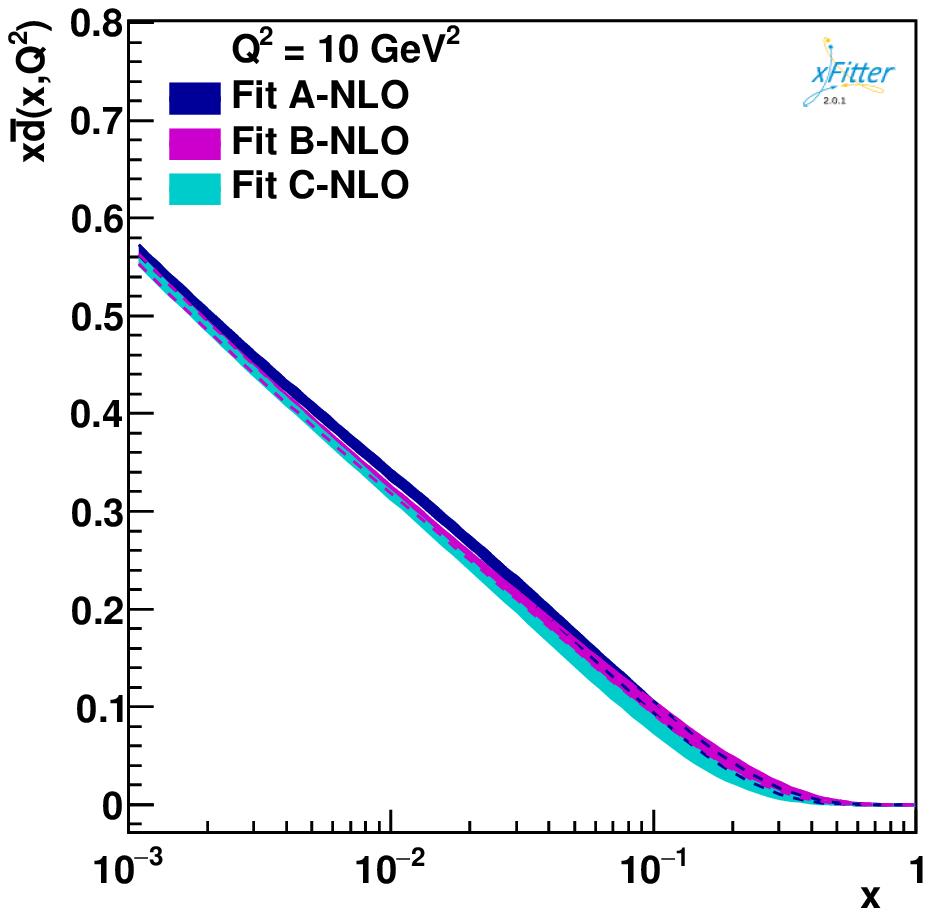}
		\includegraphics[scale = 0.4]{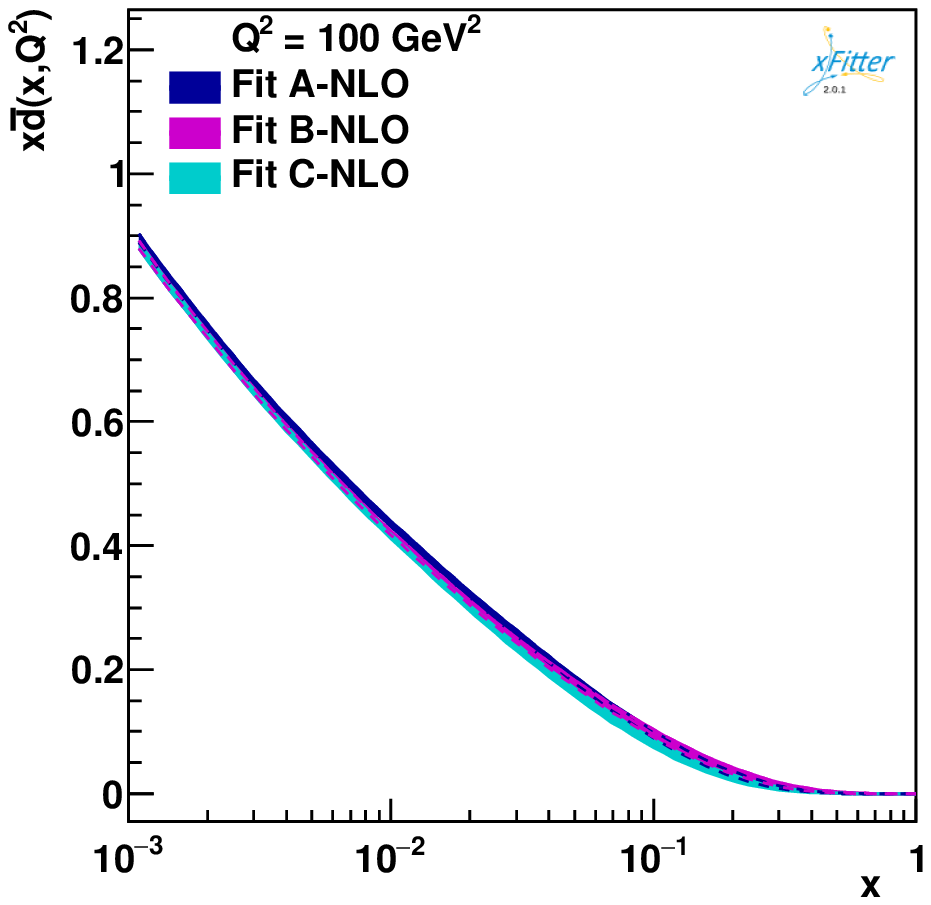}	               
		
		\includegraphics[scale = 0.4]{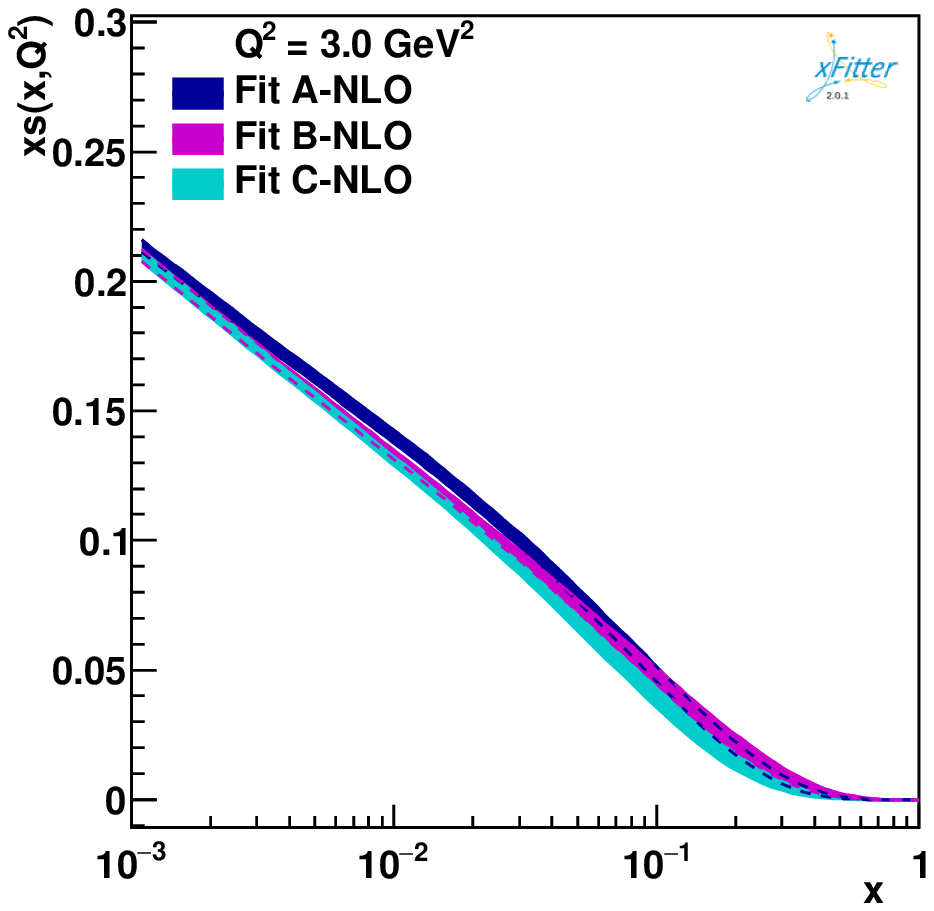}		
		\includegraphics[scale = 0.4]{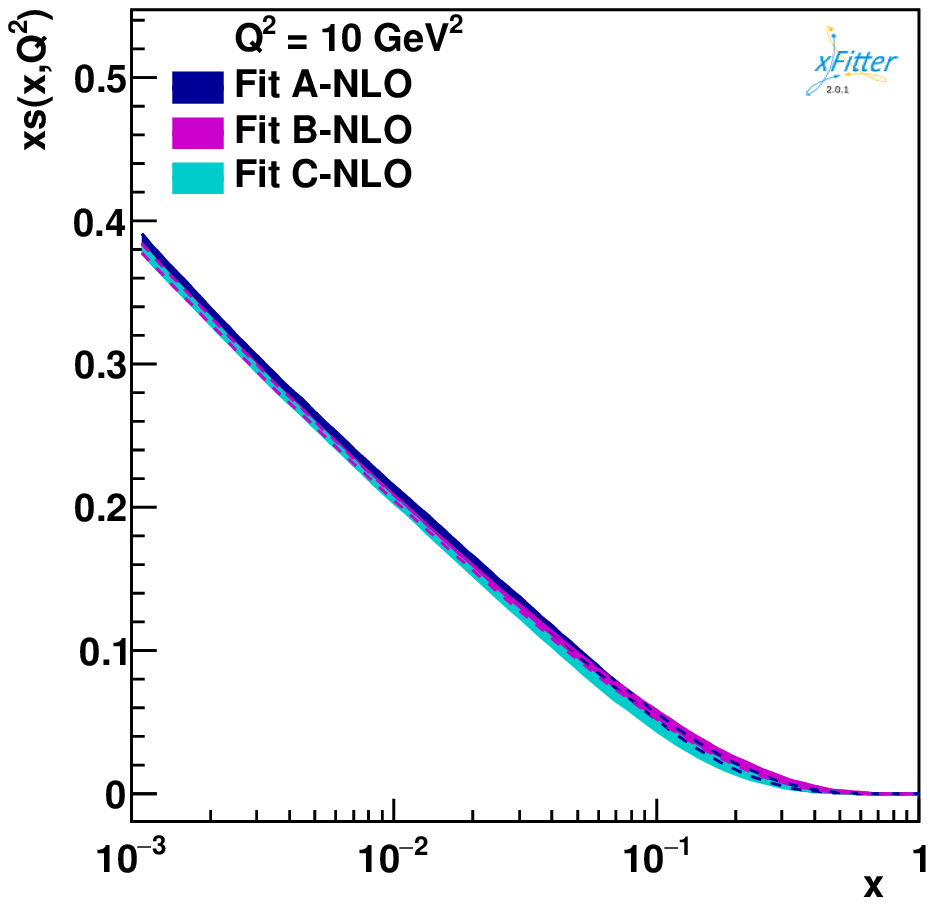}
		\includegraphics[scale = 0.4]{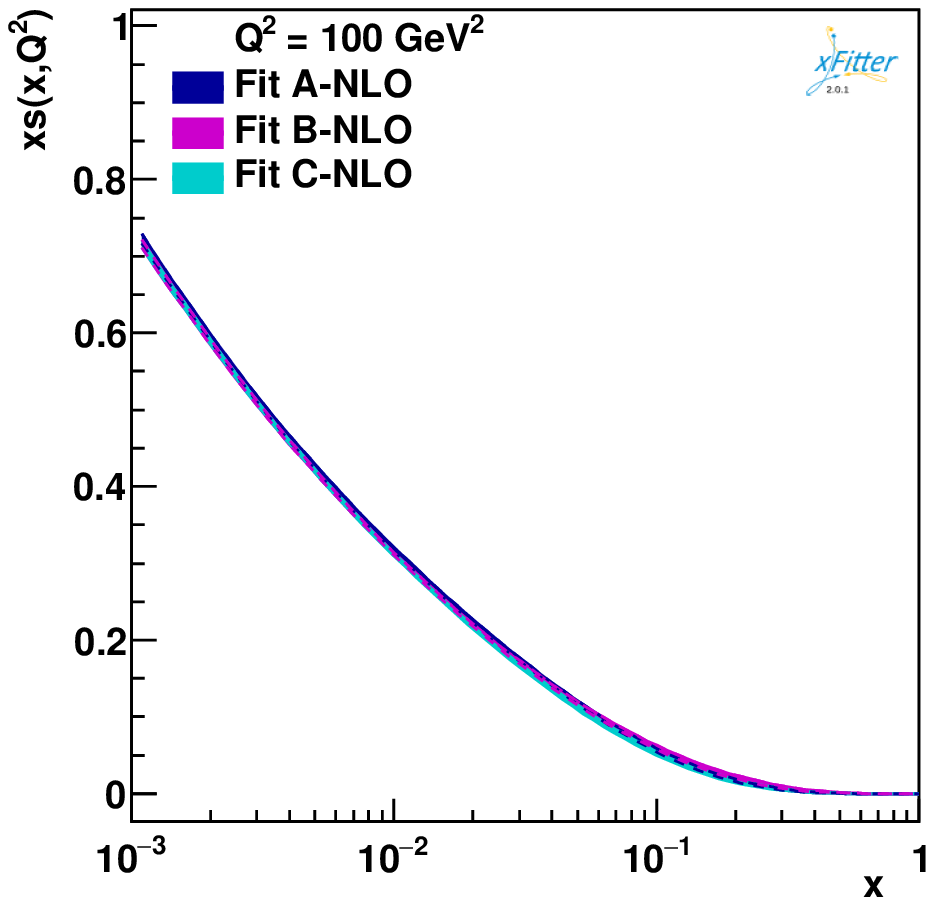}

		\includegraphics[scale = 0.4]{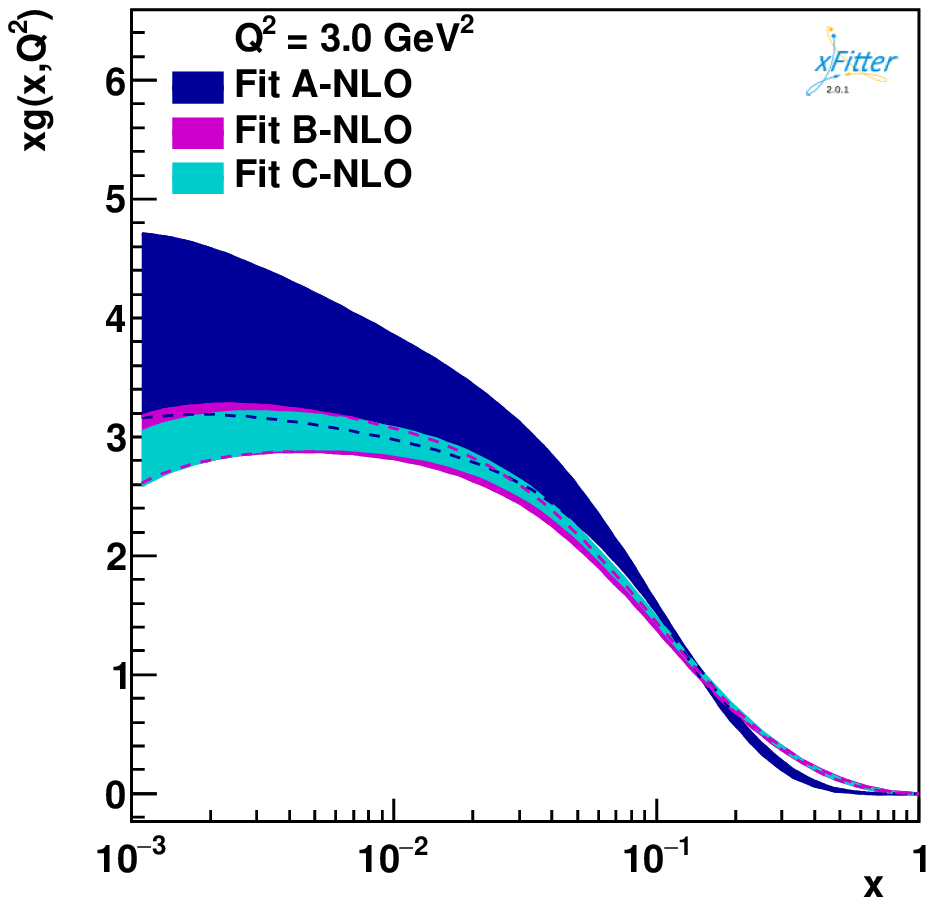}		
		\includegraphics[scale = 0.4]{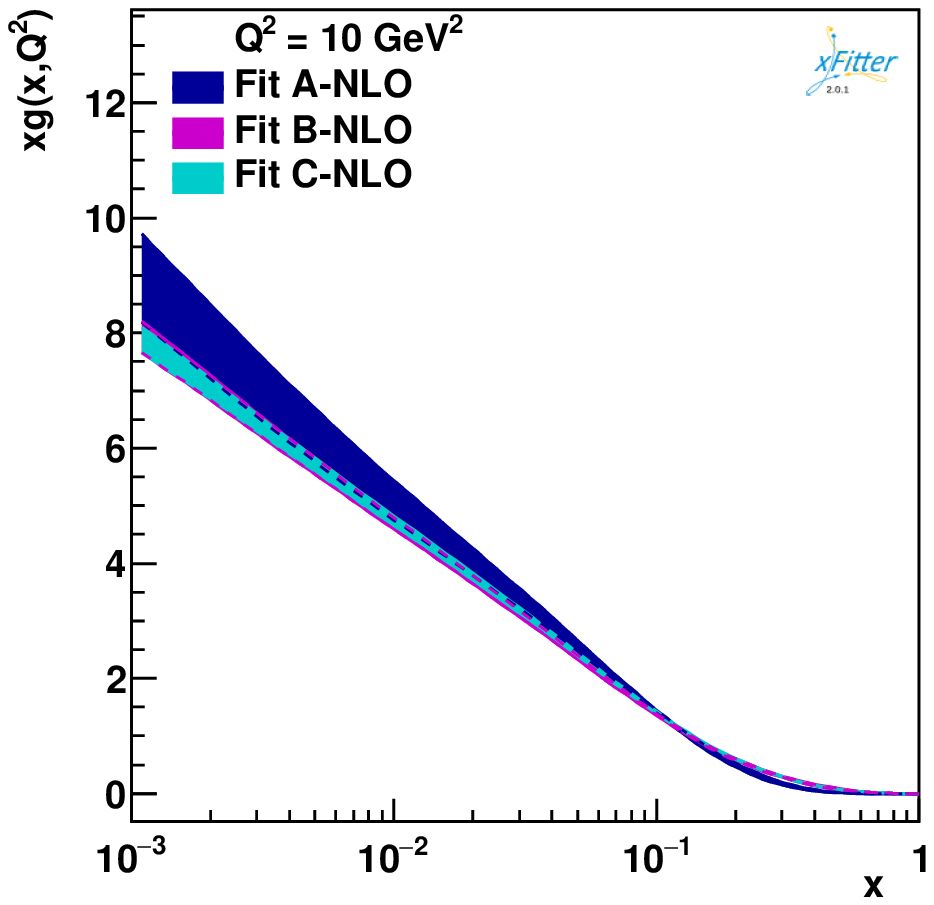}
		\includegraphics[scale = 0.4]{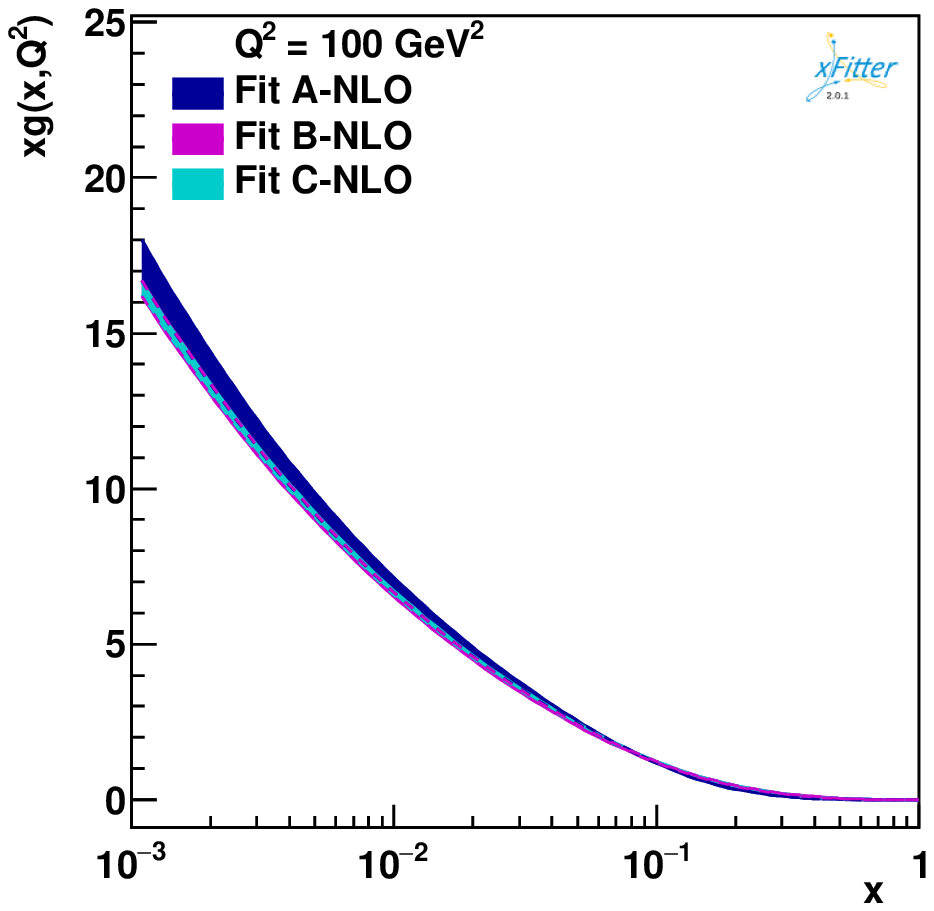}		
		\caption{The NLO parton distribution of $xu_v$, $xd_v$, $x\bar{u}$, $x\bar{d}$, $xs$, and $xg$,  as a  function of $x$ and for different values of $Q^2=$3, 10 and 100 GeV$^2$.}
		\label{fig:PDF-Qdep-NLO}
	\end{center}
\end{figure*}

\begin{figure*}[!htb]
	\begin{center}
	    \includegraphics[scale = 0.4]{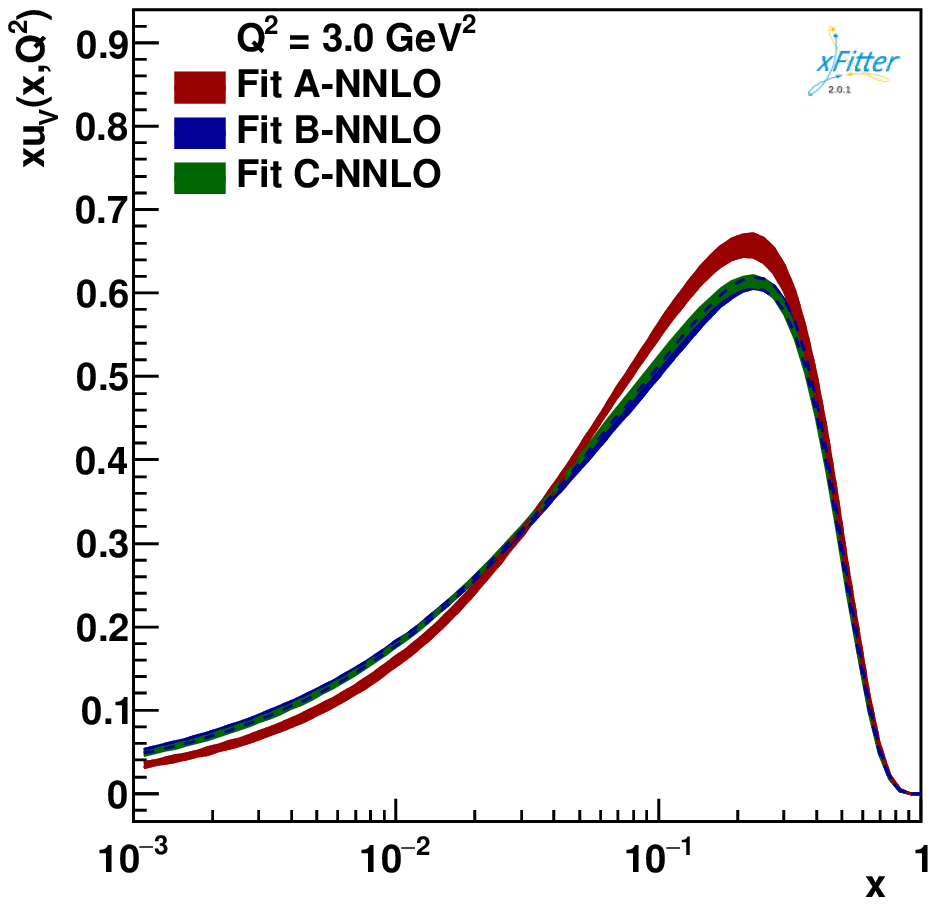}
	    \includegraphics[scale = 0.4]{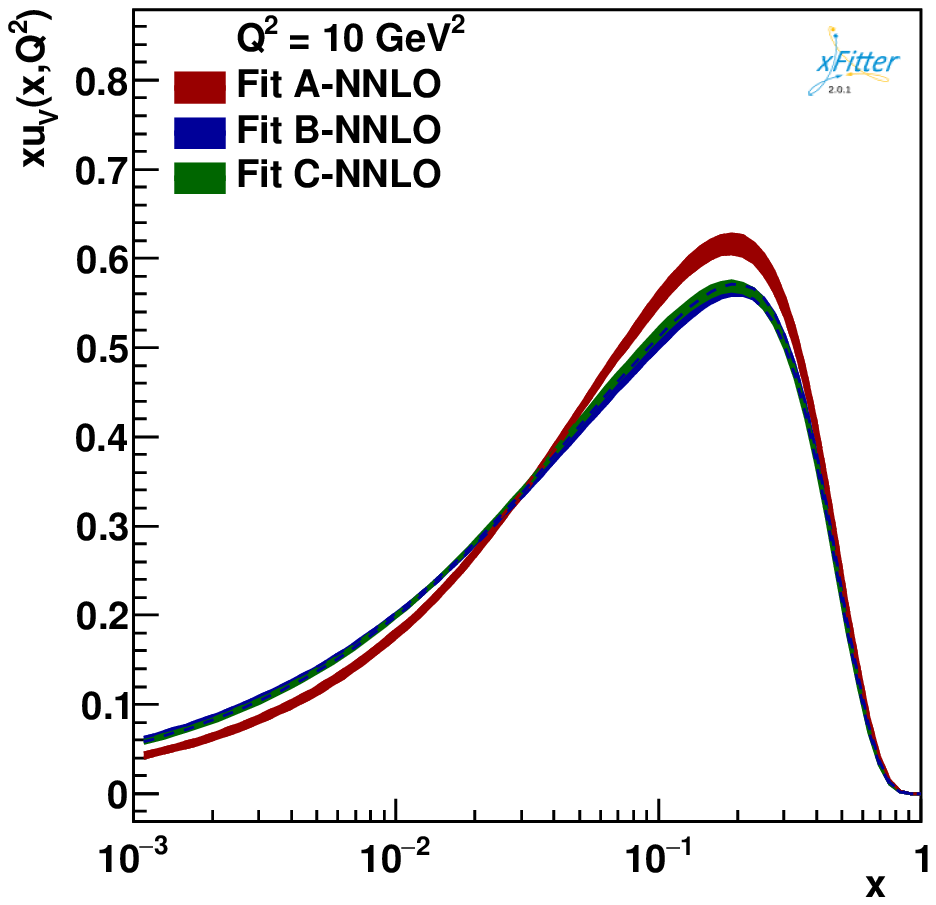}
	    \includegraphics[scale = 0.4]{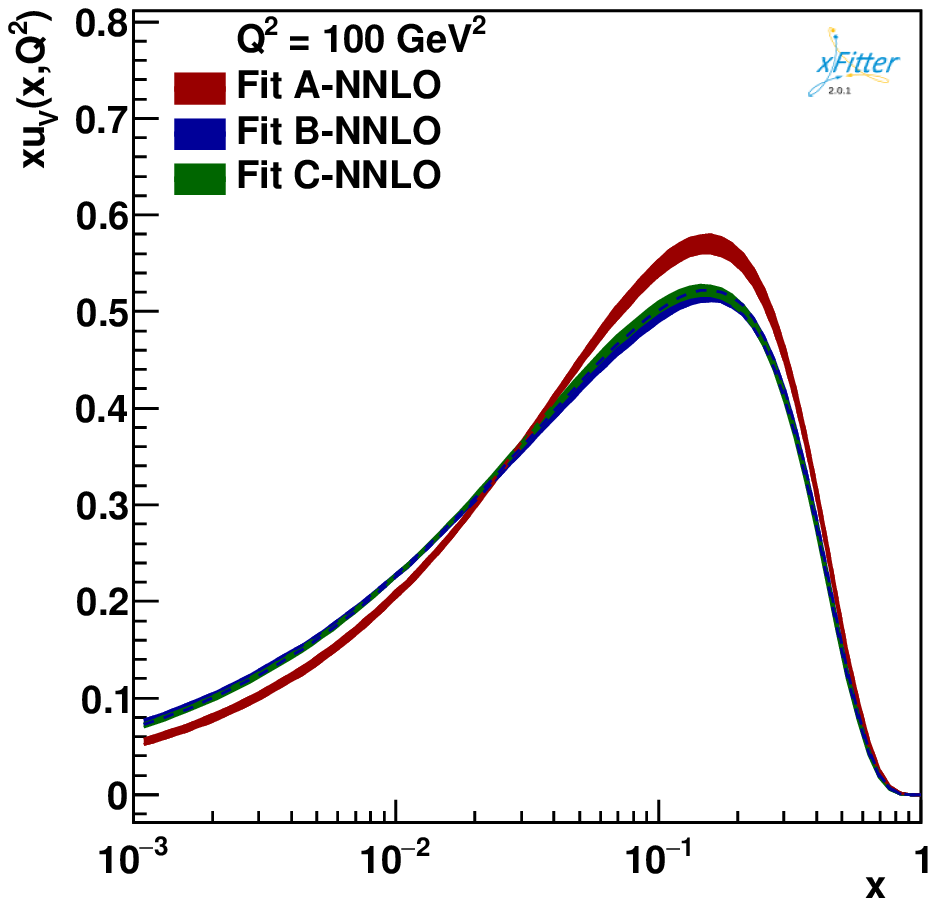}
	    	    	    	    	
	    \includegraphics[scale = 0.4]{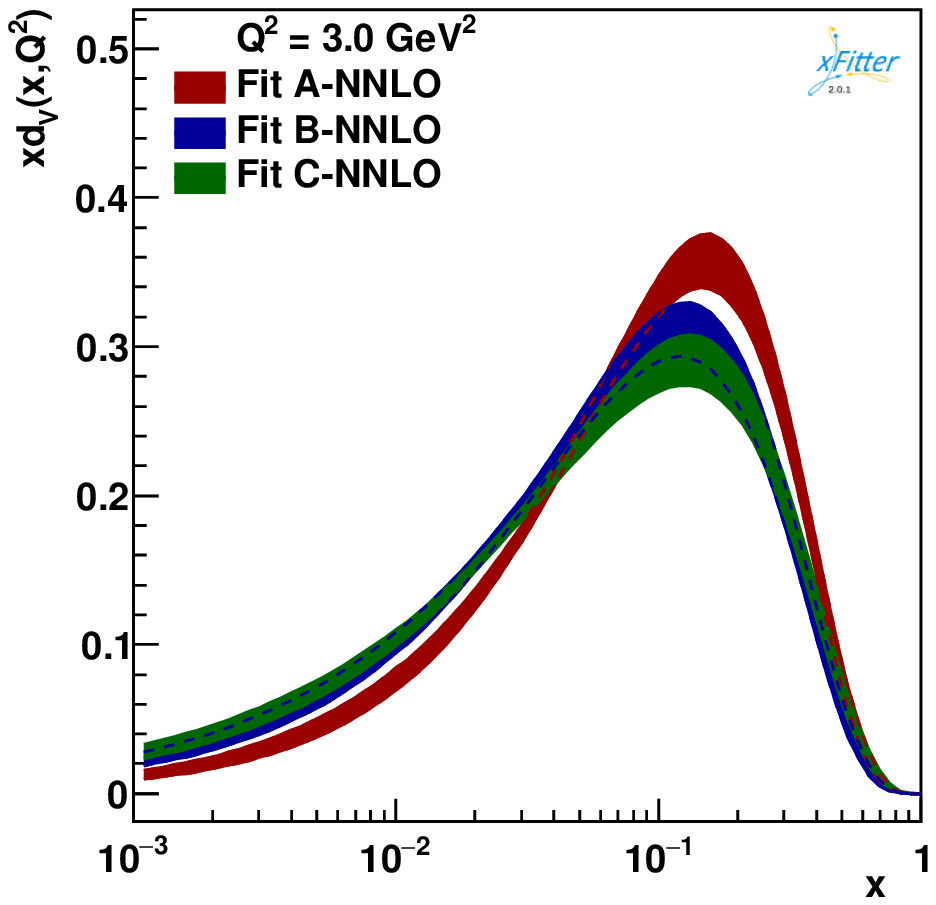}		
	    \includegraphics[scale = 0.4]{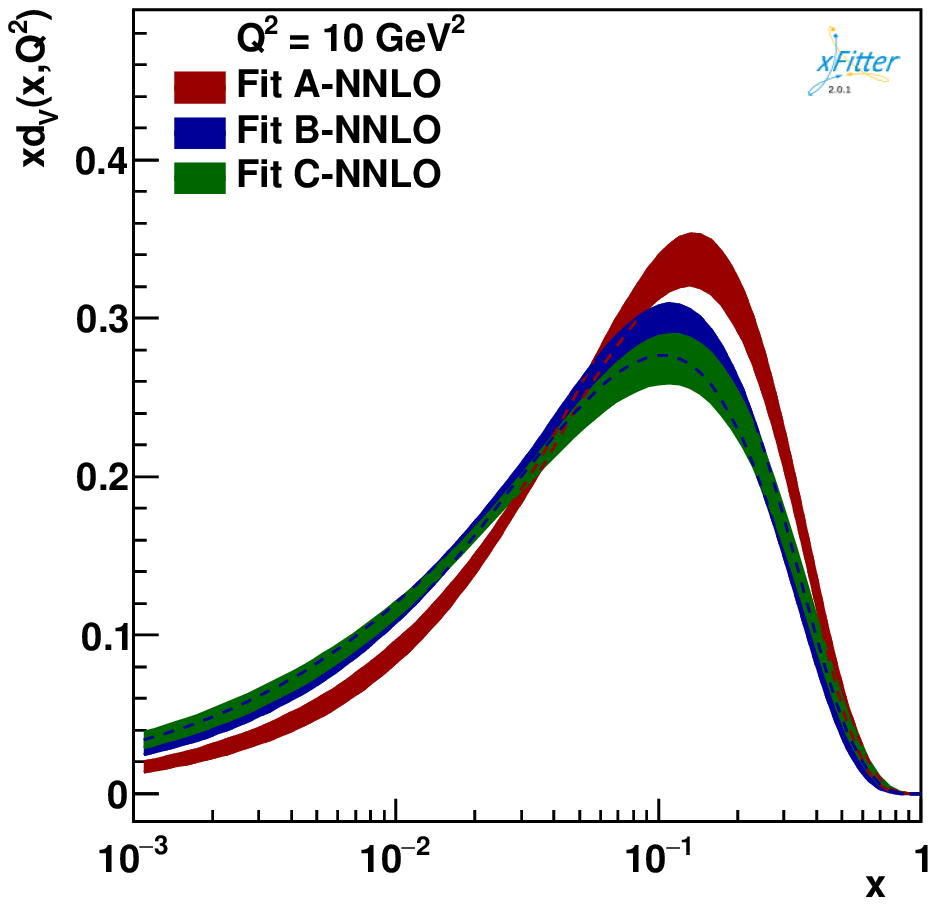}
	    \includegraphics[scale = 0.4]{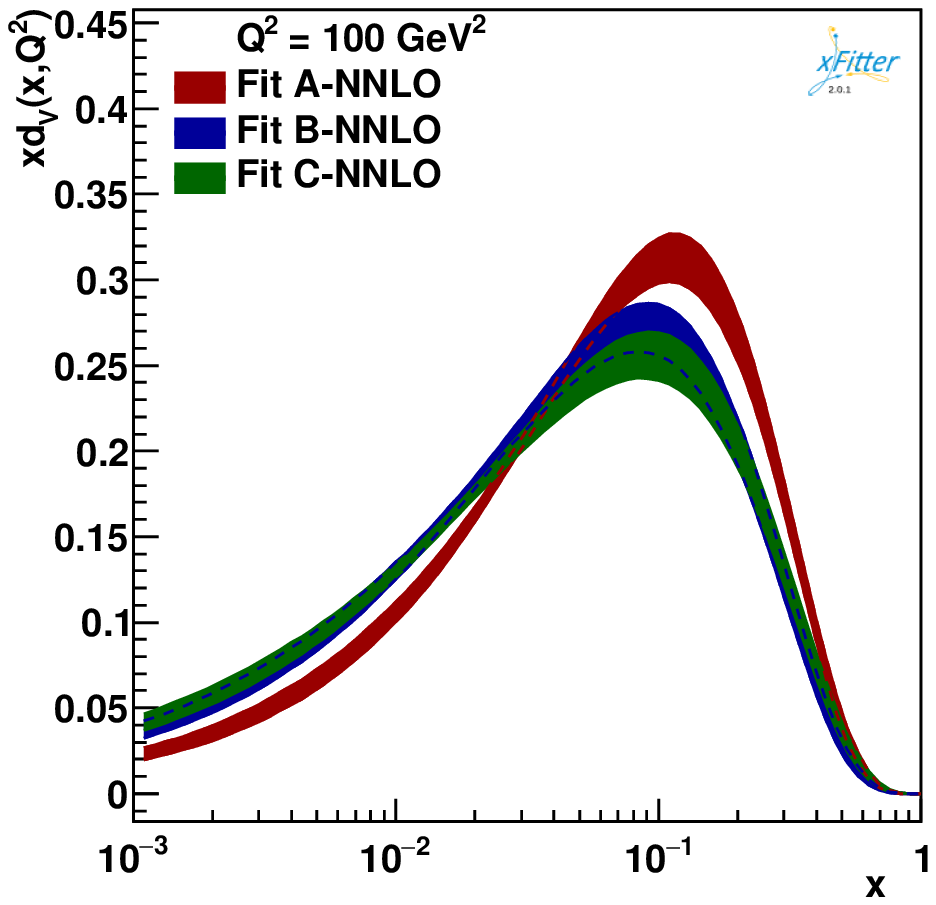}
	    
	    \includegraphics[scale = 0.4]{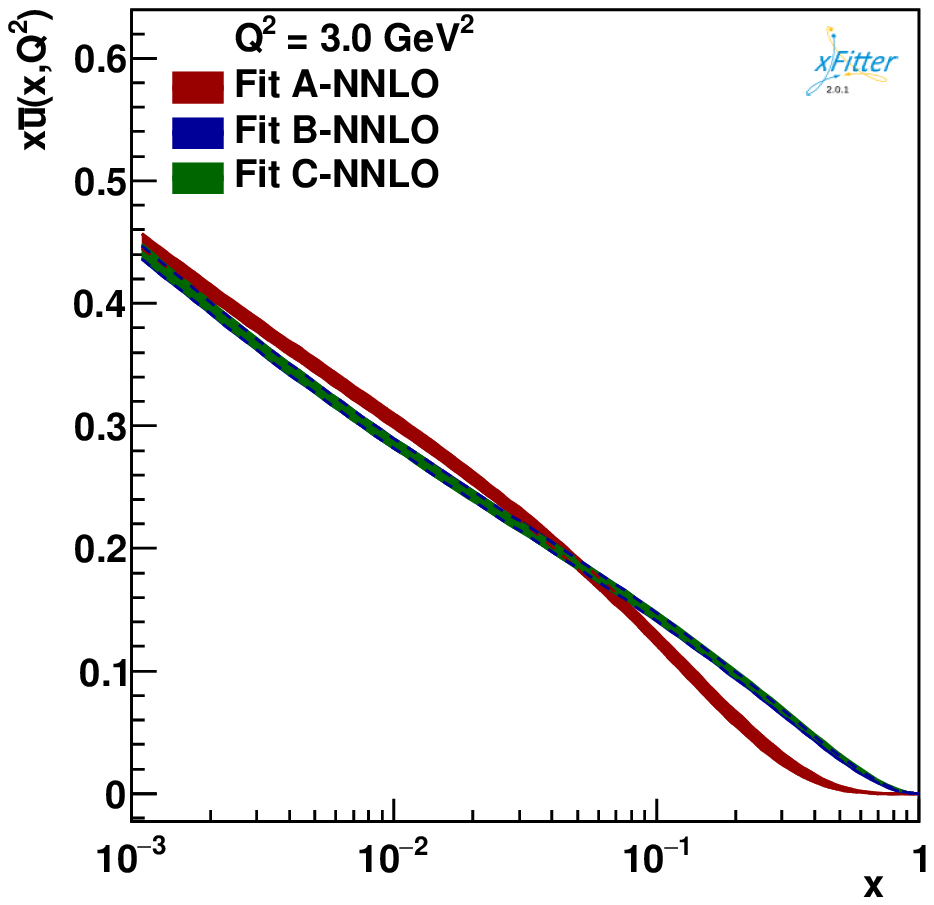}		
	    \includegraphics[scale = 0.4]{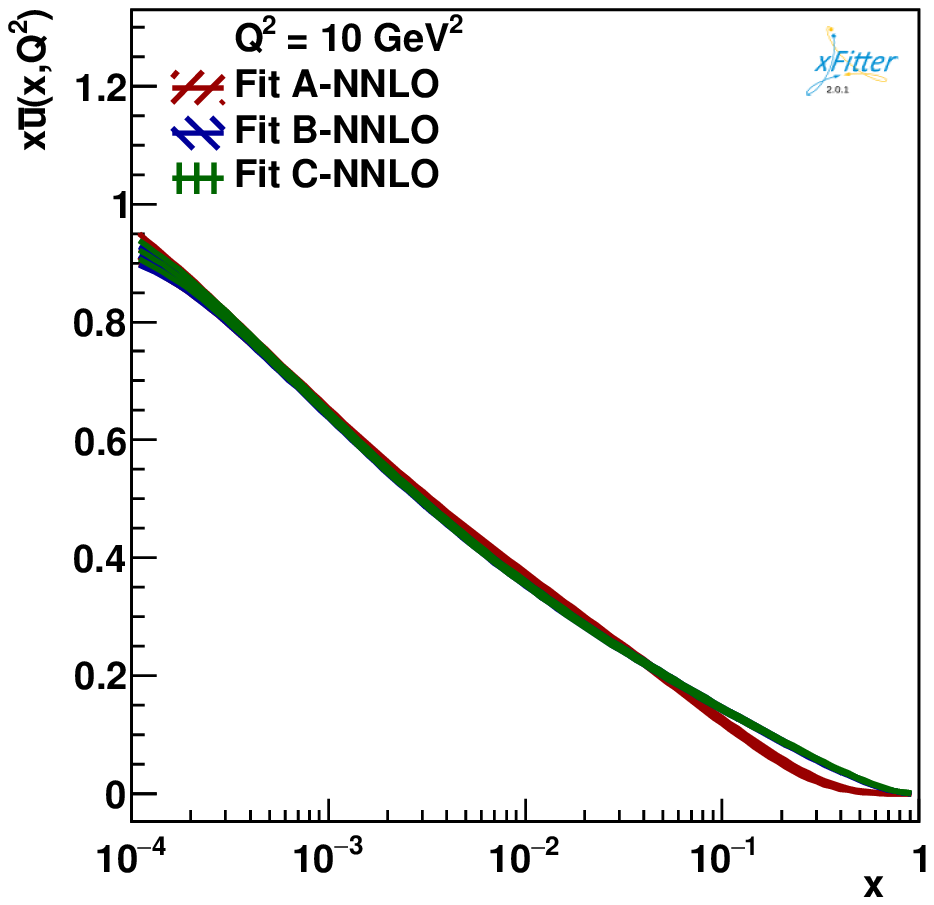}
	    \includegraphics[scale = 0.4]{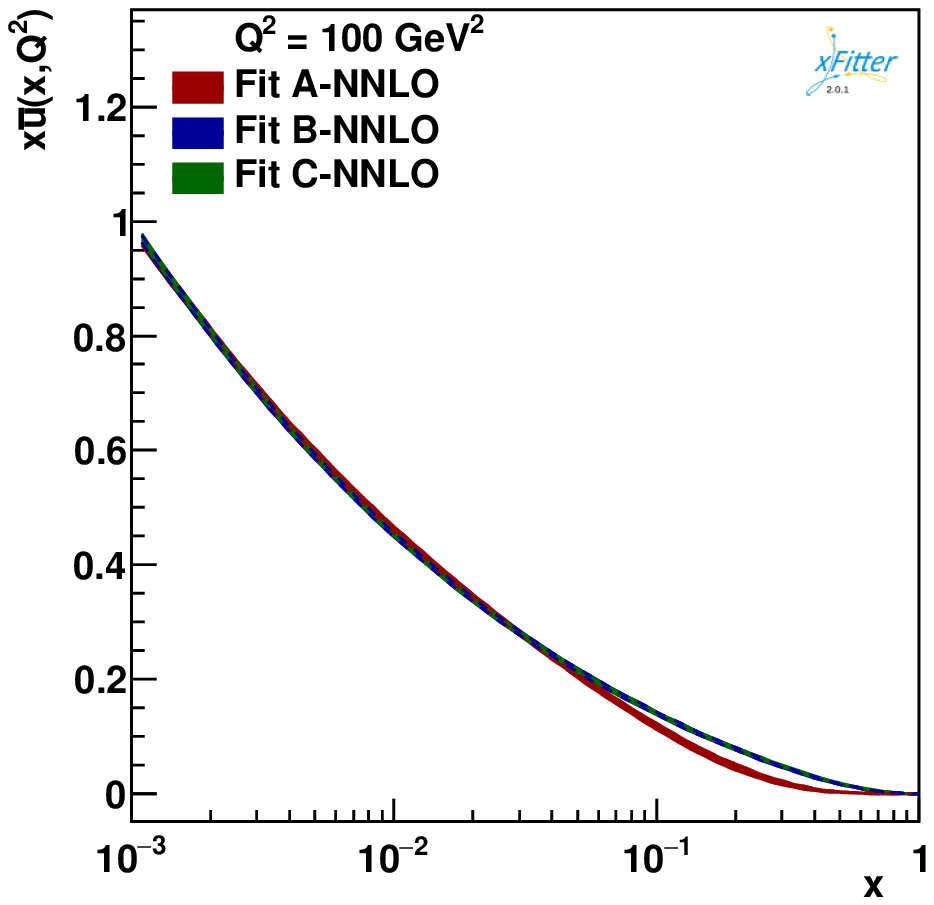}	 
	    
	    \includegraphics[scale = 0.4]{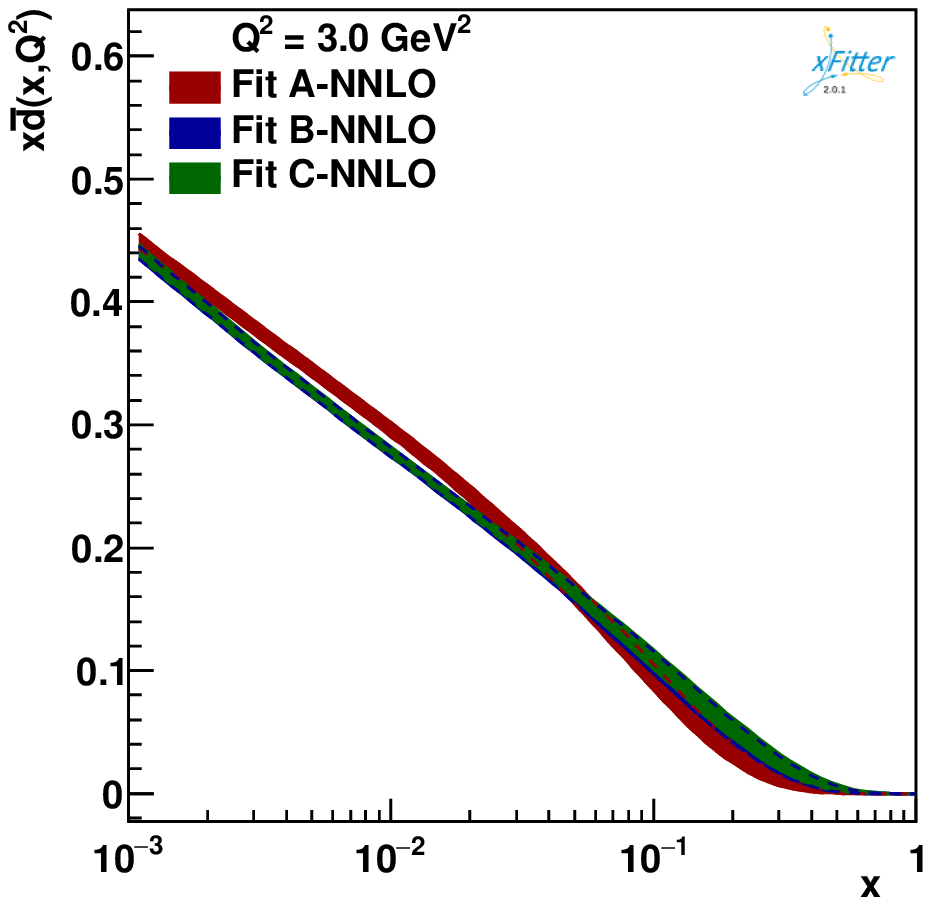}		
	    \includegraphics[scale = 0.4]{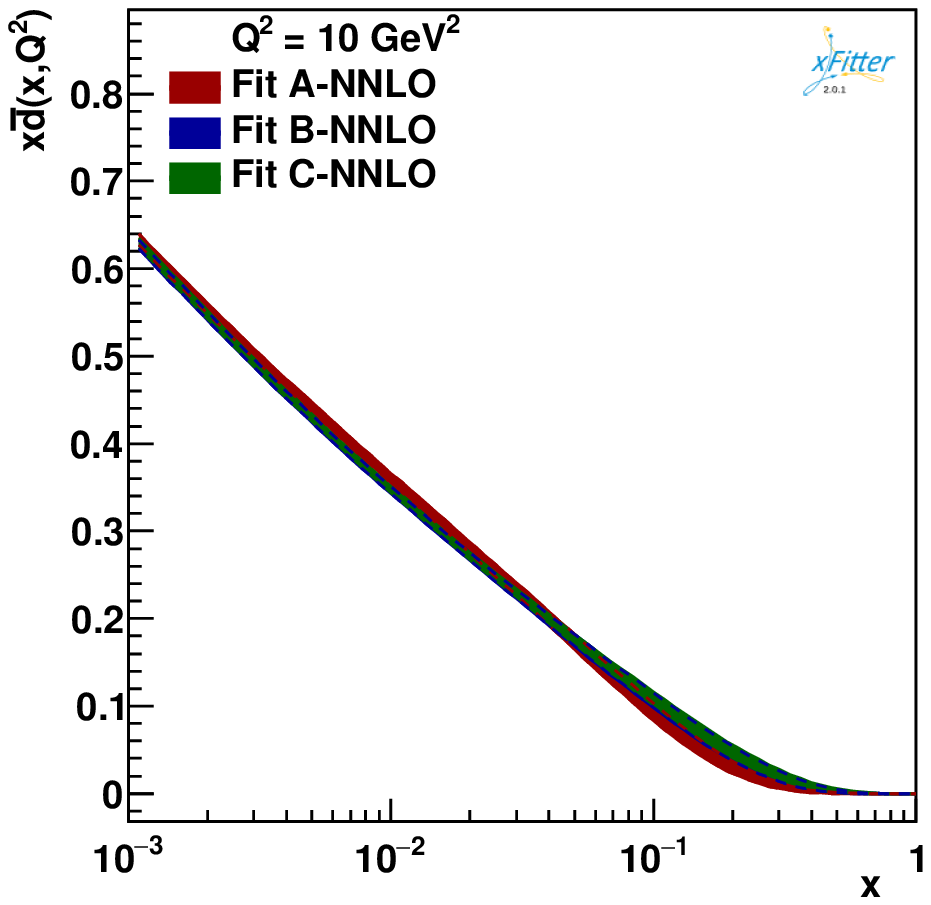}
	    \includegraphics[scale = 0.4]{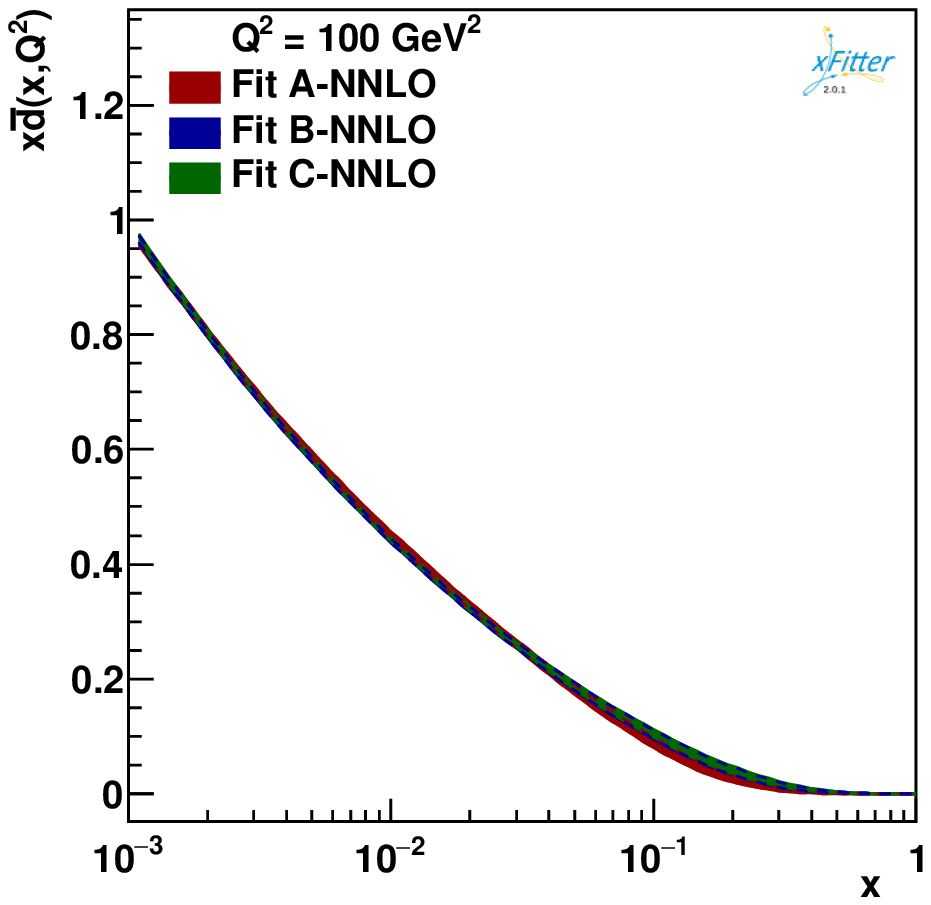}	               
	    	    	    	    	
	    \includegraphics[scale = 0.4]{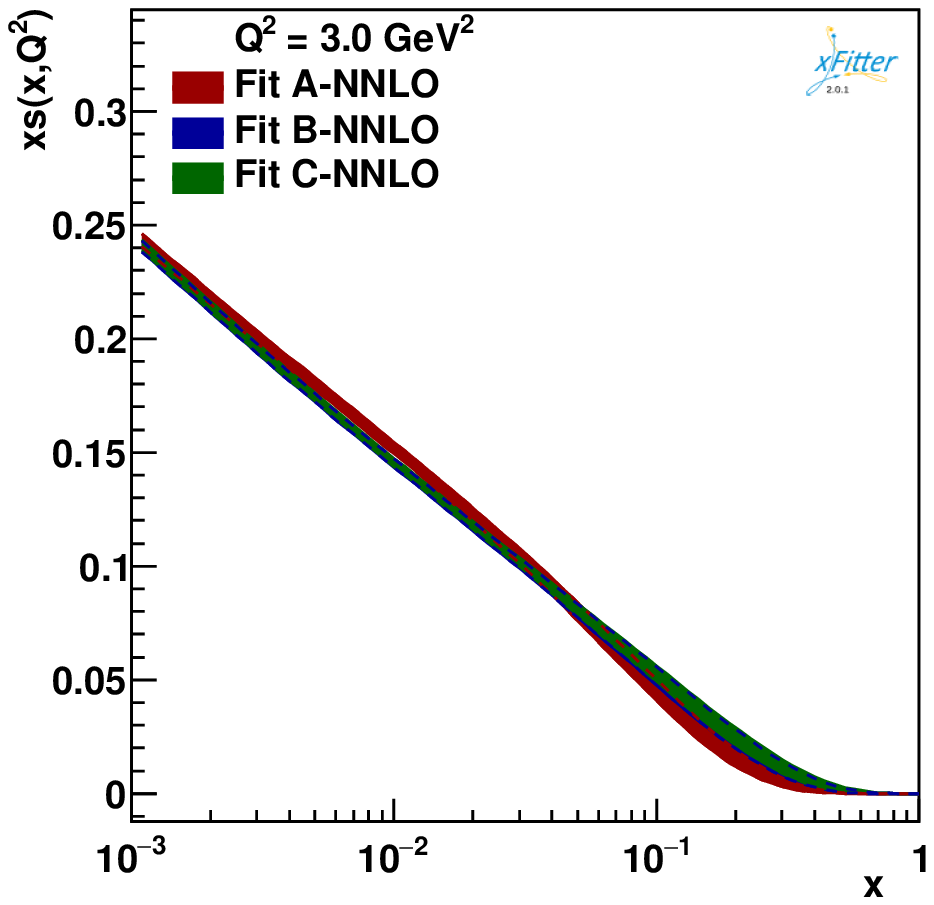}		
	    \includegraphics[scale = 0.4]{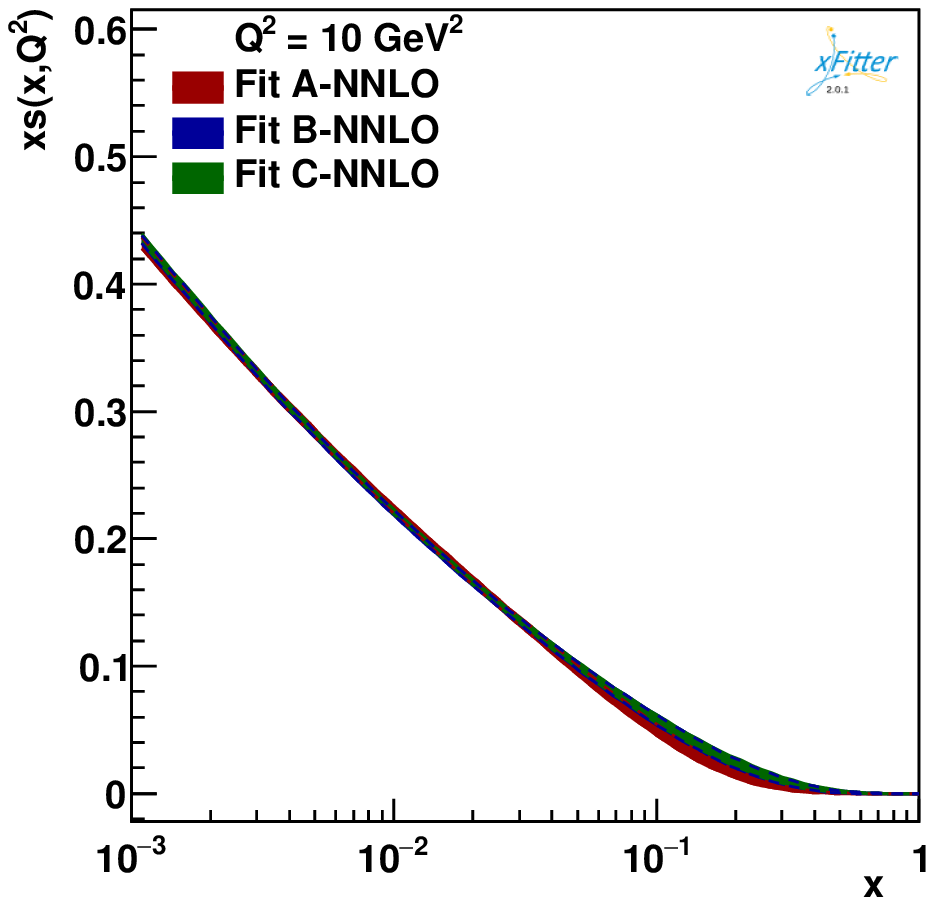}
	    \includegraphics[scale = 0.4]{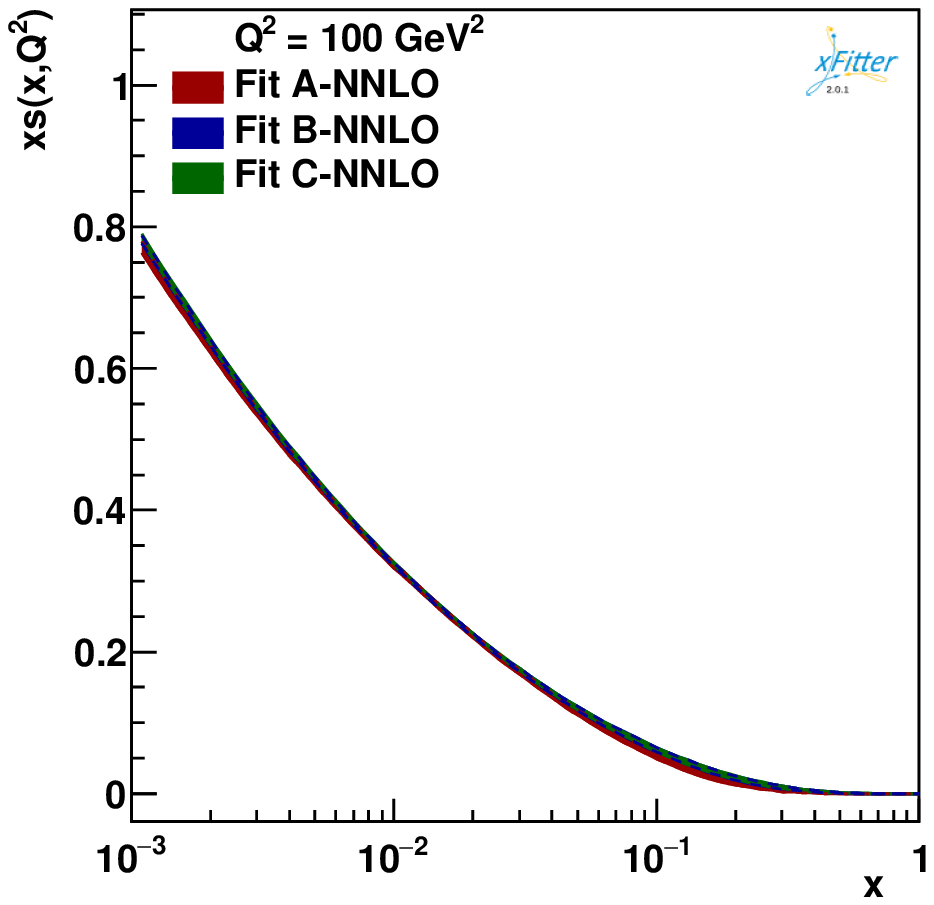}

	    \includegraphics[scale = 0.4]{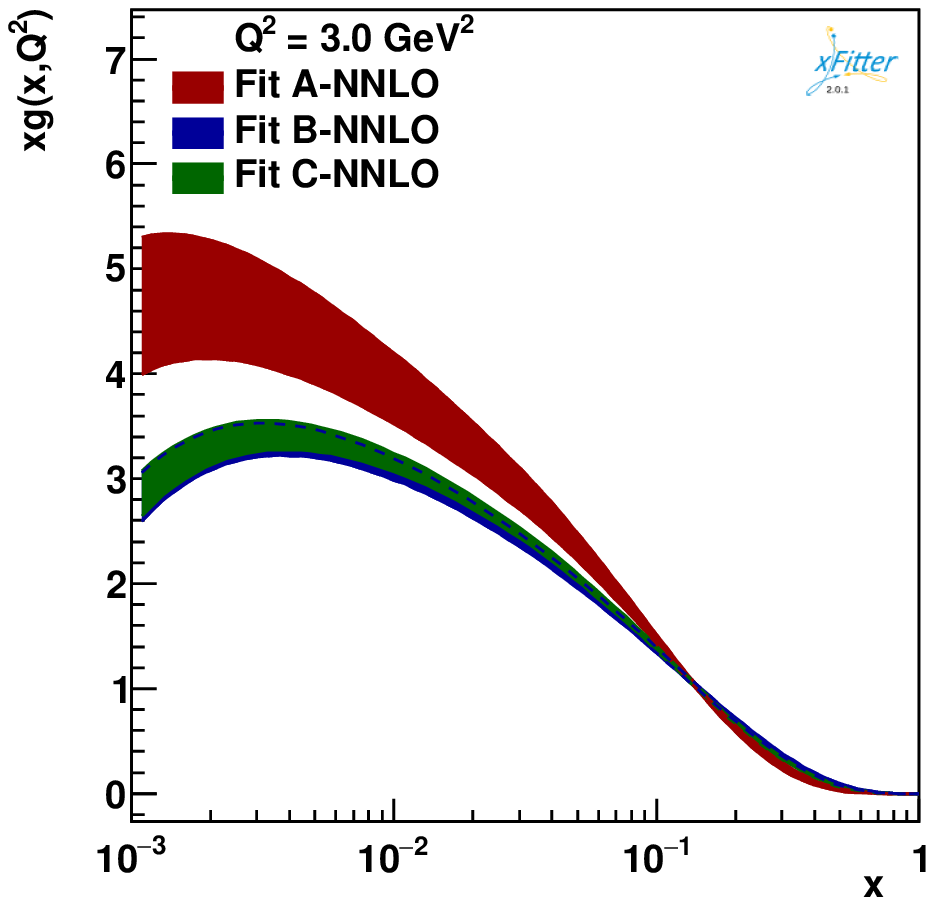}		
	    \includegraphics[scale = 0.4]{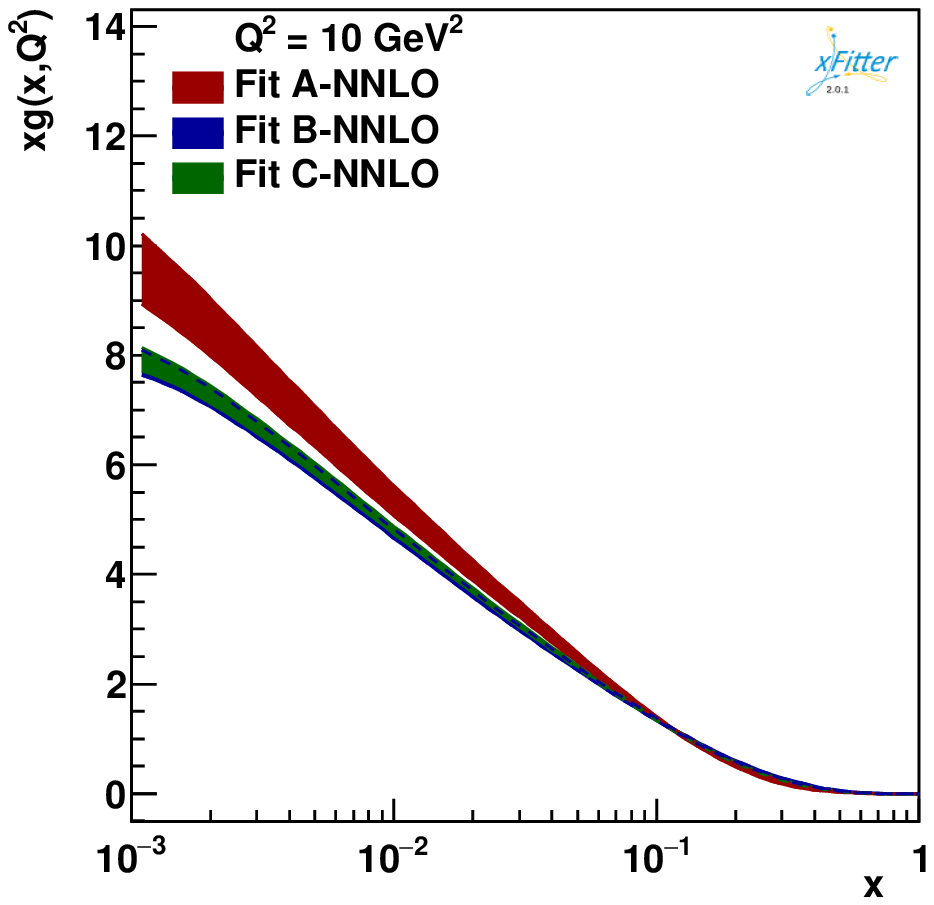}
	    \includegraphics[scale = 0.4]{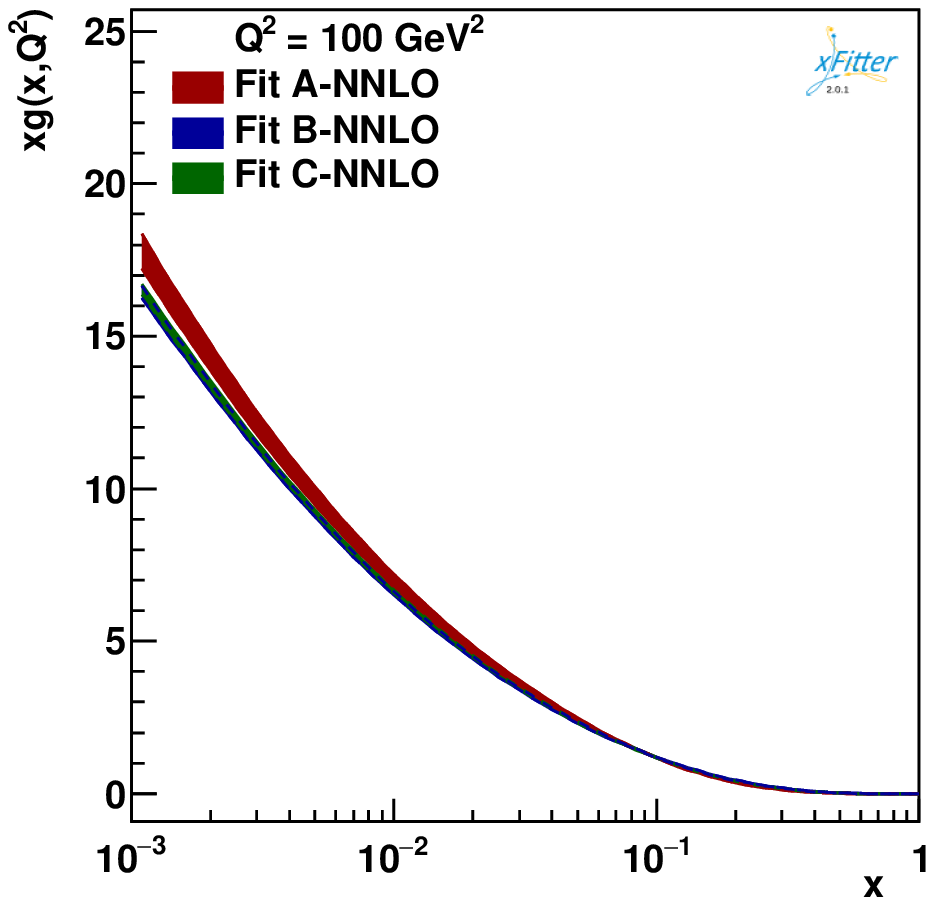}
	    	    	    	    		                        
		\caption{The NNLO parton distribution of $xu_v$, $xd_v$, $x\bar{u}$, $x\bar{d}$, $xs$, and $xg$,  as a  function of $x$ and for different values of $Q^2=$3, 10 and 100 GeV$^2$.}
		\label{fig:PDF-Qdep-NNLO}
	\end{center}
\end{figure*}

\begin{figure*}[!htb]
	\begin{center}
		\includegraphics[scale = 0.4]{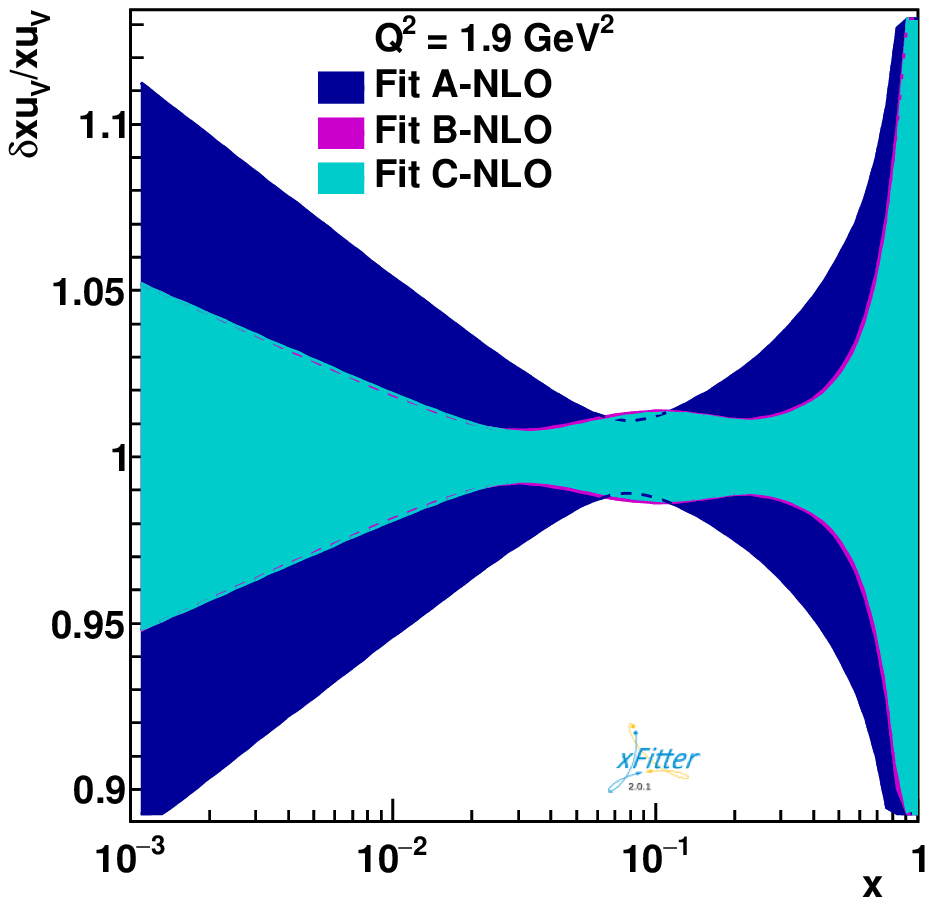}	
		\includegraphics[scale = 0.4]{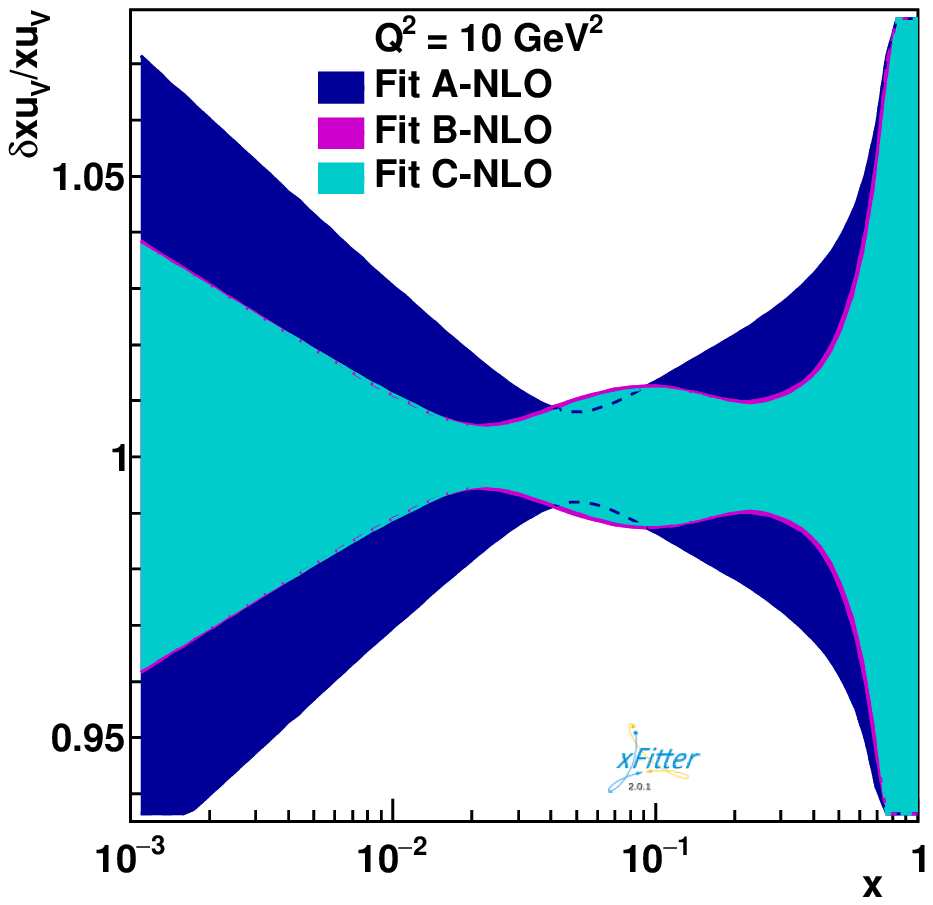}		
	    \includegraphics[scale = 0.4]{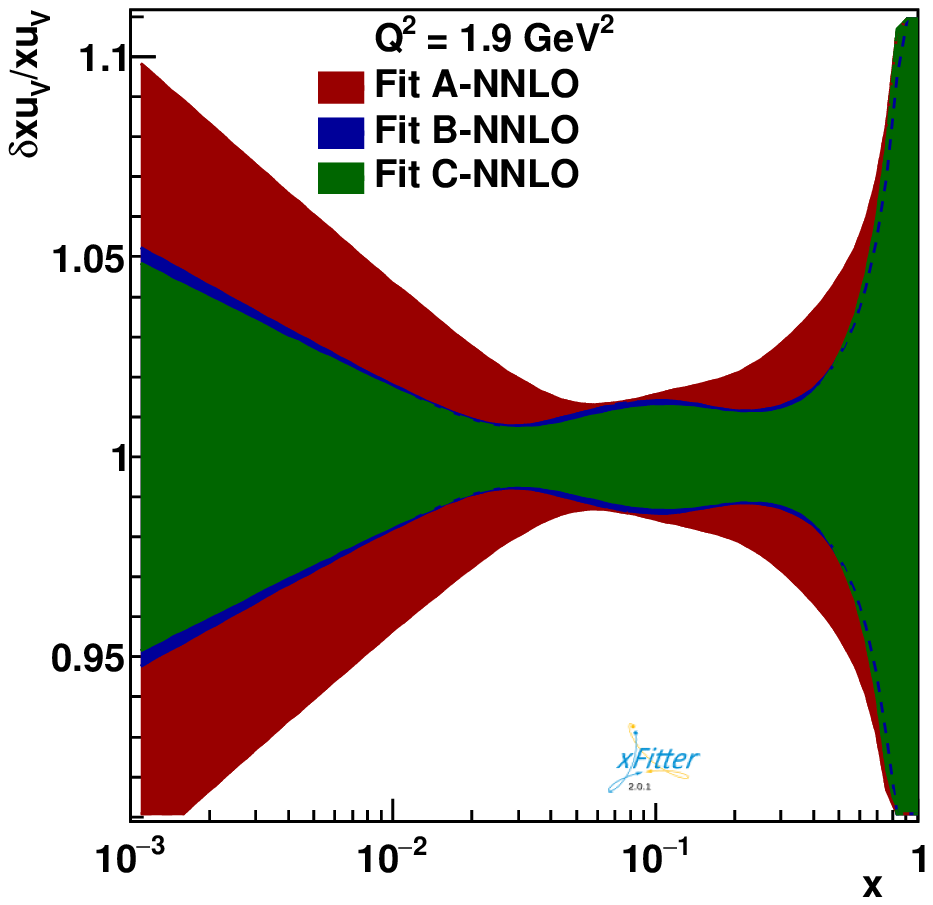}
	    \includegraphics[scale = 0.4]{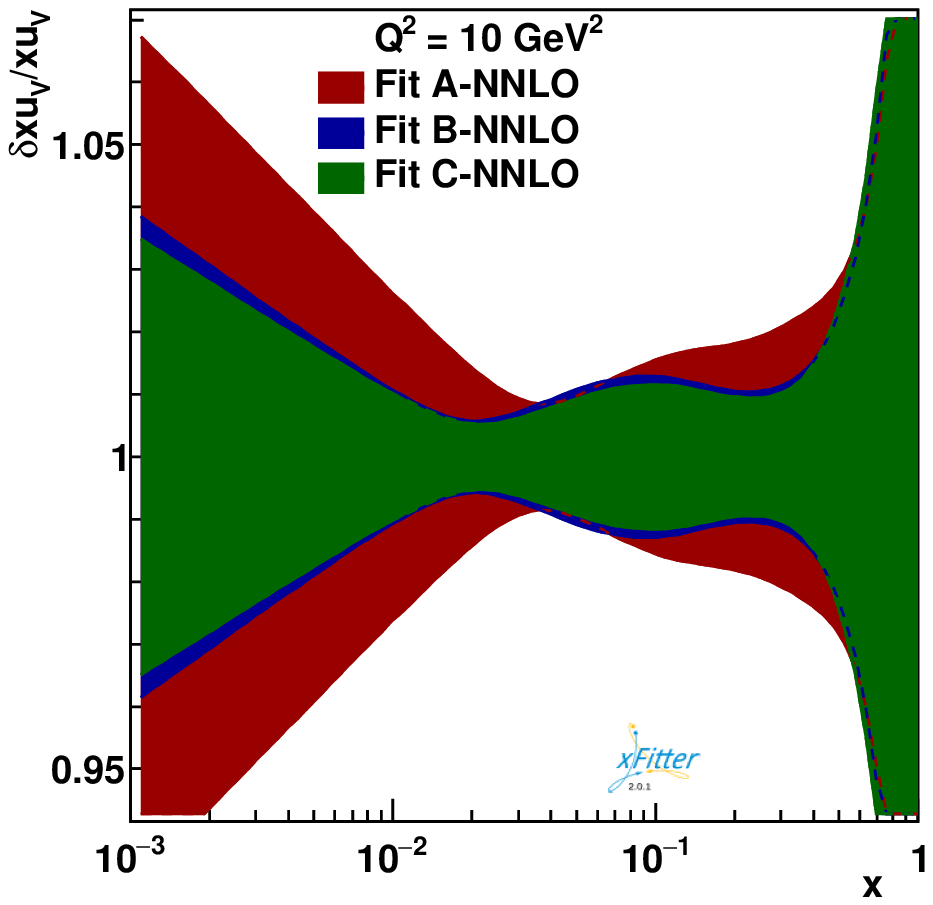}

		\includegraphics[scale = 0.4]{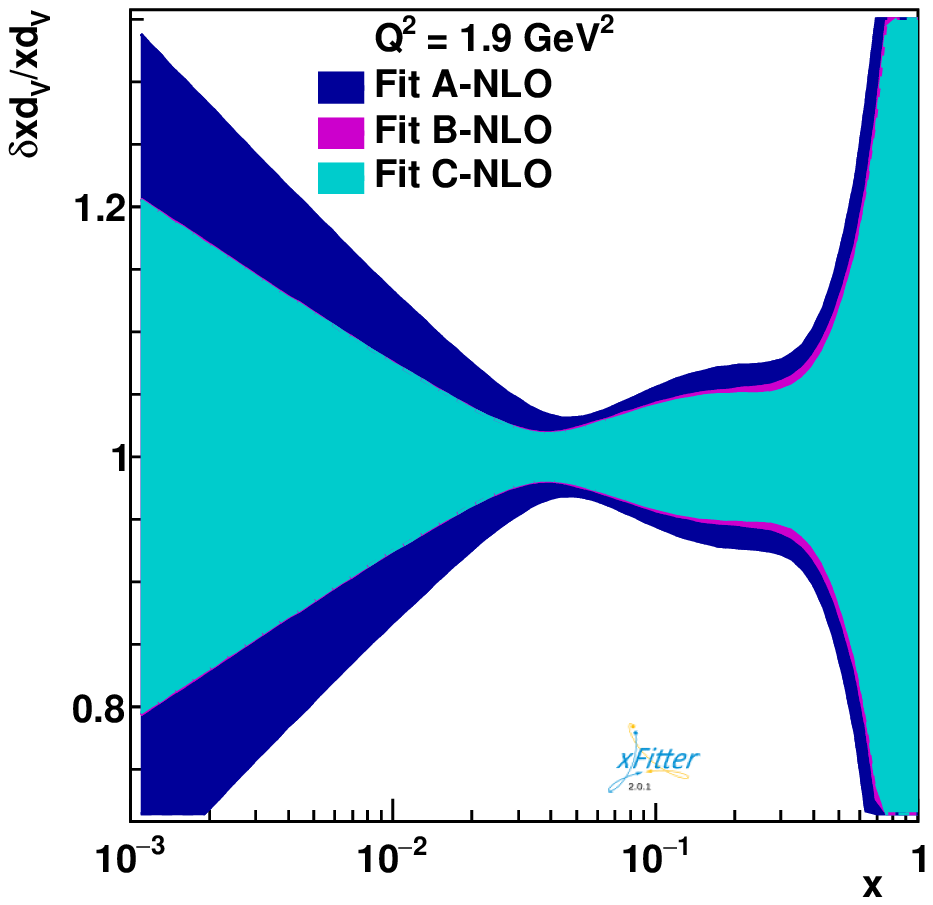}
		\includegraphics[scale = 0.4]{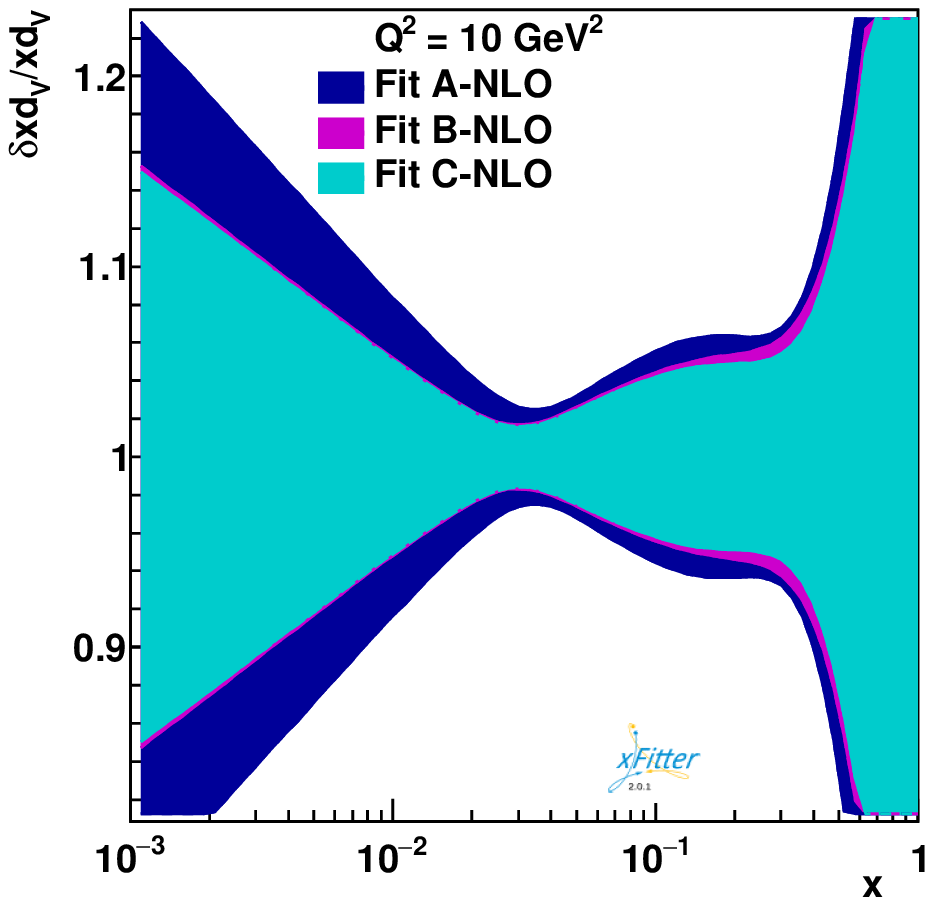}	 
	    \includegraphics[scale = 0.4]{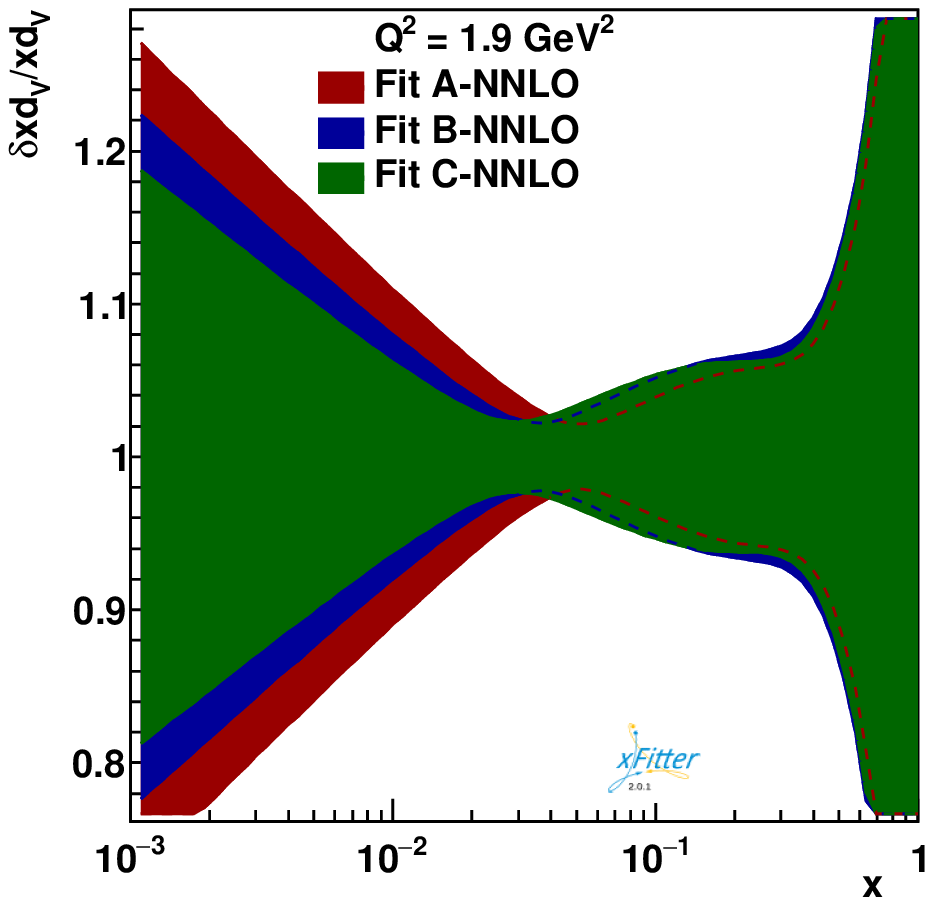}	
	    \includegraphics[scale = 0.4]{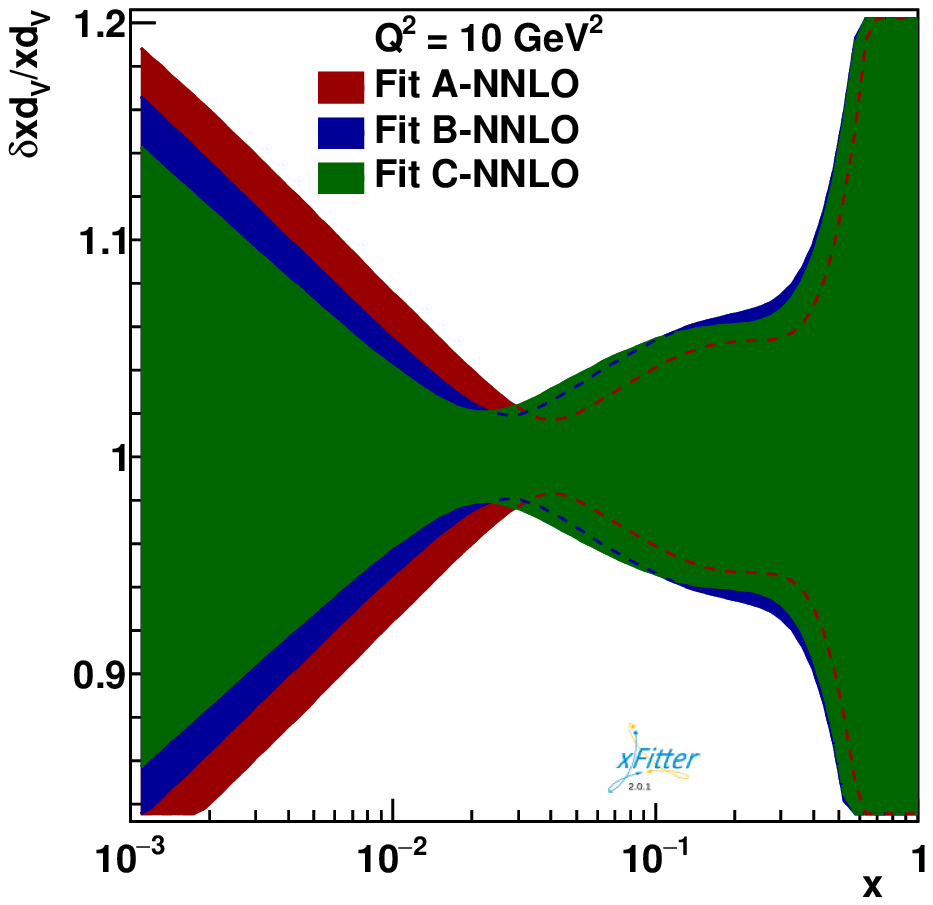}
	    
		\includegraphics[scale = 0.4]{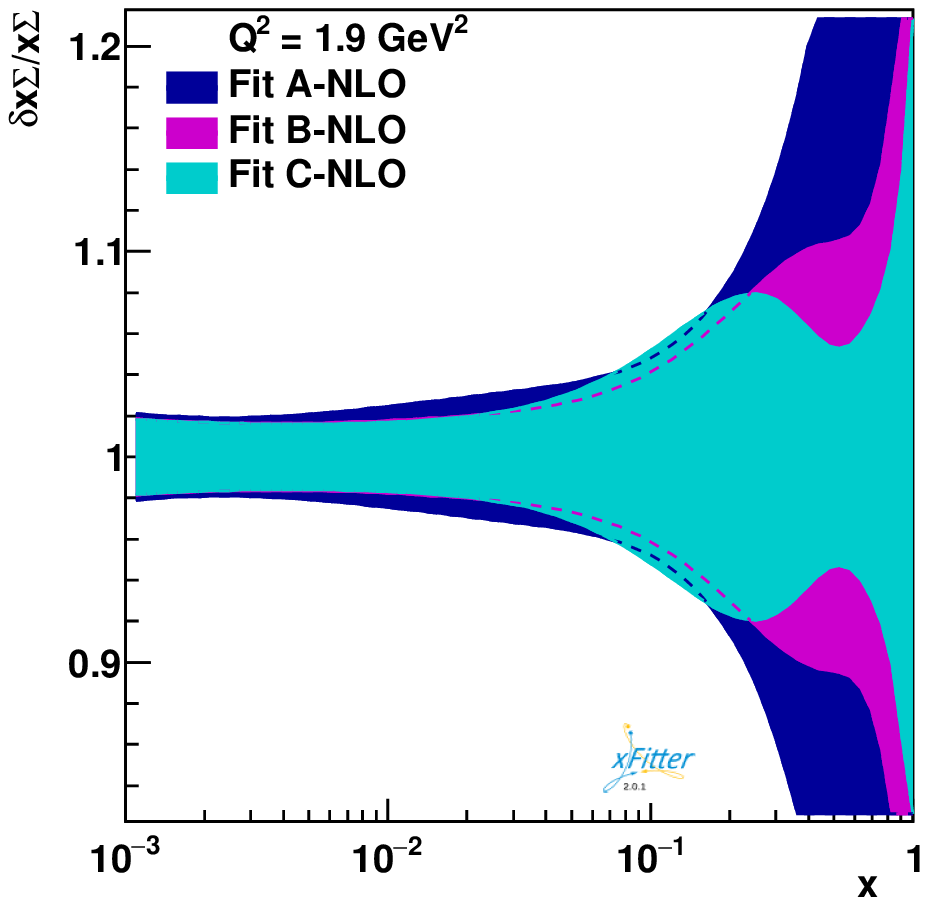}
		\includegraphics[scale = 0.4]{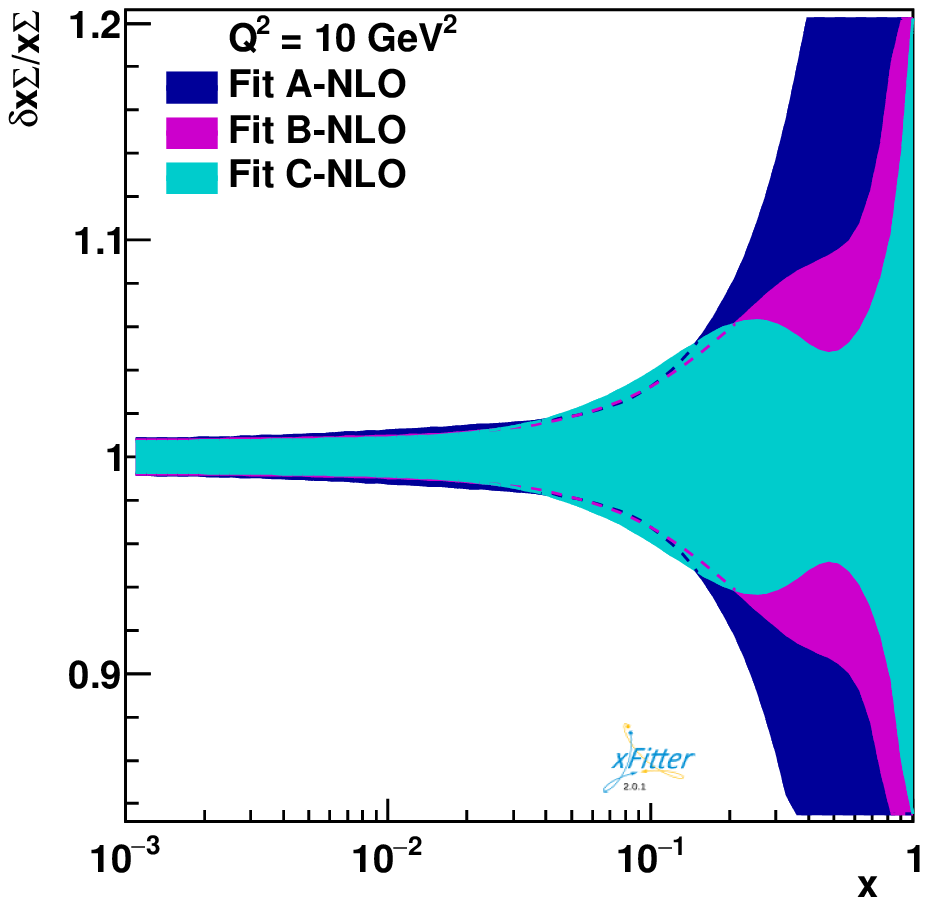}
	    \includegraphics[scale = 0.4]{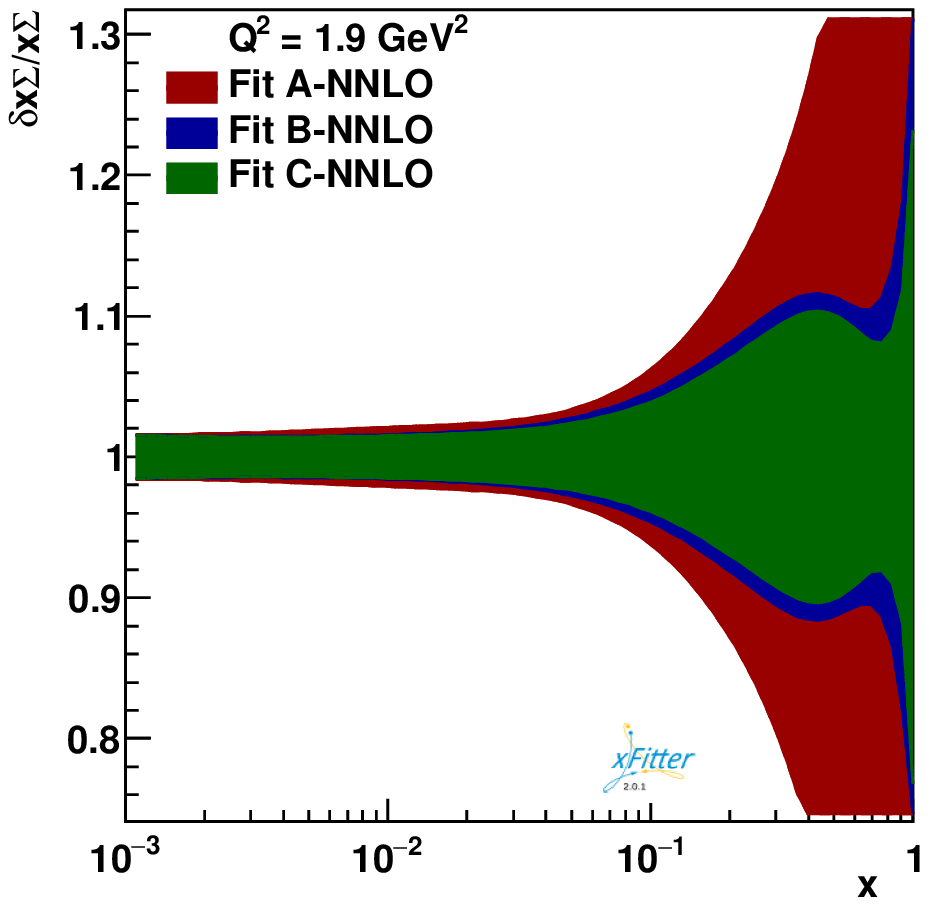}
	    \includegraphics[scale = 0.4]{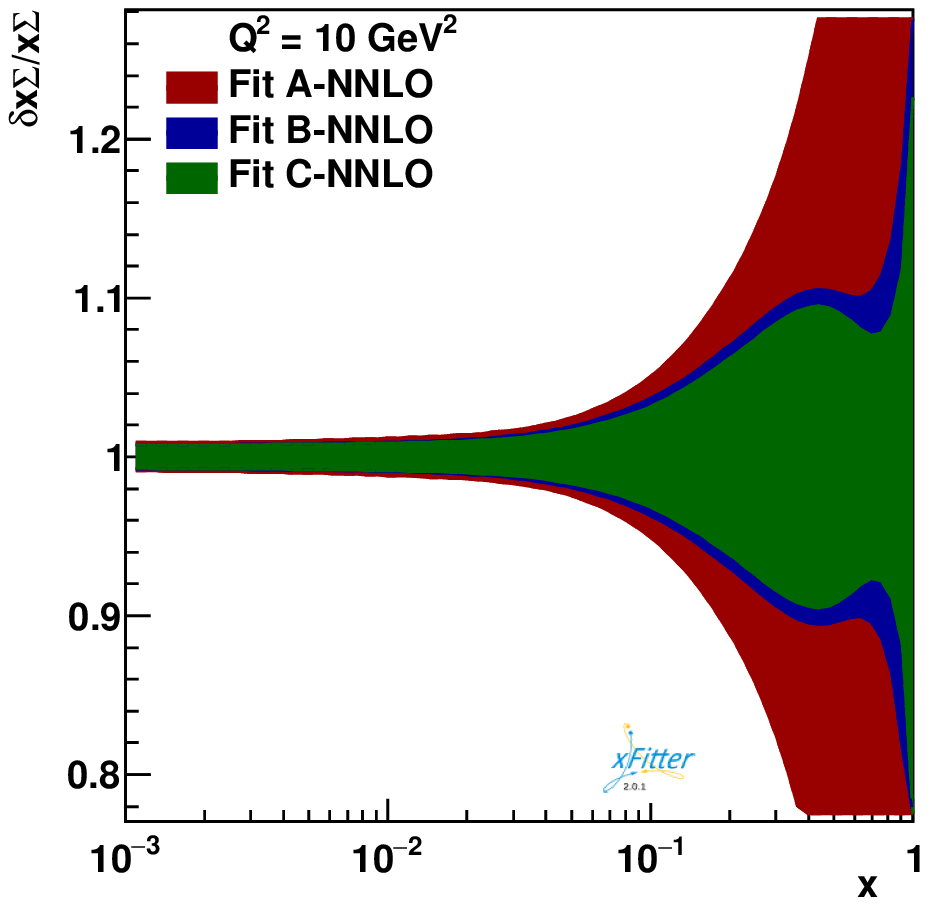} 	
	    
	    \includegraphics[scale = 0.4]{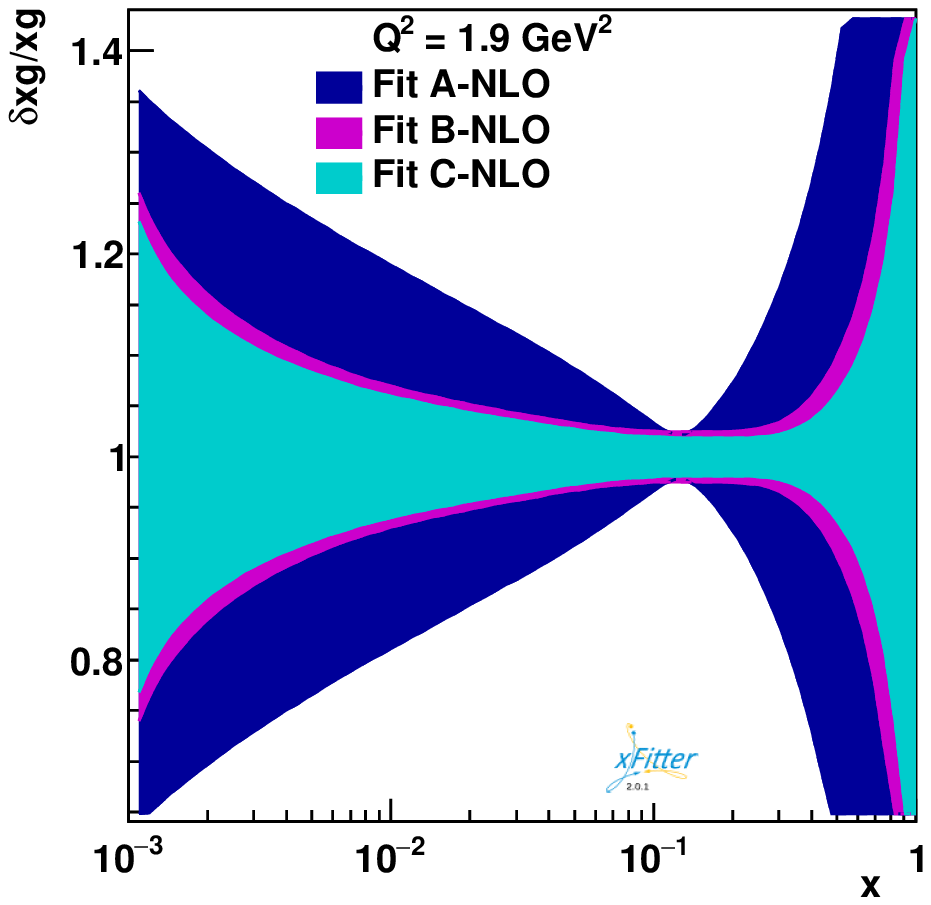}	
		\includegraphics[scale = 0.4]{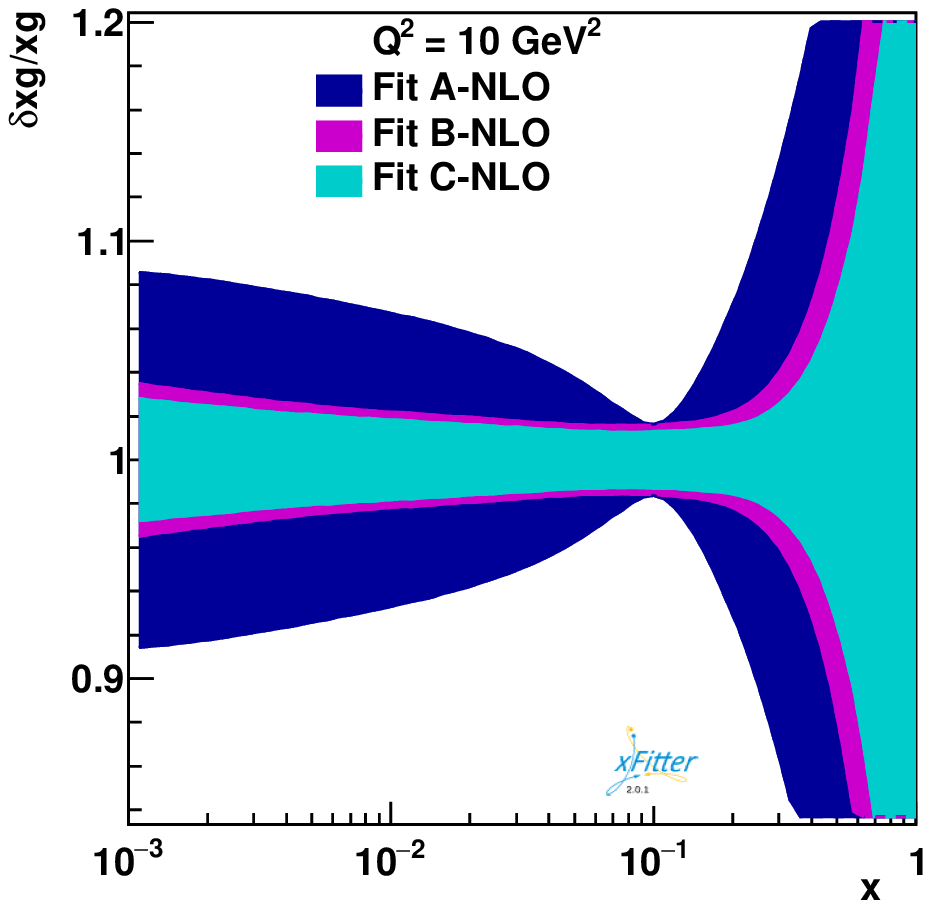}
		\includegraphics[scale = 0.4]{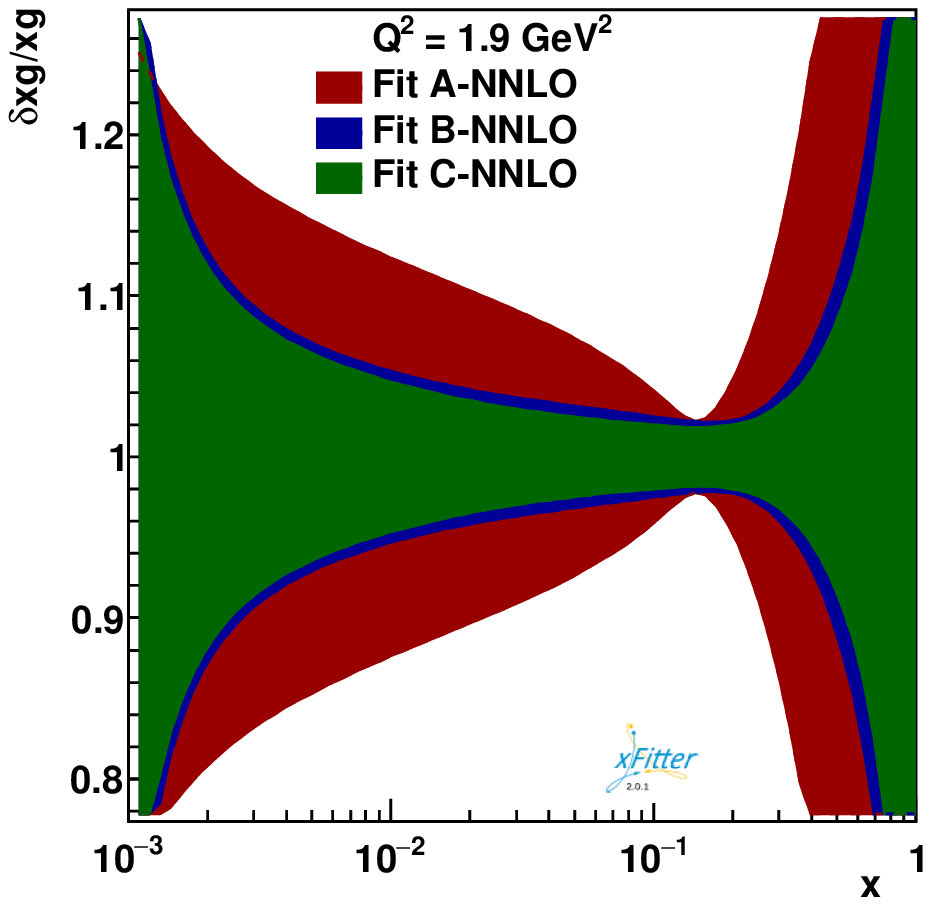}
		\includegraphics[scale = 0.4]{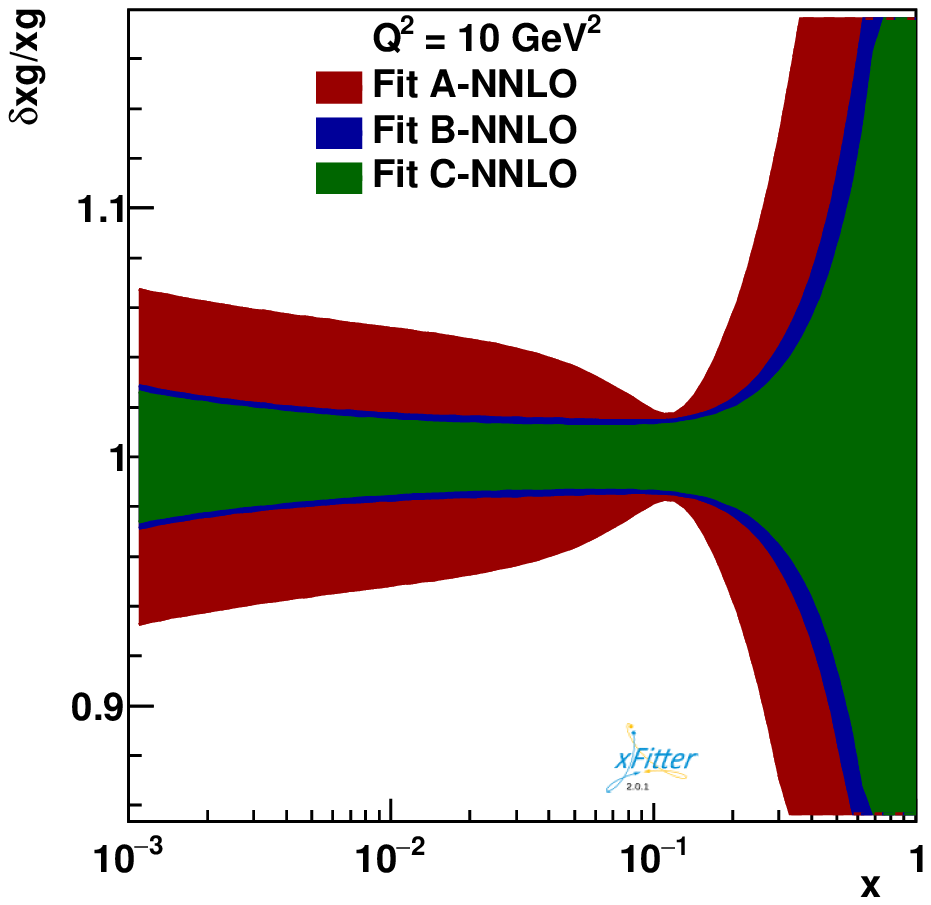}

		\caption{The NLO and NNLO results of the relative
uncertainties $\delta xq(x,Q^2)/xq(x,Q^2)$ for $q=u_v, d_v, \Sigma,$ and $g$  at the selected scales $Q^{2}$= 1.9,  10  GeV$^{2}$  as a function of $x$ and for individual Fits A,B, and C. In the left panels, we present our results for NLO, whereas the right panels are for NNLO.}
		\label{fig:PDF--RelUncer-NLO-NNLO}
	\end{center}
\end{figure*}

In Figs.~\ref{fig:PDF-Qdep-NLO} and \ref{fig:PDF-Qdep-NNLO}, the results of PDFs based on our three fits,  as a function of $x$ for $Q^{2}$= 3 GeV$^{2}$, $Q^{2}$= 10 GeV$^{2}$, and $Q^{2}$= 100 GeV$^{2}$ are presented at NLO and NNLO, respectively.  According to these figures, there are not significant changes in the valence and sea PDFs. 
To investigate the specific impact of the HERA I+II, non-LHC, and LHC data on PDFs in  Figs.~\ref{fig:PDF-Qdep-NLO} and \ref{fig:PDF-Qdep-NNLO}, 
we  need to compare our results for the relative uncertainties  of $\delta xq(x,Q^2)/xq(x,Q^2)$. 
In Fig.~\ref{fig:PDF--RelUncer-NLO-NNLO}, we present the results of the relative
uncertainties $\delta xq(x,Q^2)/xq(x,Q^2)$ for $q=u_v, d_v, \Sigma,$ and $g$  for the selected scales $Q^{2}$= 1.9,  10  GeV$^{2}$  as a function of $x$ and for our individual Fits A, B, and C. In the left panels, we present our results for NLO, whereas the right panels are for NNLO. 
According to this figure, the changes of the PDFs uncertainties are observed, especially for $\delta x\Sigma(x,Q^2)/x\Sigma(x,Q^2)$ at large $x$  and $\delta xg(x,Q^2)/xg(x,Q^2)$ for both low and large  parton $x$ values and for NLO and NNLO.  In this figure also, the relative uncertainties ratio for  valence PDFs  are decreased at low values of $x$ from Fit A to Fit C, obviously.  This figure illustrates the impact of the non-LHC and LHC data on PDFs in different ranges of $x$ values when theses data sets are added to HERA data.

\begin{figure*}[!htb]
	\begin{center}
		\includegraphics[scale = 0.4]{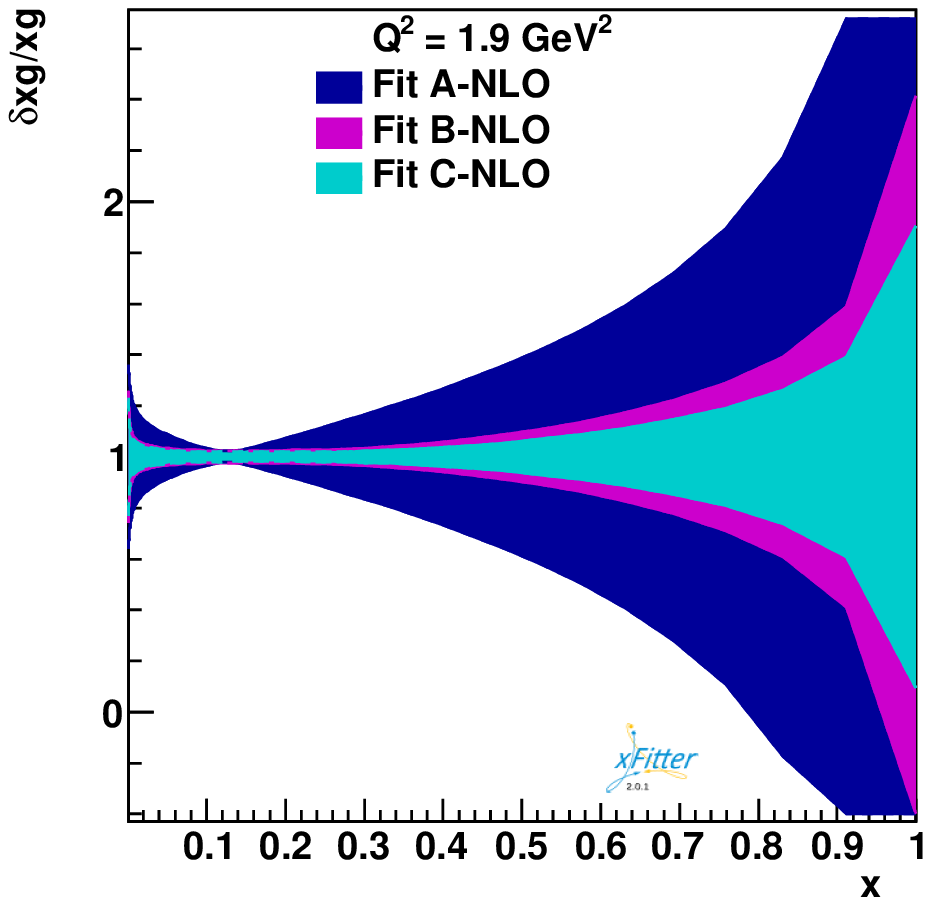}
		\includegraphics[scale = 0.4]{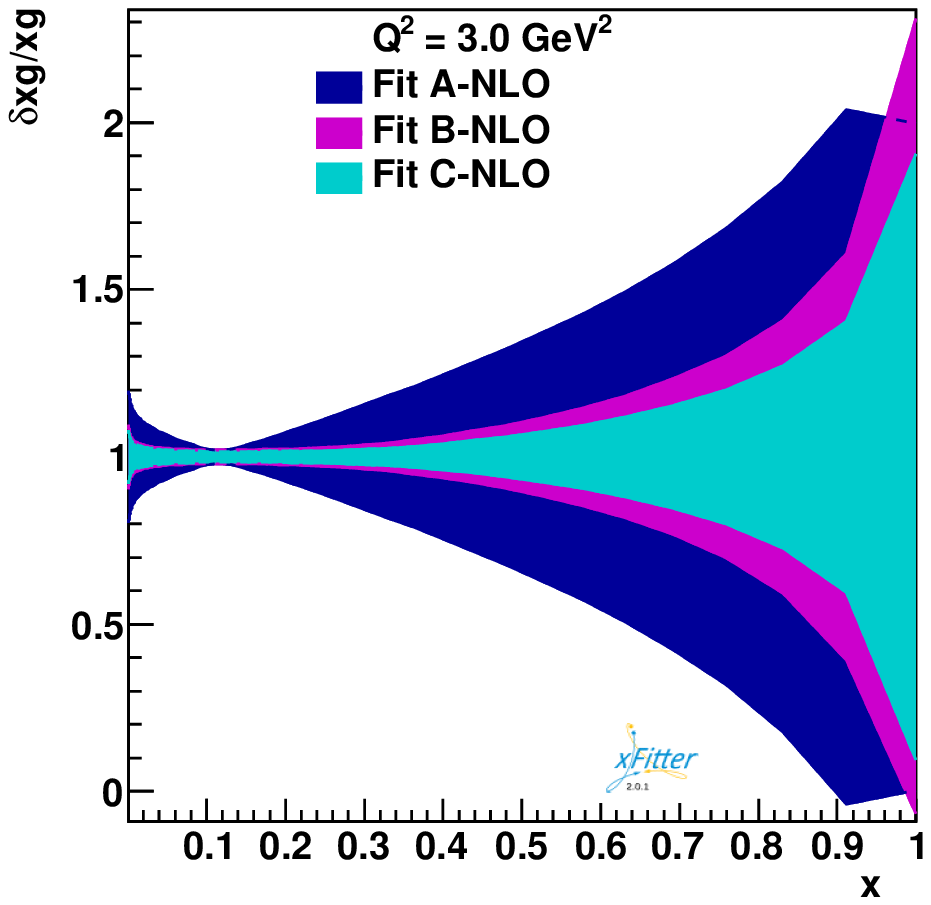}				   
		\includegraphics[scale = 0.4]{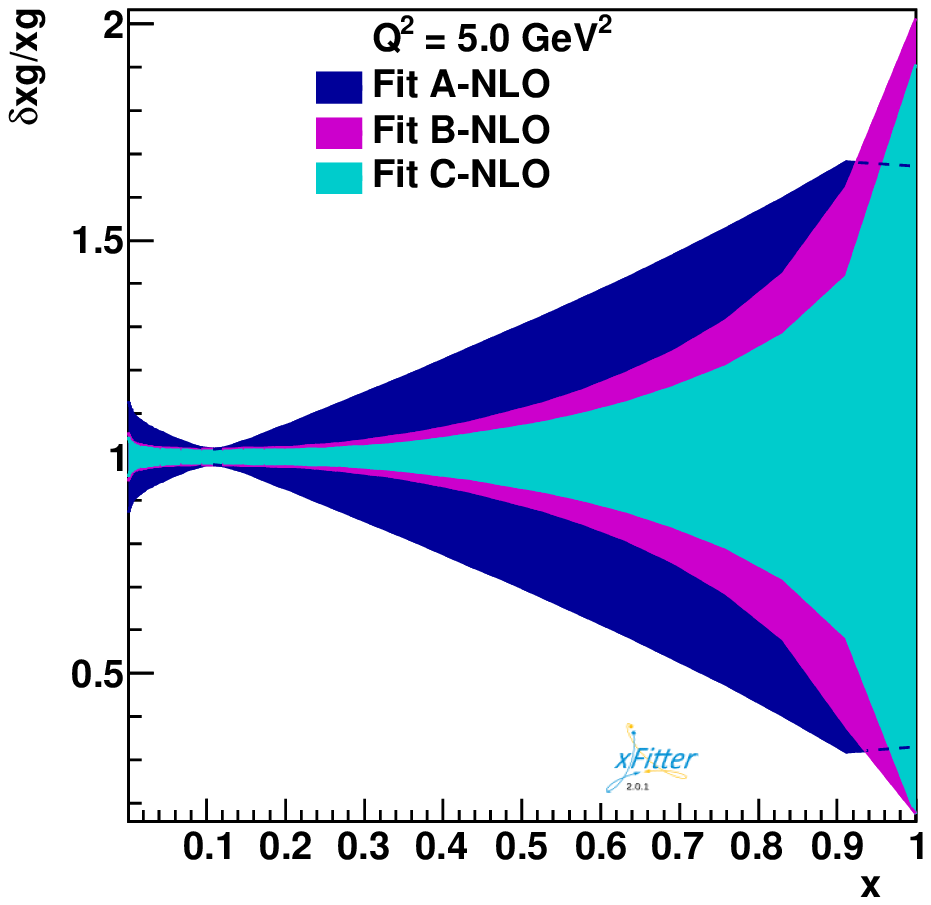}
        \includegraphics[scale = 0.4]{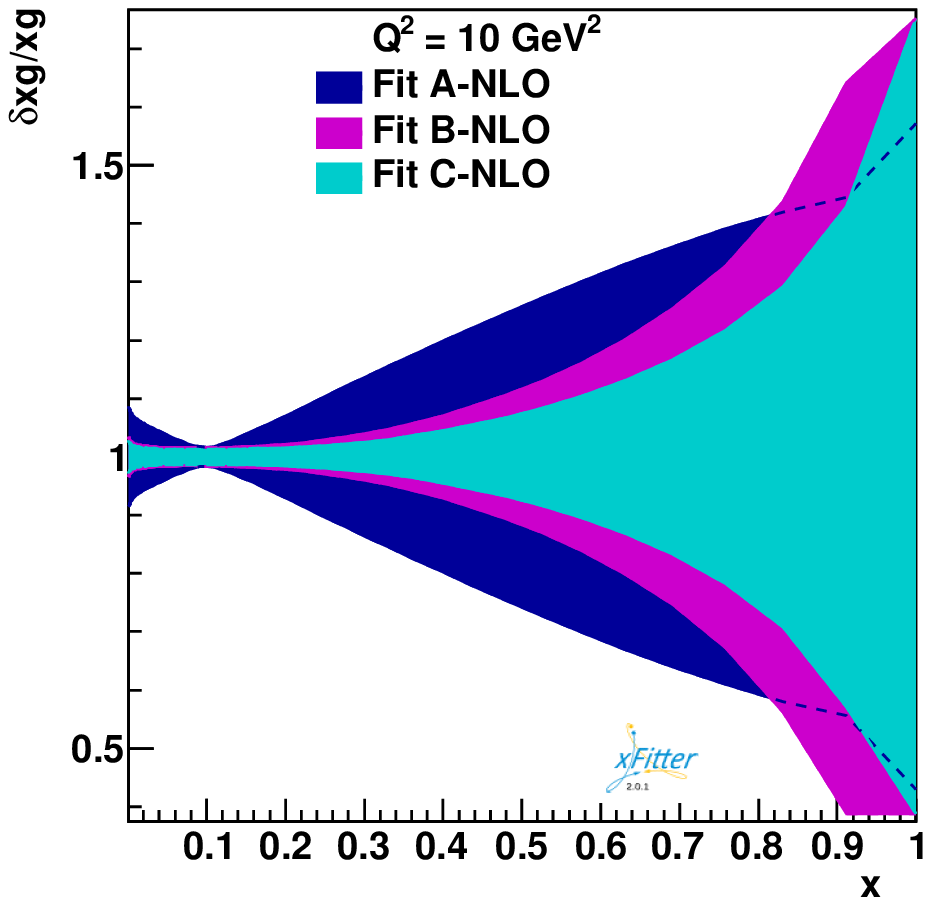}
		
		\includegraphics[scale = 0.4]{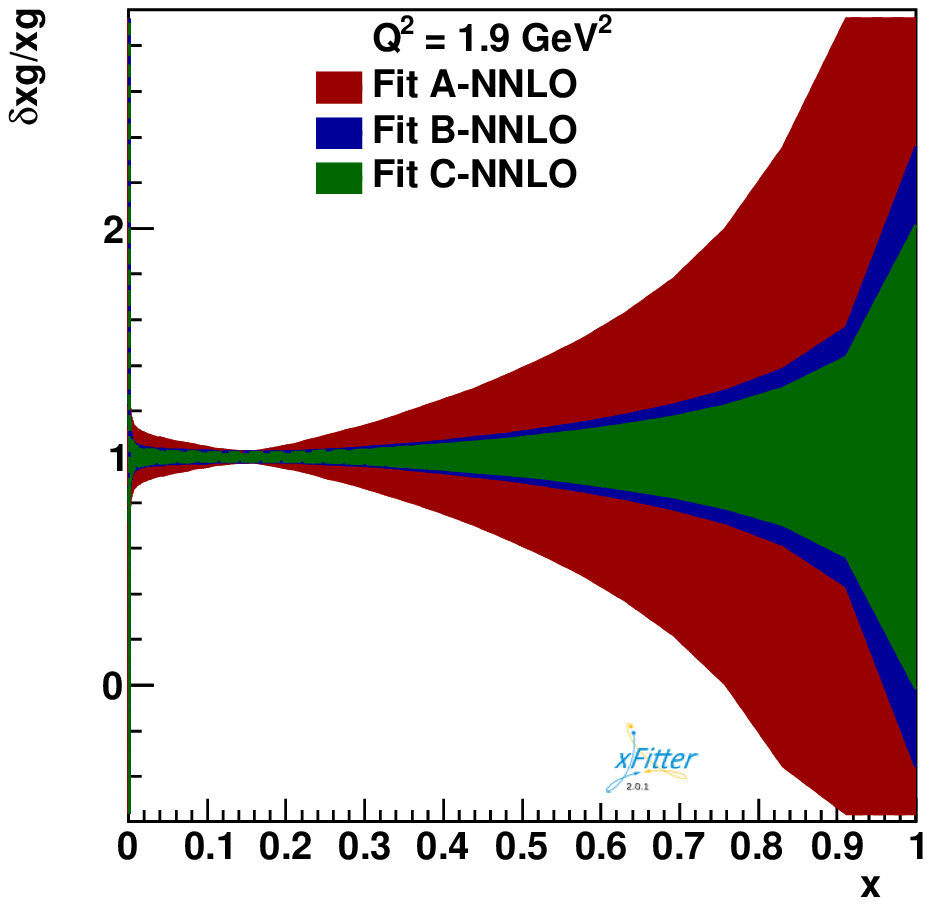}
	    \includegraphics[scale = 0.4]{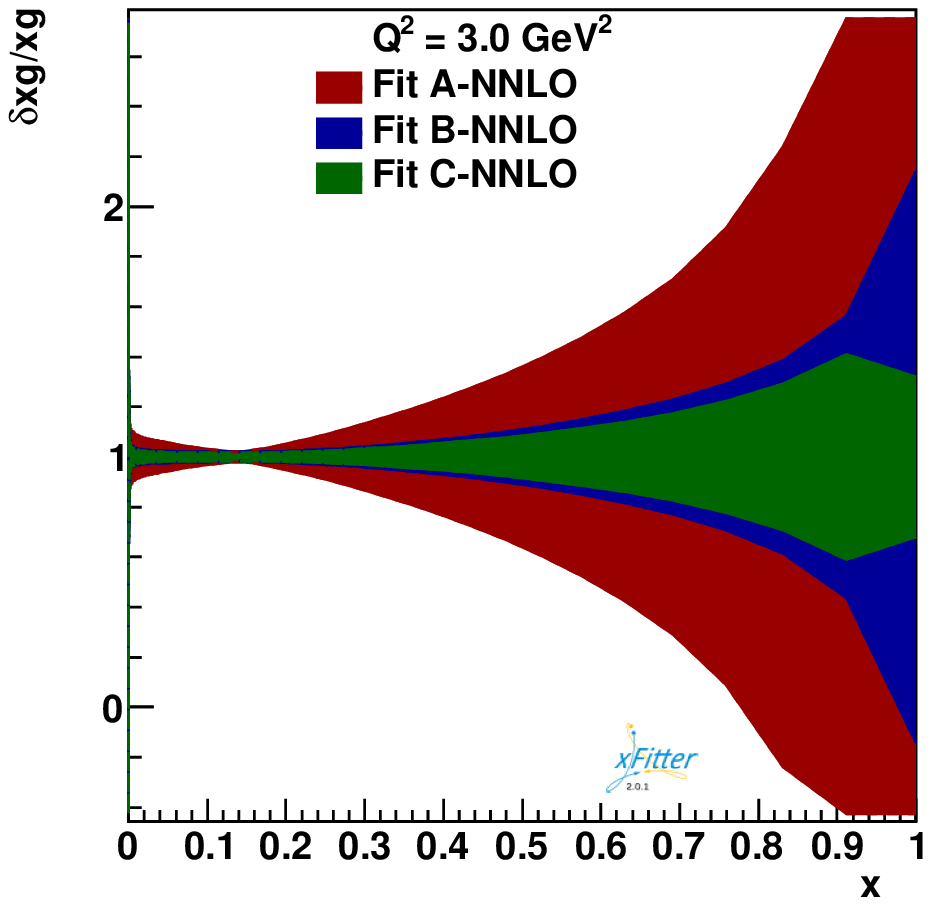}       	     
        \includegraphics[scale = 0.4]{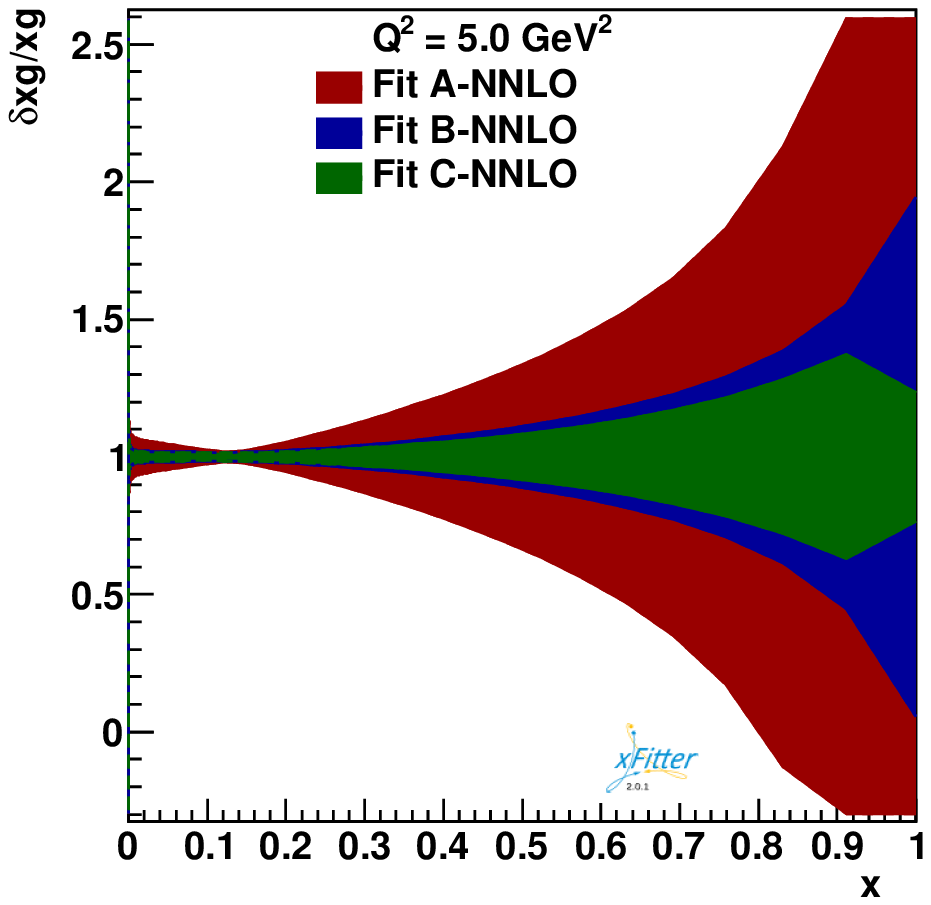}   	            
        \includegraphics[scale = 0.4]{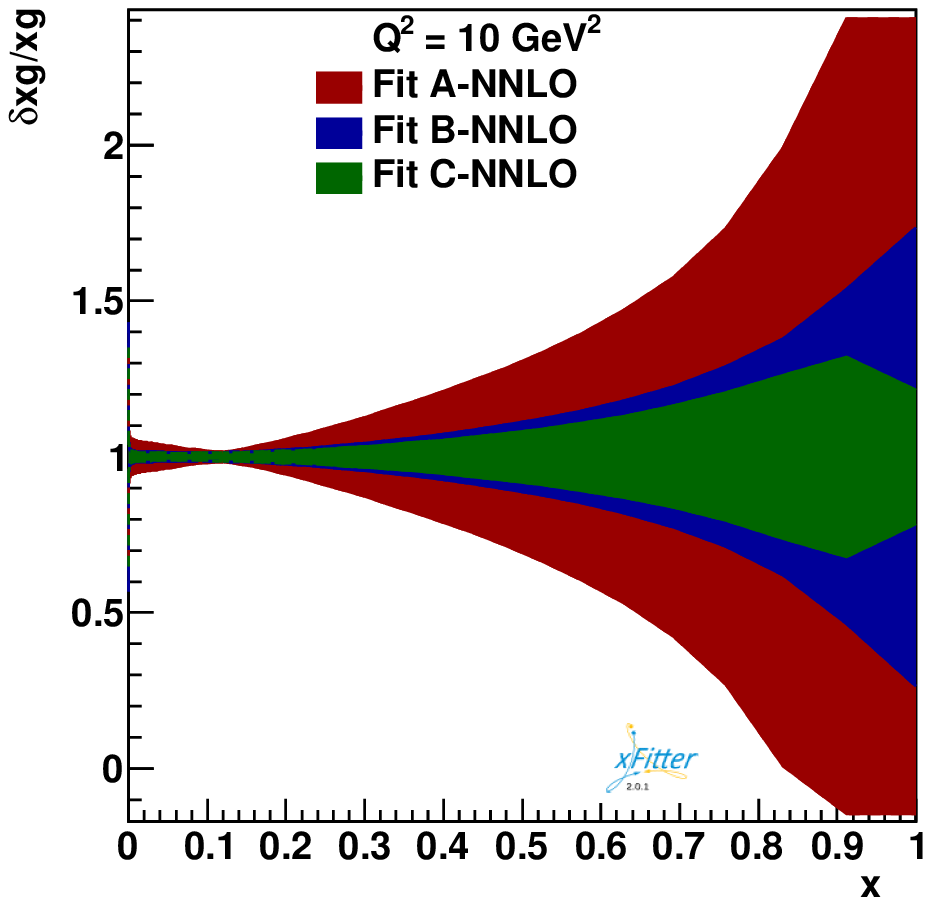}		
		
		\caption{The impact of the LHC data (Fit C) on the high-$x$  relative
uncertainties $\delta xg(x,Q^2)/xg(x,Q^2)$ at NLO and NNLO with comparison of HERA I+II (Fit A), Non-LHC data (Fit B), and LHC data (Fit C). The linear plots of gluon PDF relative ratio as a function of $x$ are presented at 1.9, 3, 5 and 10 GeV$^2$. In the up panels, we present our results for NLO, whereas the down panels are for NNLO.
}		
		\label{fig:PDF-gluon-RUR-NLO-NNLO}
	\end{center}
\end{figure*}

To investigate the specific impact of LHC data on gluon PDF at large $x$, we need also to present our results for the relative
uncertainties $\delta xg(x,Q^2)/xg(x,Q^2)$ in linear plots. Figure \ref{fig:PDF-gluon-RUR-NLO-NNLO} illustrates the NLO and NNLO  QCD fit results for gluon PDF as a function of $x$ at different scale values $Q^{2}$= 1.9,  3, 5, 10  GeV$^{2}$. In the up panels, we present our results for NLO, whereas the down panels are for NNLO.
In this figure  significant changes in the gluon PDF uncertainties at large $x$ region  in presence of HERA I+II (Fit A), non-LHC data (Fit B), and LHC data (Fit C) are obtained. Figure \ref{fig:PDF-gluon-RUR-NLO-NNLO}  illustrates clearly two significant outcomes. First, one can find how adding non-LHC and LHC data sets to HERA I + II impacts on reducing the gluon relative uncertainties at high values of $x$ significantly, as it is obvious from results of Fit C where the LHC experimental data are included in fit procedure. Second, a comparison between these diagrams  shows shifting from NLO to NNLO creates remarkable constrains on gluon PDFs of Fit C particularly at high-$x$.

\begin{figure*}[!htb]
	\begin{center}
		\includegraphics[scale = 0.4]{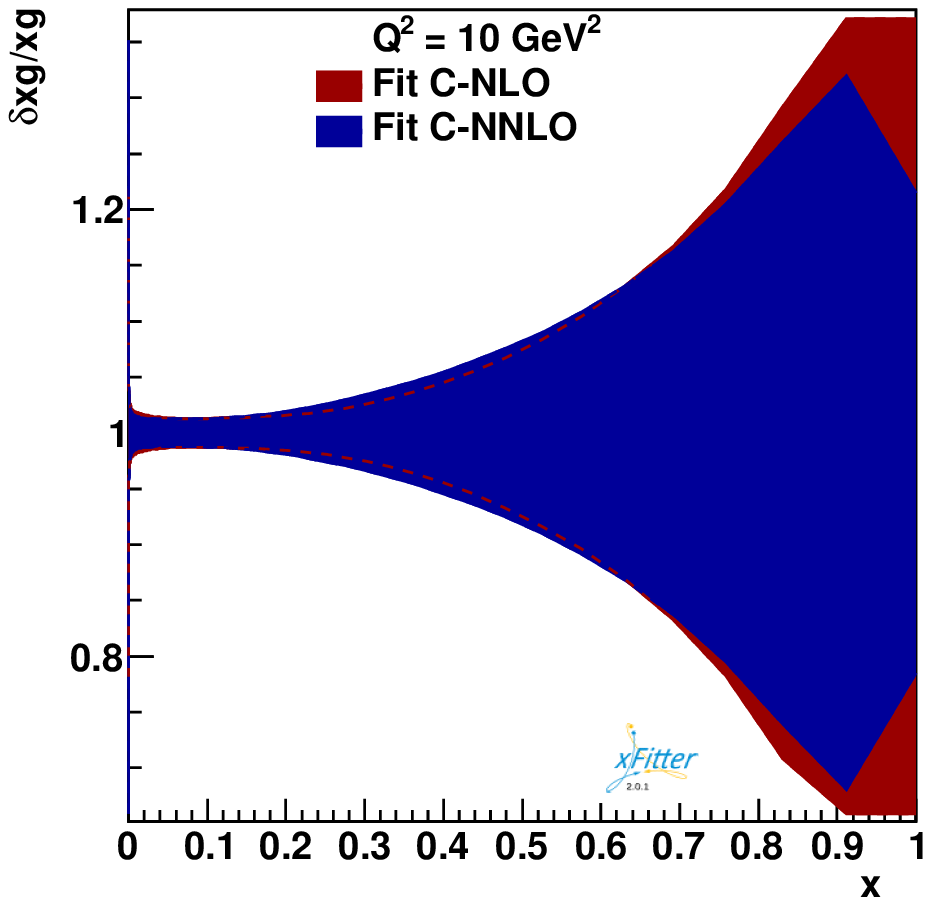}	
		\includegraphics[scale = 0.4]{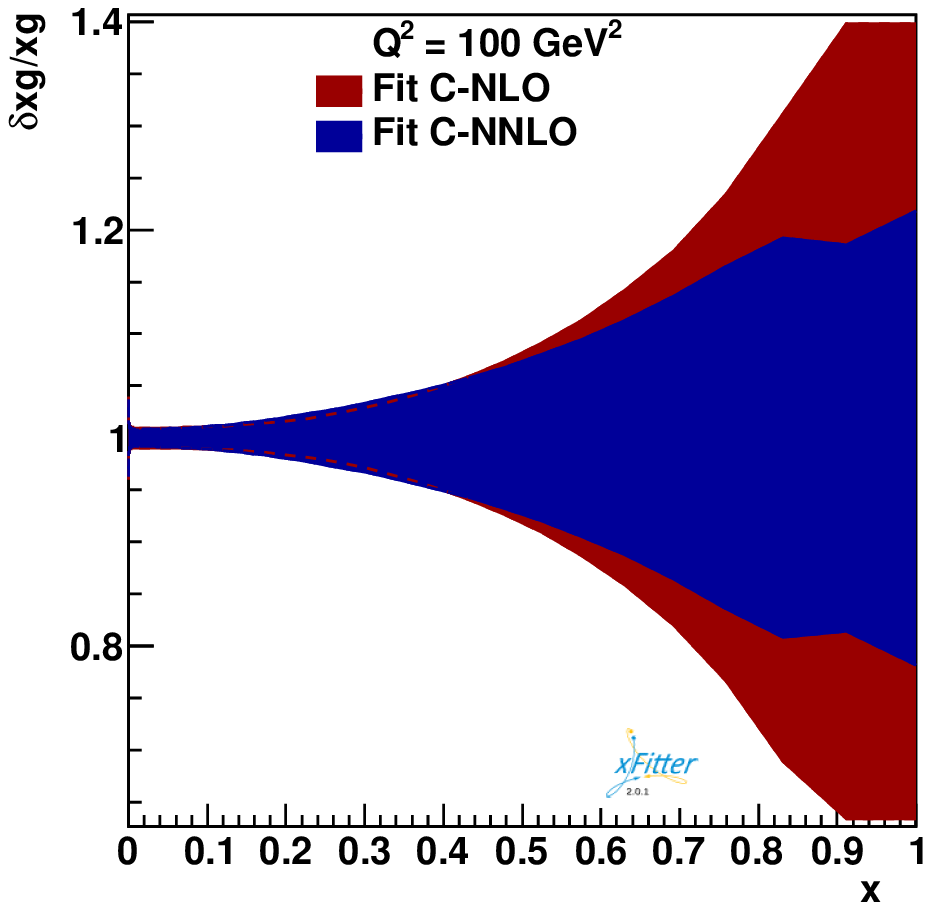}		
		\includegraphics[scale = 0.4]{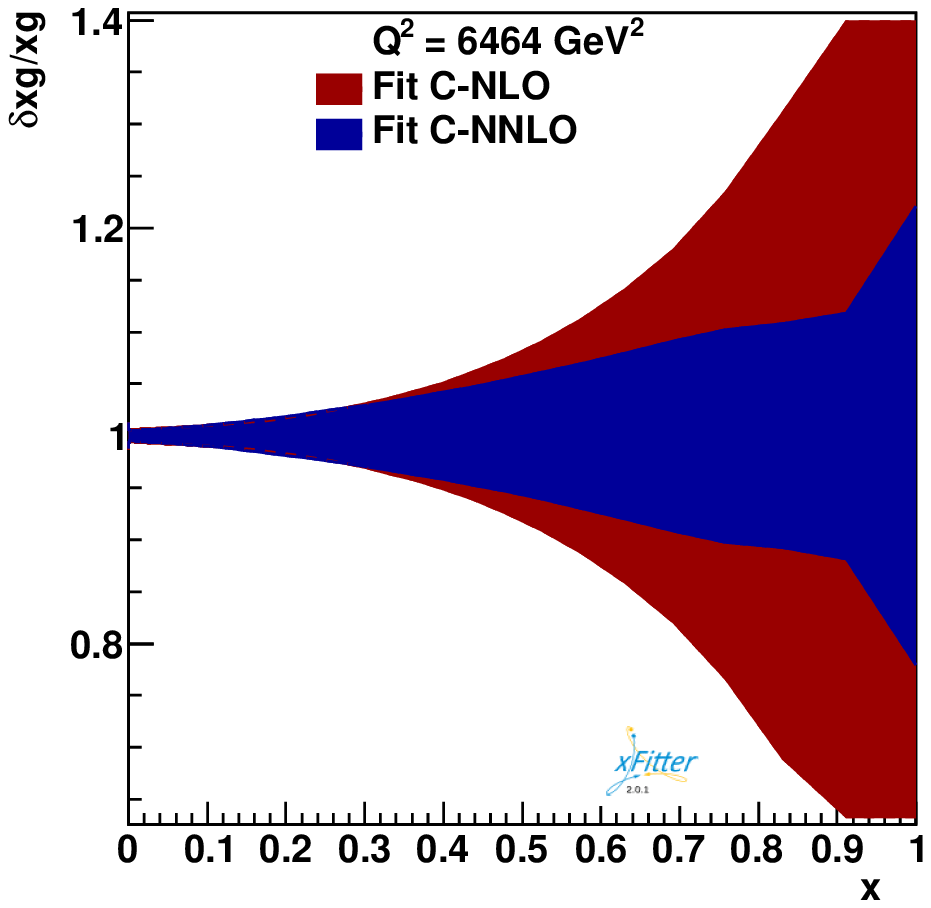}
		\includegraphics[scale = 0.4]{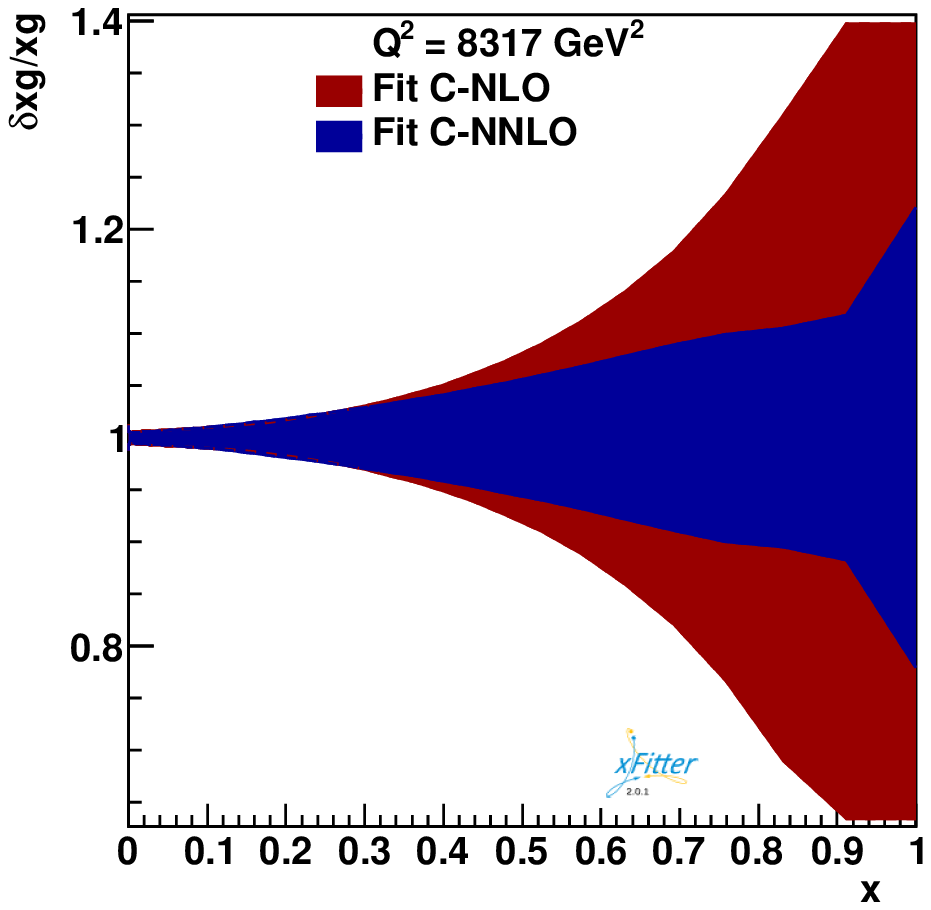}	            
		
		\caption{The comparison of NLO and NNLO gluon relative uncertainties $\delta xg(x,Q^2)/xg(x,Q^2)$ for Fit C as a function of $x$ at $Q^2$=10, 100, 6464, and 8317 GeV$^2$.}
		\label{fig:g-FitC-nlo-nnlo}
	\end{center}
\end{figure*}

For more clarification, a direct comparison between NLO and NNLO results at large $x$, the relative
uncertainties $\delta xg(x,Q^2)/xg(x,Q^2)$ for Fit C, as a completed fit procedure in this paper are presented in Fig.~\ref{fig:g-FitC-nlo-nnlo} for different $Q^2$=10, 100, 6464, and 8317 GeV$^2$ values.  

\begin{figure*}[!htb]
	\begin{center}
	    \includegraphics[scale = 0.4]{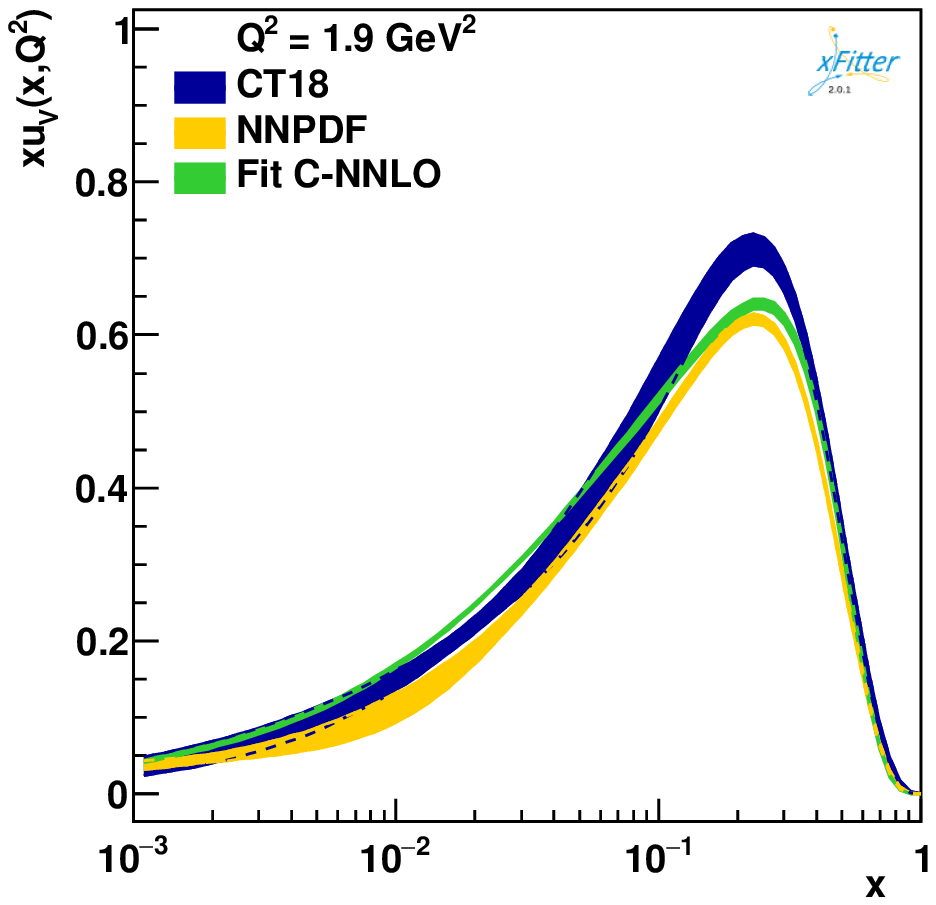}	
	    \includegraphics[scale = 0.4]{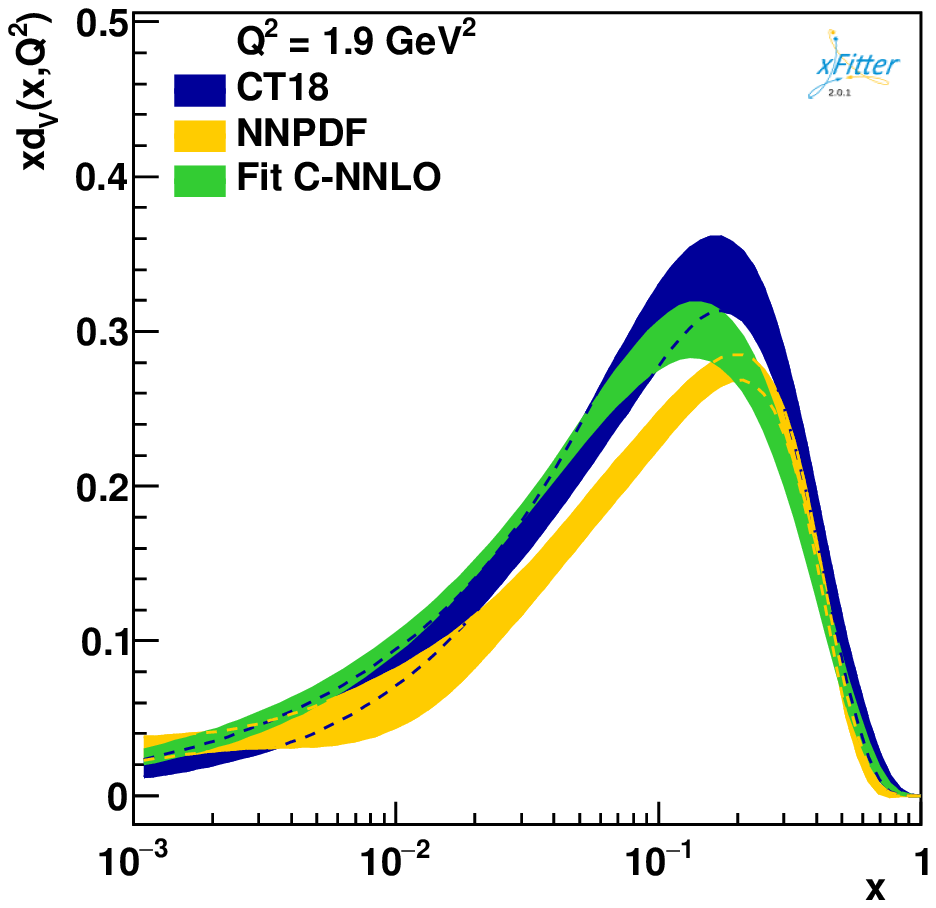}		
        \includegraphics[scale = 0.4]{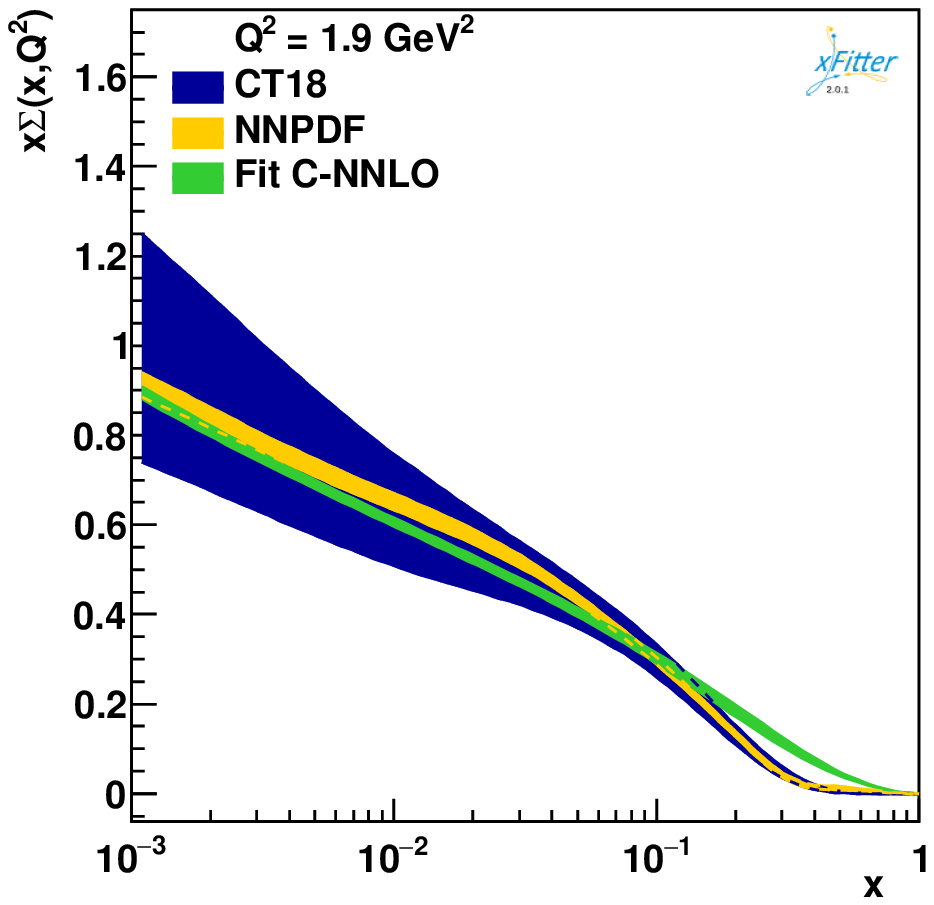}        
	    \includegraphics[scale = 0.4]{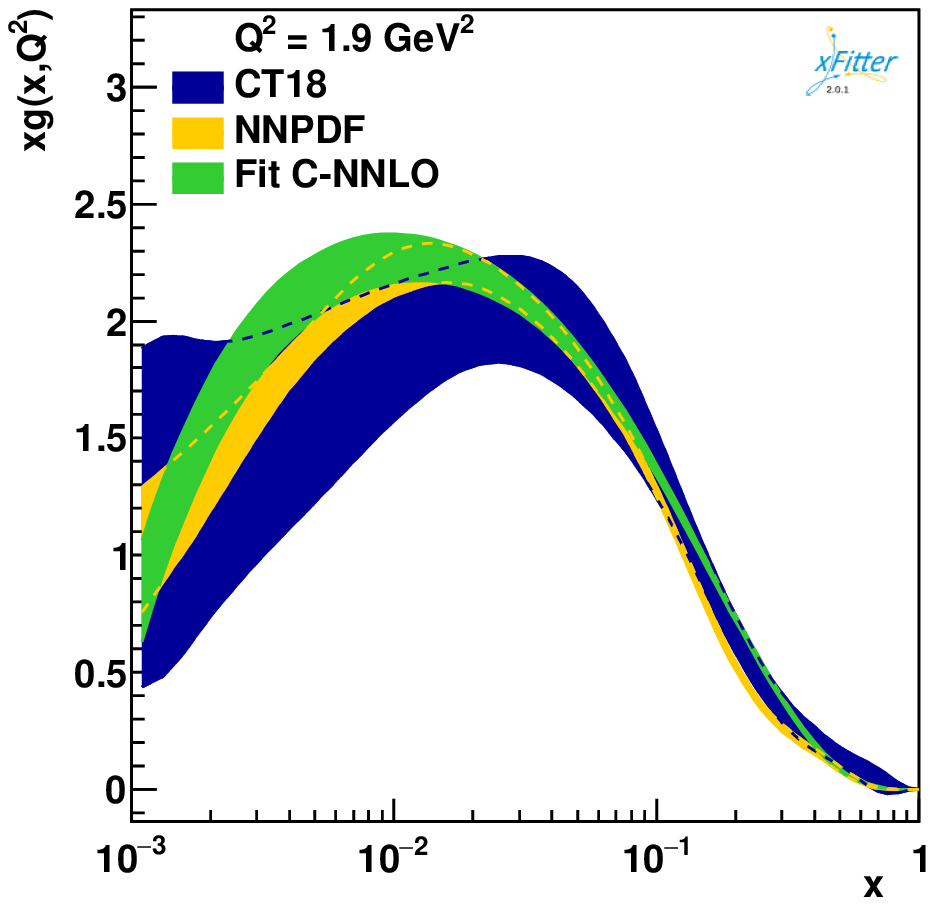}  
	    
	    \includegraphics[scale = 0.4]{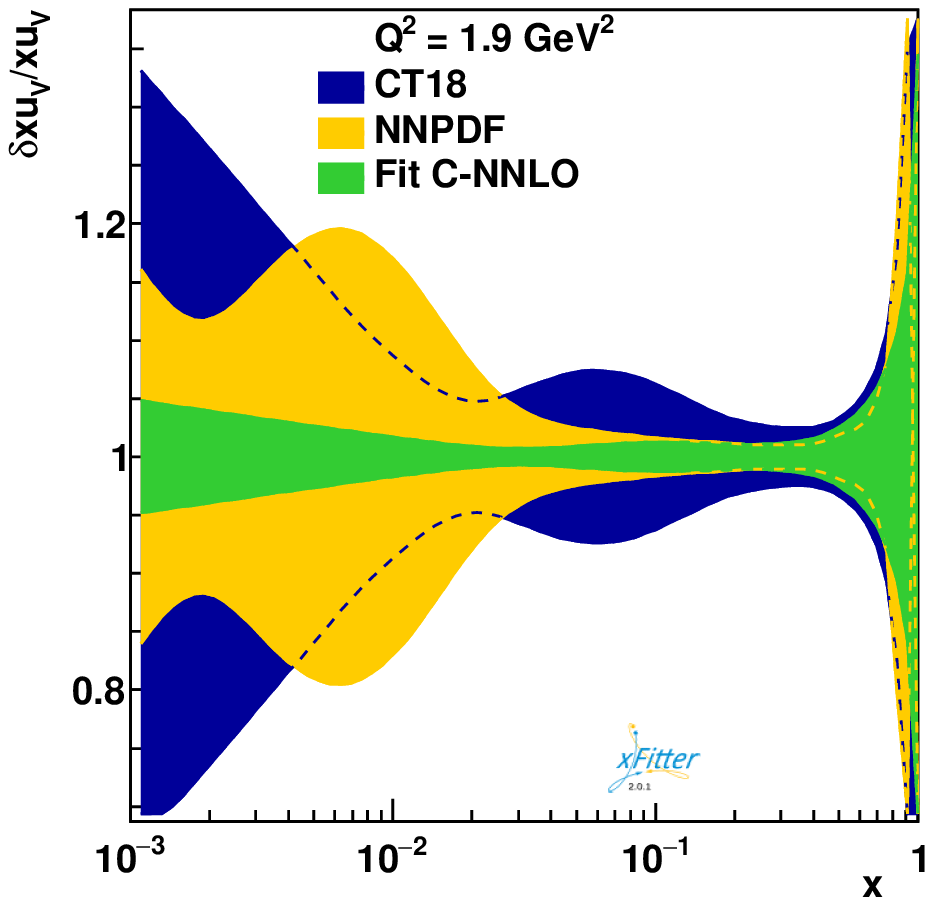}	
	    \includegraphics[scale = 0.4]{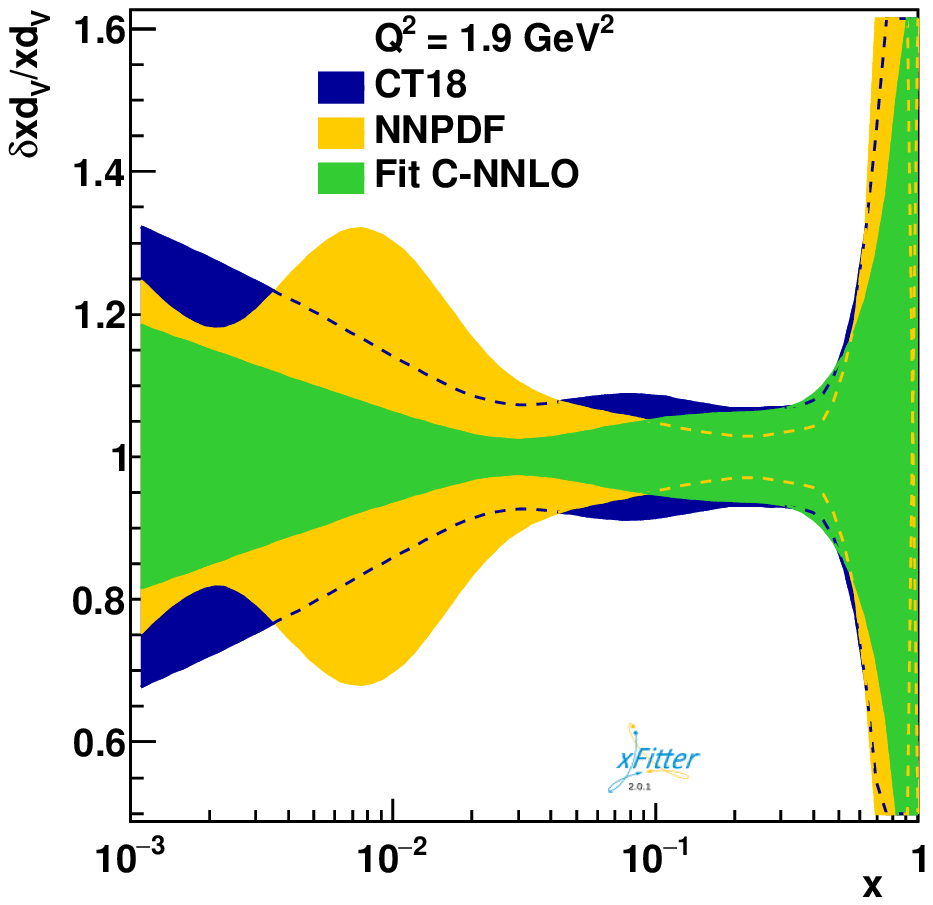}		
        \includegraphics[scale = 0.4]{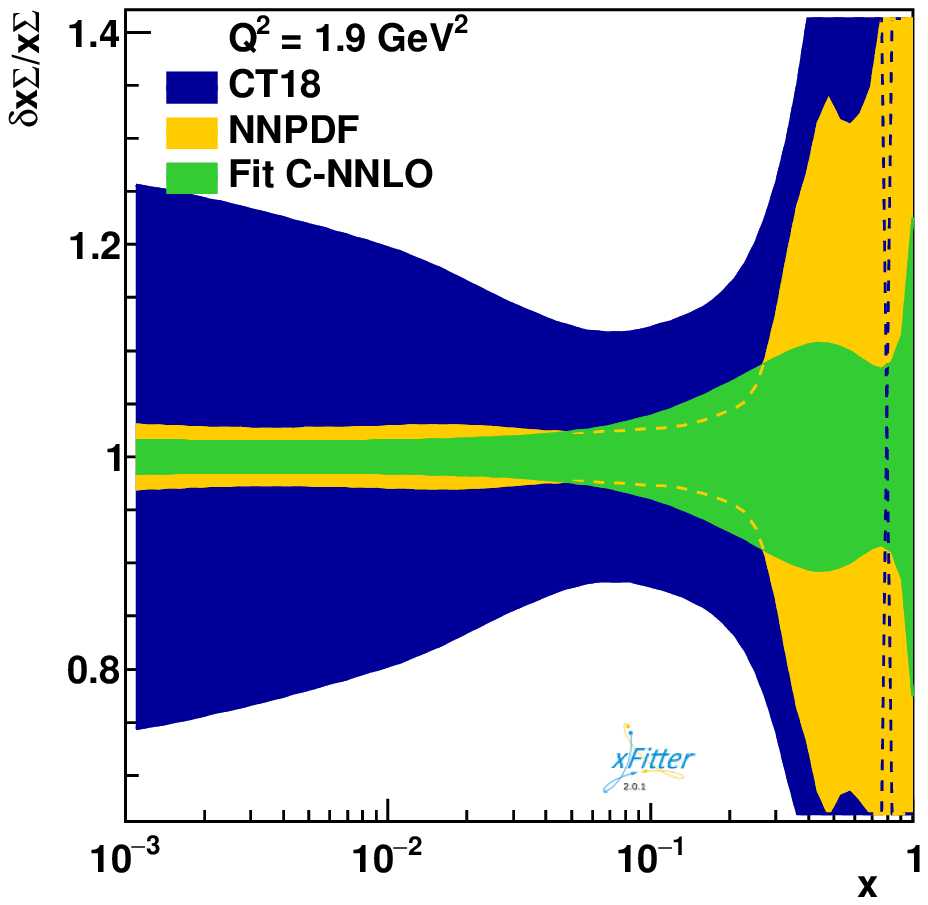}        
	    \includegraphics[scale = 0.4]{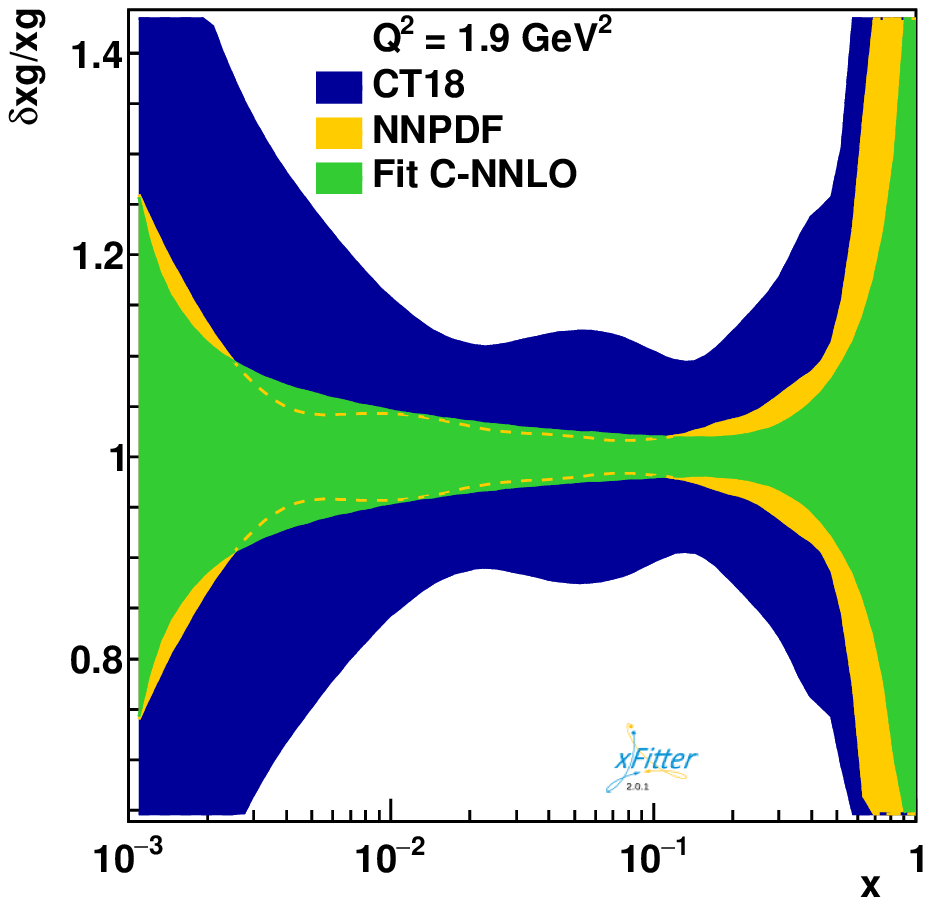}\\
	    
	    \includegraphics[scale = 0.4]{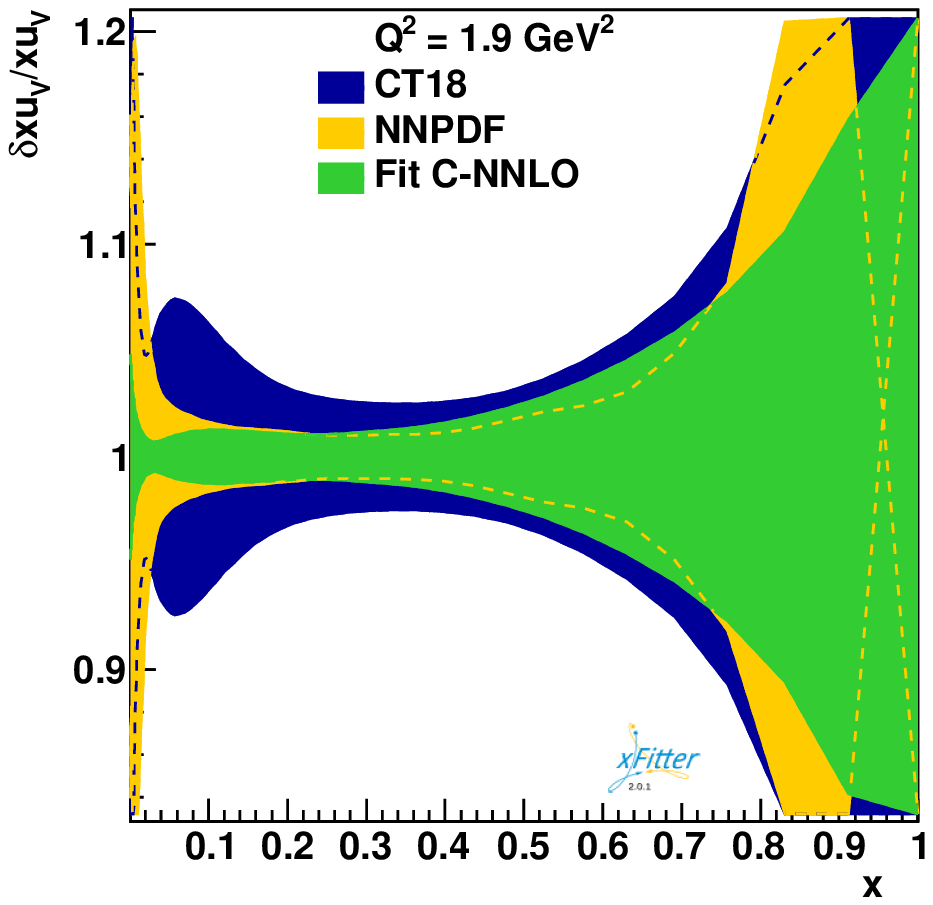}	
	    \includegraphics[scale = 0.4]{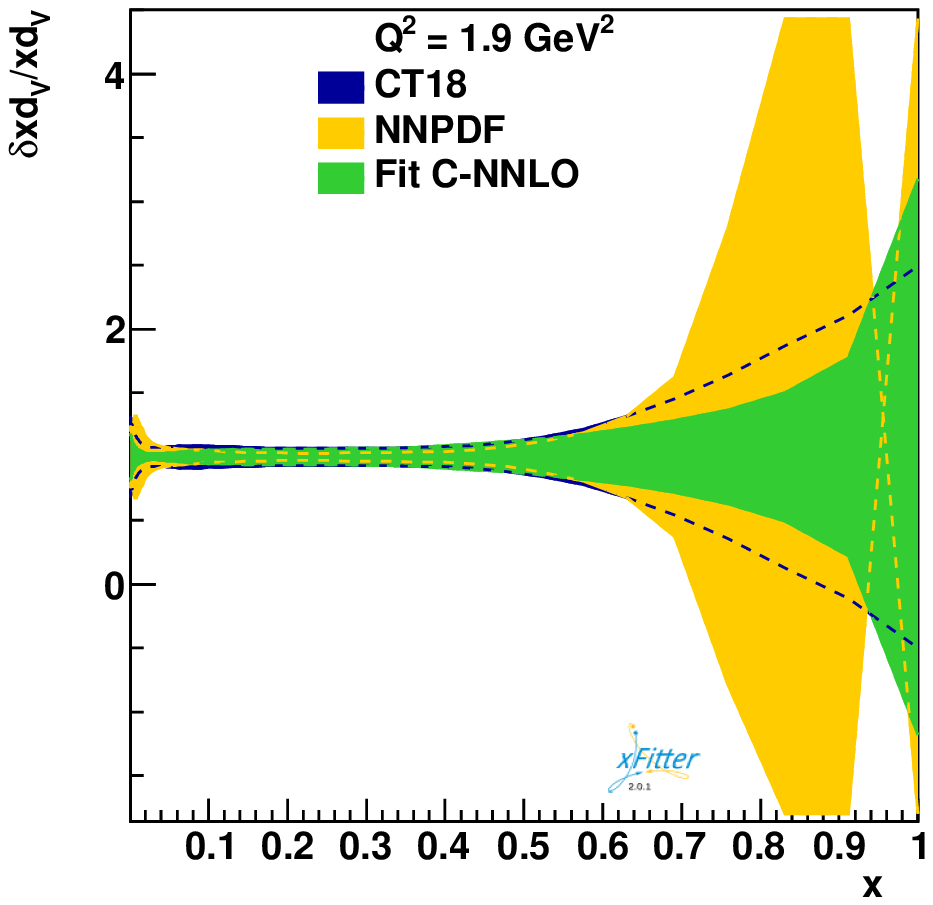}		
        \includegraphics[scale = 0.4]{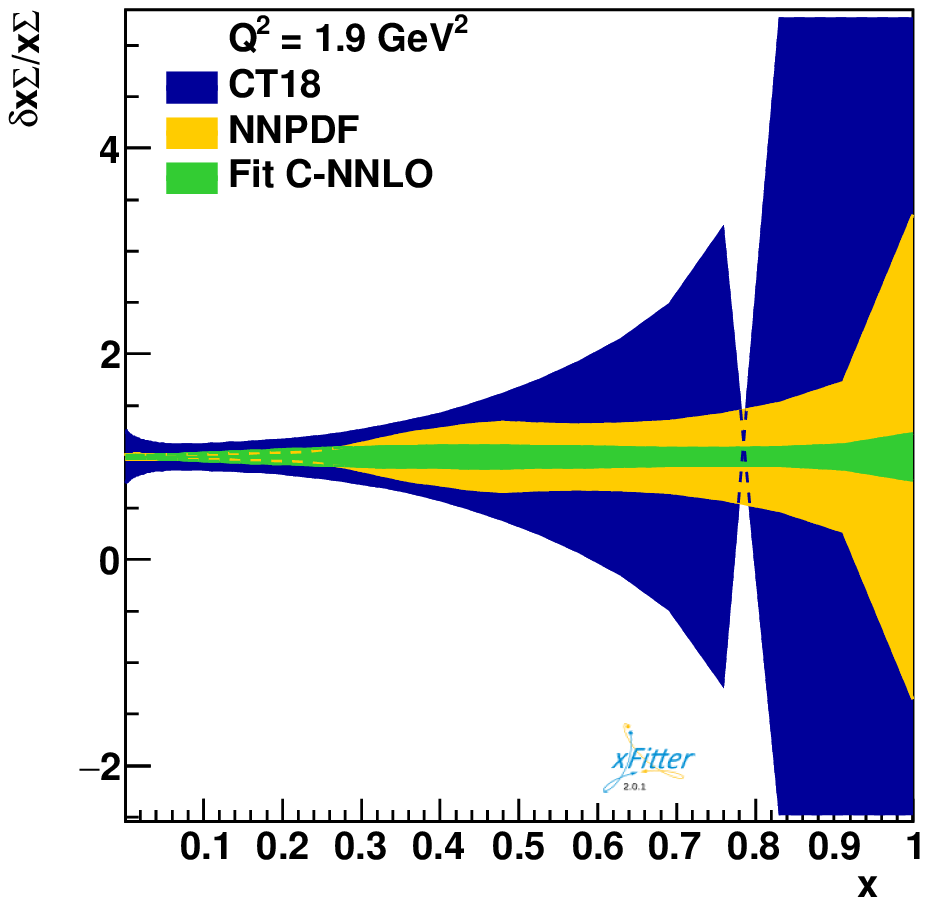}        
	    \includegraphics[scale = 0.4]{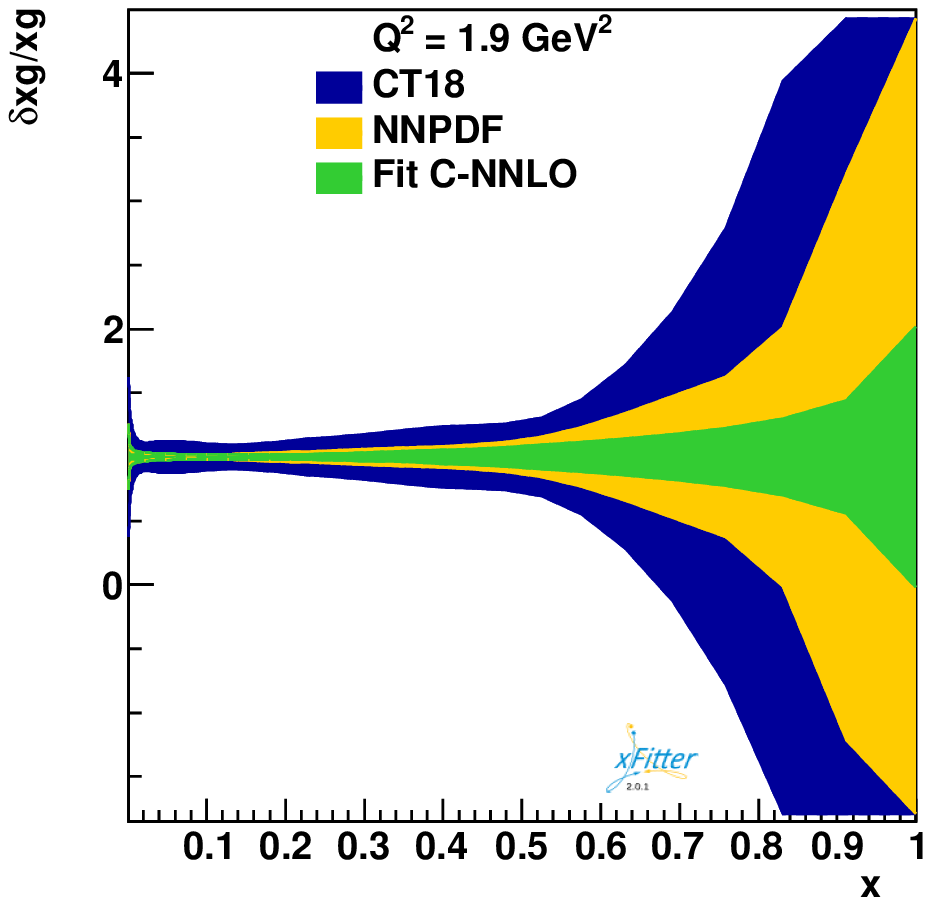}        
                
	    		\caption{The compatibility of main extracted PDF (Fit C)  at NNLO with NNPDF \cite{Ball:2021leu} and CT18 \cite{Hou:2019efy} for $xu_v$, $xd_v$, $x\Sigma$, and $xg$ distribution as a function of $x$ at 1.9 GeV$^2$. The relative uncertainties $\delta xq(x,Q^2)/xq(x,Q^2)$ PDF with both logarithmic and linear  are  also presented.}
		\label{fig:ModernPDF}
	\end{center}
\end{figure*}

The several well-known modern PDFs are also available, which have used almost the same data set in their analysis. All these groups, were looking for a common goal, which is to extract the PDFs using the experimental data. To investigate the compatibility of main extracted PDFs, $i.e.,$ Fit C at the NNLO with other modern PDF sets such as NNPDF \cite{Ball:2021leu} and CT18 \cite{Hou:2019efy} for example, we compare our results with these PDF sets that already exist at NNLO. Figure~\ref{fig:ModernPDF} displays the compatibility of our results for Fit~C with NNPDF and CT18 for $xu_v$, $xd_v$, $x\Sigma$, and $xg$ PDFs as a function of $x$ at 1.9 GeV$^2$.  In this figure, the relative PDF uncertainties $\delta xq(x,Q^2)/xq(x,Q^2)$  are shown in both logarithmic and linear plots. The difference of our analysis  with other groups is that we have 14 free parameters and 1891 data points in Fit~C, but  CT18  and  NNPDF  are extracted by utilizing 3681 and 4618 data points respectively, with very different number of parameters. It seems that, in general, an overall good agreements between our results and other reported results from modern PDFs at NNLO do exist.

\begin{figure*}[!htb]
	\begin{center}

	    \includegraphics[scale = 0.4]{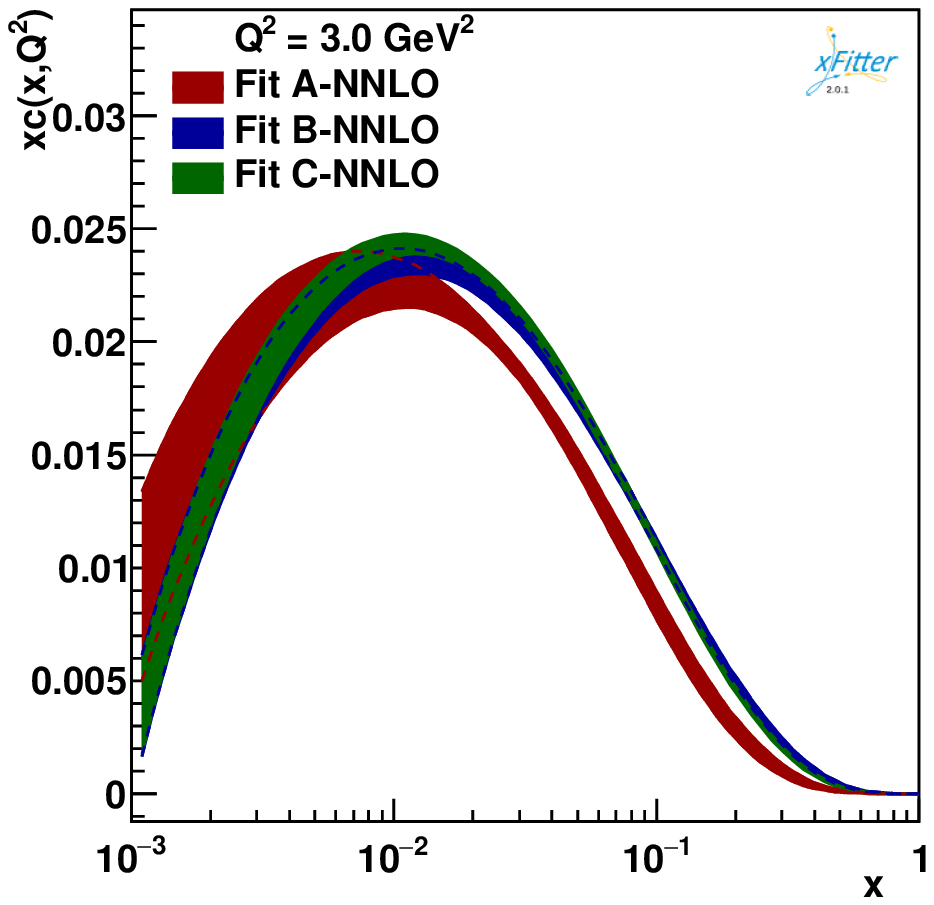}	
	    \includegraphics[scale = 0.4]{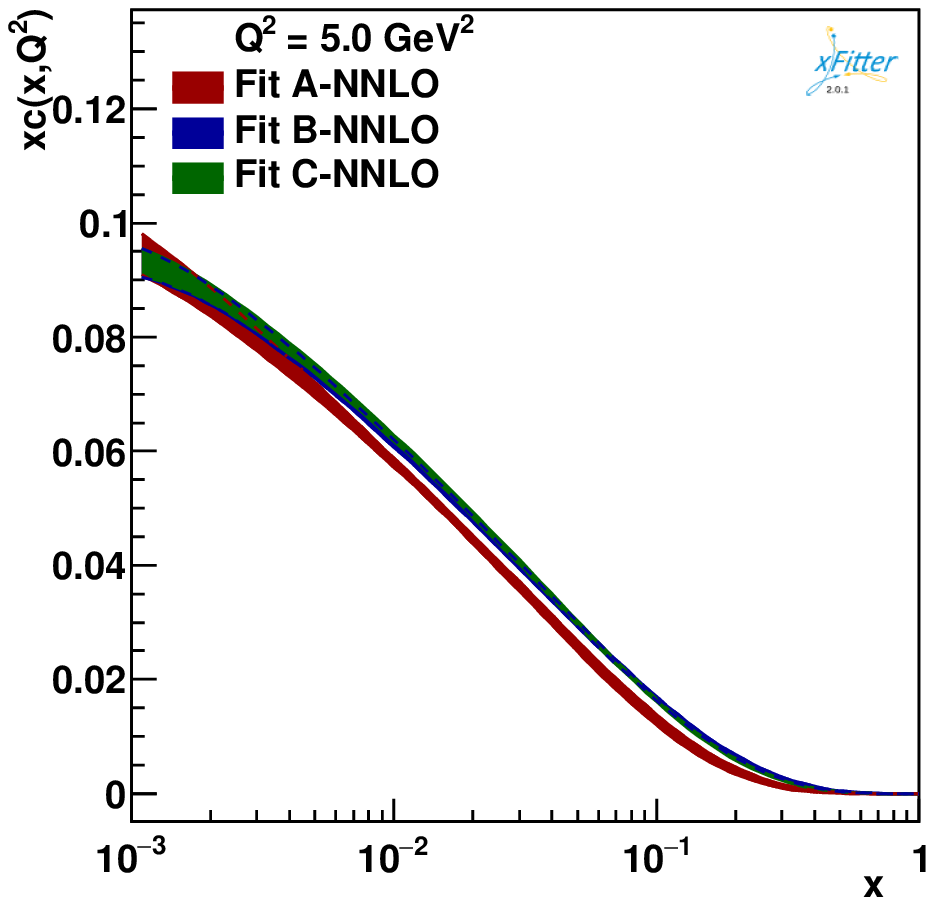}		
	    \includegraphics[scale = 0.4]{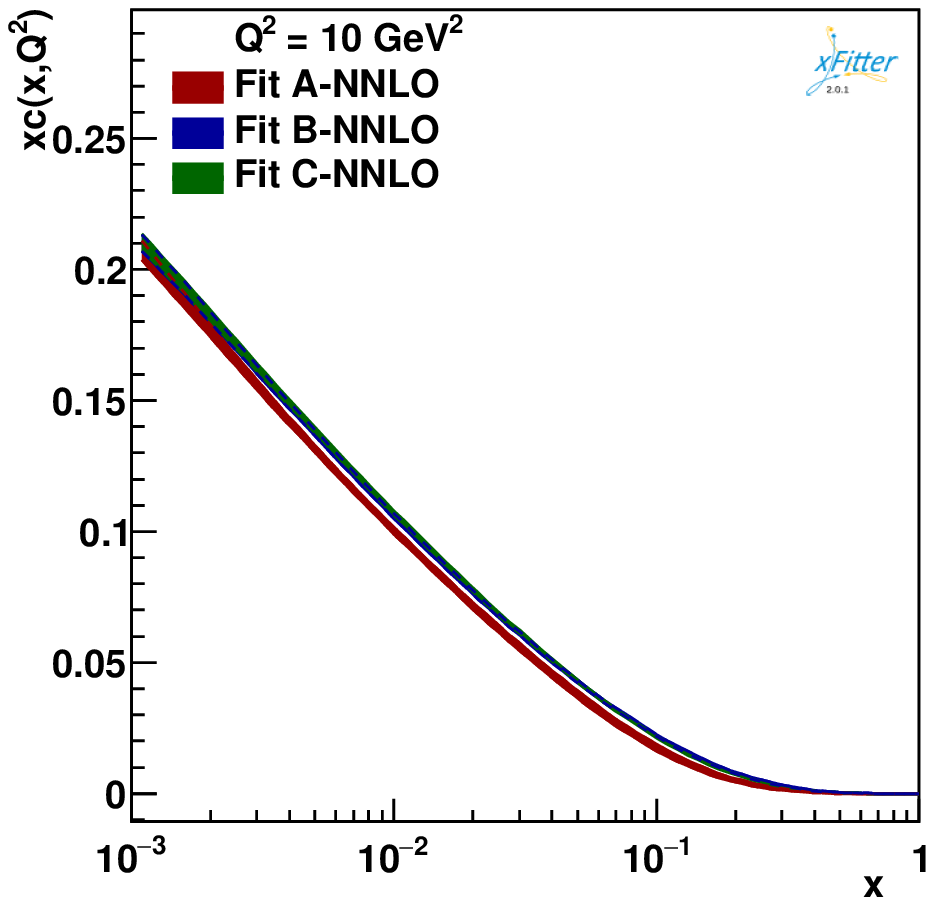}
	    \includegraphics[scale = 0.4]{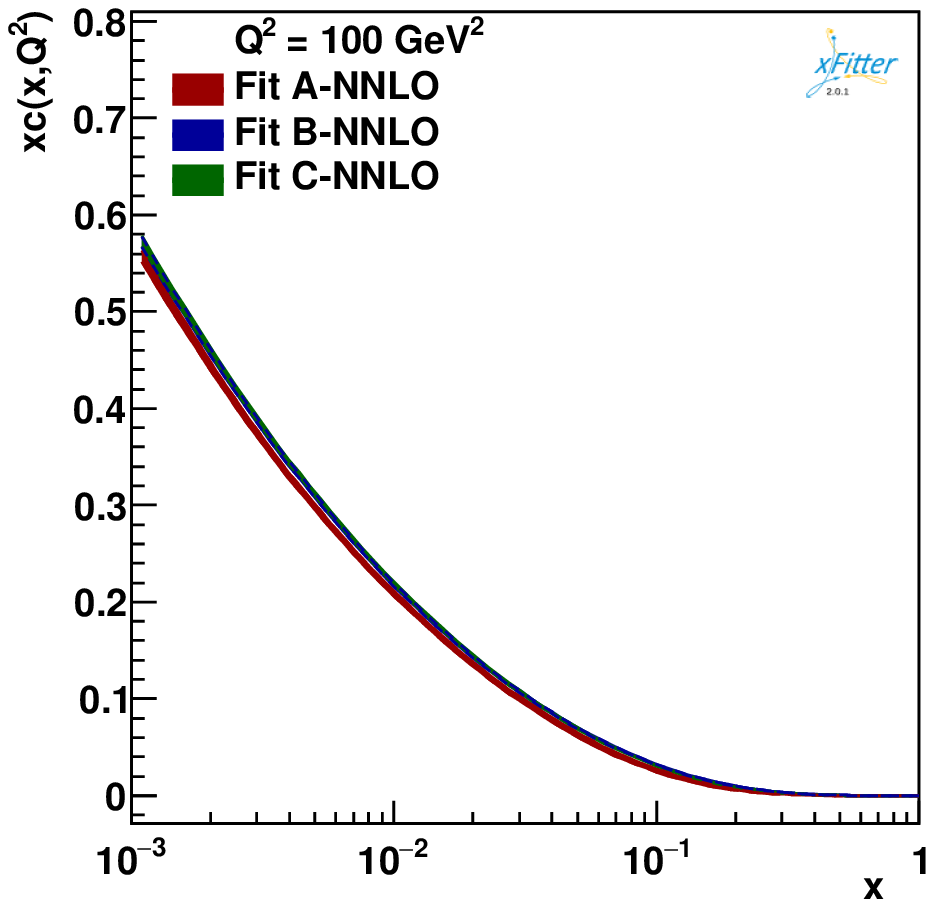}

	    \includegraphics[scale = 0.4]{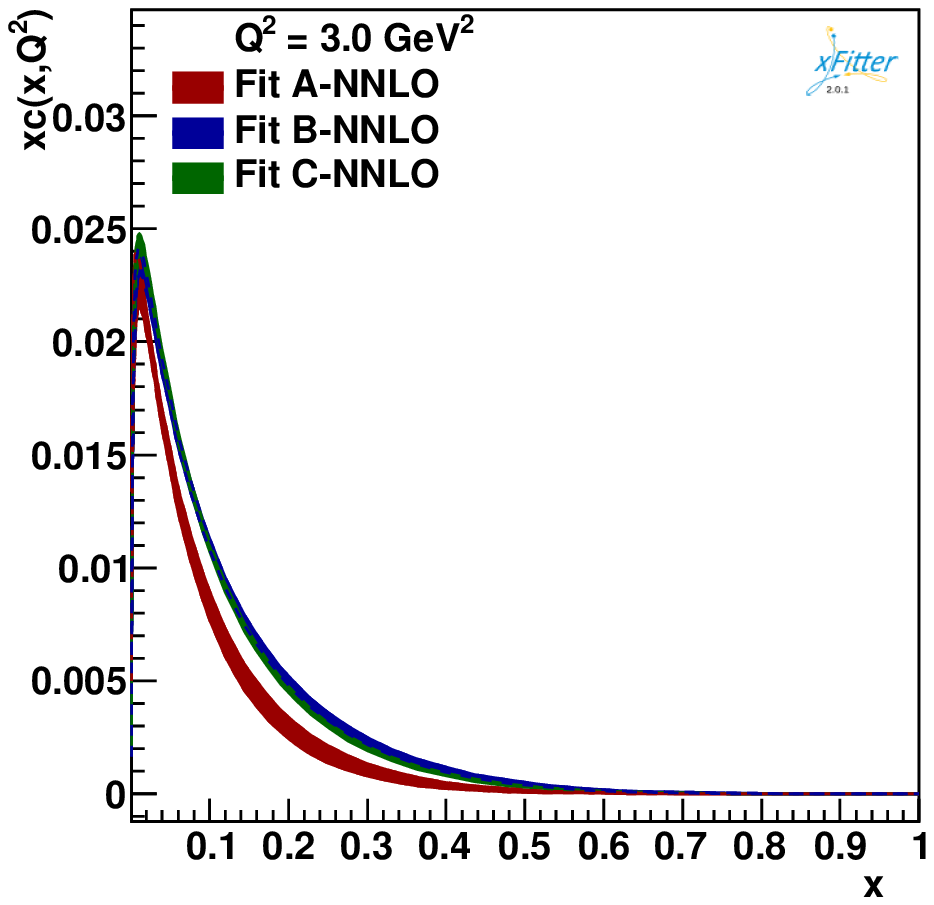}	
	    \includegraphics[scale = 0.4]{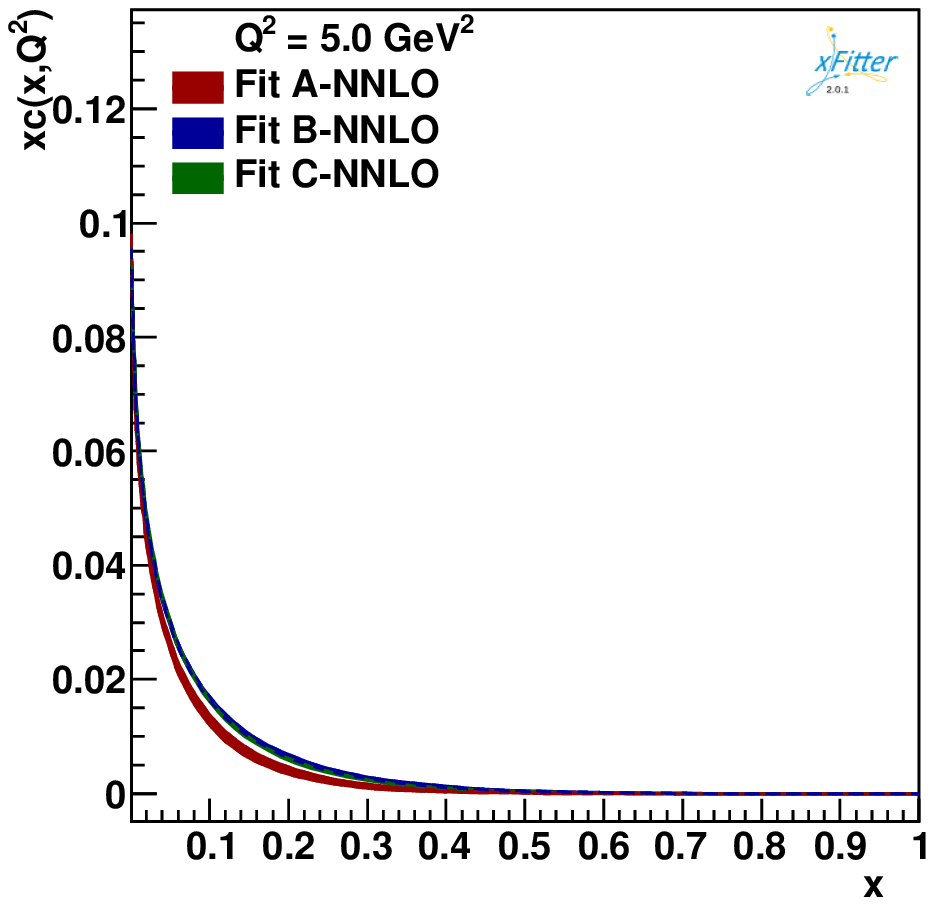}		
	    \includegraphics[scale = 0.4]{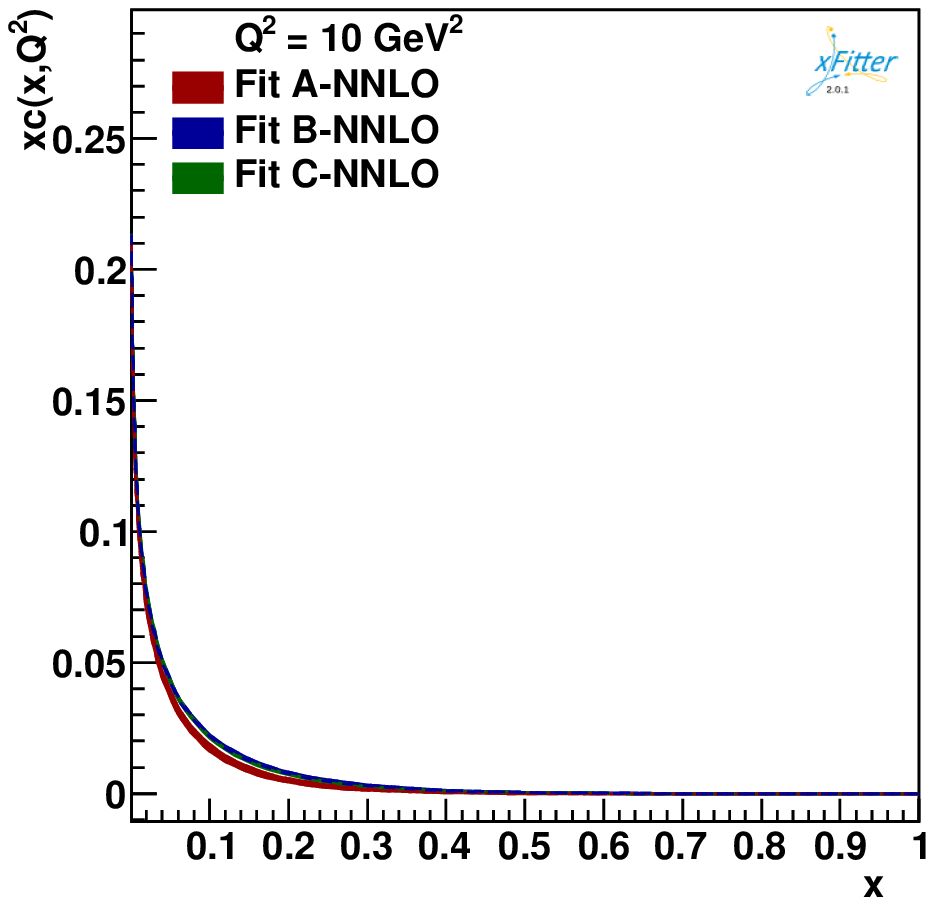}
	    \includegraphics[scale = 0.4]{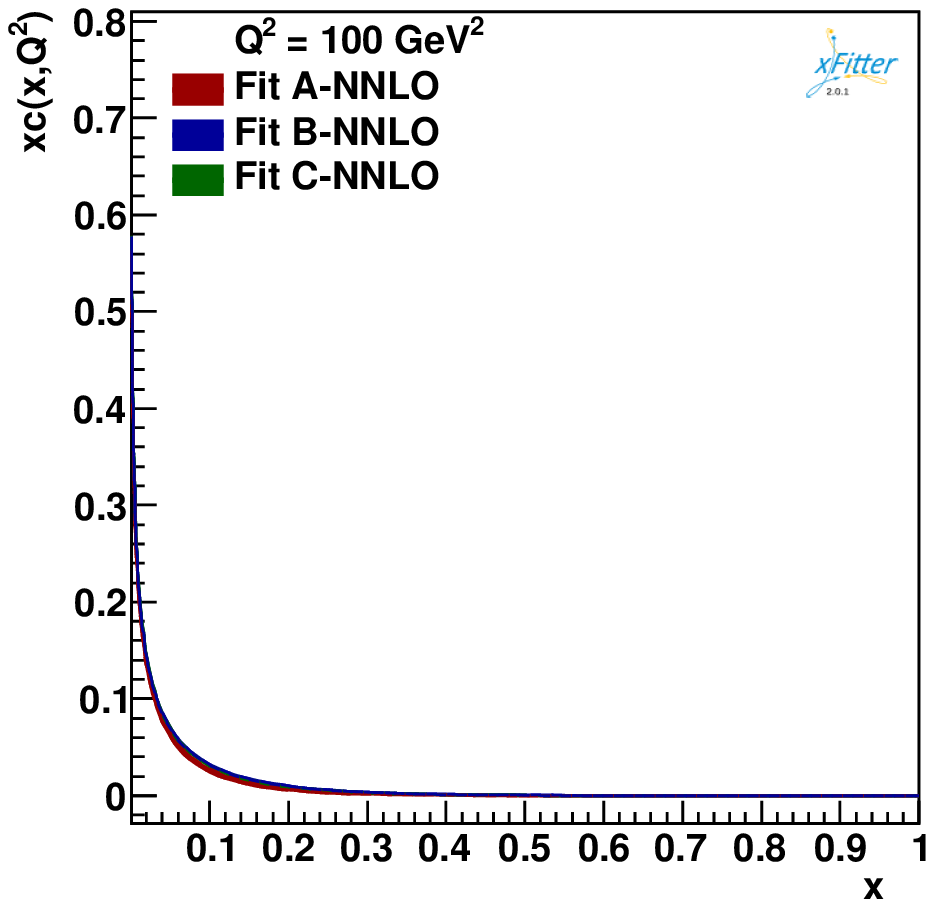}		
	
	    \includegraphics[scale = 0.4]{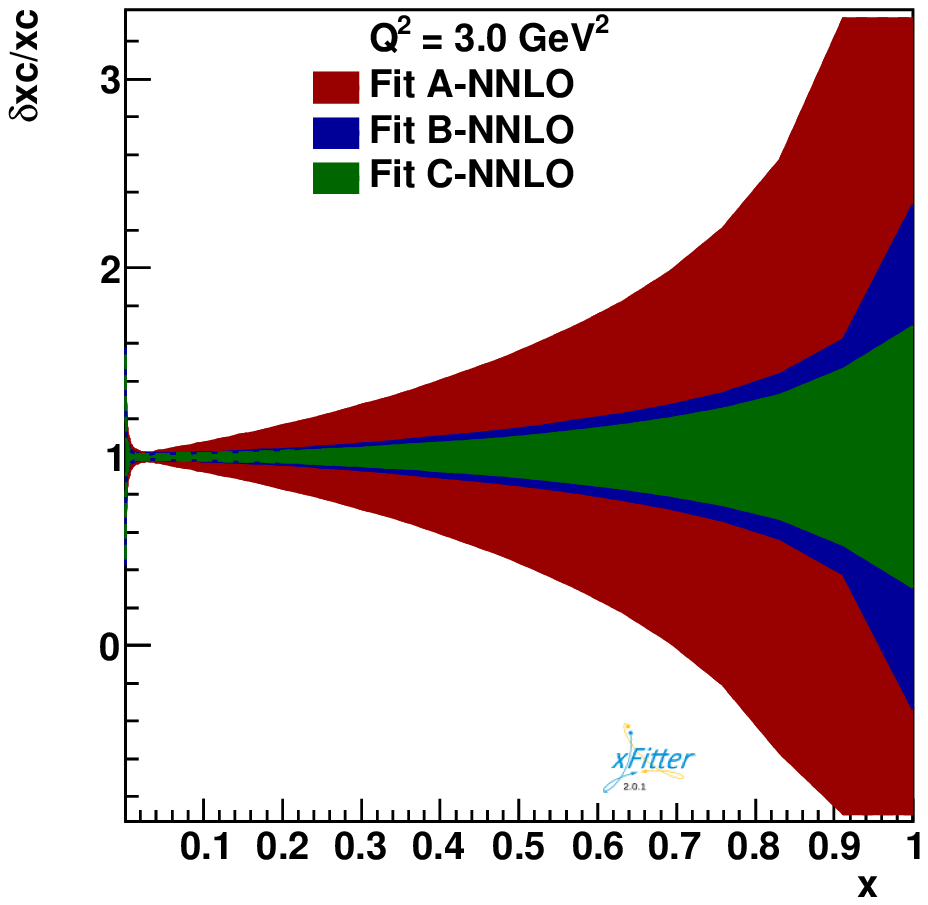}	
	    \includegraphics[scale = 0.4]{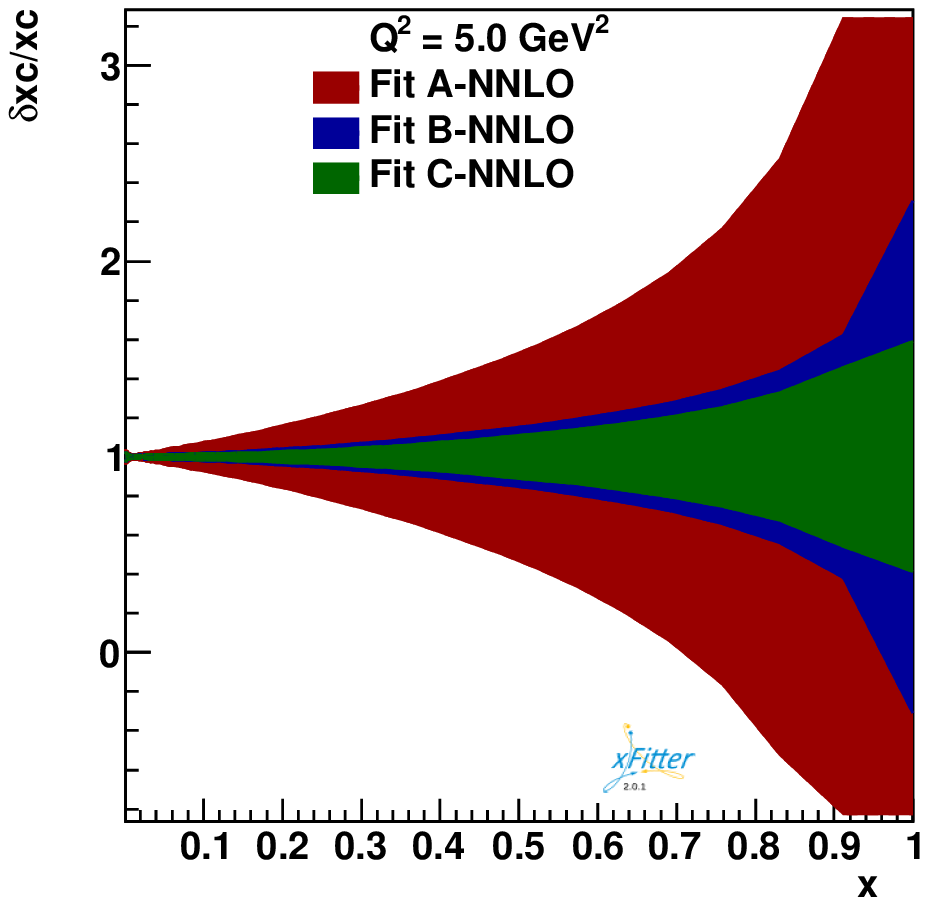}		
	    \includegraphics[scale = 0.4]{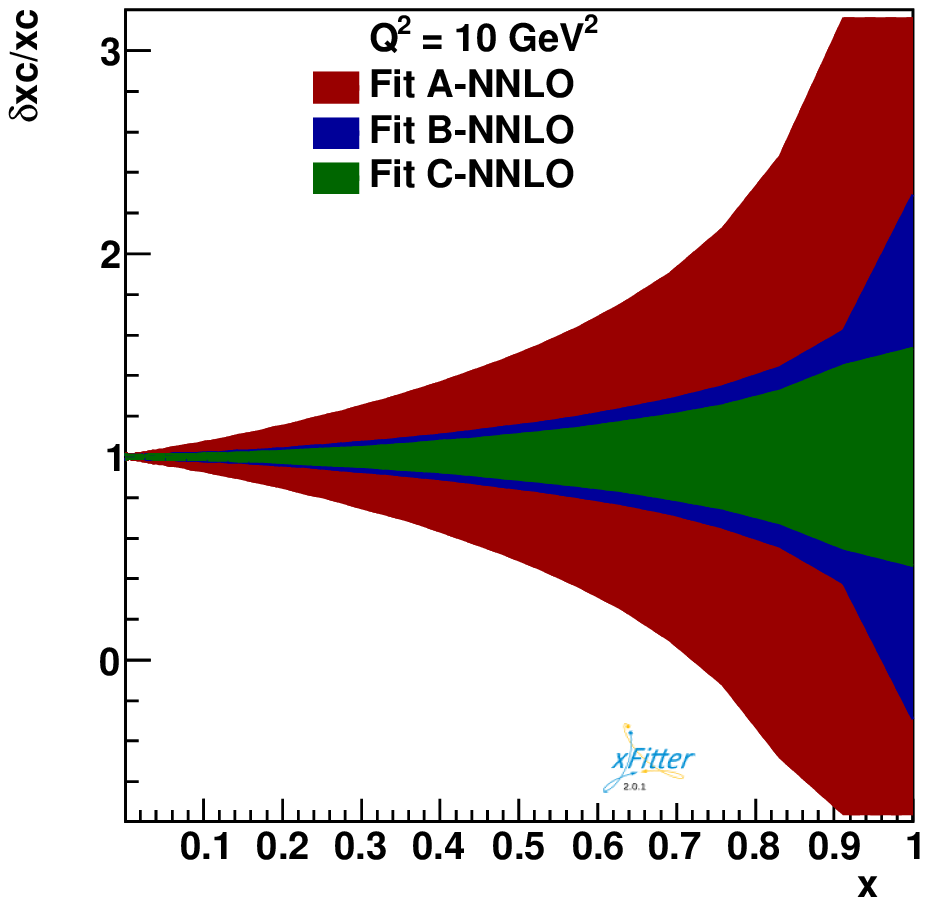}
	    \includegraphics[scale = 0.4]{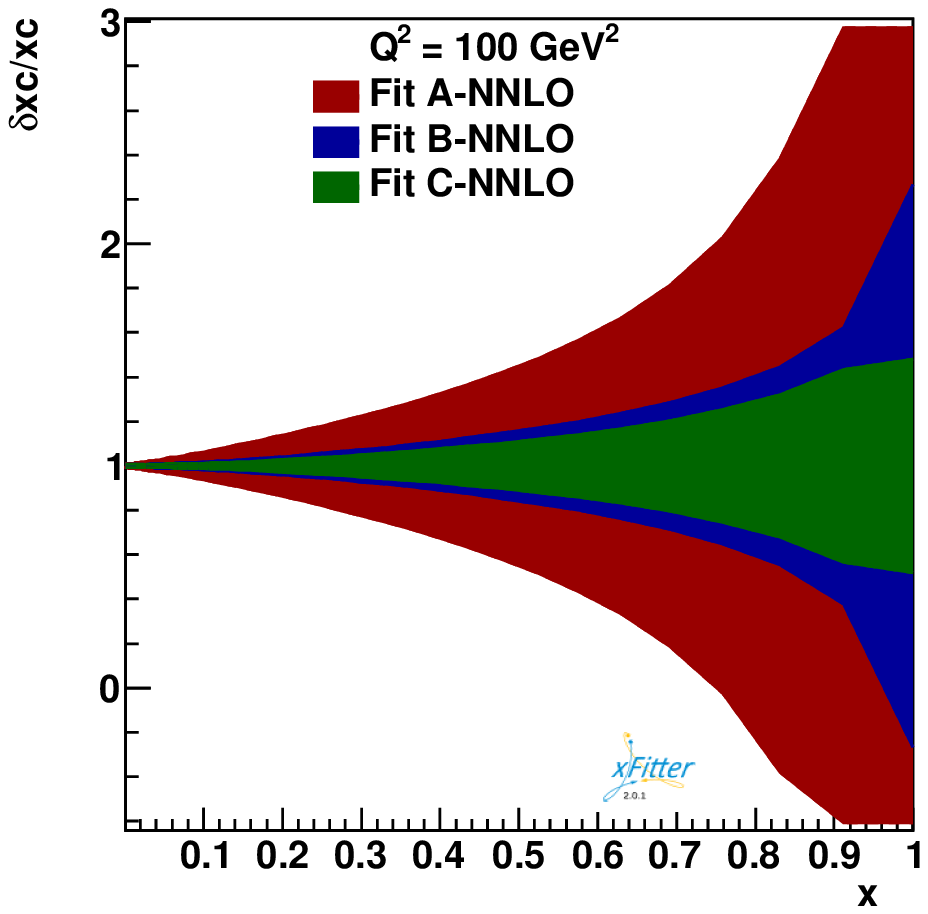}

	    		\caption{The impact of the LHC data (Fit C) on the high-$x$ charm PDF at NNLO with comparison of HERA I+II (Fit A) and Non-LHC data (Fit B). The charm PDF with both logarithmic and linear plots and also the relative uncertainties $\delta xc(x,Q^2)/xc(x,Q^2)$ as a function of $x$  at 3, 5, 10 and 100 GeV$^2$ are presented.}
		\label{fig:charm-AllQ}
	\end{center}
\end{figure*}

Because the charm PDF benefits from the accuracy of the gluon PDF, so it would be very worth to present our results for charm PDF.
Undoubtedly, a significant reduction in gluon PDF uncertainty, especially in large $x$, will reduce the charm PDF uncertainty.
 In Fig.~\ref{fig:charm-AllQ}, we present the impact of the LHC data (Fit C) on charm PDF at NNLO in comparison with HERA I+II (Fit A) and Non-LHC data (Fit B). The charm PDF with both logarithmic and linear plots and also the relative uncertainties $\delta xc(x,Q^2)/xc(x,Q^2)$ as a function of $x$  at 3, 5, 10 and 100 GeV$^2$ are shown with significant reduction on charm PDF uncertainty at large $x$.
This result, $i.e.,$ such a reduction of charm PDF uncertainty at large $x$ can be very worth for future experiments at LHC and also in some processes which we need to include the intrinsic charm (IC)  to extrinsic charm PDF. It should be noted that the IC is dominant at large $x$ region only.

\subsection{The strong coupling $\alpha_s(M_Z^2)$  }

In some PDF studies, the value of $\alpha_s(M_Z^2)$ may be considered as a fixed parameter with the world average value ~\cite{ParticleDataGroup:2020ssz}. 
In our fit procedure, we allow the value of $\alpha_s(M_Z^2)$ to vary as a free parameter. 
In fact, in the present analysis the strong coupling constant
value $\alpha_s(M_Z^2)$ is obtained simultaneously with
PDFs at both NLO and NNLO approximation in QCD.

The different data sets, may affect the PDF and strong coupling constant and also their uncertainties extraction. In the present analysis the strong coupling constant values and their uncertainties are obtained by considering different data sets.
The central value of the strong coupling constant and also its uncertainty is very sensitive to gluon PDF and its uncertainty. 
Since the large-$x$ gluon PDF benefits from an accurate determination of quark PDFs, we found that only the top-quark
production and jet cross section measurements impact on the central value and also impact significantly on the uncertainty of
gluon PDF and the strong coupling constant.

The different data sets which are listed in Table~\ref{tab:Chi2}, $i.e.,$  HERA I+II data, Non-LHC, and LHC data sets, play an important
 role in determination of the strong coupling constant value.

To study the particular significance of these three samples, we
consider Fit A without including the non-LHC and LHC data sets. In this case, the values
of
$\alpha_s(M_Z^2)=0.1167 \pm 0.0014$
and 
$\alpha_s(M_Z^2)=0.1134 \pm 0.0024$
at NLO and NNLO in QCD are obtained. However, they are  smaller than the  world average value of $\alpha_{s}(M_{Z}^{2})=0.1179\pm0.0085$~\cite{ParticleDataGroup:2020ssz}.

When the non-LHC data listed in Table~\ref{tab:Chi2} are added to HERA I+II data in Fit B, we find the values $\alpha_s(M_Z^2)=0.1198 \pm 0.0008$ for NLO and $\alpha_s(M_Z^2)=0.1187 \pm 0.0007$  for NNLO,
which are larger than the $\alpha_s(M_Z^2)$ values, which obtained from the HERA I+II DIS data only, but have a significantly small statistical error.

Finally, when the LHC data listed in Table~\ref{tab:Chi2} are included in Fit C, the  values of 
$\alpha_s(M_Z^2)=0.1188  \pm 0.0007$
and $\alpha_s(M_Z^2)=0.1179  \pm 0.0006$
are obtained at NLO and NNLO respectively, which are not very different from the extracted values from Fit B, and have a slightly smaller statistical error, but have a significantly smaller
statistical error in comparison to Fit A. 
The main origin of small uncertainty for the strong coupling constant in Fit B and C is related to the significant reduction in the gluon PDF uncertainty, especially at large $x$ in presence of the top-quark
production and jet cross section measurements.
In fact, such a reduction in the gluon PDF uncertainty 
causes to extract the small uncertainties for the strong coupling constant.

Other theoretical groups, have also determined the value of $\alpha_s(M_Z^2)$  in the PDF fits.
The NNLO values $\alpha_s(M_Z^2)=0.1164\pm0.0024$, and 
$\alpha_s(M_Z^2)=0.1156\pm0.0011$ obtained by the CTEQ \cite{Hou:2019efy} and HERAPDF \cite{H1:2021xxi} groups respectively, despite looking similar, have a different origin and data sets. Our new results are compatible with the reported results $\alpha_s(M_Z^2)= 0.1147\pm0.0008$ at NNLO \cite{Alekhin:2017kpj} and $\alpha_s(M_Z^2)=0.1191\pm0.0008$ at NLO \cite{Abdolmaleki:2019tbb}.

However, in the present analysis, different data sets that are listed in Table~\ref{tab:Chi2}, more or less are  sensitive to the variation of 	$\alpha_s(M_Z^2)$. In order to examine the sensitivity of each data set to the variation of QCD coupling we define 	$\Delta\chi^2_i(a)=\chi^2_i(a)-\chi^2_i(a_{min})$, as a function of parameter $a$. Therefore, according to this definition $\Delta\chi^2_{tot}(a_{min})=0$. Considering $a=\alpha_s(M_Z^2)$, the values of $\Delta\chi^2_i(\alpha_s(M_Z^2))$ is the difference between the values of $\chi^2$ for experiment $i$ at the fixed value of $\alpha_s(M_Z^2)$ with considering the rest of free parameters and the value of $\chi^2$ related to same experiment $i$ extracted from one of Fit A, B or C.

\begin{figure*}[!htb]
	\begin{center}
		\includegraphics[scale = 0.75]{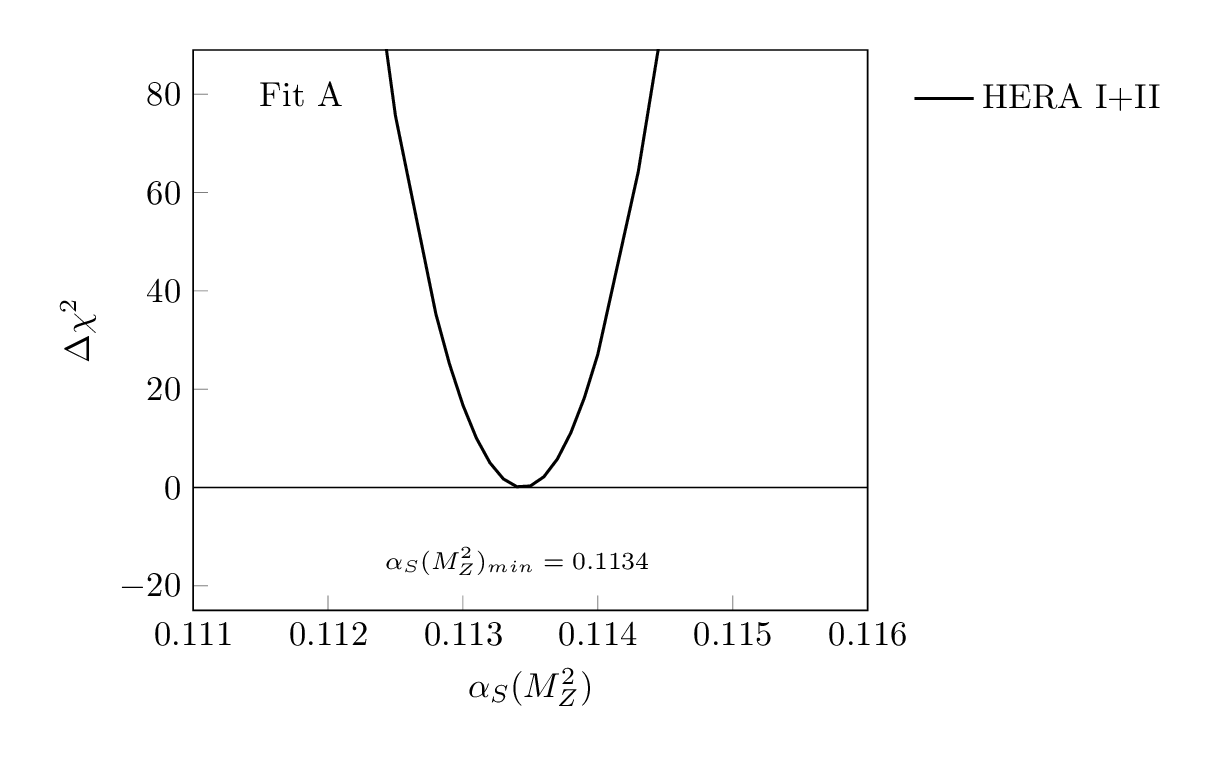}\\
		\includegraphics[scale = 0.75]{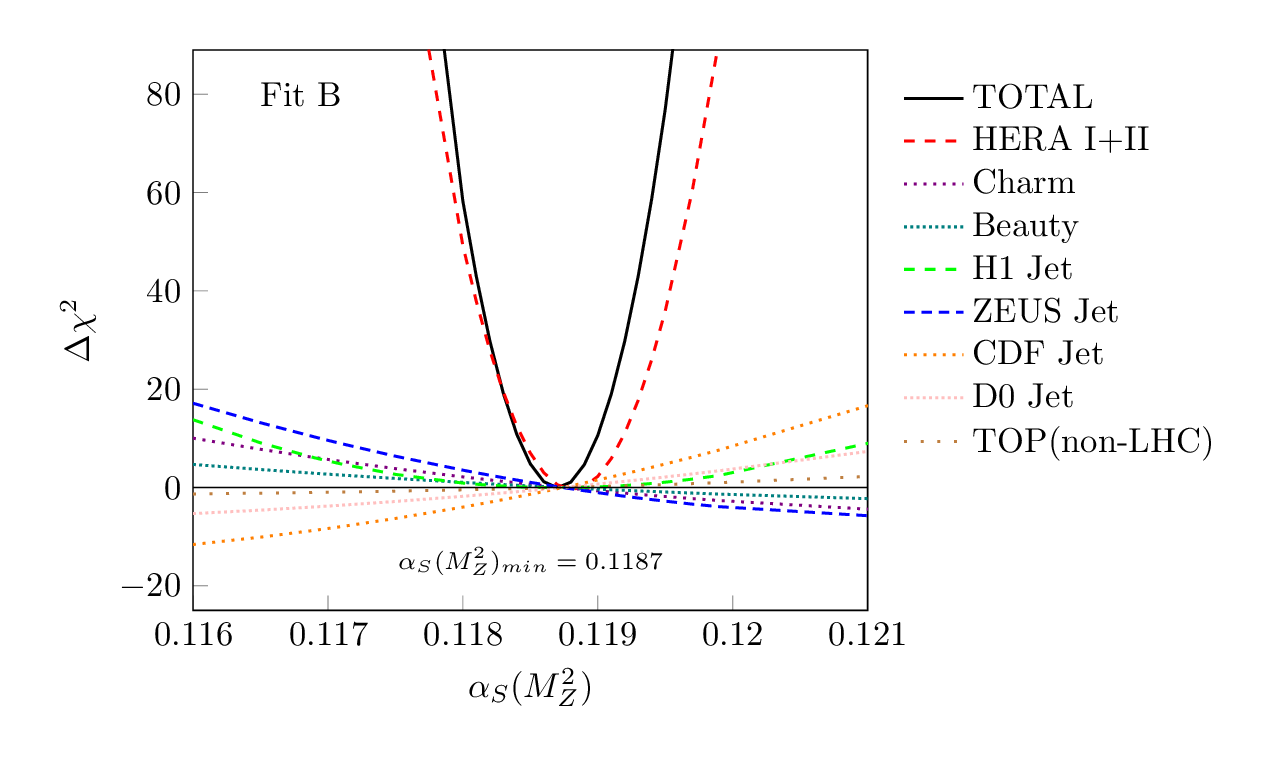}\\
		\includegraphics[scale = 0.75]{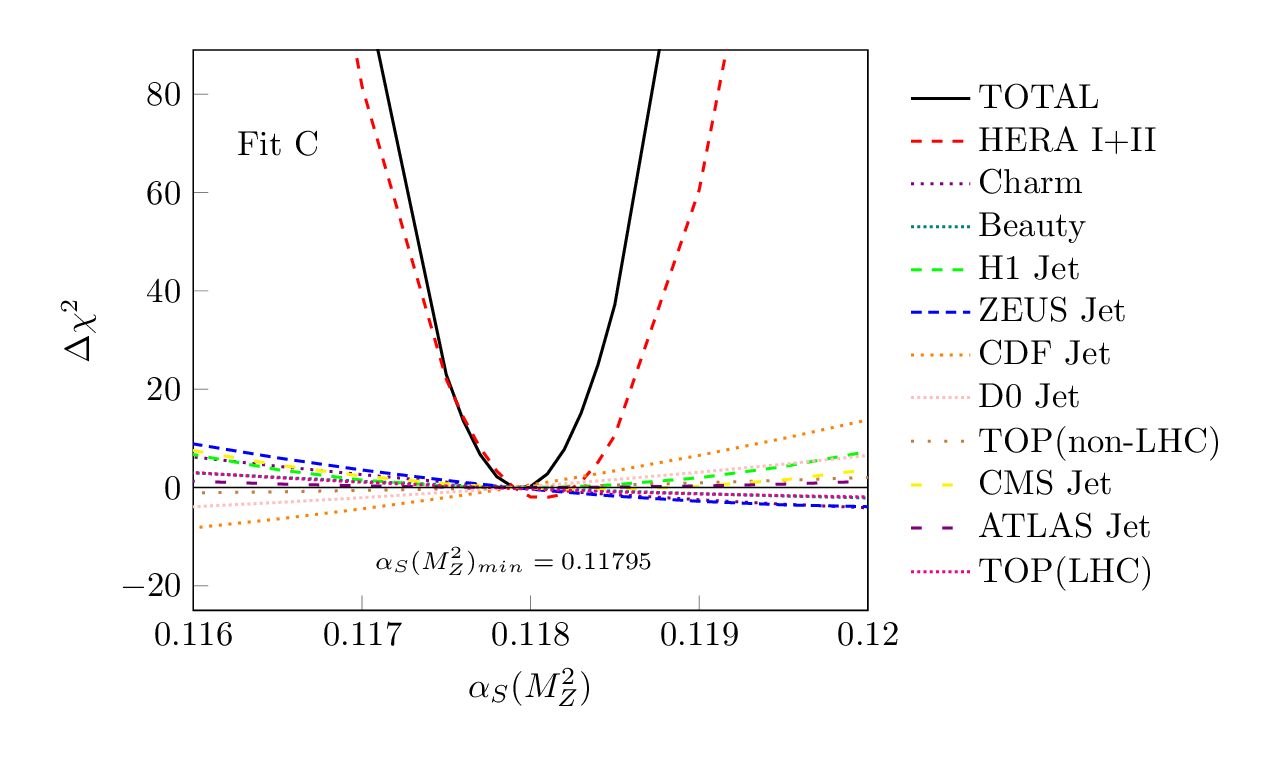}		
		\caption{ The scan of QCD strong coupling constant $\alpha_s(M_Z^2)$ at the NNLO precision. Different curves for $\Delta\chi^2$ as a function of  strong coupling $\alpha_s(M_Z^2)$ which obtained from Fit A, B and Fit C data sets at NNLO and for each type of data sets are presented. The $\Delta\chi^2_{tot}$ curves for all experiments are also shown.	
		}
		\label{fig:chi-scan}
	\end{center}
\end{figure*}

In order to show sensitivity of the individual data sets, 
we present in Fig. \ref{fig:chi-scan}  a series of curves for $\Delta\chi^2_i$ as a function of  the strong coupling $\alpha_s(M_Z^2)$
obtained from Fit A, B and Fit C at NNLO in the scan fit using the $\alpha_s(M_Z^2)$ value spanning by 0.111-0.116 range for Fit A,   the range of Fit B is 0.116-0.121 and the range of $\alpha_s(M_Z^2)$ related to Fit C is 0.116-0.120 in units of 0.001 for each type of data sets. The $\Delta\chi^2_{tot}$ curves for all experiments are also shown and Fit A, B and C prefer $\alpha_s(M_Z^2)$ to 
$0.1134 \pm 0.0024$, $0.1187 \pm 0.0007$ and  $0.1179  \pm 0.0006$  at NNLO, respectively.

 To summarize, we find that Fit C data set produce the strongest impact on QCD coupling value and prefers a little smaller value of $\alpha_s(M_Z^2)$ and uncertainty than in Fit B and also 
prefers a larger value of $\alpha_s(M_Z^2)$ and smaller uncertainty than in Fit A.

	As the last part of this section, we are interested in studying the role of tolerance parameter $T$ on the uncertainty of $\alpha_s(M_Z^2)$.
	According to Hessian formalism and in order to quantify the physical prediction uncertainties which depend on parton distribution functions, the tolerance parameter $T$  may be considered to correspond to the acceptable fits region \cite{Pumplin:2001ct}. Note that due to a combination of parameterization inflexibility, tensions between datasets and incomplete theory entering the fit, the error values increase by using a tolerance value. 
	In order to quantify the uncertainties, the variation of $\chi^2$ in the neighborhood of its minimum should be explored. In fact, the minimum will increase $\chi^2$ by an amount $\Delta\chi^2$ by moving the free parameters away from the minimum.
	In this case, one can introduce the relevant neighborhood of the minimum as $\Delta\chi^2 \leq  T^2$ \cite{Pumplin:2001ct}, where $T$ is a tolerance parameter. By taking into account $T^2\approx$16,  the uncertainties of $\alpha_s(M_Z^2)$ for Fit B and C are increased. 
	Note that all uncertainties are proportional to $T$. Using this tolerance value, we obtain the error of $\alpha_s(M_Z^2)$ as 0.0032 and 0.0028 for Fit B,  and 0.0028 and 0.0024 for Fit C, at the NLO and NNLO respectively.

\subsection{Total charm PDFs}
As we mentioned before, reducing the uncertainty of gluon distribution in large $x$ produce the strong impact on charm PDF (``extrinsic charm'') in this region. Undoubtedly, such a reduction in uncertainty of charm PDF at large $x$  in some processes which are sensitive to adding of the extrinsic  and ``intrinsic charm''  (IC) contribution would be worthwhile. Note that IC contribution is dominant only at high $x$. 

\begin{figure*}
\begin{centering}
	    \includegraphics[scale = 0.85]{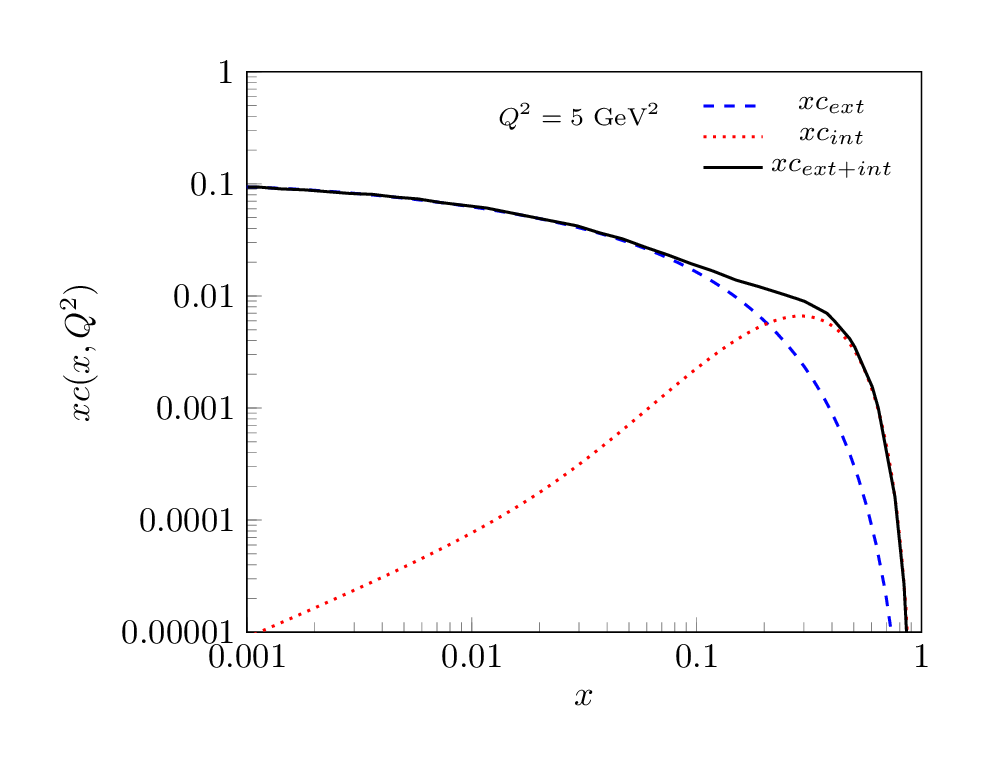}	\\	
	    \includegraphics[scale = 0.85]{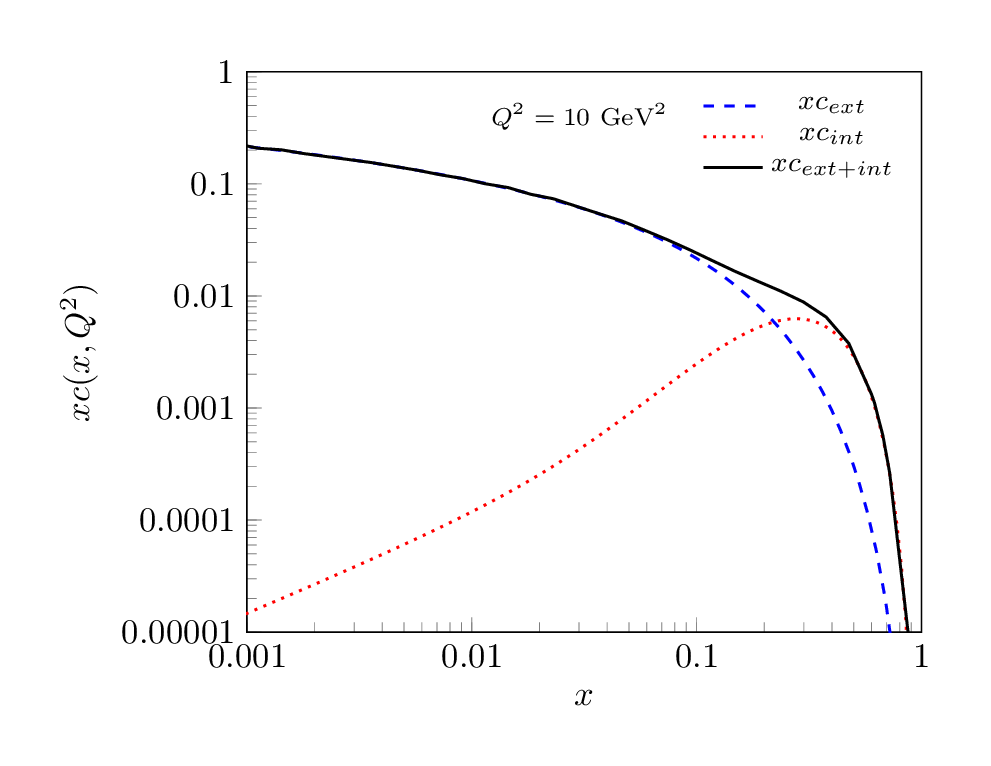}\\
	    \includegraphics[scale = 0.85]{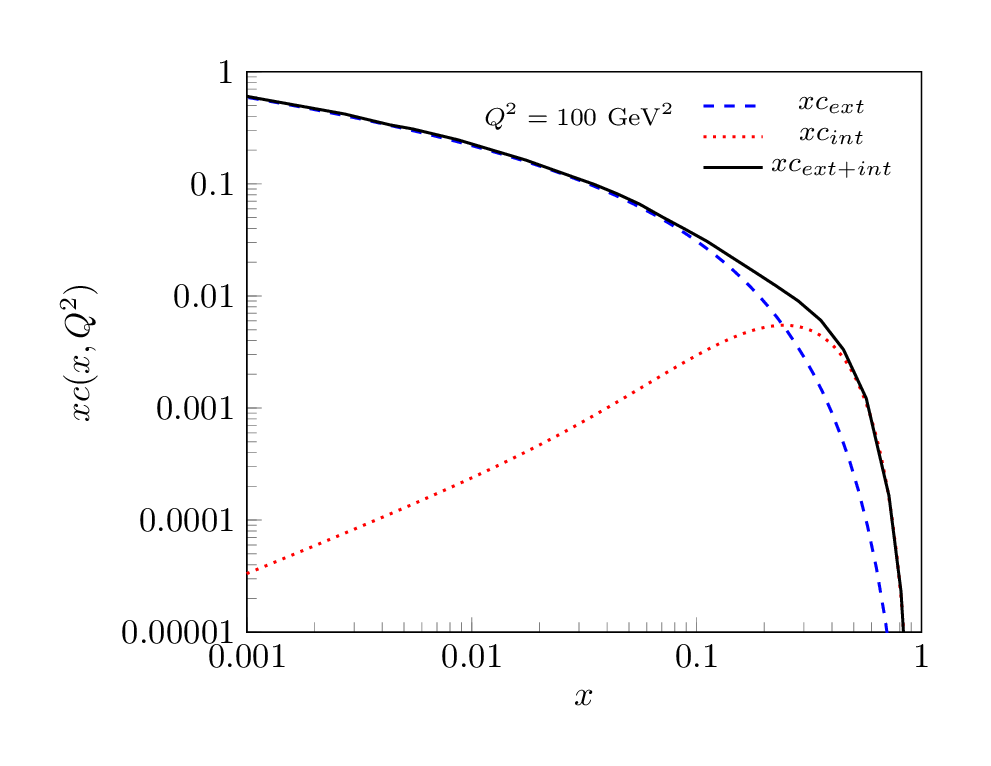}
\par\end{centering} 
 \caption{The intrinsic charm $xc_{int}(x,Q^2)$, extrinsic charm $xc_{ext}(x,Q^2)$, and total charm PDF 
 $xc(x,Q^2) = (xc_{int} + xc_{ext})(x,Q^2)$, extracted from Fit~C as a function of $x$ for $Q^2=$ 5, 10 and 100 GeV$^2$ with ${P}_{c{\bar c/p}} = 1$\%. The uncertainties for $xc_{ext}(x,Q^2)$ and total charm $xc(x,Q^2)$ are presented.
}\label{pic:xctotal2022}
\end{figure*}

In this subsection we present the full charm PDF by the sum of $xc_{ext}(x,Q^2)$ which is radiatively generated (perturbative) by the DGLAP equation in our QCD analysis and the  $xc_{int}(x,Q^2)$ (non-perturbative) with considering BHPS model \cite{BHPS1,BHPS2}. We refer the reader to see
Refs. \cite{Brodsky:2015fna} as an important review of the intrinsic heavy PDF content of the proton,
and references therein for a detailed
discussion of this model.
According to BHPS model the intrinsic heavy charm distribution is given by 
\begin{eqnarray}
\label{heavyBHPS}
c_{int}(x) & =&P_{c{\bar c}/p} 1800~x^2 \Big[ \frac{(1-x)}{3} \left( 1 + 10x + x^2 \right) \nonumber \\ && + 2 x (1+x) \ln(x) \Big]~.	
 \end{eqnarray}
Above distribution at an initial scale $Q_0 \simeq m_c$ is controlled by non-singlet evaluation equations \cite{Abdolmaleki:2019tbb,Lyonnet:2015dca} at any scale.  In Fig.~\ref{pic:xctotal2022}, we display the  intrinsic charm  $xc_{int}$ with ${P}_{c{\bar c/p}} = 1\%$, $xc_{ext}$  extrinsic  charm PDF  which extracted from our QCD Fit C,  and also the total charm $xc_{int} + xc_{ext}$  distribution as a function of $x$ and  for $Q^2=5, 10, 100$ GeV$^2$. In this figure the uncertainties for $xc_{ext}(x,Q^2)$ and total charm $xc(x,Q^2)$ are presented as well.

These results  facilitate more precise predictions to estimate the impact of heavy intrinsic contribution on heavy new physics at future facilities such as the Large Hadron-Electron Collider (LHeC), Electron Ion Collider (EIC), or AFTER@LHC. 

\subsection{Higgs boson cross section prediction using extracted PDFs}

\begin{figure*}[!htb]
	\begin{center}
	    \includegraphics[scale = 1.]{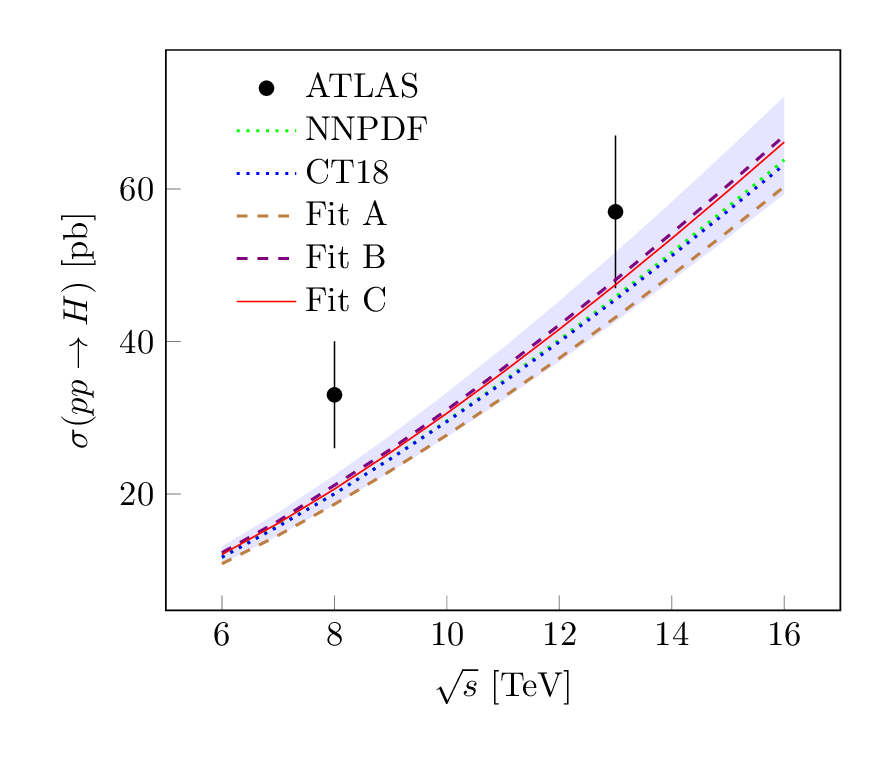}

	    		\caption{Comparison between experimental measurements of the cross section of Higgs boson  at  8 TeV and 13 TeV centre-of-mass energies which presented by ATLAS \cite{ATLAS:2015ygg,ATLAS:2018pgp} and theoretical calculation  by considering modern PDF sets NNPDF \cite{Ball:2021leu} and CT18 \cite{Hou:2019efy}, and also our PDF set from Fit A, B and Fit C at NNLO.}
		\label{fig:tth}
	\end{center}
\end{figure*}

In this subsection, we present our predictions for the cross section of Higgs boson using our extracted PDFs.
We determine this cross section at the center of mass energies from 6 TeV to 16 TeV  using the PDF sets which are extracted from Fit A to Fit C. These calculations have been done using our PDFs and also some modern PDF sets, such as  CT18 \cite{Hou:2019efy} and NNPDF \cite{Ball:2021leu} by utilizing the SusHi-1.7 \cite{Harlander:2012pb,Harlander:2016hcx} computational package.  The theoretical results with comparison of the experimental measurements for cross sections of $H$  production in $pp$ collisions at the LHC from ATLAS collaboration at the center of mass energies of 8 and 13 TeV \cite{ATLAS:2015ygg,ATLAS:2018pgp}  are shown in Fig.~\ref{fig:tth}. 
In this figure, we present the theoretical results  by considering PDF sets of NNPDF \cite{Ball:2021leu} and CT18 \cite{Hou:2019efy}, and our PDF sets which are extracted from Fit A, B and Fit C also the uncertainty related to predictions of Fit C is illustrated in Fig. 14.

The cross section predictions in this figure display a good agreement between the theory predictions at the NNLO in QCD based on our
PDFs and other theoretical models. Our results for Fit C is very close to Fit B and there are an overall good agreement between our results and other PDF sets.

 \section{Conclusion and Discussion} \label{Conc}

In this paper, we have presented the parton distribution functions (PDFs), including HERA I+II  DIS experimental data as a base data set, non-LHC and LHC data  at NLO and NNLO.  

For more clarification about the impact of various types of base, non-LHC and LHC data set on PDFs at both NLO and NNLO we present several fits. 
The first data set (Fit A) only contains HERA I+II data to prepare a fine base for investigating the impact of other data sets on PDFs. In second data set (Fit B), the non-LHC data are added to first data set.  And finally, the third data set (Fit C) contains all previous mentioned data together with LHC measurements. We extract PDFs together with their uncertainties. Also the values for QCD strong coupling constant are determined in all data sets.

In this study, one can find how adding non-LHC and LHC  data to HERA I + II impacts on reduction of the gluon relative uncertainties at high values of $x$ significantly, as it is obvious from results of Fit C where the LHC experimental data are included in fit procedure. Also, a comparison between the extracted results show that shifting from NLO to NNLO create remarkable constrains on gluon PDFs of Fit C particularly at high-$x$.

To investigate the specific impact of the HERA I+II, non-LHC, and LHC data on PDFs, we present the relative uncertainty ratios $xq(x,Q^2)/xq(x,Q^2)_{ref}$ with respect to the Fit~A at NLO and also NNLO. For both NLO and NNLO analysis,  the changes of the central values of all PDFs, their uncertainties, or both are observed. The changes of the PDFs uncertainties are observed, especially for $\delta x\Sigma(x,Q^2)/x\Sigma(x,Q^2)$ at large $x$  and $\delta xg(x,Q^2)/xg(x,Q^2)$ for both low and large $x$ values and for NLO and NNLO. Although there are no significant changes in valence and sea PDFs in all fits at NLO and NNLO, remarkable changes
in the central value of gluon PDF and their uncertainties at large value of $x$ are observed from Fit~A to Fit~C. 

To investigate the compatibility of main extracted PDFs, $i.e.,$ Fit C at the NNLO with other modern PDF sets such as NNPDF and CT18,  we compared our results with these modern PDF sets. It seems that, in general,  overall good agreements between our results and other reported results from modern PDFs at NNLO exist. 

We find that the LHC data causes even more significant impact, not only on the central value of $\alpha_s(M^2_Z)$ but also on its uncertainties in comparison to Fit A where we used only the  HERA I+II combined data. We find that Fit C data set produces the strongest impact on QCD coupling constant and prefers a little smaller value and uncertainty for $\alpha_s(M_Z^2)$ comparing to  Fit B and also prefers a larger value and smaller uncertainty for $\alpha_s(M_Z^2)$ comparing to Fit A.

Because the charm PDF benefits from the accuracy of the gluon PDF, so we present our results for charm PDF with its uncertainty. A significant reduction in gluon PDF uncertainty, especially in large $x$  impacts on reduction of charm PDF uncertainty.
The reduction of charm PDF uncertainty at large $x$ can be very worth for future experiment at LHC and also in some processes which we need to include IC  to extrinsic charm PDF. It should be noted that the IC is dominant at large $x$ region only.

In order to benchmark our PDFs, the Higgs boson cross section  at the LHC calculated using modern PDFs and extracted results from Fit A, B and C that they illustrate good compatibility with each other.

 \section{Acknowledgement} 
 M. A.  appreciate Daniel Britzger, Alexander Glazov  and Amir Shabanpour  for their  guidance about computational programs. A. K. is grateful to Stanley Brodsky for valuable discussions, comments and suggestions. 
 A. K. also appreciate Oleksandr Zenaiev for useful discussion about \textsc{xFitter} framework. M. A. also thanks Roberto Di Nardo for useful discussion on  theoretical  predictions of the Higgs boson cross section.

\newpage





\end{document}